\pdfoutput = 0 

\documentclass[fleqn]{article} 
\usepackage{textbook} \usepackage{englcomp} 
\usepackage{pt} 
\usepackage{pstricks} \usepackage{pst-node}
 \usepackage{makeidx} \usepackage{multicol} \makeindex      
\psset{unit=1cm}

\begin{document}

\author{C.A. Middelburg}
\title{A Short Introduction \\ to Process Theory}
\date{October 2016}
\maketitle
\cleardoublepage
\pagenumbering{roman}
\setcounter{page}{3}

\chapter*{Preface}
\label{ch-preface}
\markboth{Preface}{Preface}
\addcontentsline{toc}{chapter}{\numberline{}Preface}

Complex computer-based systems have become an essential part of our 
society.
These complex systems are generally composed of a number of components 
that act concurrently and interact with each other and the environment 
of the system concerned.
The complexity arises to a great extent from the many ways in which 
the components of the system can interact.
Not surprisingly, means for the description and analysis of system 
behaviour become increasingly important to discover flaws in 
computer-based systems.

When it comes to description and analysis, there is advantage in 
treating systems, the components of which they are composed, and the 
environments with which they interact, on an equal footing.
Therefore, we call them all processes.
These lecture notes concern 
\label{process theory}%
\emph{process theory}, i.e.\ the theory of process behaviour, but 
intentionally does not cover the entire field.
First of all, we do not consider all possible kinds of processes, but
only a kind of frequent occurrence.
In particular, we use the term 
\index{process}%
\emph{process} to mean any system whose behaviour is made up of 
discrete actions.
Each action of a process is either performed synchronously with an 
action of another process, in which case an interaction takes place 
between those processes, or it is performed on its own.
Moreover, we restrict ourselves to basic concepts for the 
description of process behaviour.

More concretely, we focus in these lecture notes on the concept of a
(labelled) \emph{transition system}%
\label{transition system}, a concept first introduced in~\cite{Kel76}.
The reason for this is twofold.
Firstly, the concept of a transition system can be considered to be the 
fundamental concept for the description of process behaviour.
Almost all formalisms meant for the description of process behaviour
are based on the concept of a transition system 
(see e.g.~\cite{Old86,Fok00,GV15}).
Secondly, although mathematically simple, transition systems can model 
virtually all relevant properties of processes.
The transition system describing the behaviour of a process is 
generally a suitable basis for checking properties of that process
(see e.g.~\cite{Arn94}). 

\section*{Outline of the lecture notes}

These lecture notes are organized in six chapters and an appendix in 
which the desirable background in elementary set theory is shortly 
reviewed.

Chap.~\ref{ch-basics}, which is an introductory chapter, is primarily
meant to acquire a good insight into the concept of a transition system
and its relevance to the description of process behaviour.
No attention is paid in this chapter to issues material to the
application of transition systems for the description of process 
behaviour.
These issues are treated in the subsequent chapters, which all build
on Chap.~\ref{ch-basics}.

If we have a process composed of a number of subprocesses that act 
concurrently and interact with each other, the following important
question arises.
How do we obtain a transition system describing the behaviour of the 
whole process from the transition systems describing the behaviours of 
the subprocesses?
Therefore, the issue of concurrency and interaction must be dealt with 
in the setting of transition systems.
This is done in Chap.~\ref{ch-interaction}.

Frequently, the behaviour of a process is first described at a high 
level of abstraction, and then as a process composed of several 
subprocesses that act concurrently and interact with each other.
In order to show that the high-level description is correctly refined 
by the other one, we have to abstract from the actions added for the
interactions between the subprocesses.
This issue of abstraction is treated in Chap.~\ref{ch-abstraction}.

Composing a process of subprocesses that act concurrently and interact 
with each other is only one way of combining processes.
Other ways of combining processes, especially the ones known as 
sequential composition, alternative composition and iteration, are 
useful in case of large processes to master their complexity.
Chap.~\ref{ch-composition} deals with the issue of composition in this 
wider sense.

Transition systems describing the behaviour of real-life processes are
generally very large or even infinite.
The size can be reduced strongly by using expressions representing the 
behaviour of processes instead.
The operators occurring in such process expressions correspond to ways 
of combining processes such as the ones treated in 
Chap.~\ref{ch-composition}.
Furthermore, process expressions enable us to define processes by means 
of recursive specifications.
Process expressions and recursive specifications are the subjects of 
Chap.~\ref{ch-expressions}.

There are many interesting topics related to process expressions and 
recursive specifications which are not treated in 
Chap.~\ref{ch-expressions}.
Some selected topics, including structural operational semantics and
equational laws for process expressions, are covered in
Chap.~\ref{ch-topics}.

All concepts and issues treated in these lecture notes are first 
introduced by means of simple examples, sometimes not even related to 
real-life systems, and later on illustrated by more complex examples 
based on real-life systems.
To quicken an intuitive understanding, direct connections with 
programs and automata are established wherever appropriate.
For the interested reader, direct connections with Petri nets are also 
established.
Those connections are relevant because Petri nets are basically 
generalizations of transition systems that support the direct 
description of concurrency.

In each chapter, except the last one, it shows that what has been 
dealt with so far still has certain limitations.
Each time, the next chapter is devoted to reducing the limitations
concerned.
It is worth mentioning that, as a result of this set-up, the notion of 
a transition system is first defined in Sect.~\ref{sect-basics-formal}, 
and then redefined in Sects.~\ref{sect-abstraction-formal} 
and~\ref{sect-composition-formal-1}.
  
\section*{How to use the lecture notes}

These lecture notes can be used in courses for undergraduate students in 
computer science.
Some familiarity with set theory is assumed.
The desirable background in set theory is shortly reviewed in 
App.~\ref{ch-prelims}, which also establishes the terminology and 
notation concerning sets.
Each chapter is a prerequisite for all subsequent chapters.
The examples are integrated with the text.
They should not be ignored.

\section*{History of the lecture notes}

In 2002, I was invited to write lecture notes for an introductory course
on process theory for first year undergraduate computer science students 
at Eindhoven University of Technology that could serve as a preparation 
for an undergraduate course on process algebra based on~\cite{BW90}.
This has led to an unpolished version of the current lecture notes.
They have been written while consistently applying the following three 
simple rules: 
(i)~begin with an elementary concept, 
(ii)~introduce additional concepts not until the need for them has been 
explained clearly, and
(iii)~stray from the main topic for no other reason than explanation.

The unpolished lecture notes from 2002 have been adapted in 2003 by the 
lecturer of the course in question to his ideas and preferences without 
taking the above-mentioned rules fully into account.
Those adapted lecture notes and shortened versions thereof are 
circulated since, mostly under the title ``Introduction to Process 
Theory''.

\section*{Acknowledgement}

I thank Michel Reniers from Eindhoven University of Technology, 
Department of Mechanical Engineering, for contributing most pictures in 
the lectures notes.

\vskip .9cm
\noindent Amsterdam, September 2016 
\hfill 
\emph{Kees Middelburg}

\vfill

\tableofcontents 

\cleardoublepage
\pagenumbering{arabic}

\chapter{Transition Systems}
\label{ch-basics}

The notion of a transition system can be considered to be the
fundamental notion for the description of process behaviour.
This chapter is meant to acquire a good insight into this notion and
its relevance for the description of process behaviour.
First of all, we explain informally what transition are systems and 
give some simple examples of their use in describing process behaviour
(Sect.~\ref{sect-basics-informal}).
After that, we define the notion of a transition system in a 
mathematically precise way (Sect.~\ref{sect-basics-formal}).
For a better understanding, we next investigate the connections between 
the notion of a transition system and the familiar notions of a program
(Sect.~\ref{sect-basics-conn-programs}) and an automaton
(Sect.~\ref{sect-basics-conn-automata}).
For the interested reader, we also investigate the connections with the
notion of a Petri net (Sect.~\ref{sect-basics-conn-nets}).
Finally, we discuss two equivalences on transition systems, called 
trace equivalence (Sect.~\ref{sect-basics-trace-eqv}) and bisimulation 
equivalence (Sect.~\ref{sect-basics-bisim-eqv}).
Those equivalences are useful because they allow us to abstract from  
details of transition systems that we often want to ignore.

\section{Informal explanation}
\label{sect-basics-informal}

Transition systems are often considered to be the same as automata. 
Both consist of states and labeled transitions between states.
The main difference is that automata are primarily regarded as abstract 
machines to recognize certain languages and transition systems are 
primarily regarded as a means to describe the behaviour of interacting 
processes.
In the case of transition systems, the intuition is that a transition 
is a state change caused by performing the action labeling the 
transition.
A transition from a state $s$ to a state $s'$ labeled by an action $a$
is usually written $\astep{s}{a}{s'}$.
This can be read as ``the system is capable of changing its state from
$s$ into $s'$ by performing action $a$''.
Let us give an example to illustrate that it is quite natural to look 
at real-life computer-based systems as systems that change their state 
by performing actions.
\begin{example}[Simple telephone system]\index{simple telephone system}
\label{exa-telephones}
We consider a simple telephone system.
In this telephone system each telephone is provided with a process,
called its basic call process, to establish and maintain connections
with other telephones.
Actions of this process include receiving an off-hook or on-hook signal
{from} the telephone, receiving a dialed number from the telephone,
sending a signal to start or to stop emitting a dial tone, ring tone or
ring-back tone to the telephone, and receiving an alert signal from
another telephone -- indicating an incoming call.
Suppose that a basic call process is in the idling state.
In this state, it can change its state to the initial dialing state by
receiving an off-hook signal from the telephone.
Alternatively, it can change its state to the initial ringing state by
receiving an alert signal from another telephone.
In the initial dialing state, it can change its state to another 
dialing state by sending a signal to start emitting a dial tone to the 
telephone.
In the initial ringing state, it can change its state to another
ringing state by sending a signal to start emitting a ring tone to the
telephone.
And so forth.
\end{example}

Transition systems have been devised as a means to describe the
behaviour of systems that have only discrete state changes.
Despite this underlying purpose of transition systems, they can deal
with continuous state changes as well.
However, such use of transition systems will not be treated in these
lecture notes.
Instead, we focus on acquiring a good insight into the basics of 
transition systems.
That is not for pedagogical reasons alone.
Systems that have only discrete state changes are still of utmost
importance in the practice of developing computer-based systems and 
will remain so for a long time.
Here are a couple of examples of the use of transition systems in 
describing the behaviour of systems with discrete state changes.
\begin{example}[Bounded counter]\index{counter!bounded}
\label{exa-bcounter}
We first consider a very simple system, viz.\ a bounded counter.
A bounded counter can perform increments of its value by $1$ till a
certain value $k$ is reached and can perform decrements of its value by
$1$ till the value $0$ is reached.
As states of a bounded counter, we have the natural numbers $0$ to $k$.
State $i$ is the state in which the value of the counter is $i$.
As actions, we have $\kw{inc}$ (increment) and $\kw{dec}$ (decrement).
As transitions  of a bounded counter, we have the following:
\begin{iteml}
\item
for each state $i$ that is less than $k$, a transition from state $i$
to state $i+1$ labeled with the action $\kw{inc}$, written
$\astep{i}{\kw{inc}}{i+1}$; 
\item
for each state $i$ that is less than $k$, a transition from state $i+1$ 
to state $i$ labeled with the action $\kw{dec}$, written 
$\astep{i+1}{\kw{dec}}{i}$.
\end{iteml}
If the number of states and transitions is small, a transition system
can easily be represented graphically.
The transition system describing the behaviour of the bounded counter
is represented graphically in Fig.~\ref{fig-bcounter} for the case
where $k = 3$.%
\footnote
{In graphical representations of transition systems, we use circles or
 ellipses for states and arrows for transitions.
 We indicate the initial state by an incoming unlabeled arrow.
}
\begin{figure}
\begin{pspicture}(0,0)(7,2)

 \psset{arrows=->}

 \pnode(0,1){S}
 \rput(1,1){\circlenode{0}{0}}
 \rput(3,1){\circlenode{1}{1}}
 \rput(5,1){\circlenode{2}{2}}
 \rput(7,1){\circlenode{3}{3}}

 \ncline{S}{0}
 \nccurve[angleA=30,angleB=150]{0}{1}\naput{$\kw{inc}$}
 \nccurve[angleA=210,angleB=330]{1}{0}\naput{$\kw{dec}$}
 \nccurve[angleA=30,angleB=150]{1}{2}\naput{$\kw{inc}$}
 \nccurve[angleA=210,angleB=330]{2}{1}\naput{$\kw{dec}$}
 \nccurve[angleA=30,angleB=150]{2}{3}\naput{$\kw{inc}$}
 \nccurve[angleA=210,angleB=330]{3}{2}\naput{$\kw{dec}$}
\end{pspicture}
\caption{Transition system for the bounded counter}
\label{fig-bcounter}
\end{figure}
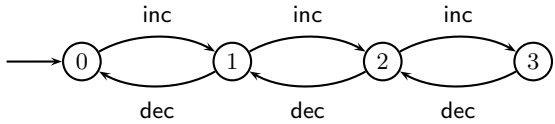

Notice that the bounded counter has a finite number of states and a 
finite number of transitions.
Furthermore, the bounded counter will never reach a terminal state, 
i.e.\ a state from which no transition is possible.
Thus, the finiteness of the bounded counter does not keep the counter
from making an infinite number of transitions.

The bounded counter can easily be adapted to become a counter modulo 
$k$, i.e.\ a counter whose value becomes $0$ by performing an increment 
by $1$ when its value is $k$ and whose value becomes $k$ by performing 
a decrement by $1$ when its value is $0$.
We have the same states and actions as before and we have two 
additional transitions:
\begin{iteml}
\item
a transition from state $k$ to state $0$ labeled with the action
$\kw{inc}$, written $\astep{k}{\kw{inc}}{0}$; 
\item
a transition from state $0$ to state $k$ labeled with the action
$\kw{dec}$, written $\astep{0}{\kw{dec}}{k}$.
\end{iteml}
\end{example}
\begin{example}[Bounded buffer]\index{buffer!bounded}
\label{exa-bbuffer}
We next consider another simple system, viz.\ a bounded buffer.
A bounded buffer can add new data to the sequence of data that it keeps
if the capacity $l$ of the buffer is not exceeded, i.e.\ if the length
of the sequence of data that it keeps is not greater than $l$.
As long as it keeps data, it can remove the data that it keeps -- in 
the order in which they were added.
As states of a bounded buffer, we have the sequences of data of which 
the length is not greater than $l$.
State $\sigma$ is the state in which the sequence of data $\sigma$ is
kept in the buffer.
As actions, we have $\kw{add}(d)$ (add $d$) and $\kw{rem}(d)$ (remove
$d$) for each datum $d$.
As transitions  of a bounded buffer, we have the following:
\begin{iteml}
\item
for each datum $d$ and each state $\sigma$ that has a length less than
$l$, a transition from state $\sigma$ to state $d\, \sigma$ labeled 
with the action $\kw{add}(d)$, written
$\astep{\sigma}{\kw{add}(d)}{d\, \sigma}$; 
\item
for each datum $d$ and each state $\sigma\, d$, a transition from state 
$\sigma\, d$ to state $\sigma$ labeled with the action $\kw{rem}(d)$, 
written $\astep{\sigma\, d}{\kw{rem}(d)}{\sigma}$.
\end{iteml}
The transition system describing the behaviour of the bounded buffer is
represented graphically in Fig.~\ref{fig-bbuffer} for the case where
$l = 2$ and the only data involved are the natural numbers $0$ and $1$.
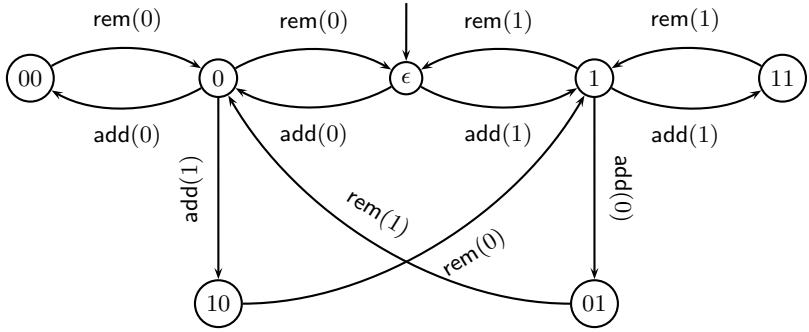
\begin{figure}
\begin{pspicture}(0,0.5)(11,5.5)

 \psset{arrows=->}

 \pnode(5.5,5){S}

 \rput(5.5,4){\circlenode{e}{$\epsilon$}}

 \rput(3,4){\circlenode{0}{0}}
 \rput(8,4){\circlenode{1}{1}}

 \rput(0.5,4){\circlenode{00}{00}}
 \rput(3,1){\circlenode{10}{10}}
 \rput(8,1){\circlenode{01}{01}}
 \rput(10.5,4){\circlenode{11}{11}}

 \ncline{S}{e}
 \nccurve[angleA=210,angleB=330]{e}{0}\naput{$\kw{add}(0)$}
 \nccurve[angleA=330,angleB=210]{e}{1}\nbput{$\kw{add}(1)$}
 \nccurve[angleA=210,angleB=330]{0}{00}\naput{$\kw{add}(0)$}
 \ncline{0}{10}\nbput[nrot=:180]{$\kw{add}(1)$}
 \ncline{1}{01}\naput[nrot=:0]{$\kw{add}(0)$}
 \nccurve[angleA=330,angleB=210]{1}{11}\nbput{$\kw{add}(1)$}

 \nccurve[angleA=30,angleB=150]{0}{e}\naput{$\kw{rem}(0)$}
 \nccurve[angleA=150,angleB=30]{1}{e}\nbput{$\kw{rem}(1)$}
 \nccurve[angleA=30,angleB=150]{00}{0}\naput{$\kw{rem}(0)$}
 \nccurve[angleA=0,angleB=240]{10}{1}\nbput[nrot=:0]{$~~~~~~\kw{rem}(0)$}
 \nccurve[angleA=180,angleB=300]{01}{0}\nbput[nrot=:180]{$\kw{rem}(1)~~~$}
 \nccurve[angleA=150,angleB=30]{11}{1}\nbput{$\kw{rem}(1)$}

\end{pspicture}
\caption{Transition system for the bounded buffer}
\label{fig-bbuffer}
\end{figure}
Although it has a finite capacity, the bounded buffer will have an 
infinite number of states and an infinite number of transitions in the 
case where the number of data involved is infinite.

The bounded buffer can easily be adapted to become unreliable, e.g.\ to 
get into an error state by adding a datum when it is full.
We have one additional state, say $\kw{err}$, no additional actions, 
and the following additional transitions: 
\begin{iteml}
\item
for each datum $d$ and each state $\sigma$ that has a length equal to
$l$, a transition from state $\sigma$ to state $\kw{err}$ labeled with
the action $\kw{add}(d)$, 
written $\astep{\sigma}{\kw{add}(d)}{\kw{err}}$.
\end{iteml}
Notice that no more transitions are possible when this unreliable 
bounded buffer has reached the state $\kw{err}$.
Thus, the additional feature of this buffer may keep it from making an
infinite number of transitions.
\end{example}

It is usual to designate one of the states of a transition system as
its initial state. 
At the start-up of a system, i.e.\ before it has performed any action,
the system is considered to be in its initial state.
The expected initial states of the bounded counter from
Example~\ref{exa-bcounter} and the bounded buffer from
Example~\ref{exa-bbuffer} are $0$ and $\epsilon$ (the empty sequence), 
respectively.

Although bounded counters and buffers arise frequently as basic
components in computer-based systems, they are not regarded as typical
examples of real-life computer-based systems.
In the following example, we consider a simplified version of a small
real-life computer-based system, viz.\ a calculator.
\begin{example}[Calculator]\index{calculator}
\label{exa-calculator}
We consider a calculator that can perform simple arithmetical 
operations on integers.
It can only perform addition, subtraction, multiplication and division
on integers between a certain values, say $\nm{min}$ and $\nm{max}$.
As states of the calculator, we have pairs $\tup{i,o}$, where 
$\nm{min} \leq i \leq \nm{max}$ or $i = \und$ and 
$o \in
 \set{\kw{add},\kw{sub},\kw{mul},\kw{div},\kw{eq},\kw{clr},\und}$.
State $\tup{i,o}$ is roughly the state in which the result of the 
preceding calculations is $i$ and the operator that must be applied 
next is $o$.
If $o = \und$, the operator that must be applied next is not 
available; and if in addition $i= \und$, the result of the 
preceding calculations is not available either. 
As initial state, we have the pair $\tup{\und,\und}$.
As actions, we have 
$\kw{rd}(i)$ (read operand $i$) and $\kw{wr}(i)$ (write result $i$), 
both for $\nm{min} \leq i \leq \nm{max}$, and 
$\kw{rd}(o)$ (read operator $o$), 
for $o \in \set{\kw{add},\kw{sub},\kw{mul},\kw{div},\kw{eq},\kw{clr}}$.
As transitions of the calculator, we have the following:
\begin{iteml}
\item
for each $i$ with $\nm{min} \leq i \leq \nm{max}$: 
\begin{iteml}
\item 
a transition 
$\astep{\tup{\und,\und}}{\kw{rd}(i)}{\tup{i,\und}}$, 
\item
a transition 
$\astep{\tup{i,\kw{clr}}}{\kw{wr}(0)}{\tup{\und,\und}}$,
\item
a transition 
$\astep{\tup{i,\kw{eq}}}{\kw{wr}(i)}{\tup{i,\und}}$; 
\end{iteml}
\item
for each $i$ with $\nm{min} \leq i \leq \nm{max}$ and  
$o \in \set{\kw{add},\kw{sub},\kw{mul},\kw{div},\kw{eq},\kw{clr}}$: 
\begin{iteml}
\item 
a transition $\astep{\tup{i,\und}}{\kw{rd}(o)}{\tup{i,o}}$; 
\end{iteml}
\item
for each $i$ with $\nm{min} \leq i \leq \nm{max}$ and  
$j$ with $\nm{min} \leq j \leq \nm{max}$: 
\begin{iteml}
\item
a transition 
$\astep{\tup{i,\kw{add}}}{\kw{rd}(j)}{\tup{i+j,\und}}$
if $\nm{min} \leq i+j \leq \nm{max}$, 
\item
a transition 
$\astep{\tup{i,\kw{sub}}}{\kw{rd}(j)}{\tup{i-j,\und}}$ 
if $\nm{min} \leq i-j \leq \nm{max}$, 
\item
a transition 
$\astep{\tup{i,\kw{mul}}}{\kw{rd}(j)}{\tup{i \cdot j,\und}}$
if $\nm{min} \leq i \cdot j \leq \nm{max}$,  
\item
a transition 
$\astep{\tup{i,\kw{div}}}{\kw{rd}(j)}{\tup{i \div j,\und}}$
if $\nm{min} \leq i \div j \leq \nm{max}$ and $j \neq 0$. 
\end{iteml}
\end{iteml}
The transition system describing the behaviour of the calculator is
represented graphically in Fig.~\ref{fig-calculator} for the case where
$\nm{min} = 0$ and $\nm{max} = 1$.
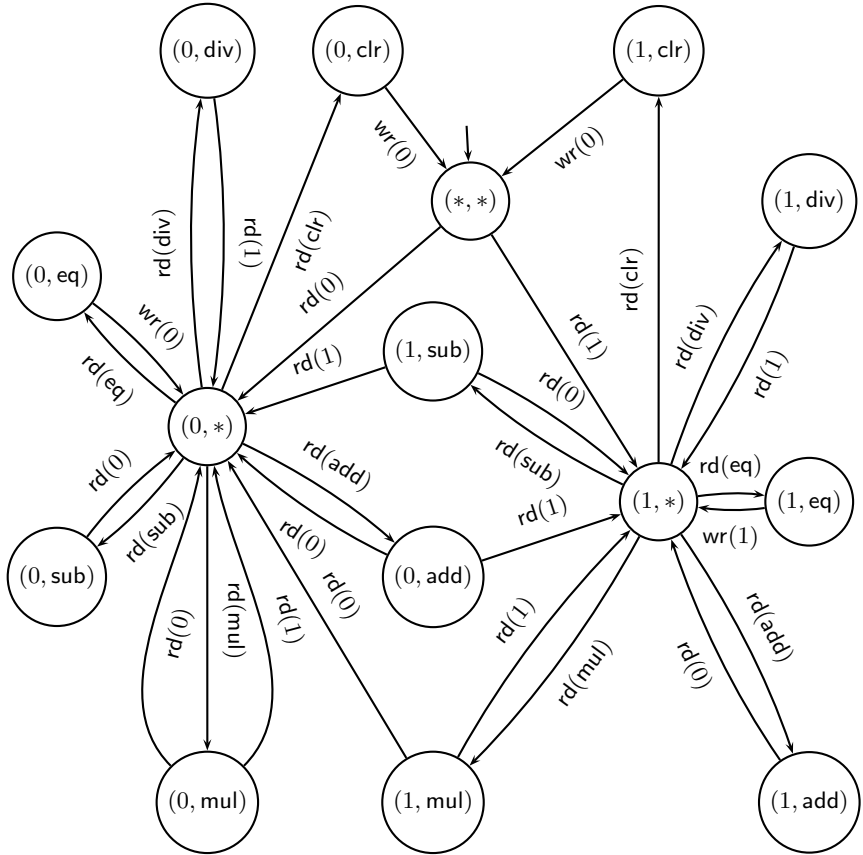
\begin{figure}
\begin{pspicture}(1,-1)(12,12)

 \psset{arrows=->}

 \pnode(7.5,9){S}
 \rput(7.5,8){\circlenode{**}{$(*,*)$}}
 \rput(4,5){\circlenode{0*}{$(0,*)$}}

 \rput(6,10){\circlenode{0clr}{$(0,\kw{clr})$}}
 \rput(2,7){\circlenode{0eq}{$(0,\kw{eq})$}}
 \rput(7,3){\circlenode{0add}{$(0,\kw{add})$}}
 \rput(2,3){\circlenode{0sub}{$(0,\kw{sub})$}}
 \rput(4,0){\circlenode{0mul}{$(0,\kw{mul})$}}
 \rput(4,10){\circlenode{0div}{$(0,\kw{div})$}}

 \rput(10,4){\circlenode{1*}{$(1,*)$}}
 \rput(10,10){\circlenode{1clr}{$(1,\kw{clr})$}}
 \rput(12,4){\circlenode{1eq}{$(1,\kw{eq})$}}
 \rput(12,0){\circlenode{1add}{$(1,\kw{add})$}}
 \rput(7,6){\circlenode{1sub}{$(1,\kw{sub})$}}
 \rput(7,0){\circlenode{1mul}{$(1,\kw{mul})$}}
 \rput(12,8){\circlenode{1div}{$(1,\kw{div})$}}

 \ncline{S}{**}

 \ncline{**}{0*}\nbput[nrot=:180]{$\kw{rd}(0)$}
 \ncline{**}{1*}\naput[nrot=:0]{$\kw{rd}(1)$}

 \ncline{0clr}{**}\nbput[nrot=:0]{$\kw{wr}(0)$}
 \ncline{1clr}{**}\naput[nrot=:180]{$\kw{wr}(0)$}

 \ncarc{0eq}{0*}\naput[nrot=:0]{$\kw{wr}(0)$}
 \ncarc{1eq}{1*}\naput[nrot=:180]{$\kw{wr}(1)$}

 \ncarc{0*}{0add}\naput[nrot=:0]{$\kw{rd}(\kw{add})$}
 \ncarc{1*}{1add}\naput[nrot=:0]{$\kw{rd}(\kw{add})$}
 \ncarc{0*}{0sub}\naput[nrot=:180]{$\kw{rd}(\kw{sub})\quad$}
 \ncarc{1*}{1sub}\naput[nrot=:180]{$\kw{rd}(\kw{sub})$}
 \ncline{0*}{0mul}\naput[nrot=:0]{$\quad\kw{rd}(\kw{mul})$}
 \ncarc{1*}{1mul}\naput[nrot=:180]{$\kw{rd}(\kw{mul})$}
 \ncarc{0*}{0div}\naput[nrot=:0]{$\kw{rd}(\kw{div})$}
 \ncarc{1*}{1div}\naput[nrot=:0]{$\kw{rd}(\kw{div})$}
 \ncarc{0*}{0eq}\naput[nrot=:180]{$\kw{rd}(\kw{eq})$}
 \ncarc{1*}{1eq}\naput[nrot=:0]{$\kw{rd}(\kw{eq})$}
 \ncline{0*}{0clr}\nbput[nrot=:0]{$\quad\kw{rd}(\kw{clr})$}
 \ncline{1*}{1clr}\naput[nrot=:0]{$\kw{rd}(\kw{clr})$}

 \ncarc{0add}{0*}\naput[nrot=:180]{$\quad\kw{rd}(0)$}
 \ncline{0add}{1*}\naput[nrot=:0]{$\kw{rd}(1)$}
 \ncarc{1add}{1*}\naput[nrot=:180]{$\kw{rd}(0)$}

 \ncarc{0sub}{0*}\naput[nrot=:0]{$\kw{rd}(0)$}
 \ncarc{1sub}{1*}\naput[nrot=:0]{$\kw{rd}(0)\quad$}
 \ncline{1sub}{0*}\nbput[nrot=:180]{$\quad\kw{rd}(1)$}

 \nccurve[angleA=135,angleB=260]{0mul}{0*}\nbput[nrot=:0]{$\kw{rd}(0)$}
 \nccurve[angleA=45,angleB=280]{0mul}{0*}\nbput[nrot=:180]{$\kw{rd}(1)$}
 \ncline{1mul}{0*}\nbput[nrot=:180]{$\kw{rd}(0)$}
 \ncarc{1mul}{1*}\naput[nrot=:0]{$\kw{rd}(1)$}

 \ncarc{0div}{0*}\naput[nrot=:0]{$\kw{rd}(1)$}
 \ncarc{1div}{1*}\naput[nrot=:180]{$\kw{rd}(1)$}

\end{pspicture}
\caption{Transition system for the calculator}
\label{fig-calculator}
\end{figure}
Although the extremely small range of integers makes this case actually
useless, it turns out to be difficult to represent the transition 
system graphically.
The textual description given above is still intelligible.
However, it is questionable whether this would be the case for a more 
realistic calculator.
\end{example}

Examples like Example~\ref{exa-calculator} indicate that in the case of
real-life systems we probably need a way to describe process behaviour 
more concisely than by directly giving a transition system.
This is one of the issues treated in the remaining chapters of these 
lecture notes.

\section{Formal definition}
\label{sect-basics-formal}

With the previous section, we have prepared the way for the formal 
definition of the notion of a transition system.

\begin{definition}[Transition system]\index{transition system}
A \emph{transition system} $T$ is a quadruple $\tup{S,A,\step{},s_0}$ 
where
\begin{iteml}
\item
$S$ is a set of \emph{states}\index{state};
\item
$A$ is a set of \emph{actions}\index{action};
\item 
${\step{}} \subseteq S \x A \x S$ is a set of 
\emph{transitions}\index{transition};
\item
$s_0 \in S$ is the \emph{initial state}\index{state!initial}.
\end{iteml}
If $S$ and $A$ are finite, $T$ is called a \emph{finite}
transition system.
We write $\astep{s}{a}{s'}$ instead of $\tup{s,a,s'} \in {\step{}}$.
We write $\acts(T)$ for $A$, i.e.\ the set of actions of $T$.
The set  ${\gstep{}{}{}} \subseteq S \x \seqof{A} \x S$ of
\emph{generalized transitions}\index{transition!generalized} of $T$ is 
the smallest subset of $S \x \seqof{A} \x S$ satisfying:
\begin{iteml}
\item
$\astep{s}{\epsilon}{s}$ for each $s \in S$;
\item
if $\astep{s}{a}{s'}$, then $\gstep{s}{a}{s'}$;
\item
if $\gstep{s}{\sigma}{s'}$ and $\gstep{s'}{\sigma'}{s''}$, then 
$\gstep{s}{\sigma\, \sigma'}{s''}$.
\end{iteml}
A state $s \in S$ is called a \emph{reachable}\index{state!reachable} 
state of $T$ if there is a $\sigma \in \seqof{A}$ such that 
${\gstep{s_0}{\sigma}{s}}$.
A state $s \in S$ is called a \emph{terminal}\index{state!terminal} 
state of $T$ if there is no $a \in A$ and $s' \in S$ such that 
$\astep{s}{a}{s'}$.
\end{definition}

When a system has reached one of its terminal states, no more 
transitions are possible.
Sometimes, certain terminal states are designated as final states.
The convention is to do so if there is a need to make a distinction
between terminal states in which the system is considered to terminate 
successfully and terminal states in which the system is considered not 
to terminate successfully.
In that case, the final states are the terminal states in which the
system is considered to terminate successfully.
Final states are also loosely called successfully terminating states.
A system that reaches a terminal state different from a final state
is said to become inactive. 
With certain terminal states designated as final states, a transition 
system is a quintuple $\tup{S,A,\step{},\term{},s_0}$, where $S$, $A$, 
$\step{}$, $s_0$ are as before and the set ${\term{}} \subseteq S$ of 
\emph{final states} or \emph{successfully terminating states} consists 
of terminal states only.
We will return to such transition systems in Chap.~\ref{ch-composition}.

We will now return to some of the transition systems introduced
informally in the previous section. 
\begin{example}[Bounded counter]\index{counter!bounded}
\label{exa-bcounter-formal}
We look again at the bounded counter from Example~\ref{exa-bcounter}.
Formally, the behaviour of a bounded counter with bound $k$ is 
described by the transition system $\tup{S,A,\step{},s_0}$ where
\[
\begin{aeqns}
S & = & \set{i \in \Nat \where i \leq k}\enspace, \\
A & = & \set{\kw{inc},\kw{dec}}\enspace, \\
{\step{}} & = & 
 \set{\tup{i,\kw{inc},i+1} \where i \in \Nat, i < k} \union
 \set{\tup{i+1,\kw{dec},i} \where i \in \Nat, i < k}\enspace, \\
s_0 & = & 0\enspace.
\end{aeqns}
\]
All states of this finite transition system are reachable.
It does not have terminal states.
\end{example}
\begin{example}[Unreliable bounded buffer]%
\index{buffer!bounded!unreliable}
\label{exa-bbuffer-formal}
We also look at the unreliable bounded buffer mentioned in 
Example~\ref{exa-bbuffer}.
Formally, the behaviour of the unreliable  bounded buffer with 
capacity $l$ is described by the transition system 
$\tup{S,A,\step{},s_0}$ where
\[
\begin{aeqns}
S & = & \set{\sigma \in \seqof{D} \where |\sigma| \leq l} \union
        \set{\kw{err}}\enspace, \\
A & = & \set{\kw{add}(d) \where d \in D} \union 
        \set{\kw{rem}(d) \where d \in D}\enspace, \\
{\step{}} & = & 
 \set{\tup{\sigma,\kw{add}(d),d\, \sigma} \where 
     \sigma \in \seqof{D}, |\sigma| < l} \\ & {} \union & 
 \set{\tup{\sigma,\kw{add}(d),\kw{err}} \where 
     \sigma \in \seqof{D}, |\sigma| = l} \\ & {} \union &
 \set{\tup{\sigma\, d,\kw{rem}(d),\sigma} \where 
     \sigma \in \seqof{D}, |\sigma| < l}\enspace, \\
s_0 & = & \epsilon\enspace.
\end{aeqns}
\]
All states of this transition system are reachable.
It has one terminal state, viz.\ $\kw{err}$.
\end{example}

Henceforth, we will only occasionally introduce transition systems in
this formal style.

After the informal explanation and formal definition of the notion of
a transition system, we are now in the position to relate it to the 
notions of a program and an automaton.

\section{Programs and transition systems}
\label{sect-basics-conn-programs}

For a better understanding of the notion of a transition system, we now
look into its connections with the familiar notion of a program.

The behaviour of a program upon execution can be regarded as a
transition system.
In doing so, we can abstract from how the actions performed by a
program are processed by a machine, and hence from how the values
assigned to the program variables are maintained.
In that case, we focus on the flow of control.
The states of the transition system only serve as the control points of
the program and its actions are merely requests to perform actions such 
as assignments, tests, etc.
What we have in view here will be called the behaviour of a program
upon \emph{abstract} execution to distinguish it clearly from the 
behaviour of a program upon execution on a machine, which applies to 
the processing by a machine of the actions performed by the program.
Here is an example of the use of transition systems in describing the 
behaviour of programs upon abstract execution. 
\begin{example}[Factorial program]\index{factorial program}
\label{exa-factorial}
We consider the following PASCAL~\cite{Wir71} program to calculate factorials:
\begin{verbatim}
PROGRAM factorial(input,output);
VAR i,n,f: 0..maxint;
BEGIN
   read(n); 
   i := 0; f := 1; 
   WHILE i < n DO 
      BEGIN i := i + 1; f := f * i END;
   write(f)
END
\end{verbatim}
The behaviour of this program upon abstract execution can be described 
by a transition system as follows.
As states of the factorial program, we have the natural numbers $0$ to 
$7$, with $0$ as initial state.
The states can be viewed as the values of a ``program counter''.
As actions, we have an action corresponding to each atomic statement of
the program as well as each test of the program and its opposite.
As transitions, we have the following:
\begin{ldispl}
\astep{0}{\kw{read(n)}}{1},\;
\astep{1}{\kw{i\,:=\,0}}{2},\;
\astep{2}{\kw{f\,:=\,1}}{3}, \\
\astep{3}{\kw{i\,<\,n}}{4},\; 
\astep{4}{\kw{i\,:=\,i\,+\,1}}{5},\;
\astep{5}{\kw{f\,:=\,f\,*\,i}}{3}, \\
\astep{3}{\kw{NOT\,i\,<\,n}}{6},\;
\astep{6}{\kw{write(f)}}{7}.
\end{ldispl}
The transition system for the factorial program is represented
graphically in Fig.~\ref{fig-factorial}.
\begin{figure}
\begin{pspicture}(0,0)(12,3)

 \psset{arrows=->}

 \pnode(1,3){S}
 \rput(1,2){\circlenode{0}{0}}
 \rput(3,2){\circlenode{1}{1}}
 \rput(5,2){\circlenode{2}{2}}
 \rput(7,2){\circlenode{3}{3}}
 \rput(8,0){\circlenode{4}{4}}
 \rput(6,0){\circlenode{5}{5}}
 \rput(9,2){\circlenode{6}{6}}
 \rput(11,2){\circlenode{7}{7}}

 \ncline{S}{0}
 \ncline{0}{1}\naput{$\kw{read(n)}$}
 \ncline{1}{2}\naput{$\kw{i\,:=\,0}$}
 \ncline{2}{3}\naput{$\kw{f\,:=\,1}$}
 \ncline{3}{4}\naput[nrot=:0]{$\kw{i\,<\,n}$}
 \ncline{4}{5}\naput{$\kw{i\,:=\,i\,+\,1}$}
 \ncline{5}{3}\naput[nrot=:0]{$\kw{f\,:=\,f\,*\,i}$}
 \ncline{3}{6}\naput{$\kw{NOT\,i\,<\,n}$}
 \ncline{6}{7}\naput{$\kw{write(f)}$}

\end{pspicture}
\caption{Transition system for the factorial program}
\label{fig-factorial}
\end{figure}
\end{example}
Here is another example.
\begin{example}[Greatest common divisor program]%
\index{greatest common divisor program}
\label{exa-gcd}
We consider the following PASCAL program to calculate greatest common 
divisors:
\begin{verbatim}
PROGRAM gcd(input,output);
VAR m,n: 0..maxint;
BEGIN
   read(m); read(n); 
   REPEAT
      WHILE m > n DO m := m - n;
      WHILE n > m DO n := n - m
   UNTIL m = n;
   write(m)
END
\end{verbatim}
The behaviour of this program upon abstract execution can be described 
by a transition system as follows.
As states of the greatest common divisor program, we have the natural
numbers $0$ to $8$, with $0$ as initial state.
As actions, we have an action corresponding to each atomic statement of
the program as well as each test of the program and its opposite.
As transitions, we have the following:
\begin{ldispl}
\astep{0}{\kw{read(m)}}{1},\;
\astep{1}{\kw{read(n)}}{2}, \\
\astep{2}{\kw{m\,>\,n}}{3},\; 
\astep{3}{\kw{m\,:=\,m\,-\,n}}{2}, \\
\astep{2}{\kw{NOT\,m\,>\,n}}{4},\; 
\astep{4}{\kw{n\,>\,m}}{5},\; 
\astep{5}{\kw{n\,:=\,n\,-\,m}}{4},\;
\astep{4}{\kw{NOT\,n\,>\,m}}{6},\; 
\astep{6}{\kw{NOT\,m\,=\,n}}{2}, \\
\astep{6}{\kw{m\,=\,n}}{7},\; 
\astep{7}{\kw{write(f)}}{8}.
\end{ldispl}
The transition system for the greatest common divisor program is 
represented graphically in Fig.~\ref{fig-gcd}.
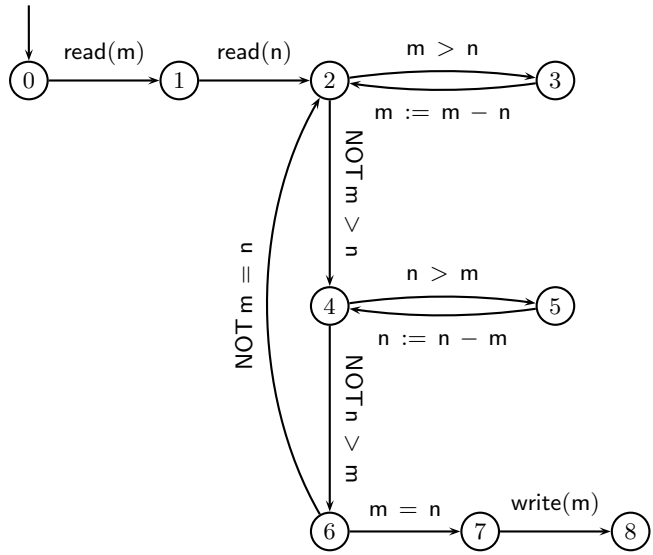
\begin{figure}
\begin{pspicture}(0,-.5)(10,7)

 \psset{arrows=->}

 \pnode(1,7){S}
 \rput(1,6){\circlenode{0}{0}}
 \rput(3,6){\circlenode{1}{1}}
 \rput(5,6){\circlenode{2}{2}}
 \rput(8,6){\circlenode{3}{3}}
 \rput(5,3){\circlenode{4}{4}}
 \rput(8,3){\circlenode{5}{5}}
 \rput(5,0){\circlenode{6}{6}}
 \rput(7,0){\circlenode{7}{7}}
 \rput(9,0){\circlenode{8}{8}}

 \ncline{S}{0}
 \ncline{0}{1}\naput{$\kw{read(m)}$}
 \ncline{1}{2}\naput{$\kw{read(n)}$}
 \ncarc{2}{3}\naput{$\kw{m\,>\,n}$}
 \ncarc{3}{2}\naput{$\kw{m\,:=\,m\,-\,n}$}
 \ncline{2}{4}\naput[nrot=:0]{$\kw{NOT\,m\,>\,n}$}
 \ncarc{4}{5}\naput{$\kw{n\,>\,m}$}
 \ncarc{5}{4}\naput{$\kw{n\,:=\,n\,-\,m}$}
 \ncline{4}{6}\naput[nrot=:0]{$\kw{NOT\,n\,>\,m}$}
 \ncline{6}{7}\naput{$\kw{m\,=\,n}$}
 \nccurve[angleA=120,angleB=240]{6}{2}\naput[nrot=:0]{$\kw{NOT\,m\,=\,n}$}
 \ncline{7}{8}\naput{$\kw{write(m)}$}

\end{pspicture}
\caption{Transition system for the greatest common divisor program}
\label{fig-gcd}
\end{figure}
\end{example}

Notice that the transition systems described in 
Examples~\ref{exa-factorial} and~\ref{exa-gcd} have a single terminal 
state.
In both cases, the program is considered to terminate successfully 
in its terminal state.

A transition system derived from a program in the way described and
illustrated above is reminiscent of a flowchart.
However, the underlying idea is that the transition system describes 
the behaviour of the program upon execution in such a way that it can 
act concurrently and interact with a machine that processes the actions
performed by the program.
If it does so, the combined behaviour can be regarded as the behaviour
of the program upon execution on a machine.
Interaction between processes is one of the issues treated in the
remaining chapters of these lecture notes.
We can also directly give a transition system describing the behaviour 
of the program upon execution on a machine.
In that case, we have to take into account that an assignment changes 
the value of a program variable, the values of the program variables
determine whether a test succeeds, etc.
This is illustrated in the following couple of examples, which are 
concerned with the same programs as the previous two examples.
\begin{example}[Factorial program]\index{factorial program}
\label{exa-factorial-mach}
We consider again the program from Example~\ref{exa-factorial}.
The intended behaviour of this program upon execution on a machine can
be described by a transition system as follows.
As states of the program, we have pairs $\tup{l,s}$, where 
$l \in \Nat$ with $0 \leq l \leq 7$ and 
$s = \tup{i,n,f}$ with 
$i,n,f \in 
 \set{i \in \Nat \where i \leq \nm{maxint}} \union \set{\und}$.
These states can be viewed as follows:  $l$ is the value of the program
counter and $s = \tup{i,n,f}$ is the storage that keeps the values of 
the program variables $\kw{i}$, $\kw{n}$, and $\kw{f}$ in that order.
The special value $\und$ is used to indicate that a value has not 
yet been assigned to a program variable.
The initial state is $\tup{0,\tup{\und,\und,\und}}$.
As actions, we have again an action corresponding to each atomic 
statement of the program as well as each test of the program and its 
opposite.
As transitions, we have the following:
\begin{iteml}
\item
for each $n$:
\begin{iteml}
\item
a transition
$\astep{\tup{0,\tup{\und,\und,\und}}}
  {\kw{read(n)}}{\tup{1,\tup{\und,n,\und}}}$, 
\item
a transition
$\astep{\tup{1,\tup{\und,n,\und}}}{\kw{i\,:=\,0}}
  {\tup{2,\tup{0,n,\und}}}$, 
\item
a transition
$\astep{\tup{2,\tup{0,n,\und}}}{\kw{f\,:=\,1}}
  {\tup{3,\tup{0,n,1}}}$;
\end{iteml}
\item
for each $i,n,f$ such that $i < n$ and $f = i\,!$:
\begin{iteml}
\item
a transition
$\astep{\tup{3,\tup{i,n,f}}}{\kw{i\,<\,n}}
  {\tup{4,\tup{i,n,f}}}$, 
\item
a transition
$\astep{\tup{4,\tup{i,n,f}}}{\kw{i\,:=\,i\,+\,1}}
  {\tup{5,\tup{i + 1,n,f}}}$,
\item
a transition
$\astep{\tup{5,\tup{i + 1,n,f}}}{\kw{f\,:=\,f\,*\,i}}
  {\tup{3,\tup{i + 1,n,f \cdot (i + 1)}}}$;
\end{iteml}
\item
for each $i,n,f$ such that $i = n$ and $f = i\,!$:
\begin{iteml}
\item
a transition
$\astep{\tup{3,\tup{i,n,f}}}{\kw{NOT\,i\,<\,n}}{\tup{6,\tup{i,n,f}}}$,
\item
a transition
$\astep{\tup{6,\tup{i,n,f}}}{\kw{write(f)}}{\tup{7,\tup{i,n,f}}}$.
\end{iteml}
\end{iteml}
There are some noticeable differences between this transition system 
and the transition system from Example~\ref{exa-factorial}.
The two relevant intuitions are as follows.
In the same state, reading different numbers does not cause the same
state change.
In the same state, a test and its opposite do not succeed both.

Not all states are reachable.
For example, states $\tup{l,\tup{i,n,f}}$ with $i \neq \und$ and 
$n \neq \und$ for which $i > n$ holds are not reachable.
We did not bother to restrict the transition system to the reachable
states: we will see later that the resulting transition system would 
describe essentially the same behaviour.
The transition system for the factorial program is represented
graphically in Fig.~\ref{fig-factorial-mach} for the case where
$\nm{maxint} = 2$.
\begin{figure}
\begin{pspicture}(0,0)(12,18)

 \psset{arrows=->}

 \pnode(5,17.5){S}
 \rput(5,16.5){\ovalnode{0***}{(0,(*,*,*))}}

 \rput(1,15){\ovalnode{1*0*}{(1,(*,0,*))}}
 \rput(1,13.5){\ovalnode{200*}{(2,(0,0,*))}}
 \rput(1,12){\ovalnode{3001}{(3,(0,0,1)}}
 \rput(1,10.5){\ovalnode{6001}{(6,(0,0,1)}}
 \rput(1,9){\ovalnode{7001}{(7,(0,0,1))}}

 \rput(5,15){\ovalnode{1*1*}{(1,(*,1,*))}}
 \rput(5,13.5){\ovalnode{201*}{(2,(0,1,*))}}
 \rput(5,12){\ovalnode{3011}{(3,(0,1,1))}}
 \rput(5,10.5){\ovalnode{4011}{(4,(0,1,1))}}
 \rput(5,9){\ovalnode{5111}{(5,(1,1,1))}}
 \rput(5,7.5){\ovalnode{3111}{(3,(1,1,1))}}
 \rput(5,6){\ovalnode{6111}{(6,(1,1,1))}}
 \rput(5,4.5){\ovalnode{7111}{(7,(1,1,1))}}

 \rput(9,15){\ovalnode{1*2*}{(1,(*,2,*))}}
 \rput(9,13.5){\ovalnode{202*}{(2,(0,2,*))}}
 \rput(9,12){\ovalnode{3021}{(3,(0,2,1))}}
 \rput(9,10.5){\ovalnode{4021}{(4,(0,2,1))}}
 \rput(9,9){\ovalnode{5121}{(5,(1,2,1))}}
 \rput(9,7.5){\ovalnode{3121}{(3,(1,2,1))}}
 \rput(9,6){\ovalnode{4121}{(4,(1,2,1))}}
 \rput(9,4.5){\ovalnode{5221}{(5,(2,2,1))}}
 \rput(9,3){\ovalnode{3222}{(3,(2,2,2))}}
 \rput(9,1.5){\ovalnode{6222}{(6,(2,2,2))}}
 \rput(9,0){\ovalnode{7222}{(7,(2,2,2))}}

 \ncline{S}{0***}
 \ncline{0***}{1*0*}\nbput[nrot=:180]{$\kw{read(n)}$}
 \ncline{1*0*}{200*}\naput{$\kw{i\,:=\,0}$}
 \ncline{200*}{3001}\naput{$\kw{f\,:=\,1}$}
 \ncline{3001}{6001}\naput{$\kw{NOT\,i\,<\,n}$}
 \ncline{6001}{7001}\naput{$\kw{write(f)}$}

 \ncline{0***}{1*1*}\naput{$\kw{read(n)}$}
 \ncline{1*1*}{201*}\naput{$\kw{i\,:=\,0}$}
 \ncline{201*}{3011}\naput{$\kw{f\,:=\,1}$}
 \ncline{3011}{4011}\naput{$\kw{i\,<\,n}$}
 \ncline{4011}{5111}\naput{$\kw{i\,:=\,i+1}$}
 \ncline{5111}{3111}\naput{$\kw{f\,:=\,f*i}$}
 \ncline{3111}{6111}\naput{$\kw{NOT\,i\,<\,n}$}
 \ncline{6111}{7111}\naput{$\kw{write(f)}$}

 \ncline{0***}{1*2*}\naput[nrot=:0]{$\kw{read(n)}$}
 \ncline{1*2*}{202*}\naput{$\kw{i\,:=\,0}$}
 \ncline{202*}{3021}\naput{$\kw{f\,:=\,1}$}
 \ncline{3021}{4021}\naput{$\kw{i\,<\,n}$}
 \ncline{4021}{5121}\naput{$\kw{i\,:=\,i+1}$}
 \ncline{5121}{3121}\naput{$\kw{f\,:=\,f*i}$}
 \ncline{3121}{4121}\naput{$\kw{i\,<\,n}$}
 \ncline{4121}{5221}\naput{$\kw{i\,:=\,i+1}$}
 \ncline{5221}{3222}\naput{$\kw{f\,:=\,f*i}$}
 \ncline{3222}{6222}\naput{$\kw{NOT\,i\,<\,n}$}
 \ncline{6222}{7222}\naput{$\kw{write(f)}$}

\end{pspicture}
\caption{Another transition system for the factorial program}
\label{fig-factorial-mach}
\end{figure}
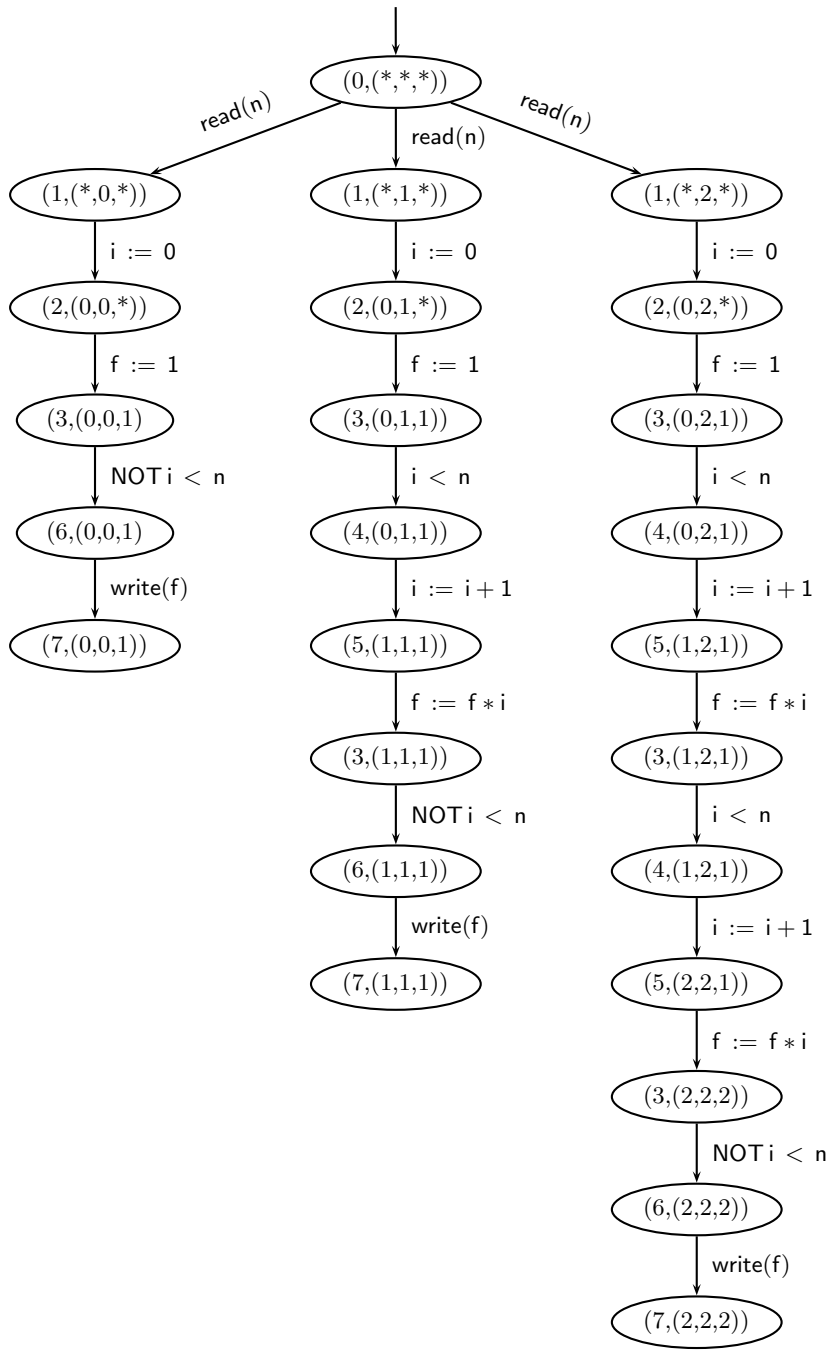
\end{example}
\begin{example}[Greatest common divisor program]%
\index{greatest common divisor program}
\label{exa-gcd-mach}
We also consider again the program from Example~\ref{exa-gcd}.
The intended behaviour of this program upon execution on a machine can
be described by a transition system as follows.
As states of the program, we have pairs $\tup{l,s}$, 
where $l \in \Nat$ with $0 \leq l \leq 8$ 
and $s = \tup{m,n}$ with 
$m,n \in
 \set{i \in \Nat \where i \leq \nm{maxint}} \union \set{\und}$.
These states are like in Example~\ref{exa-factorial-mach}.
The initial state is $\tup{0,\tup{\und,\und}}$.
As actions, we have again an action corresponding to each atomic 
statement of the program as well as each test of the program and its 
opposite.
As transitions, we have the following:
\begin{iteml}
\item
for each $m$:
\begin{iteml}
\item
a transition
$\astep{\tup{0,\tup{\und,\und}}}{\kw{read(m)}}
  {\tup{1,\tup{m,\und}}}$; 
\end{iteml}
\item
for each $m,n$:
\begin{iteml}
\item
a transition
$\astep{\tup{1,\tup{m,\und}}}{\kw{read(n)}}
  {\tup{2,\tup{m,n}}}$; 
\end{iteml}
\item
for each $m,n$ such that $m > n$:
\begin{iteml}
\item
a transition
$\astep{\tup{2,\tup{m,n}}}{\kw{m\,>\,n}}
  {\tup{3,\tup{m,n}}}$, 
\item
a transition
$\astep{\tup{3,\tup{m,n}}}{\kw{m\,:=\,m\,-\,n}}
  {\tup{2,\tup{m - n,n}}}$;
\end{iteml}
\item
for each $m,n$ such that $m \leq n$:
\begin{iteml}
\item
a transition
$\astep{\tup{2,\tup{m,n}}}{\kw{NOT\,m\,>\,n}}
  {\tup{4,\tup{m,n}}}$;
\end{iteml}
\item
for each $m,n$ such that $m < n$:
\begin{iteml}
\item
a transition
$\astep{\tup{4,\tup{m,n}}}{\kw{n\,>\,m}}
  {\tup{5,\tup{m,n}}}$, 
\item
a transition
$\astep{\tup{5,\tup{m,n}}}{\kw{n\,:=\,n\,-\,m}}
  {\tup{4,\tup{m,n - m}}}$;
\end{iteml}
\item
for each $m,n$ such that $m \geq n$:
\begin{iteml}
\item
a transition
$\astep{\tup{4,\tup{m,n}}}{\kw{NOT\,n\,>\,m}}
  {\tup{6,\tup{m,n}}}$;
\end{iteml}
\item
for each $m,n$ such that $m \neq n$:
\begin{iteml}
\item
a transition
$\astep{\tup{6,\tup{m,n}}}{\kw{NOT\,m\,=\,n}}
  {\tup{2,\tup{m,n}}}$; 
\end{iteml}
\item
for each $m,n$ such that $m = n$:
\begin{iteml}
\item
a transition
$\astep{\tup{6,\tup{m,n}}}{\kw{m\,=\,n}}
  {\tup{7,\tup{m,n}}}$, 
\item
a transition
$\astep{\tup{7,\tup{m,n}}}{\kw{write(m)}}{\tup{8,\tup{m,n}}}$.
\end{iteml}
\end{iteml}
The differences between this transition system and the transition 
system given in Example~\ref{exa-gcd} are of the same kind as between
the transition systems given for factorial program.
Like in Example~\ref{exa-factorial-mach}, not all states are reachable.
The transition system for the greatest common divisor program is 
represented graphically in Fig.~\ref{fig-gcd-mach} for the case where
$\nm{maxint} = 2$.
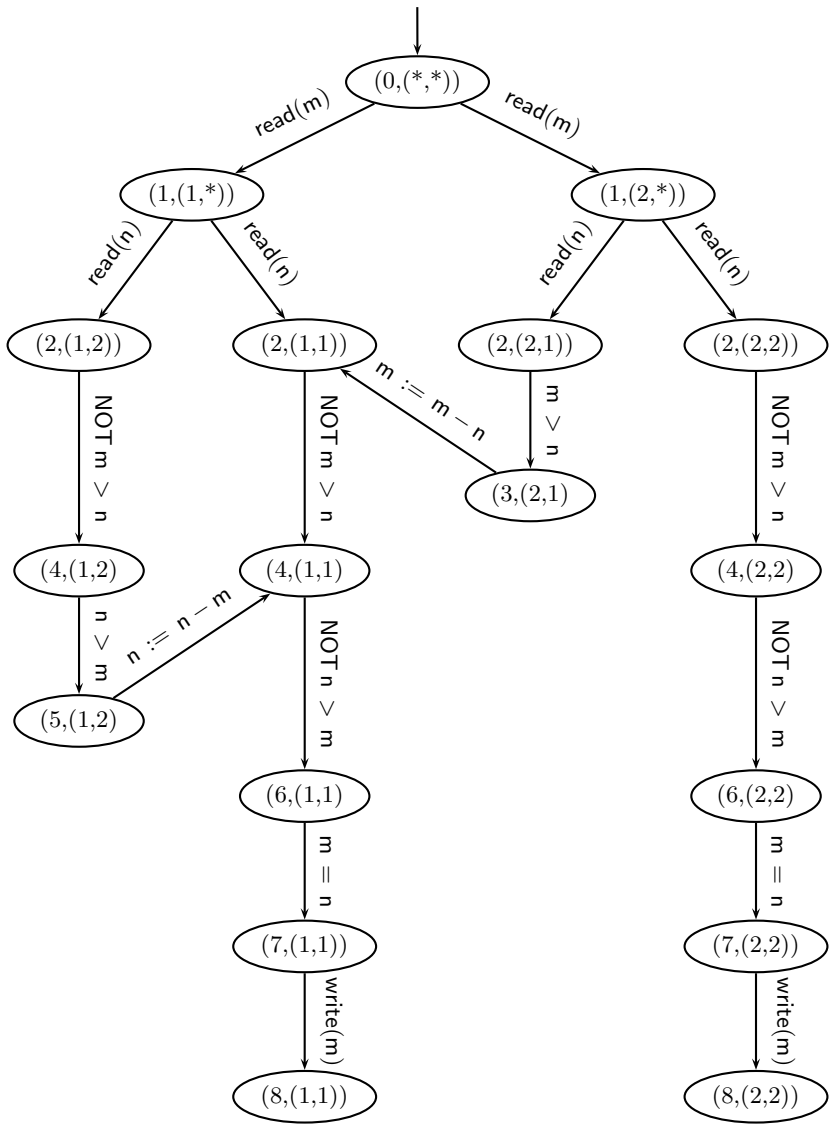
\begin{figure}
\begin{pspicture}(0,3)(12,18)

 \psset{arrows=->}

 \pnode(5.5,17.5){S}

 \rput(5.5,16.5){\ovalnode{0**}{(0,(*,*))}}

 \rput(2.5,15){\ovalnode{11*}{(1,(1,*))}}
 \rput(8.5,15){\ovalnode{12*}{(1,(2,*))}}

 \rput(4,13){\ovalnode{211}{(2,(1,1))}}
 \rput(1,13){\ovalnode{212}{(2,(1,2))}}
 \rput(7,13){\ovalnode{221}{(2,(2,1))}}
 \rput(10,13){\ovalnode{222}{(2,(2,2))}}

 \rput(7,11){\ovalnode{321}{(3,(2,1)}}

 \rput(4,10){\ovalnode{411}{(4,(1,1)}}
 \rput(1,10){\ovalnode{412}{(4,(1,2)}}
 \rput(10,10){\ovalnode{422}{(4,(2,2)}}

 \rput(1,8){\ovalnode{512}{(5,(1,2)}}

 \rput(4,7){\ovalnode{611}{(6,(1,1)}}
 \rput(10,7){\ovalnode{622}{(6,(2,2)}}

 \rput(4,5){\ovalnode{711}{(7,(1,1))}}
 \rput(10,5){\ovalnode{722}{(7,(2,2))}}

 \rput(4,3){\ovalnode{811}{(8,(1,1))}}
 \rput(10,3){\ovalnode{822}{(8,(2,2))}}

 \ncline{S}{0**}
 \ncline{0**}{11*}\nbput[nrot=:180]{$\kw{read(m)}$}
 \ncline{0**}{12*}\naput[nrot=:0]{$\kw{read(m)}$}
 \ncline{11*}{211}\naput[nrot=:0]{$\kw{read(n)}$}
 \ncline{11*}{212}\nbput[nrot=:180]{$\kw{read(n)}$}
 \ncline{12*}{221}\nbput[nrot=:180]{$\kw{read(n)}$}
 \ncline{12*}{222}\naput[nrot=:0]{$\kw{read(n)}$}
 \ncline{211}{411}\naput[nrot=:0]{$\kw{NOT\,m\,>\,n}$}
 \ncline{212}{412}\naput[nrot=:0]{$\kw{NOT\,m\,>\,n}$}
 \ncline{222}{422}\naput[nrot=:0]{$\kw{NOT\,m\,>\,n}$}
 \ncline{221}{321}\naput[nrot=:0]{$\kw{m\,>\,n}$}
 \ncline{321}{211}\nbput[nrot=:180]{$\kw{m\,:=\,m-n}$}
 \ncline{411}{611}\naput[nrot=:0]{$\kw{NOT\,n\,>\,m}$}
 \ncline{422}{622}\naput[nrot=:0]{$\kw{NOT\,n\,>\,m}$}
 \ncline{412}{512}\naput[nrot=:0]{$\kw{n\,>\,m}$}
 \ncline{512}{411}\naput[nrot=:0]{$\kw{n\,:=\,n-m}$}
 \ncline{611}{711}\naput[nrot=:0]{$\kw{m\,=\,n}$}
 \ncline{711}{811}\naput[nrot=:0]{$\kw{write(m)}$}
 \ncline{622}{722}\naput[nrot=:0]{$\kw{m\,=\,n}$}
 \ncline{722}{822}\naput[nrot=:0]{$\kw{write(m)}$}

\end{pspicture}
\caption
  {Another transition system for the greatest common divisor program}
\label{fig-gcd-mach}
\end{figure}
\end{example}

For a given programming language, the behaviour of its programs upon 
execution on a machine is called its operational semantics.
It is usually described in a style known as structural operational 
semantics.
This means that the behaviour of a compound language construct is 
described in terms of the behaviour of its constituents.
The transition systems from the previous two examples were not formally
based on a given (structural) operational semantics.

\section{Automata and transition systems}
\label{sect-basics-conn-automata}

For a better understanding of the notion of a transition system, we 
looked in the previous section into its connections with the familiar 
notion of a program.
For the same reason, we now look into its connections with the familiar 
notion of an automaton from automata theory (see e.g.~\cite{HMU14} for 
an introduction).

Automata can be regarded as a specialized kind of transition systems. 
In this section, we restrict ourselves to the kind of automata known as
\emph{non-deterministic finite accepters}.
They are illustrative for almost any kind of automata.
If no confusion can arise, we will call them simply \emph{automata}.
The difference between automata and transition systems is mainly a
matter of intended use.
As mentioned in Section~\ref{sect-basics-informal}, transition systems 
are primarily regarded as a means to describe the behaviour of processes 
and automata are primarily regarded as abstract machines to recognize 
certain languages.
Because of the different intended use, final states are indispensable
in the case of automata: reaching a final state means that a complete
sentence has been recognized.
The final states of automata are usually not required to satisfy the
restriction that they are terminal states.
This restriction would be harmless in the sense that it would not have 
any influence on the languages that automata are able to recognize.
Automata that satisfy the restriction can be regarded as finite
transition systems with designated final states.
We do not give the standard definition of the notion of an automaton.
Our definition underlines the resemblance to transition systems 
mentioned above.
\begin{definition}[Automaton]\index{automaton}
An \emph{automaton} $M$ is a quintuple 
$\tup{S,A,\linebreak[2]\step{},\linebreak[2]s_0,F}$
where
\begin{iteml}
\item
$S$ is a finite set of \emph{internal states};
\item
$A$ is a finite set of \emph{symbols}, called the \emph{input alphabet};
\item 
${\step{}} \subseteq S \x A \x S$ is a set of \emph{transitions};
\item
$s_0 \in S$ is the \emph{initial state};
\item
$F \subseteq S$ is a set of \emph{final states}.
\end{iteml}
A state $s \in S$ is called a \emph{terminal} state of $M$ if there is
no $a \in A$ and $s' \in S$ such that $\astep{s}{a}{s'}$, just as in 
the case of transition systems.
The set ${\gstep{}{}{}} \subseteq S \x \seqof{A} \x S$ of
\emph{generalized transitions} of $M$ is also defined exactly as for
transition systems.
The \emph{language} accepted by $M$, written $\Lang(M)$, is the 
set
$\set{\sigma \in \seqof{A} \where 
      \gstep{s_0}{\sigma}{s} \mathrm{\;for\;some\;} s \in F}$.
\end{definition}
In the standard definition of the notion of an automaton, we have a
\emph{transition function} $\funct{\delta}{S \x A}{\setof{(S)}}$
instead of a set ${\step{}} \subseteq S \x A \x S$ of transitions.
If we take $\delta$ such that $s' \in \delta(s,a)$ if and only if $\astep{s}{a}{s'}$, then we get an automaton according to the standard
definition.

If we regard symbols as actions of reading the symbols, automata are 
simply transition systems with designated final states.
An automaton can be considered to accept certain sequences of symbols 
as follows.
A transition of an automaton is regarded as a state change caused by 
reading a symbol.
A sequence of symbols $a_1\, \ldots\, a_n$ is accepted if a sequence of 
consecutive state changes from the initial state to one of the final 
states can be obtained by reading the symbols $a_1$, \ldots, $a_n$ in 
turn.
This informal explanation can be made more precise as follows.

Let $M$ be the automaton $\tup{S,A,\step{},s_0,F}$ and 
let $A'$ be the set of actions $\set{\kw{read}(a) \where a \in A}$.
Suppose that each state in $F$ is a terminal state of $M$.
Now consider the transition system $T = \tup{S,A',\step{}',s_0}$ where 
$\astepp{s_1}{\kw{read}(a)}{s_2}$ iff $\astep{s_1}{a}{s_2}$.
The sentences of the language accepted by $M$ are exactly the sequences 
of symbols that can be consecutively read by $T$ till a terminal state 
is reached that is contained in $F$.

Let us look at a simple example of the use of automata in recognizing a
language.
\begin{example}[Pidgingol]\index{pidgingol}
\label{exa-pidgingol}
We consider a very simple language.
A sentence of the language consists of a noun clause followed by a verb
followed by a noun clause.
A noun clause consists of an article followed by a noun.
A noun is either $\kw{man}$ or $\kw{machine}$.
A verb is either $\kw{simulates}$ or $\kw{mimics}$.
An example sentence is \textsf{the man mimics a machine}.
This language is accepted by the following automaton.
As internal states of the automaton, we have pairs $\tup{p,i}$, 
where $p \in \set{\kw{left},\kw{right}}$ and $i \in \Nat$ with 
$0 \leq i \leq 2$. 
The choice of states is not really relevant.
We could have taken the natural numbers $0$ to $5$ equally well, but
the choice made here allows for a short presentation of the automaton.
The initial state is $\tup{\kw{left},0}$ and the only final state is 
$\tup{\kw{right},2}$.
The input alphabet consists of $\kw{a}$, $\kw{the}$, $\kw{man}$,
$\kw{machine}$, $\kw{simulates}$ and $\kw{mimics}$.
As transitions, we have the following:
\begin{iteml}
\item
for $p = \kw{left},\kw{right}$:
\begin{iteml}
\item
a transition 
$\astep{\tup{p,0}}{\kw{a}}{\tup{p,1}}$,
\item
a transition 
$\astep{\tup{p,0}}{\kw{the}}{\tup{p,1}}$,
\item
a transition 
$\astep{\tup{p,1}}{\kw{man}}{\tup{p,2}}$,
\item
a transition 
$\astep{\tup{p,1}}{\kw{machine}}{\tup{p,2}}$;
\end{iteml}
\item
a transition 
$\astep{\tup{\kw{left},2}}{\kw{simulates}}
  {\tup{\kw{right},0}}$;
\item
a transition 
$\astep{\tup{\kw{left},2}}{\kw{mimics}}
  {\tup{\kw{right},0}}$.
\end{iteml}
The automaton for our very simple language is represented
graphically in Fig.~\ref{fig-pidging}.
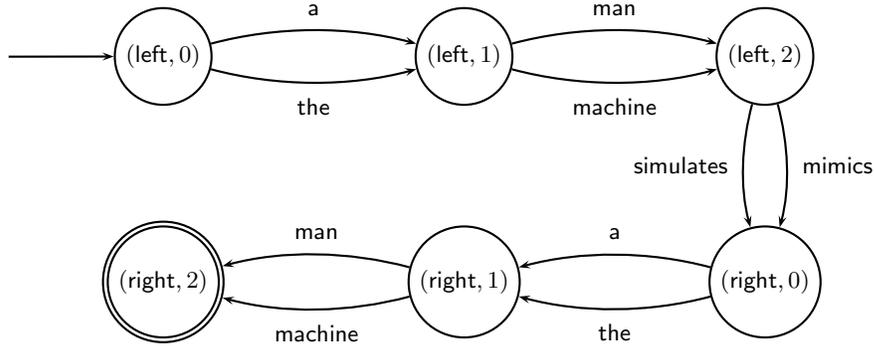
\begin{figure}
\begin{pspicture}(0,0)(12,5)

 \psset{arrows=->}

 \pnode(0,4){S}
 \rput(2,4){\circlenode{left0}{$(\kw{left},0)$}}
 \rput(6,4){\circlenode{left1}{$(\kw{left},1)$}}
 \rput(10,4){\circlenode{left2}{$(\kw{left},2)$}}
 \rput(10,1){\circlenode{right0}{$(\kw{right},0)$}}
 \rput(6,1){\circlenode{right1}{$(\kw{right},1)$}}
 \rput(2,1){\circlenode[doubleline=true]{right2}{$(\kw{right},2)$}}

 \ncline{S}{left0}
 \nccurve[angleA=15,angleB=165]{left0}{left1}\naput{$\kw{a}$}
 \nccurve[angleA=345,angleB=195]{left0}{left1}\nbput{$\kw{the}$}
 \nccurve[angleA=15,angleB=165]{left1}{left2}\naput{$\kw{man}$}
 \nccurve[angleA=345,angleB=195]{left1}{left2}\nbput{$\kw{machine}$}
 \nccurve[angleA=255,angleB=105]{left2}{right0}\nbput{$\kw{simulates}$}
 \nccurve[angleA=285,angleB=75]{left2}{right0}\naput{$\kw{mimics}$}
 \nccurve[angleA=165,angleB=15]{right0}{right1}\nbput{$\kw{a}$}
 \nccurve[angleA=195,angleB=345]{right0}{right1}\naput{$\kw{the}$}
 \nccurve[angleA=165,angleB=15]{right1}{right2}\nbput{$\kw{man}$}
 \nccurve[angleA=195,angleB=345]{right1}{right2}\naput{$\kw{machine}$}

\end{pspicture}
\caption{Automaton accepting a very simple language}
\label{fig-pidging}
\end{figure}
It is obvious that this automaton accepts the same sequences of symbols
as the finite transition system obtained from this automaton by 
replacing the symbols $\kw{a}$, $\kw{the}$, $\kw{man}$, $\kw{machine}$, 
$\kw{simulates}$ and $\kw{mimics}$ by actions of reading these symbols. 
\end{example}

Conversely, we can also view any finite transition system as an 
automaton by regarding its actions as symbols and its terminal states 
as final states.
This is interesting because the sequences of actions it can 
consecutively perform are an important aspect of the behaviour of a
process.
We will get back to that later in Section~\ref{sect-basics-trace-eqv}.
Here is an example that illustrates the potential usefulness of 
focussing on the sequences of actions that a system can consecutively 
perform.
\begin{example}[Unreliable bounded counter]%
\index{counter!bounded!unreliable}
\label{exa-bcounter-automaton}
We consider an unreliable version of the bounded counter with bound 
$k$ from Example~\ref{exa-bcounter}.
It gets into an error state by performing an increment by $1$ when its 
bound is reached.
We have one additional state, $\kw{err}$, and the additional transition
$\astep{k}{\kw{inc}}{\kw{err}}$.
More precisely, the behaviour of the unreliable bounded counter with 
bound $k$ is described by the transition system 
$\tup{S,A,\step{},s_0}$ where
\[
\begin{aeqns}
S & = & \set{i \in \Nat \where i \leq k} \union 
        \set{\kw{err}}\enspace, \\
A & = & \set{\kw{inc},\kw{dec}}\enspace, \\
{\step{}} & = & 
 \set{\tup{i,\kw{inc},i+1} \where i \in \Nat, i < k} \union
 \set{\tup{k,\kw{inc},\kw{err}}} \\ & {} \union &
 \set{\tup{i+1,\kw{dec},i} \where i \in \Nat, i < k}\enspace, \\
s_0 & = & 0
\end{aeqns}
\]
This transition system has only one terminal state, viz.\ $\kw{err}$.
The sequences of actions that lead to this state are exactly the
sequences $w$ that satisfy the following conditions:
\begin{iteml}
\item
$n_{\kw{inc}}(w) - n_{\kw{dec}}(w) = k + 1$,
\item
for all proper prefixes $v$ of $w$, 
$0 \leq n_{\kw{inc}}(v) - n_{\kw{dec}}(v) \leq k$;
\end{iteml}
where $n_a(u)$ stands for the number of occurrences of action $a$ in 
sequence $u$.
This description of the sequences of actions that lead to its terminal 
state may be regarded as the specification of the intended system.

If we designate the terminal state as final state, the transition
system can be viewed as an automaton recognizing the language on the 
alphabet $\set{\kw{inc},\kw{dec}}$ that consists of the sequences 
$w \in \seqof{\set{\kw{inc},\kw{dec}}}$ satisfying the conditions
just mentioned.
When viewing the transition system as an automaton, the point is that 
$\kw{inc}$ and $\kw{dec}$ are considered to be symbols to be read
instead of actions to be performed.
\end{example}

The following is known from automata theory.
The languages that can be accepted by an automaton as defined here,
i.e.\ a non-deterministic finite accepter, are exactly the regular 
languages.
Intuitively, a regular language has a structure simple enough that a 
limited memory is sufficient to accept all its sentences.
Many actual languages are not regular.
Broader language categories include the context-free languages and
the context-sensitive languages.
They can be accepted by automata of more powerful kinds:
non-deterministic pushdown accepters for context-free languages and
linear bounded accepters for context-sensitive languages.
Those kinds of automata are in turn closely related to restricted
kinds of infinite transition systems.

\section{Petri nets and transition systems}
\label{sect-basics-conn-nets}

For a better understanding of the notion of a transition system, we 
looked in the previous two sections into its connections with the 
familiar notions of a program and an automaton.
For the interested reader, we now look into its connections with the  
notion of a Petri net.
Sometimes, the notion of a Petri net is considered to be the fundamental
notion for the description of process behaviour.
We believe that it is too complicated to be acceptable as a fundamental
notion.
However, there are many applications of Petri nets in a wide variety of 
areas.
The central developments of more than fifty years of Petri net theory
and practice are presented in~\cite{Rei13}.

The notion of a Petri net is essentially a generalization of the notion 
of a transition system.
In this section, we restrict our attention to the kind of Petri nets 
known as \emph{place/transition nets} with arc weight $1$.
They are illustrative for almost any other kind of Petri nets.
If no confusion can arise, we will call them simply \emph{nets}.
The crucial difference between nets and transition systems is the 
following.
In transition systems, choices between behaviours and sequentiality of 
behaviours are regarded as the basic aspects of process behaviour,
whereas in nets, concurrency of behaviours is also regarded as a basic 
aspect of process behaviour.
How concurrency can be dealt with in the setting of transition systems 
is treated in Chap.~\ref{ch-interaction}.
Nets support the direct description of concurrency because they can 
deal with states that are distributed over several places.
We do not give the standard definition of the notion of a net.
Our definition, which is taken from~\cite{Old87}, underlines the 
similarities between transition systems and nets.

\begin{definition}[Net]\index{net}
A \emph{net} $N$ is a quadruple $\tup{P,A,\step{},m_0}$ 
where
\begin{iteml}
\item
$P$ is a set of \emph{places};
\item
$A$ is a set of \emph{actions};
\item 
${\step{}} \subseteq 
 (\fsetof{(P)} \diff \emptyset) \x A \x (\fsetof{(P)} \diff \emptyset)$
is a set of \emph{transitions};
\item
$m_0 \in \fsetof{(P)} \diff \emptyset$ is the \emph{initial marking}.
\end{iteml}
Let $t$ be the transition $\astep{Q}{a}{Q'}$.
Then the \emph{preset} of $t$, written $\nm{pre}(t)$, is $Q$; 
the \emph{postset} of $t$, written $\nm{post}(t)$, is $Q'$; and
the \emph{action} of $t$, written $\nm{act}(t)$, is $a$.
\end{definition}
In the standard definition of the notion of a place/transition net, a
net has a set $T$ of transitions which are not necessarily composed of
their preset, postset and action.
The pre- and postsets of each transition is in the standard definition
given by a \emph{flow relation} $F \subseteq (P \x T) \union (T \x P)$
and the action of each transition by a \emph{labeling function}
$\funct{\ell}{T}{A}$.
Moreover, there is a \emph{arc weight function} $\funct{W}{F}{\Nat}$ in
the standard definition.
Because, we restrict ourselves to the case where the arc weight is 
invariably $1$, the arc weight function is superfluous.
If we take $T = {\step{}}$, $F$ such that 
$\tup{p,\astep{Q}{a}{Q'}} \in F$ if and only if $p \in Q$ and
$\tup{\astep{Q}{a}{Q'},p} \in F$ if and only if $p \in Q'$, and 
$\ell$ such that $\ell(\astep{Q}{a}{Q'}) = a$, then we get a 
place/transition net according to the standard definition.

If we regard singleton sets of places as states, transition systems are
nets where the presets, postsets and initial marking are singleton 
sets.
A net can be considered to distribute the states of a transition system
over several places as follows.
Each place contains zero, one or more \emph{tokens}.
The numbers of tokens contained in the different places make up the 
states of a net, also called markings. 
A transition $t$ is firable in a marking if there is at least one 
token in each place from the preset of $t$.
By firing $t$, one token is removed from each place from the preset of 
$t$ and one token is inserted in each place from the postset of $t$.
This informal explanation can be made more precise as follows. 

Let $N$ be the net $\tup{P,A,\step{},m_0}$.
Then a \emph{marking} of $N$ is a multiset of places, i.e.\ a function 
$\funct{m}{P}{\Nat}$.
A transition $t$ of $N$ is \emph{firable} in a marking $m$ if 
$m(p) > 0$ for all $p \in \nm{pre}(t)$. 
If transition $t$ is firable in marking $m$, the \emph{firing} of $t$ 
in $m$ produces the unique marking $m'$ such that for all $p \in P$:
\[
m'(p) = 
\left\{
\begin{array}[c]{ll}
m(p) - 1 & 
\mathrm{if}\; p \in \nm{pre}(t) \;\mathrm{and}\; p \not\in \nm{post}(t),
\\
m(p) + 1 & 
\mathrm{if}\; p \not\in \nm{pre}(t) \;\mathrm{and}\; p \in \nm{post}(t),
\\
m(p)     & \mathrm{otherwise}.
\end{array}
\right.
\]
The notation $\astep{m}{t}{m'}$ is used to indicate that firing 
transition $t$ in marking $m$ produces marking $m'$.
A set $Q \subseteq P$ is identified with the unique marking $m$ such 
that $m(p) = 1$ if $p \in Q$ and $m(p) = 0$ otherwise.

Let $N$ be the net $\tup{P,A,\step{},m_0}$ and $\sigma \in \seqof{A}$.
The notation $\gstep{m}{\sigma}{m'}$ is used to indicate that there are
markings $m_1,\ldots,m_{n+1}$ and transitions $t_1,\ldots,t_n$ such
that $\astep{m_1}{t_1}{m_2}$, \ldots, $\astep{m_n}{t_n}{m_{n+1}}$,
$m_1 = m$, $m_{n+1} = m'$ and 
$\sigma = \nm{act}(t_1)\,\ldots\,\nm{act}(t_n)$.
A marking $m$ of $N$ is called a \emph{reachable} marking of $N$ if 
there is a $\sigma \in \seqof{A}$ such that $\gstep{m_0}{\sigma}{m}$.
Reachable markings make an important link between nets and transition
systems.

The transition system describing the behaviour of a net is defined as 
follows.
Let $N$ be the net $\tup{P,A,\step{},m_0}$ 
and $M$ be the set of reachable markings of $N$.
Then the transition system associated with $N$ is the transition 
system $\TrSy(N) = \tup{M,A,\step{}',m_0}$ where $\astepp{m_1}{a}{m_2}$
iff there exists a transition $t$
of $N$ such that $\astep{m_1}{t}{m_2}$ and $\nm{act}(t) = a$.

Let us look at an example of the use of nets in describing process 
behaviour.
\begin{example}[Binary memory cell]\index{binary memory cell}
\label{exa-binvar}
We consider a binary memory cell.
A binary memory cell holds at any moment either the value $0$ or the 
value $1$. 
Initially, it holds the value $0$.
The binary memory cell can store a value and retrieve its value. 
Its behaviour can be described by a net as follows.
As places of the binary memory cell, we have the pairs 
$\tup{b,\kw{rtr}}$, $\tup{b,\kw{sto}}$ for $b = 0,1$. 
If its marking includes the place $\tup{b,\kw{rtr}}$, the cell can 
retrieve the value $b$. 
If its marking includes the place $\tup{b,\kw{sto}}$, the cell can 
store the value $b$.
If its marking includes both $\tup{b,\kw{rtr}}$ and $\tup{b,\kw{sto}}$,
the cell can store the value $1 - b$.
As initial marking, we have $\set{\tup{0,\kw{rtr}},\tup{0,\kw{sto}}}$.
As actions, we have $\kw{sto}(b)$ (store $b$) and $\kw{rtr}(b)$ 
(retrieve $b$) for $b = 0,1$.
As transitions, we have the following (for $b = 0,1$):
\begin{ldispl}
\astep{\set{\tup{b,\kw{rtr}}}}{\kw{rtr}(b)}{\set{\tup{b,\kw{rtr}}}}, \\
\astep{\set{\tup{b,\kw{sto}}}}{\kw{sto}(b)}{\set{\tup{b,\kw{sto}}}}, \\
\astep{\set{\tup{b,\kw{rtr}},\tup{b,\kw{sto}}}}{\kw{sto}(1 - b)}
  {\set{\tup{1 - b,\kw{rtr}},\tup{1 - b,\kw{sto}}}}.
\end{ldispl}

The transition system associated with this net is as follows.
As states, we have the markings 
$\set{\tup{b,\kw{rtr}},\tup{b,\kw{sto}}}$ for $b = 0,1$, with
$\set{\tup{0,\kw{rtr}},\tup{0,\kw{sto}}}$ as the initial state.
As actions, we still have $\kw{sto}(b)$ and $\kw{rtr}(b)$ for 
$b = 0,1$.
As transitions, we have the following (for $b = 0,1$):
\begin{ldispl}
\astep{\set{\tup{b,\kw{rtr}},\tup{b,\kw{sto}}}}{\kw{rtr}(b)}
  {\set{\tup{b,\kw{rtr}},\tup{b,\kw{sto}}}}, \\
\astep{\set{\tup{b,\kw{rtr}},\tup{b,\kw{sto}}}}{\kw{sto}(b)}
  {\set{\tup{b,\kw{rtr}},\tup{b,\kw{sto}}}}, \\
\astep{\set{\tup{b,\kw{rtr}},\tup{b,\kw{sto}}}}{\kw{sto}(1 - b)}
  {\set{\tup{1 - b,\kw{rtr}},\tup{1 - b,\kw{sto}}}}.
\end{ldispl}
The transition system for the binary memory cell does not indicate that
if both $\kw{rtr}(b)$ and $\kw{sto}(b)$ can occur, they can also occur 
simultaneously.
This can be covered as well if we generalize transition systems by
taking multisets of actions as labels of transitions.
We will not discuss this generalization in these lecture notes.
\end{example}
Let us look at one more example of the use of nets in describing 
process behaviour.
\begin{example}[Milner's scheduling problem]%
\index{Milner's scheduling problem}
\label{exa-scheduler}
We consider the system of scheduled processes from Milner's scheduling
problem (see~\cite{Mil89}).
It consists of processes $P_1,\ldots,P_n$ ($n > 1$), each wishing to 
perform a certain task repeatedly, and a scheduler ensuring that they 
start their task in cyclic order, beginning with $P_1$.
The behaviour of this system can be described by a net as 
follows.
As places of the system, we have the pairs 
$\tup{i,\kw{idle}}$, $\tup{i,\kw{busy}}$, $\tup{i,\kw{sch}}$ for 
$1 \leq i \leq n$. 
If its marking includes both $\tup{i,\kw{idle}}$ and $\tup{i,\kw{sch}}$,
process $P_i$ can start performing its task.
If its marking includes $\tup{i,\kw{busy}}$, process $P_i$ can finish
performing its task. 
As initial marking, we have 
$\set{\tup{1,\kw{idle}},\ldots,\tup{n,\kw{idle}},\tup{1,\kw{sch}}}$.
As actions, we have $\kw{start}(i)$ (start task $i$) and 
$\kw{finish}(i)$ (finish task $i$) for $1 \leq i \leq n$.
As transitions, we have the following (for $1 \leq i \leq n$):
\begin{ldispl}
\astep{\set{\tup{i,\kw{idle}},\tup{i,\kw{sch}}}}{\kw{start}(i)}
  {\set{\tup{i,\kw{busy}},\tup{\nm{nxt}(i),\kw{sch}}}}, \\
\astep{\set{\tup{i,\kw{busy}}}}{\kw{finish}(i)}
  {\set{\tup{i,\kw{idle}}}},
\end{ldispl}
where $\nm{nxt}(i) = i + 1$ if $i < n$ and $\nm{nxt}(n) = 1$.
The behaviour of the system is much easier to grasp from this net than 
from the transition system associated with the net because the 
structure of the system is clearly reflected in the net.
The net for the system of scheduled processes is represented graphically
in Fig.~\ref{fig-scheduler} for the case where $n = 3$.
\begin{figure}
\begin{pspicture}(-0.5,0.5)(11.5,5.5)

 \psset{arrows=->,radius=.3}

 \Cnode(1,2){1idle}\nput{90}{1idle}{$(1,\kw{idle})$}
 \cnode[fillstyle=solid,fillcolor=black](1,2){.1}{T1}
 \Cnode(0,4){1sch}\nput{90}{1sch}{$(1,\kw{sch})$}
 \cnode[fillstyle=solid,fillcolor=black](0,4){.1}{T4}
 \Cnode(3,2){1busy}\nput{90}{1busy}{$(1,\kw{busy})$}

 \Cnode(5,2){2idle}\nput{90}{2idle}{$(2,\kw{idle})$}
 \cnode[fillstyle=solid,fillcolor=black](5,2){.1}{T2}
 \Cnode(4,4){2sch}\nput{90}{2sch}{$(2,\kw{sch})$}
 \Cnode(7,2){2busy}\nput{90}{2busy}{$(2,\kw{busy})$}

 \Cnode(9,2){3idle}\nput{90}{3idle}{$(3,\kw{idle})$}
 \cnode[fillstyle=solid,fillcolor=black](9,2){.1}{T3}
 \Cnode(8,4){3sch}\nput{90}{3sch}{$(3,\kw{sch})$}
 \Cnode(11,2){3busy}\nput{90}{3busy}{$(3,\kw{busy})$}

 \fnode(2,3){start1}\nput{90}{start1}{$\kw{start}(1)$}
 \fnode(6,3){start2}\nput{90}{start2}{$\kw{start}(2)$}
 \fnode(10,3){start3}\nput{90}{start3}{$\kw{start}(3)$}
 \fnode(2,1){finish1}\nput{270}{finish1}{$\kw{finish}(1)$}
 \fnode(6,1){finish2}\nput{270}{finish2}{$\kw{finish}(2)$}
 \fnode(10,1){finish3}\nput{270}{finish3}{$\kw{finish}(3)$}

 \ncline{1idle}{start1}
 \ncline{1sch}{start1}
 \ncline{start1}{1busy}
 \ncline{start1}{2sch}
 \ncline{1busy}{finish1}
 \ncline{finish1}{1idle}

 \ncline{2idle}{start2}
 \ncline{2sch}{start2}
 \ncline{start2}{2busy}
 \ncline{start2}{3sch}
 \ncline{2busy}{finish2}
 \ncline{finish2}{2idle}

 \ncline{3idle}{start3}
 \ncline{3sch}{start3}
 \ncline{start3}{3busy}
 \nccurve[angleA=45,angleB=45]{start3}{1sch}
 \ncline{3busy}{finish3}
 \ncline{finish3}{3idle}

\end{pspicture}
\caption{Net for the system of scheduled processes}
\label{fig-scheduler}
\end{figure}
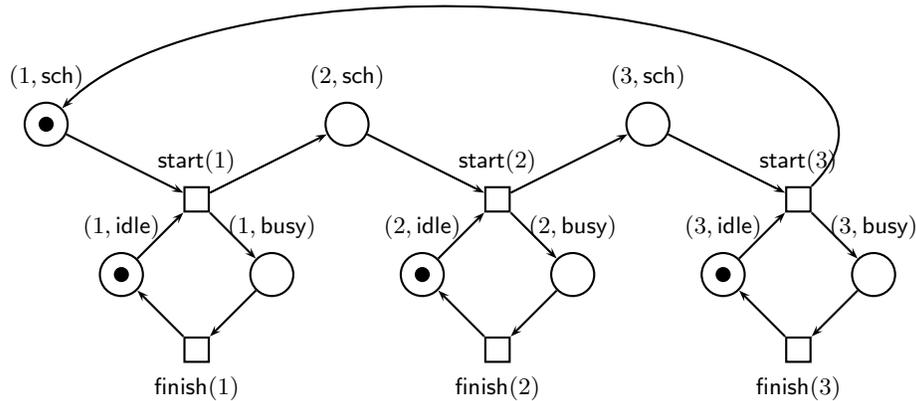
The places and transitions are represented as follows.
Places $p$ are represented as circles and transitions $t$ as boxes
labeled with $\nm{act}(t)$ and connected via directed arcs to the
circles representing the places in $\nm{pre}(t)$ and $\nm{post}(t)$.
The initial marking is represented by putting a bullet into the circles 
representing the places that are in the initial marking.
\end{example}

\section{Equivalences on transition systems}
\label{sect-basics-trace-eqv}
\label{sect-basics-bisim-eqv}

In this section, we look at a couple of notions that are taken up to 
abstract from those details of transition systems that are often 
supposed to be irrelevant.

Usually, transition systems show details that are not considered to be 
relevant to the behaviour of processes.
There are, for example, applications of transition systems where only 
the sequences of actions that can be performed consecutively starting 
from the initial state of a transition system, called the traces of the
transition system, matter.
Here is a simple example of a case where only the traces matter.
\begin{example}[Bounded counter]\index{counter!bounded}
\label{exa-bcounter-trace-eqv}
We consider again the bounded counter with bound $k$ from
Example~\ref{exa-bcounter}.
Its traces are exactly the traces $w$ for which the following condition 
holds: for all prefixes $v$ of $w$, 
$0 \leq n_{\kw{inc}}(v) - n_{\kw{dec}}(v) \leq k$.
This description of its traces expresses all we expect from the bounded 
counter: we regard any transition system that has those traces as a
bounded counter.
For this reason, only the traces are relevant in this case.
\end{example}
Notice that in all cases where a transition system is used to accept a 
language, as described in Section~\ref{sect-basics-conn-automata}, only
the traces are relevant.

In all those cases where only the traces of the transition system 
matter, it is useful to ignore all other details.
This is done by identifying transition systems that have the same 
set of traces.
Such transition systems are called trace equivalent.
Here is a precise definition.
\begin{definition}[Trace]\index{trace}
Let $T = \tup{S,A,\step{},s_0}$ be a transition system.
A \emph{trace} of $T$ is a sequence $\sigma \in \seqof{A}$ such that
$\gstep{s_0}{\sigma}{s}$ for some $s \in S$.
We write $\tr(T)$ for the set of all traces of $T$.
Then two transition systems $T$ and $T'$ are 
\emph{trace equivalent}\index{trace equivalence},
written $T \treqv T'$, if $\tr(T) = \tr(T')$.
\end{definition}

We will see below that there are also cases where not only the traces 
of the transition system matter.
In those cases, trace equivalence is obviously not the right 
equivalence to make use of.

There exist different viewpoints on what should be considered relevant
to the behaviour of processes.
The equivalence known as bisimulation equivalence is based on the idea
that not only the traces of equivalent transition systems should
coincide, but also the stages at which the choices of different 
possibilities occur.
Therefore, bisimulation equivalence is said to preserve the branching
structure of transition systems.
Here is an example of a case where apparently not only the traces 
matter, but also the stages at which the choices of different 
possibilities occur.
\begin{example}[Split connection]\index{split connection}
\label{exa-split}
We consider a split connection between nodes in a network 
(see e.g.~\cite{BMS96,Ste00}).
A split connection has one input port and two output ports.
A datum that has been consumed at the input port can be delivered at 
either of the output ports.
That is, the choice of the output ports is resolved after the datum  
has been consumed.
The behaviour of a split connection with input port $k$ and output 
ports $l$ and $m$ can be described as follows.
We assume a set of data $D$.
As states of the split connection, we have $\und$ and the data 
$d \in D$, with $\und$ as initial state.
As actions, we have $\kw{s}_i(d)$ (send $d$ at port $i$) and 
$\kw{r}_i(d)$ (receive $d$ at port $i$) for $i = k,l,m$ and $d \in D$.
As transitions, we have the following:
\begin{iteml}
\item
for each $d \in D$, a transition $\astep{\und}{\kw{r}_k(d)}{d}$; 
\item
for each $i \in \set{l,m}$ and $d \in D$, 
a transition $\astep{d}{\kw{s}_i(d)}{\und}$.
\end{iteml}
The transition system for the split connection is represented 
graphically in Fig.~\ref{fig-split} for the case where $D = \set{0,1}$.
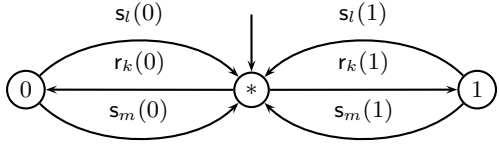
\begin{figure}
\begin{pspicture}(0,0)(7,3)

 \psset{arrows=->}

 \pnode(4,2){S}
 \rput(4,1){\circlenode{*}{$*$}}
 \rput(1,1){\circlenode{0}{$0$}}
 \rput(7,1){\circlenode{1}{$1$}}

 \ncline{S}{*}
 \ncline{*}{0}\nbput{$\kw{r}_k(0)$}
 \nccurve[angleA=45,angleB=135]{0}{*}\naput{$\kw{s}_l(0)$}
 \nccurve[angleA=315,angleB=225]{0}{*}\naput{$\kw{s}_m(0)$}
 \ncline{*}{1}\naput{$\kw{r}_k(1)$}
 \nccurve[angleA=135,angleB=45]{1}{*}\nbput{$\kw{s}_l(1)$}
 \nccurve[angleA=225,angleB=315]{1}{*}\nbput{$\kw{s}_m(1)$}

\end{pspicture}
\caption{Transition system for the split connection}
\label{fig-split}
\end{figure}
Next we consider a transition system that is trace equivalent to the
one just presented.
As states, we have the pairs $\tup{i,d}$ for $i =k,l,m$ and 
$d \in D \union \set{\und}$, with $\tup{k,\und}$ as initial 
state.
As actions, we still have $\kw{s}_i(d)$ and $\kw{r}_i(d)$ for 
$i = k,l,m$ and $d \in D$.
As transitions, we have the following:
\begin{iteml}
\item
for each $i \in \set{l,m}$ and $d \in D$: 
$\astep{\tup{k,\und}}{\kw{r}_k(d)}{\tup{i,d}}$,
$\astep{\tup{i,d}}{\kw{s}_i(d)}{\tup{k,\und}}$.
\end{iteml}
This transition system is represented graphically in 
Fig.~\ref{fig-split-like} for the case where $D = \set{0,1}$.
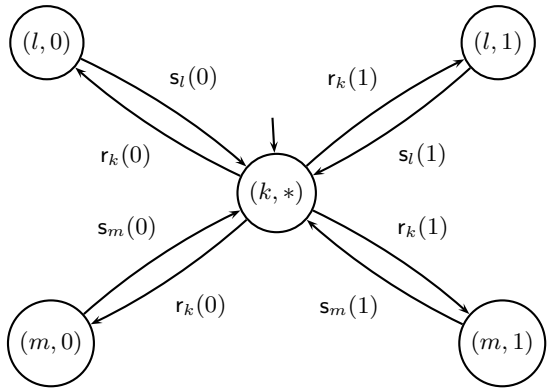
\begin{figure}
\begin{pspicture}(0,0)(7,6)

 \psset{arrows=->}

 \pnode(4,4){S}
 \rput(4,3){\circlenode{k*}{$(k,*)$}}
 \rput(1,5){\circlenode{l0}{$(l,0)$}}
 \rput(7,5){\circlenode{l1}{$(l,1)$}}
 \rput(1,1){\circlenode{m0}{$(m,0)$}}
 \rput(7,1){\circlenode{m1}{$(m,1)$}}

 \ncline{S}{k*}
 \ncarc{k*}{l0}\naput{$\kw{r}_k(0)$}
 \ncarc{k*}{m0}\naput{$\kw{r}_k(0)$}
 \ncarc{l0}{k*}\naput{$\kw{s}_l(0)$}
 \ncarc{m0}{k*}\naput{$\kw{s}_m(0)$}
 \ncarc{k*}{l1}\naput{$\kw{r}_k(1)$}
 \ncarc{k*}{m1}\naput{$\kw{r}_k(1)$}
 \ncarc{l1}{k*}\naput{$\kw{s}_l(1)$}
 \ncarc{m1}{k*}\naput{$\kw{s}_m(1)$}

\end{pspicture}
\caption{Transition system for the split-like connection}
\label{fig-split-like}
\end{figure}
This transition system does not describe the intended behaviour of the
split connection correctly.
A datum that has been consumed cannot be delivered at either of the 
output ports because the choice of the output ports is resolved at the 
instant that the datum is consumed.
So, we do not want to identify this transition system with the previous
one.
They are not identified by bisimulation equivalence.
\end{example}

What is exactly meant by ``the stages at which the choices of different 
possibilities occur'' in our intuitive explanation of bisimulation
equivalence becomes clear in the following informal definition.
Two transition systems $T$ and $T'$ are bisimulation equivalent if 
their states can be related such that: 
\begin{iteml}
\item
the initial states are related; 
\item
if states $s_1$ and $s_1'$ are related and in $T$ a transition with 
label $a$ is possible from $s_1$ to some $s_2$, then in $T'$ a 
transition with label $a$ is possible from $s_1'$ to some $s_2'$ such 
that $s_2$ and $s_2'$ are related;
\item
likewise, with the role of $T$ and $T'$ reversed.
\end{iteml}
This means that, starting from any pair of related states, $T$ can
simulate $T'$ and conversely $T'$ can simulate $T$.

Bisimulation equivalence can also be characterized as follows:
it identifies transition systems if they cannot be distinguished by any 
conceivable experiment with an experimenter that is only able to detect 
which actions are performed at any stage.
The kind of identifications made by bisimulation equivalence is 
illustrated with the following example.    
\begin{example}[Merge connection]\index{merge connection}
\label{exa-merge}
We consider a merge connection between nodes in a network 
(see e.g.~\cite{BMS96,Ste00}).
A merge connection has two input ports and one output port.
Each datum that has been consumes at one of the input ports is 
delivered at the output port.
The behaviour of a merge connection with input ports $k$ and $l$ and 
output port $m$ can be described as follows.
We assume a set of data $D$.
As states, we have the pairs $\tup{i,d}$ for $i =k,l,m$ and 
$d \in D \union \set{\und}$, with $\tup{m,\und}$ as initial 
state.
As actions, we have again $\kw{s}_i(d)$ and $\kw{r}_i(d)$ for 
$i = k,l,m$ and $d \in D$.
As transitions, we have the following:
\begin{iteml}
\item
for each $i \in \set{k,l}$ and $d \in D$: 
$\astep{\tup{m,\und}}{\kw{r}_i(d)}{\tup{i,d}}$,
$\astep{\tup{i,d}}{\kw{s}_m(d)}{\tup{m,\und}}$.
\end{iteml}
This transition system for the merge connection is represented 
graphically in Fig.~\ref{fig-merge} for the case where $D = \set{0,1}$.
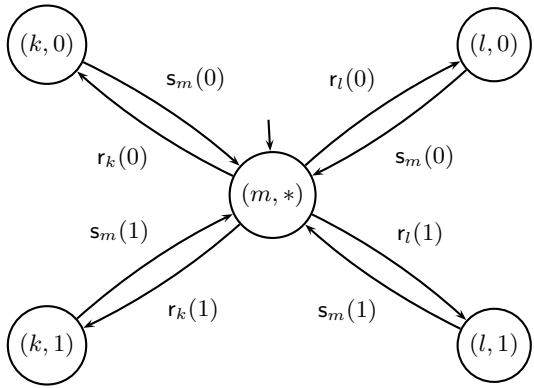
\begin{figure}
\begin{pspicture}(0,0)(7,6)

 \psset{arrows=->}

 \pnode(4,4){S}
 \rput(4,3){\circlenode{m*}{$(m,*)$}}
 \rput(1,5){\circlenode{k0}{$(k,0)$}}
 \rput(7,5){\circlenode{l0}{$(l,0)$}}
 \rput(1,1){\circlenode{k1}{$(k,1)$}}
 \rput(7,1){\circlenode{l1}{$(l,1)$}}

 \ncline{S}{m*}
 \ncarc{m*}{k0}\naput{$\kw{r}_k(0)$}
 \ncarc{m*}{l0}\naput{$\kw{r}_l(0)$}
 \ncarc{k0}{m*}\naput{$\kw{s}_m(0)$}
 \ncarc{l0}{m*}\naput{$\kw{s}_m(0)$}
 \ncarc{m*}{k1}\naput{$\kw{r}_k(1)$}
 \ncarc{m*}{l1}\naput{$\kw{r}_l(1)$}
 \ncarc{k1}{m*}\naput{$\kw{s}_m(1)$}
 \ncarc{l1}{m*}\naput{$\kw{s}_m(1)$}

\end{pspicture}
\caption{Transition system for the merge connection}
\label{fig-merge}
\end{figure}
Next we consider the following transition system.
As states, we have $\und$ and the data 
$d \in D$, with $\und$ as initial state.
As actions, we still have $\kw{s}_i(d)$ and $\kw{r}_i(d)$ for 
$i = k,l,m$ and $d \in D$.
As transitions, we have the following:
\begin{iteml}
\item
for each $i \in \set{k,l}$ and $d \in D$,  
a transition $\astep{\und}{\kw{r}_i(d)}{d}$;
\item 
for each $d \in D$, a transition $\astep{d}{\kw{s}_m(d)}{\und}$.
\end{iteml}
This transition system is represented graphically in 
Fig.~\ref{fig-merge-alt} for the case where $D = \set{0,1}$.
\begin{figure}
\begin{pspicture}(0,0)(7,3)

 \psset{arrows=->}

 \pnode(4,2){S}
 \rput(4,1){\circlenode{*}{$*$}}
 \rput(1,1){\circlenode{0}{$0$}}
 \rput(7,1){\circlenode{1}{$1$}}

 \ncline{S}{*}
 \ncline{0}{*}\naput{$\kw{s}_m(0)$}
 \nccurve[angleA=45,angleB=135]{*}{1}\naput{$\kw{r}_k(1)$}
 \nccurve[angleA=315,angleB=225]{*}{1}\naput{$\kw{r}_l(1)$}
 \ncline{1}{*}\nbput{$\kw{s}_m(1)$}
 \nccurve[angleA=135,angleB=45]{*}{0}\nbput{$\kw{r}_k(0)$}
 \nccurve[angleA=225,angleB=315]{*}{0}\nbput{$\kw{r}_l(0)$}

\end{pspicture}
\caption{Another transition system for the merge connection}
\label{fig-merge-alt}
\end{figure}
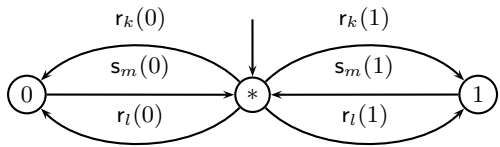
This transition system describes the intended behaviour of the merge 
connection correctly as well.
Is this transition system identified with the previous one by 
bisimulation equivalence?
Yes, it is: relate state $\tup{m,\und}$ to state $\und$ and, for 
each $i \in \set{k,l}$ and $d \in D$, state $\tup{i,d}$ to state $d$.
\end{example}

Let us now give the formal definition of bisimulation equivalence.
\begin{definition}[Bisimulation]\index{bisimulation} 
Let $T = \tup{S,A,\step{},s_0}$ 
and $T' = \tup{S',A',\linebreak[2]\step{}',\linebreak[2]s_0'}$ 
be transition systems such that $A = A'$.
Then a \emph{bisimulation} $B$ between $T$ and $T'$ is a binary 
relation $B \subseteq S \x S'$ such that the following conditions hold:
\begin{enuml}
\item
$B(s_0,s_0')$;
\item
whenever $B(s_1,s_1')$ and $\astep{s_1}{a}{s_2}$, then  there is a 
state $s_2'$ such that $\astepp{s_1'}{a}{s_2'}$ and $B(s_2,s_2')$;
\item
whenever $B(s_1,s_1')$ and $\astepp{s_1'}{a}{s_2'}$, then there is a 
state $s_2$ such that $\astep{s_1}{a}{s_2}$ and $B(s_2,s_2')$.
\end{enuml}
The two transition systems $T$ and $T'$ are 
\emph{bisimulation equivalent}\index{bisimulation equivalence}, written 
$T \bisim T'$, if there exists a bisimulation $B$ between $T$ and $T'$.
A bisimulation between $T$ and $T$ is called an 
\emph{autobisimulation}\index{autobisimulation} on $T$.
\end{definition}
Restriction to relations $B$ between the reachable states of $T$ and 
the reachable states of $T'$ does not change the notion of bisimulation 
equivalence.

Let us return to the experimenter that is only able to detect 
which actions are performed at any stage.
If performing the same experiment on a system more than once leads to 
the same outcome for all his (or her) experiments, the system behaves 
predictably.
Such a system is called determinate.
This is an important notion in the design of a system.
In many case, we have to arrive at a determinate system from components
of which some are not determinate.
This is, for example, the case with the simple data communication 
protocol treated in the next chapter.
Here is the precise definition of determinacy.
\begin{definition}[Determinacy]\index{transition system!determinacy of} 
Let $T = \tup{S,A,\step{},s_0}$ be a transition system.
Then  $T$ is \emph{determinate} if the following condition holds:
\begin{enuml}
\item[]
whenever $\gstep{s_0}{\sigma}{s}$ and $\gstep{s_0}{\sigma}{s'}$, then
there is an autobisimulation $B$ on $T$ such that $B(s,s')$.
\end{enuml}
\end{definition}
For determinate transition systems trace equivalence and bisimulation 
equivalence coincide.
\begin{property}[Determinacy]\index{transition system!determinacy of} 
Let $T = \tup{S,A,\step{},s_0}$ 
and $T' = \tup{S',A',\linebreak[2]\step{}',\linebreak[2]s_0'}$ 
be transition systems such that $A = A'$.
Then the following holds:
\begin{enuml}
\item[]
if $T$ and $T'$ are determinate, then 
$T \bisim T'$ if and only if $T \treqv T'$.
\end{enuml}
\end{property}
The notion of determinism of a transition system is closely related
to the notion of determinacy of a transition system.
\begin{definition}[Determinism]\index{transition system!determinism of} 
Let $T = \tup{S,A,\step{},s_0}$ be a transition system.
Then  $T$ is \emph{deterministic} if the following condition holds:
\begin{enuml}
\item[]
whenever $\gstep{s_0}{\sigma}{s}$ and $\gstep{s_0}{\sigma}{s'}$, then
$s = s'$.
\end{enuml}
\end{definition}
It is easy to see that all deterministic transition systems are 
determinate, but not all determinate transition systems are 
deterministic.
One could say that a determinate transition system is deterministic up
to bisimulation.

In this section, we have shortly introduced the use of equivalences for 
abstraction from details of transition systems that we want to ignore.
This plays a prominent part in techniques for the analysis of process
behaviour.
We will come back to trace and bisimulation equivalence later.


\chapter{Concurrency and Interaction}
\label{ch-interaction}

Complex systems are generally composed of a number of components that
act concurrently and interact with each other.
This chapter deals with the issue of concurrency and interaction
by introducing the notion of parallel composition of transition 
systems.
First of all, we explain informally what parallel composition of 
transition systems is and give a simple example of its use in 
describing process behaviour (Sect.~\ref{sect-interaction-informal}).
After that, we define the notion of parallel composition of transition 
systems in a mathematically precise way 
(Sect.~\ref{sect-interaction-formal}).
For a better understanding, we next investigate the connections between 
the notion of parallel composition of transition systems and the more
familiar notion of parallel execution of programs
(Sect.~\ref{sect-interaction-conn-programs}).
We also describe a typical example of a real-life system composed of
components that act concurrently and interact with each other, viz.\ 
a simple data communication protocol, using parallel composition of
transition systems (Sect.~\ref{sect-interaction-abp}).
For the interested reader, we relate the notion of parallel composition
of transition systems with the notion of parallel composition of nets
(Sect.~\ref{sect-interaction-conn-nets}).
Finally, we have another look at trace equivalence and bisimulation 
equivalence (Sect.~\ref{sect-interaction-eqv}).

\section{Informal explanation}
\label{sect-interaction-informal}

Sending a message to another component and receiving a message from 
another component are typical examples of the kinds of actions that are 
performed by a component of a system in order to interact with other 
components that act concurrently.
Synchronous communication of a message between two components is a 
typical example of an interaction that takes place when a send action 
of one component and a matching receive action of the other component 
are performed synchronously.
When two actions are performed synchronously, those actions cannot be 
observed separately. 
Therefore, the intuition is that only one action is left when two 
actions are performed synchronously.
For instance, when a send action and a matching receive action are 
performed synchronously, only a communication action can be observed.
It does not have to be the case that any two actions can be performed
synchronously.
Usually, two action can be performed synchronously only if they can 
establish an interaction.
That is, for example, not the case for two send actions.

Now consider the use of transition systems in describing the behaviour
of systems.
In the case where a system is composed of components that act
concurrently and interact which each other, we would like to reflect 
the composition in the description of the behaviour of the system.
That is, we would like to use transition systems to describe the 
behaviour of the components and to be able to describe the behaviour of
the whole system by expressing that its transition system is obtained 
from the transition systems describing the behaviour of the components 
by applying a certain operation to those transition systems. 
Parallel composition of transition systems as introduced in this 
chapter serves this purpose.
The intuition is that the parallel composition of two transition 
systems $T$ and $T'$ can perform at each stage any action that
$T$ can perform next, any action that $T'$ can perform next, and any 
action that results from synchronously performing an action that $T$
can perform next and an action that $T'$ can perform next.
Parallel composition does not prevent actions that can be performed
synchronously from being performed on their own.
In order to prevent certain actions from being performed on their own,
we introduce a seperate operation on transition systems, called
encapsulation.  The reason why parallel composition and encapsulation
are not combined in a single operation will be explained later at the 
end of Sect.~\ref{sect-interaction-formal}.
Here is an example of the use of parallel composition and encapsulation
in describing the behaviour of systems composed of components that act 
concurrently and interact which each other.
\begin{example}[Bounded buffers]\index{buffer!bounded}
\label{exa-bbuffer-parallel}
We consider the system composed of two bounded buffers, buffer $1$ and
buffer $2$, where each datum removed from the data kept in buffer $1$ 
is simultaneously added to the data kept in buffer $2$.
In this way, data from buffer $1$ is transferred to buffer $2$.
We start from the bounded buffers from Example~\ref{exa-bbuffer}.
In the case of buffer $1$, we rename the actions $\kw{add}(d)$ and 
$\kw{rem}(d)$ into $\kw{add}_1(d)$ and $\kw{rem}_1(d)$, respectively.
In the case of buffer $2$, we rename the actions $\kw{add}(d)$ and 
$\kw{rem}(d)$ into $\kw{add}_2(d)$ and $\kw{rem}_2(d)$, respectively.
In this way, we can distinguish between the action of adding a datum 
to the data kept in one buffer and the action of adding the same datum 
to the data kept in the other buffer, as well as between the action of 
removing a datum from the data kept in one buffer and the action of 
removing the same datum from the data kept in the other buffer.

The renamings yield the following.
As states of bounded buffer $i$, $i = 1,2$, with capacity $l_i$, we 
have the sequences of data of which the length is not greater than 
$l_i$.
As initial state, we have the empty sequence.
As actions, we have $\kw{add}_i(d)$ and $\kw{rem}_i(d)$ for each datum 
$d$.
As transitions of bounded buffer $i$, we have the following:
\begin{iteml}
\item
for each datum $d$ and each state $\sigma$ that has a length less than
$l_i$, a transition $\astep{\sigma}{\kw{add}_i(d)}{d\, \sigma}$; 
\item
for each datum $d$ and each state $\sigma\, d$, a transition 
$\astep{\sigma\, d}{\kw{rem}_i(d)}{\sigma}$.
\end{iteml}
In the case where, for each datum $d$, the actions $\kw{rem}_1(d)$ and 
$\kw{add}_2(d)$ can be performed synchronously, and $\kw{trf}(d)$ 
(transfer $d$) is the action left when these actions are performed 
synchronously, parallel composition of buffer $1$ and buffer $2$ 
results in the following transition system.
As states, we have pairs $\tup{\sigma_1,\sigma_2}$ where $\sigma_i$
($i = 1,2$) is a sequence of data of which the length is not greater 
than $l_i$. 
State $\tup{\sigma_1,\sigma_2}$ is the state in which the sequence of 
data $\sigma_i$ ($i = 1,2$) is kept in buffer $i$.
As initial state, we have $\tup{\epsilon,\epsilon}$.
As actions, we have $\kw{add}_i(d)$, $\kw{rem}_i(d)$ and $\kw{trf}(d)$
for each datum $d$ and $i = 1,2$.
As transitions, we have the following:
\begin{iteml}
\item
for each datum $d$ and each state $\tup{\sigma_1,\sigma_2}$ with the 
length of $\sigma_1$ less than $l_1$, a transition 
$\astep{\tup{\sigma_1,\sigma_2}}{\kw{add}_1(d)}
       {\tup{d\, \sigma_1,\sigma_2}}$; 
\item
for each datum $d$ and each state $\tup{\sigma_1,\sigma_2}$ with the 
length of $\sigma_2$ less than $l_2$, a transition 
$\astep{\tup{\sigma_1,\sigma_2}}{\kw{add}_2(d)}
       {\tup{\sigma_1,d\, \sigma_2}}$; 
\item
for each datum $d$ and each state $\tup{\sigma_1\, d,\sigma_2}$, a 
transition 
$\astep{\tup{\sigma_1\, d,\sigma_2}}{\kw{rem}_1(d)}
       {\tup{\sigma_1,\sigma_2}}$;
\item
for each datum $d$ and each state $\tup{\sigma_1,\sigma_2\, d}$, a 
transition 
$\astep{\tup{\sigma_1,\sigma_2\, d}}{\kw{rem}_2(d)}
       {\tup{\sigma_1,\sigma_2}}$;
\item
for each datum $d$ and each state $\tup{\sigma_1\, d,\sigma_2}$ with 
the length of $\sigma_2$ less than $l_2$, a 
transition 
$\astep{\tup{\sigma_1\, d,\sigma_2}}{\kw{trf}(d)}
       {\tup{\sigma_1,d\, \sigma_2}}$.
\end{iteml}
This transition system is represented graphically in 
Fig.~\ref{fig-bbuffer-parallel} for the case where $l_1 = l_2 = 1$ and 
the only data involved are the natural numbers $0$ and $1$.
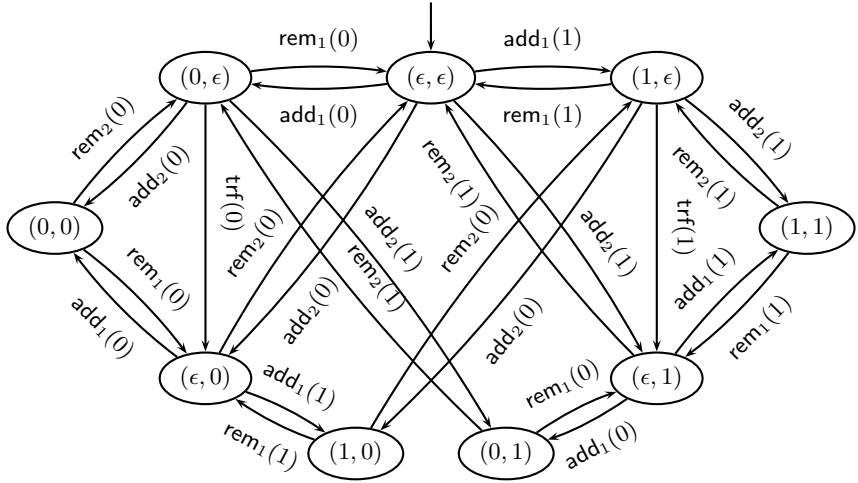
\begin{figure}
\begin{pspicture}(0,0)(12,6)

 \psset{arrows=->}

 \pnode(6,6){S}
 \rput(6,5){\ovalnode{ee}{$(\epsilon,\epsilon)$}}
 \rput(3,5){\ovalnode{0e}{$(0,\epsilon)$}}
 \rput(9,5){\ovalnode{1e}{$(1,\epsilon)$}}
 \rput(3,1){\ovalnode{e0}{$(\epsilon,0)$}}
 \rput(9,1){\ovalnode{e1}{$(\epsilon,1)$}}
 \rput(1,3){\ovalnode{00}{$(0,0)$}}
 \rput(7,0){\ovalnode{01}{$(0,1)$}}
 \rput(5,0){\ovalnode{10}{$(1,0)$}}
 \rput(11,3){\ovalnode{11}{$(1,1)$}}

 \ncline{S}{ee}

 \ncarc{ee}{0e}\naput[nrot=:180]{$\kw{add}_1(0)$}
 \ncarc{e0}{00}\naput[nrot=:180]{$\kw{add}_1(0)$}
 \ncarc{e1}{01}\naput[nrot=:180]{$\kw{add}_1(0)$}
 \ncarc{ee}{1e}\naput[nrot=:0]{$\kw{add}_1(1)$}
 \ncarc{e0}{10}\naput[nrot=:0]{$\kw{add}_1(1)$}
 \ncarc{e1}{11}\naput[nrot=:0]{$\kw{add}_1(1)$}

 \ncarc{0e}{ee}\naput[nrot=:0]{$\kw{rem}_1(0)$}
 \ncarc{00}{e0}\naput[nrot=:0]{$\kw{rem}_1(0)$}
 \ncarc{01}{e1}\naput[nrot=:0]{$\kw{rem}_1(0)$}
 \ncarc{1e}{ee}\naput[nrot=:180]{$\kw{rem}_1(1)$}
 \ncarc{10}{e0}\naput[nrot=:180]{$\kw{rem}_1(1)$}
 \ncarc{11}{e1}\naput[nrot=:180]{$\kw{rem}_1(1)$}

 \ncarc{ee}{e0}\naput[nrot=:180]{$\kw{add}_2(0)\qquad\qquad\qquad$}
 \ncarc{0e}{00}\naput[nrot=:180]{$\kw{add}_2(0)$}
 \ncarc{1e}{10}\naput[nrot=:180]{$\kw{add}_2(0)\qquad\qquad$}
 \ncarc{ee}{e1}\naput[nrot=:0]{$\qquad\qquad\kw{add}_2(1)$}
 \ncarc{0e}{01}\naput[nrot=:0]{$\kw{add}_2(1)$}
 \ncarc{1e}{11}\naput[nrot=:0]{$\kw{add}_2(1)$}

 \ncarc{e0}{ee}\naput[nrot=:0]{$\kw{rem}_2(0)\qquad\qquad$}
 \ncarc{00}{0e}\naput[nrot=:0]{$\kw{rem}_2(0)$}
 \ncarc{10}{1e}\naput[nrot=:0]{$\kw{rem}_2(0)$}
 \ncarc{e1}{ee}\naput[nrot=:180]{$\kw{rem}_2(1)\qquad\qquad\qquad\qquad$}
 \ncarc{01}{0e}\nbput[nrot=:180]{$\qquad\kw{rem}_2(1)$}
 \ncarc{11}{1e}\naput[nrot=:180]{$\kw{rem}_2(1)$}

 \ncline{0e}{e0}\naput[nrot=:0]{$\kw{trf}(0)\qquad$}
 \ncline{1e}{e1}\naput[nrot=:0]{$\kw{trf}(1)$}

\end{pspicture}
\caption{Transition system for parallel composition of bounded buffers}
\label{fig-bbuffer-parallel}
\end{figure}
For each datum $d$, actions $\kw{rem}_1(d)$ and $\kw{add}_2(d)$ can 
still be performed on their own.
Encapsulation with respect to these actions prevents them from being
performed on their own, i.e.\ it results in the following transition 
system.
We have the same states as before.
As actions, we have $\kw{add}_1(d)$, $\kw{rem}_2(d)$ and $\kw{trf}(d)$
for each datum $d$.
As transitions, we have the following:
\begin{iteml}
\item
for each datum $d$ and each state $\tup{\sigma_1,\sigma_2}$ with the 
length of $\sigma_1$ less than $l_1$, a transition 
$\astep{\tup{\sigma_1,\sigma_2}}{\kw{add}_1(d)}
       {\tup{d\, \sigma_1,\sigma_2}}$; 
\item
for each datum $d$ and each state $\tup{\sigma_1,\sigma_2\, d}$, a 
transition 
$\astep{\tup{\sigma_1,\sigma_2\, d}}{\kw{rem}_2(d)}
       {\tup{\sigma_1,\sigma_2}}$;
\item
for each datum $d$ and each state $\tup{\sigma_1\, d,\sigma_2}$ with 
the length of $\sigma_2$ less than $l_2$, a 
transition 
$\astep{\tup{\sigma_1\, d,\sigma_2}}{\kw{trf}(d)}
       {\tup{\sigma_1,d\, \sigma_2}}$.
\end{iteml}
This transition system is represented graphically in 
Fig.~\ref{fig-bbuffer-encap} for the case where $l_1 = l_2 = 1$ and 
the only data involved are the natural numbers $0$ and $1$.
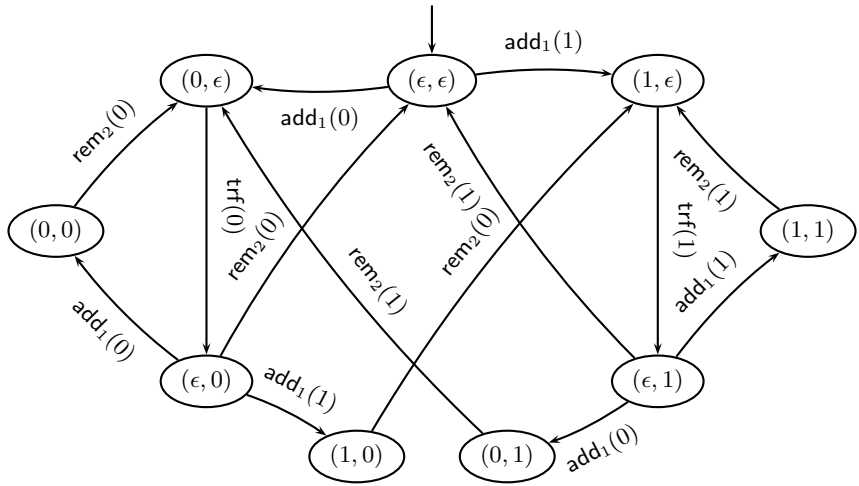
\begin{figure}
\begin{pspicture}(0,0)(12,6)

 \psset{arrows=->}

 \pnode(6,6){S}
 \rput(6,5){\ovalnode{ee}{$(\epsilon,\epsilon)$}}
 \rput(3,5){\ovalnode{0e}{$(0,\epsilon)$}}
 \rput(9,5){\ovalnode{1e}{$(1,\epsilon)$}}
 \rput(3,1){\ovalnode{e0}{$(\epsilon,0)$}}
 \rput(9,1){\ovalnode{e1}{$(\epsilon,1)$}}
 \rput(1,3){\ovalnode{00}{$(0,0)$}}
 \rput(7,0){\ovalnode{01}{$(0,1)$}}
 \rput(5,0){\ovalnode{10}{$(1,0)$}}
 \rput(11,3){\ovalnode{11}{$(1,1)$}}

 \ncline{S}{ee}

 \ncarc{ee}{0e}\naput[nrot=:180]{$\kw{add}_1(0)$}
 \ncarc{e0}{00}\naput[nrot=:180]{$\kw{add}_1(0)$}
 \ncarc{e1}{01}\naput[nrot=:180]{$\kw{add}_1(0)$}
 \ncarc{ee}{1e}\naput[nrot=:0]{$\kw{add}_1(1)$}
 \ncarc{e0}{10}\naput[nrot=:0]{$\kw{add}_1(1)$}
 \ncarc{e1}{11}\naput[nrot=:0]{$\kw{add}_1(1)$}

 \ncarc{e0}{ee}\naput[nrot=:0]{$\kw{rem}_2(0)\qquad\qquad$}
 \ncarc{00}{0e}\naput[nrot=:0]{$\kw{rem}_2(0)$}
 \ncarc{10}{1e}\naput[nrot=:0]{$\kw{rem}_2(0)$}
 \ncarc{e1}{ee}\naput[nrot=:180]{$\kw{rem}_2(1)\qquad\qquad\qquad\qquad$}
 \ncarc{01}{0e}\nbput[nrot=:180]{$\qquad\kw{rem}_2(1)$}
 \ncarc{11}{1e}\naput[nrot=:180]{$\kw{rem}_2(1)$}

 \ncline{0e}{e0}\naput[nrot=:0]{$\kw{trf}(0)\qquad$}
 \ncline{1e}{e1}\naput[nrot=:0]{$\kw{trf}(1)$}

\end{pspicture}
\caption{Transition system for encapsulation of two parallel bounded 
  buffers}
\label{fig-bbuffer-encap}
\end{figure}
So encapsulation is needed to prevent that the actions $\kw{rem}_1(d)$ 
and $\kw{add}_2(d)$ do not lead to transfer of datum $d$ from buffer 
$1$ to buffer $2$. 
The transition system obtained from the two bounded buffers by parallel 
composition and encapsulation would be bisimulation equivalent 
(see Sect.~\ref{sect-basics-bisim-eqv}) to a bounded buffer with 
capacity $l_1 + l_2$ if we could abstract from the internal transfer 
actions $\kw{trf}(d)$.
Abstraction from internal actions is one of the issues treated in the
remaining chapters of these lecture notes.
\end{example}

Although systems composed of bounded buffers that act concurrently and 
interact with each other as described above actually arise in 
computer-based systems, they are not regarded as typical examples of 
real-life computer-based systems composed of components that act 
concurrently and interact with each other.
Later, in Sect.~\ref{sect-interaction-abp}, we give a fairly typical
example, viz.\ a simple data communication protocol known as the ABP
(Alternating Bit Protocol).

\section{Formal definitions}
\label{sect-interaction-formal}

With the previous section, we have prepared the way for the formal 
definitions of the notions of parallel composition of transition 
systems and encapsulation of a transition system.

Whether two actions can be performed synchronously, and if so what 
action is left when they are performed synchronously, is mathematically
represented by a communication function.
Here is the definition of a communication function.
\begin{definition}[Communication function]\index{communication function}
\label{def-commf}
Let $A$ be a set of actions.
A \emph{communication function} on $A$ is a partial function 
$\funct{\commf}{A \x A}{A}$ satisfying for $a,b,c \in A$:
\begin{iteml}
\item
if $\commf(a,b)$ is defined, 
then $\commf(b,a)$ is defined and $\commf(a,b) = \commf(b,a)$;
\item
if $\commf(a,b)$ and $\commf(\commf(a,b),c)$ are defined,
then $\commf(b,c)$ and $\commf(a,\commf(b,c))$ are defined and
$\commf(\commf(a,b),c) = \commf(a,\commf(b,c))$.
\end{iteml}
\end{definition}
The reason for the first condition is evident: there should be no 
difference between performing $a$ and $b$ synchronously and performing 
$b$ and $a$ synchronously.
The reason for the second condition is essentially the same, but for 
the case where more than two actions can be performed synchronously.
Let us give an example to illustrate that it is straightforward to 
define the communication function needed.
\begin{example}[Bounded buffers]\index{buffer!bounded}
\label{exa-bbuffer-comm-fun}
We consider again the parallel composition of bounded buffers from
Example~\ref{exa-bbuffer-parallel}.
In that example, for each datum $d$, the actions $\kw{rem}_1(d)$ and 
$\kw{add}_2(d)$ can be performed synchronously, and $\kw{trf}(d)$ is 
the action left when these actions are performed synchronously.
This is simply represented by the communication function $\commf$ 
defined such that 
$\commf(\kw{rem}_1(d),\kw{add}_2(d)) = 
 \commf(\kw{add}_2(d),\kw{rem}_1(d)) = \kw{trf}(d)$ for each datum $d$,
and it is undefined otherwise.
\end{example}

Let us now look at the formal definitions of parallel composition and
encapsulation.
\begin{definition}[Parallel composition]\index{parallel composition}
Let $T = \tup{S,A,\step{},s_0}$ and $T' = \tup{S',A',\step{}',s'_0}$ be 
transition systems.
Let $\commf$ be a communication function on a set of actions that
includes $A \union A'$.
The \emph{parallel composition} of $T$ and $T'$ under $\commf$,
written $T \parcs{\commf} T'$, is the transition system 
$\tup{S'',A'',\step{}'',s''_0}$ where
\begin{iteml}
\item
$S'' = S \x S'$;
\item
$A'' = 
A \union A' \union 
\set{\commf(a,a') \where 
     a \in A, a' \in A', \commf(a,a') \;\mathrm{is}\;\mathrm{defined}}$;
\item
$\step{}''$ is the smallest subset of $S'' \x A'' \x S''$ such that: 
\begin{iteml}
\item
if $\astep{s_1}{a}{s_2}$ and $s' \in S'$, then
$\asteppp{\tup{s_1,s'}}{a}{\tup{s_2,s'}}$;
\item
if $\astepp{s'_1}{b}{s'_2}$ and $s \in S$, then
$\asteppp{\tup{s,s'_1}}{b}{\tup{s,s'_2}}$;
\item
if $\astep{s_1}{a}{s_2}$, $\astepp{s'_1}{b}{s'_2}$ and 
$\commf(a,b)$ is defined, then
$\asteppp{\tup{s_1,s'_1}}{\commf(a,b)}{\tup{s_2,s'_2}}$;
\end{iteml}
\item
$ s''_0 = \tup{s_0,s'_0}$.
\end{iteml}
\end{definition}
We use the convention of association to the left for parallel
composition to reduce the number of parentheses, e.g.\ we write 
$T_1 \parcs{\commf} T_2 \parcs{\commf} T'_3$ for 
$(T_1 \parcs{\commf} T_2) \parcs{\commf} T_3$.

\begin{definition}[Encapsulation]\index{encapsulation}
Let $T = \tup{S,A,\step{},s_0}$ be a transition system.
Let $H \subseteq A$.
The \emph{encapsulation} of $T$ with respect to $H$, 
written $\encap{H}(T)$, is the transition system 
$\tup{S',A',\step{}',s_0}$ where
\begin{iteml}
\item
$S' = 
 \set{s \where 
      \Exists{\sigma \in \seqof{(A \diff H)}}{\gstep{s_0}{\sigma}{s}}}$;
\item
$A' = 
 \set{a \in A \diff H \where 
      \Exists{s_1,s_2 \in S'}{\astep{s_1}{a}{s_2}}}$;
\item
$\step{}'$ is the smallest subset of $S' \x A' \x S'$ such that: 
\begin{iteml}
\item
if $\astep{s_1}{a}{s_2}$, $s_1 \in S'$ and $a \not\in H$, 
then $\astepp{s_1}{a}{s_2}$.
\end{iteml}
\end{iteml}
\end{definition}

In many applications, $\commf(\commf(a,b),c)$ is undefined for
all $a,b,c \in A$.
That case is called \emph{handshaking communication}%
\index{communication!handshaking}.
We introduce some standardized terminology and notation for handshaking
communication.
Transition systems send, receive and communicate data at \emph{ports}.
If a port is used for communication between two transition systems, it 
is called \emph{internal}.
Otherwise, it is called \emph{external}.
We write:
\begin{iteml}
\item[]
$\kw{s}_i(d)$ for the action of sending datum $d$ at port $i$;
\item[]
$\kw{r}_i(d)$ for the action of receiving datum $d$ at port $i$;
\item[]
$\kw{c}_i(d)$ for the action of communicating datum $d$ at port $i$.
\end{iteml}
Assuming a set of data $D$, the communication function is defined such
that
\begin{ldispl}
 \begin{geqns}
 \commf(\kw{s}_i(d),\kw{r}_i(d)) = \commf(\kw{r}_i(d),\kw{s}_i(d)) = 
 \kw{c}_i(d)
 \end{geqns}
\end{ldispl}
for all $d \in D$, and it is undefined otherwise.

It is important to remember that handshaking communication is just one
kind of communication.
It is not required that $\commf(\commf(a,b),c)$ is undefined for all 
$a,b,c \in A$.
Here is an example of another kind of communication.
\begin{example}[Non-handshaking communication]%
\index{communication!non-handshaking}
\label{exa-commf}
We consider a kind of communication in which three transition systems
participate.
A communication of this kind takes place by synchronously performing 
one send action and two matching receive actions.
Using a notation which is reminiscent of the standardized notation for 
handshaking communication, this ternary kind of communication can be
represented by a communication function as follows.
Assuming a set of data $D$, the communication function is defined such
that
\begin{ldispl}
 \begin{geqns}
 \commf(\kw{r}_i(d),\kw{r}_i(d)) =  
 \kw{rr}_i(d)\;,
\\
 \commf(\kw{s}_i(d),\kw{r}_i(d)) = \commf(\kw{r}_i(d),\kw{s}_i(d)) = 
 \kw{sr}_i(d)\;,
\\
 \commf(\kw{s}_i(d),\kw{rr}_i(d)) = \commf(\kw{rr}_i(d),\kw{s}_i(d)) = 
 \kw{c}_i(d)\;,
\\
 \commf(\kw{sr}_i(d),\kw{r}_i(d)) = \commf(\kw{r}_i(d),\kw{sr}_i(d)) = 
 \kw{c}_i(d)\;,
 \end{geqns}
\end{ldispl}
for all $d \in D$, and it is undefined otherwise.
The actions $\kw{sr}_i(d)$ and $\kw{rr}_i(d)$ represent the possible 
partial communications.
\end{example}
An important thing to note about the kind of communication treated
in the preceding example is the following.
If parallel composition and encapsulation were combined in a single
operation that prevents actions that can be performed synchronously
from being performed on their own, this kind of communication would be
excluded.

\section{Programs and parallel composition}
\label{sect-interaction-conn-programs}

For about thirty five years, there are programming languages in which it 
can be expressed that a number of (sequential) subprograms must be 
executed in parallel.
What exactly does that mean?
Can it be described in a straightforward way by means of transition
systems using parallel composition?
It turns out that the answers to these questions do not only depend on
whether one abstracts from the processing of actions by a machine, but
also on the way in which the programming language used supports
interaction between subprograms executed in parallel.
Roughly speaking, the basic ways of interaction are:
\begin{iteml}
\item
by synchronous communication, i.e.\ communication where the sending
subprogram must wait till each receiving subprogram (usually one) is
ready to participate in the communication;
\item
by asynchronous communication, i.e.\ communication where the sending
subprogram does not have to wait till each receiving subprogram
(usually one) is ready to participate in the communication;
\item
via shared variables, i.e.\ program variables to which more than one
subprogram has access.
\end{iteml}
Some programming languages support a combination of these basic ways.
An important thing to note is that, in virtually all programming
languages that support synchronous or asynchronous communication, the
data communicated may depend on the values of program variables. 

In this section, we will look at the questions posed above in more
detail.
We do so primarily to acquire a better understanding of the notion of
parallel composition of transition systems.
In line with Sect.~\ref{sect-basics-conn-programs}, we like to abstract
initially from how the actions performed by subprograms are processed 
by a machine.
That is, we like to focus initially on the flow of control.

Let $T_{P_1}$ and $T_{P_2}$ be transition systems describing the
behaviour of two subprograms $P_1$ and $P_2$ upon abstract execution.
If the programming language does not support synchronous communication,
then the behaviour of $P_1$ and $P_2$ upon parallel abstract execution 
can be described by
\begin{ldispl}
 \begin{geqns}
T_{P_1} \parcs{\commf} T_{P_2}\;,
 \end{geqns}
\end{ldispl}
where $\commf$ is undefined for any two actions.

In order to illustrate by an example how this works, we have to choose 
a programming language first.
Our choice is a simple extension of PASCAL introcuced by Ben-Ari back
in 1982 (see~\cite{Ben82,Ben06}).
The extension concerned simply permits to write statements of the form
\verb#COBEGIN P1; ...; Pn COEND#, where \verb#P1#, \ldots, \verb#Pn#
are procedures defined in the program, in the program body to express
that those procedures must be executed in parallel.
Moreover, assignments and tests are indivisible and nothing else is
indivisible.
That is all.
The extension does not support communication in a direct way.
Interaction is only possible via shared variables.
Let us now turn to the promised example.

\begin{example}[Peterson's protocol]\index{Peterson's protocol}
\label{exa-peterson}
We consider a program implementing a simple mutual exclusion protocol.
A mutual exclusion protocol concerns the exclusive access by components
of a system to a shared resource while using that shared resource.
As the saying is, a component is in its critical section while it is
using the shared resource.
We consider Peterson's protocol for guaranteeing that at most one
component of a system is in its critical section (see~\cite{Pet81}).
The protocol assumes that there are three shared variables \verb#c0#,
\verb#c1# and \verb#t#, with initial value $\False$, $\False$ and $0$,
respectively, and that all assignments and tests concerning these 
variables are indivisible. 

The idea behind the protocol is as follows. 
The components have sequence numbers $0$ and $1$.
The value of \verb#t# is the sequence number of the component that last
started an attempt to enter its critical section.
That the value of \verb#c0# is $\False$ signifies that component $0$
is not in its critical section; and that the value of \verb#c1# is 
$\False$ signifies that component $1$ is not in its critical section.
If component $0$ intends to enter its critical section it must assign
the value $\True$ to \verb#c0# before it checks the value of \verb#c1#,
to prevent situations in which the value of both variables is $\False$.
Analogously for component $1$.
This may lead to situations in which the value of both \verb#c0# and 
\verb#c1# is $\True$.
In order to prevent that the system becomes inactive in that case, each 
component checks whether the other last started an attempt to enter its 
critical section, and the one of which the check succeeds actually 
enters its critical section.

In the program that we will give below, we have taken the most simple
critical sections for which the mutual exclusion problem is not
trivial: a sequence of two indivisible statements.
Here is the program.
\begin{verbatim}
PROGRAM peterson;
VAR 
   c0, c1: boolean; 
   t: 0..1;
   
PROCEDURE p0;
BEGIN
   WHILE true DO
      BEGIN
         c0 := true;
         t := 0;
         REPEAT UNTIL c1 = false OR t = 1;
         enter0;  {enter critical section}
         leave0;  {leave critical section}
         c0 := false;
      END
END

PROCEDURE p1;
BEGIN
   WHILE true DO
      BEGIN
         c1 := true;
         t := 1;
         REPEAT UNTIL c0 = false OR t = 0;
         enter1;  {enter critical section}
         leave1;  {leave critical section}
         c1 := false;
      END
END

.
.
.

BEGIN
   c0 := false;
   c1 := false;
   t := 0;
   COBEGIN p1; p2 COEND
END
\end{verbatim}
Actually, \verb#enter0#, \verb#leave0#, \verb#enter1# and \verb#leave1#
are no real statements.
They stand for arbitrary indivisible statements that use the shared
resource.

The behaviour of the procedures \verb#p0# and \verb#p1# upon abstract 
execution can be described by transition systems in the same way as in 
Examples~\ref{exa-factorial} and~\ref{exa-gcd}.
As states, we have in either case the natural numbers $0$ to $7$, with
$0$ as initial state.
As actions, we have in either case an action corresponding to each
atomic statement of the procedure as well as each test of the procedure
and its opposite.
As transitions, we have the following in the case of \verb#p0#:
\begin{ldispl}
\astep{0}{\kw{true}}{1},\;
\astep{1}{\kw{c0\,:=\,true}}{2},\;
\astep{2}{\kw{t\,:=\,0}}{3}, \\
\astep{3}{\kw{NOT\,(c1\,=\,false\,OR\,t\,=\,1)}}{3},\;
\astep{3}{\kw{c1\,=\,false\,OR\,t\,=\,1}}{4}, \\
\astep{4}{\kw{enter0}}{5},\;
\astep{5}{\kw{leave0}}{6},\;
\astep{6}{\kw{c0\,:=\,false}}{0}, \\
\astep{0}{\kw{NOT\,true}}{7};
\end{ldispl}
and the following in the case of \verb#p1#:
\begin{ldispl}
\astep{0}{\kw{true}}{1},\;
\astep{1}{\kw{c1\,:=\,true}}{2},\;
\astep{2}{\kw{t\,:=\,1}}{3}, \\
\astep{3}{\kw{NOT\,(c0\,=\,false\,OR\,t\,=\,0)}}{3},\;
\astep{3}{\kw{c0\,=\,false\,OR\,t\,=\,0}}{4}, \\
\astep{4}{\kw{enter1}}{5},\;
\astep{5}{\kw{leave1}}{6},\;
\astep{6}{\kw{c1\,:=\,false}}{0}, \\
\astep{0}{\kw{NOT\,true}}{7}
\end{ldispl}
Here, $\kw{enter0}$, $\kw{leave0}$, $\kw{enter1}$ and $\kw{leave1}$ are 
no real actions.
They stand for the actions corresponding to the statements that
\verb#enter0#, \verb#leave0#, \verb#enter1# and \verb#leave1# stand 
for.

The transition systems for the procedures \verb#p0# and \verb#p1# are
represented graphically in Fig.~\ref{fig-peterson}.
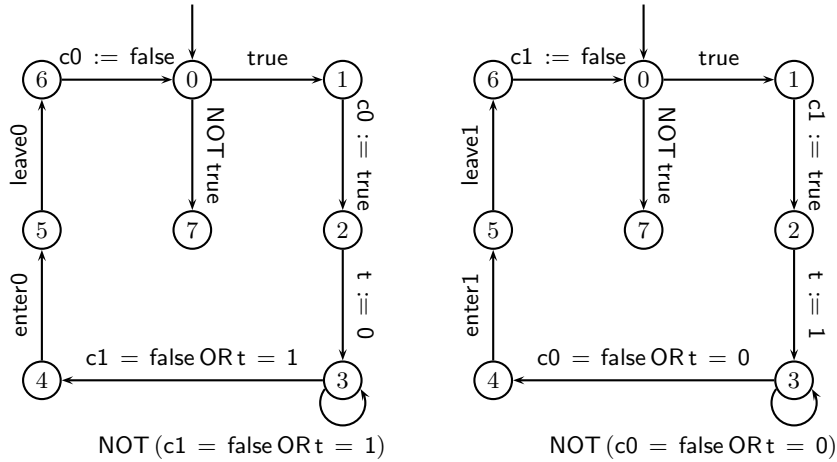
\begin{figure}
\begin{pspicture}(0,0)(12,7)

 \psset{arrows=->}

 \pnode(3,6){S}
 \rput(3,5){\circlenode{0}{$0$}}
 \rput(5,5){\circlenode{1}{$1$}}
 \rput(5,3){\circlenode{2}{$2$}}
 \rput(5,1){\circlenode{3}{$3$}}
 \rput(1,1){\circlenode{4}{$4$}}
 \rput(1,3){\circlenode{5}{$5$}}
 \rput(1,5){\circlenode{6}{$6$}}
 \rput(3,3){\circlenode{7}{$7$}}

 \ncline{S}{0}
 \ncline{0}{1}\naput[nrot=:0]{$\kw{true}$}
 \ncline{1}{2}\naput[nrot=:0]{$\kw{c0\,:=\,true}$}
 \ncline{2}{3}\naput[nrot=:0]{$\kw{t\,:=\,0}$}
 \nccircle[angleA=180]{3}{0.3}\uput[dl](5.7,0.4){$\kw{NOT\,(c1\,=\,false\,OR\,t\,=\,1)}$}
 \ncline{3}{4}\nbput[nrot=:180]{$\kw{c1\,=\,false\,OR\,t\,=\,1}$}
 \ncline{4}{5}\naput[nrot=:0]{$\kw{enter0}$}
 \ncline{5}{6}\naput[nrot=:0]{$\kw{leave0}$}
 \ncline{6}{0}\naput[nrot=:0]{$\kw{c0\,:=\,false}$}
 \ncline{0}{7}\naput[nrot=:0]{$\kw{NOT\,true}$}

 \pnode(9,6){S}
 \rput(9,5){\circlenode{0}{$0$}}
 \rput(11,5){\circlenode{1}{$1$}}
 \rput(11,3){\circlenode{2}{$2$}}
 \rput(11,1){\circlenode{3}{$3$}}
 \rput(7,1){\circlenode{4}{$4$}}
 \rput(7,3){\circlenode{5}{$5$}}
 \rput(7,5){\circlenode{6}{$6$}}
 \rput(9,3){\circlenode{7}{$7$}}

 \ncline{S}{0}
 \ncline{0}{1}\naput[nrot=:0]{$\kw{true}$}
 \ncline{1}{2}\naput[nrot=:0]{$\kw{c1\,:=\,true}$}
 \ncline{2}{3}\naput[nrot=:0]{$\kw{t\,:=\,1}$}
 \nccircle[angleA=180]{3}{0.3}\uput[dl](11.7,0.4){$\kw{NOT\,(c0\,=\,false\,OR\,t\,=\,0)}$}
 \ncline{3}{4}\nbput[nrot=:180]{$\kw{c0\,=\,false\,OR\,t\,=\,0}$}
 \ncline{4}{5}\naput[nrot=:0]{$\kw{enter1}$}
 \ncline{5}{6}\naput[nrot=:0]{$\kw{leave1}$}
 \ncline{6}{0}\naput[nrot=:0]{$\kw{c1\,:=\,false}$}
 \ncline{0}{7}\naput[nrot=:0]{$\kw{NOT\,true}$}
\end{pspicture}
\caption{Transition systems for Peterson's protocol}
\label{fig-peterson}
\end{figure}
We call these transition systems $T_{\mathtt{p0}}$ and
$T_{\mathtt{p1}}$, respectively.
The behaviour of the procedures \verb#p0# and \verb#p1# upon parallel
abstract execution can be described as follows:
\begin{ldispl}
 \begin{geqns}
  T_{\mathtt{p0}} \parcs{\commf} T_{\mathtt{p1}}
 \end{geqns}
\end{ldispl}
where the communication function $\commf$ is undefined for any two 
actions.
\end{example}
Notice that the preceding example is based on the idea that the 
parallel abstract execution of two subprograms can be reduced to 
arbitrary interleaving only, i.e.\ to performing again and again an 
action that one or the other of the two can perform next.
This is obviously problematic in the presence of synchronous
communication: simultaneously performing actions is not taken into
account.
However, if one abstracts from the processing of actions by a machine, 
there is also no alternative in the general case where the data
communicated may depend on the values of program variables. 

Let us now, like in Sect.~\ref{sect-basics-conn-programs}, take into
account how the actions performed by subprograms are processed by a
machine and turn to the behaviour of subprograms upon parallel 
execution on a machine.
We can describe the behaviour of machines on which subprograms are
executed by transition systems as well.
We will give a simple example illustrating this later.
Let $T_{P_1}$ and $T_{P_2}$ be transition systems describing the 
behaviour of two subprograms $P_1$ and $P_2$ upon abstract execution.
If we suppose that we also have the transition systems of the
appropriate machines available, the behaviour of $P_1$ and $P_2$ upon
parallel execution on a machine can in many cases best be described
in one of the following ways, depending on the way in which the
programming language used supports interaction between subprograms
executed in parallel:
\begin{ldispl}
 \begin{geqns}
\encap{H}((T_{P_1} \parcs{\commf'} T_{P_2}) \parcs{\commf} T_M)
 \end{geqns}
\end{ldispl}
or
\begin{ldispl}
 \begin{geqns}
\encap{H}
 (\encap{H_1}(T_{P_1} \parcs{\commf} T_{M_1}) \parcs{\commf} 
  \encap{H_2}(T_{P_2} \parcs{\commf} T_{M_2}))\;,
 \end{geqns}
\end{ldispl}
where the communication function $\commf'$ is undefined for any two
actions, and the communication function $\commf$, the sets of actions
$H$, $H_1$ and $H_2$, and the transition systems $T_M$, $T_{M_1}$ and
$T_{M_2}$ all depend on the way in which the programming language used
supports interaction between subprograms executed in parallel.
The transition systems $T_M$, $T_{M_1}$ and $T_{M_2}$ are supposed to
describe the behaviour of appropriate machines.

The first way of description applies if the programming language only
supports shared variables as a means to interact. 
The second way of description applies if the programming language
supports synchronous communication or asynchronous communication, but 
does not support shared variables.
Synchronous communication can be fully represented by the communication
function $\commf$, while asynchronous communication cannot be fully
represented by the communication function (as explained below).
In the case where only shared variables are supported, $P_1$ and $P_2$
are executed on the same machine: $T_M$.
In the cases where shared variables are not supported, $P_1$ and $P_2$
are executed on different machines: $T_{M_1}$ and $T_{M_2}$,
respectively.
The machines process the actions performed by the subprograms. 
In the case of asynchronous communication, they are also involved in 
the communication between subprograms.
In that case, each machine buffers the data sent to the subprogram that
the machine executes till the subprogram consumes the data.
Actually, the first way of description can be applied in the case of
asynchronous communication as well, but it is rather clumsy.

The second way of description shows that, in the case where no
abstraction from the processing of actions by a machine is made,
parallel execution of subprograms corresponds directly to 
(encapsulated) parallel composition if synchronous communication or
asynchronous communication is supported by the programming language
used, and moreover shared variables are not supported.
This makes it a compositional way of description, which has advantages 
in analysis. 
The compositionality is missing in the first way of description, which
applies if only shared variables are supported.

Here is an example that illustrates how the behaviour of machines on
which subprograms are executed can be described by transition systems.
\begin{example}[Peterson's protocol]\index{Peterson's protocol}
\label{exa-machine}
We consider again the program from Example~\ref{exa-peterson} 
concerning Peterson's mutual exclusion protocol.
The behaviour of a machine on which the procedures \verb#p0# and
\verb#p1# can be executed in parallel, after initialization of the
program variables $\kw{c0}$, $\kw{c1}$, and $\kw{t}$, is described by a
transition system as follows.
As states of the machine, we have triples $\tup{c0,c1,t}$, where 
$c0,c1 \in \Bool$ and $t \in \set{0,1}$.
These states can be viewed as follows: 
$\tup{c0,c1,t}$ is the storage that keeps the values of the program
variables $\kw{c0}$, $\kw{c1}$, and $\kw{t}$ in that order.
The initial state is $\tup{\False,\False,0}$.
As actions, we have an action corresponding to each atomic statement of
the procedures as well as each test of the procedures and its opposite.
However, these actions differ from the actions of the transition system
describing the behaviour of the procedures upon abstract execution: the
former actions are actions of processing the latter actions.
The difference is indicated by overlining the former actions.
As transitions, we have the following:
\begin{iteml}
\item
for each $c0,c1,t$:
\begin{iteml}
\item
a transition
$\astep{\tup{c0,c1,t}}{\ol{\kw{c0\,:=\,false}}}{\tup{\False,c1,t}}$, 
\item
a transition
$\astep{\tup{c0,c1,t}}{\ol{\kw{c0\,:=\,true}}}{\tup{\True,c1,t}}$, 
\item
a transition
$\astep{\tup{c0,c1,t}}{\ol{\kw{c1\,:=\,false}}}{\tup{c0,\False,t}}$, 
\item
a transition
$\astep{\tup{c0,c1,t}}{\ol{\kw{c1\,:=\,true}}}{\tup{c0,\True,t}}$, 
\item
a transition
$\astep{\tup{c0,c1,t}}{\ol{\kw{t\,:=\,0}}}{\tup{c0,c1,0}}$, 
\item
a transition
$\astep{\tup{c0,c1,t}}{\ol{\kw{t\,:=\,1}}}{\tup{c0,c1,1}}$, 
\item
a transition
$\astep{\tup{c0,c1,t}}{\ol{\kw{true}}}{\tup{c0,c1,t}}$, 
\item
a transition
$\astep{\tup{c0,c1,t}}
  {\ol{\kw{c1\,=\,false\,OR\,t\,=\,1}}}{\tup{c0,c1,t}}$
if $c1 = \False$ or $t = 1$,
\item
a transition
$\astep{\tup{c0,c1,t}}
  {\ol{\kw{NOT\,(c1\,=\,false\,OR\,t\,=\,1)}}}{\tup{c0,c1,t}}$
if $c1 \neq \False$ and $t \neq 1$,
\item
a transition
$\astep{\tup{c0,c1,t}}
  {\ol{\kw{c0\,=\,false\,OR\,t\,=\,0}}}{\tup{c0,c1,t}}$
if $c0 = \False$ or $t = 0$,
\item
a transition
$\astep{\tup{c0,c1,t}}
  {\ol{\kw{NOT\,(c0\,=\,false\,OR\,t\,=\,0)}}}{\tup{c0,c1,t}}$
if $c0 \neq \False$ and $t \neq 0$).
\end{iteml}
\end{iteml}

The transition system for the machine is
represented graphically in Fig.~\ref{fig-peterson-mach}.
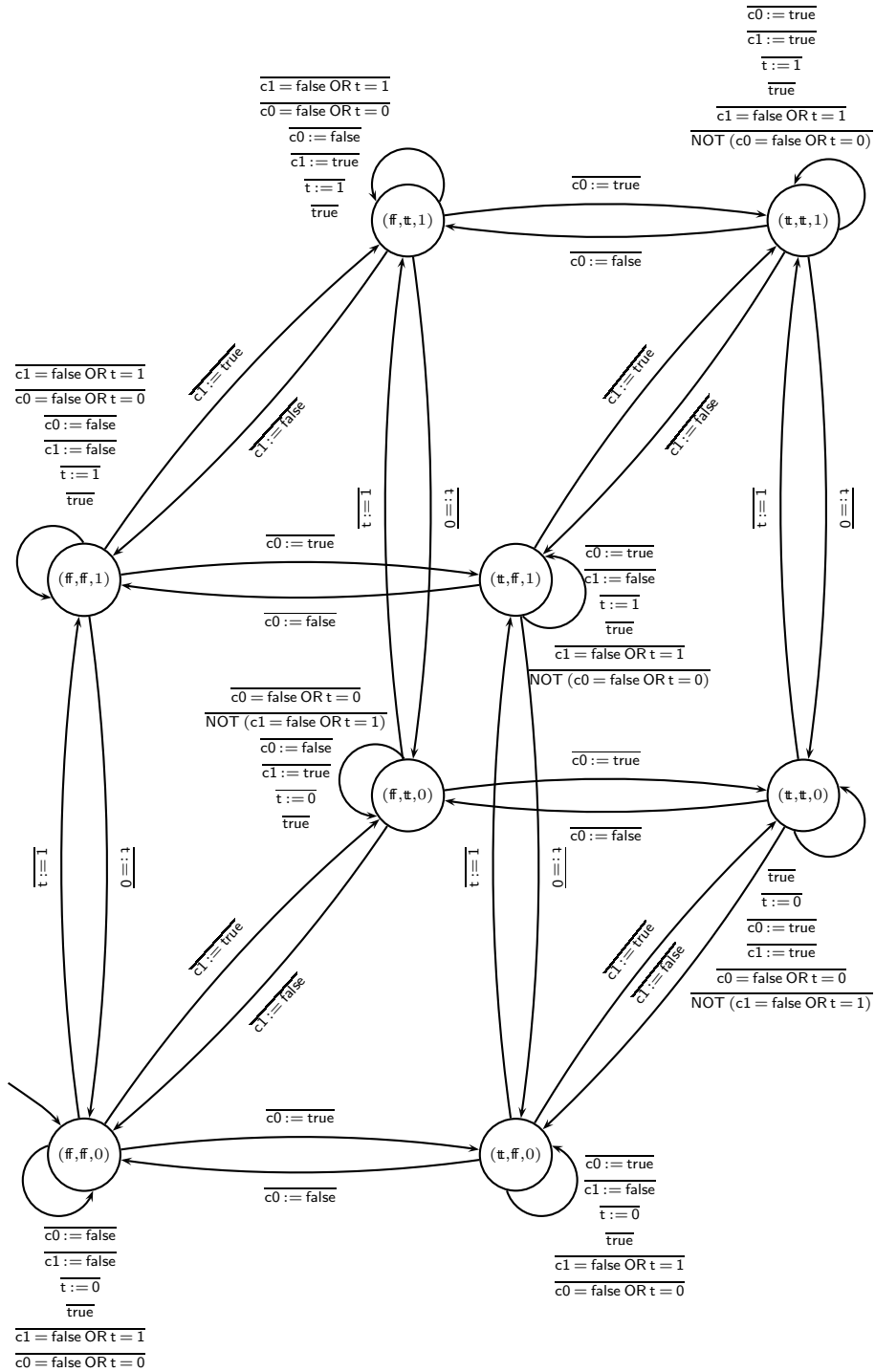
\begin{figure}
\begin{pspicture}(0,-2)(12,17)

 \psset{arrows=->}

 \pnode(0,2){S}
 \rput(1,1){\circlenode{ff0}{$\scriptstyle(\False,\False,0)$}}
 \rput(7,1){\circlenode{tf0}{$\scriptstyle(\True,\False,0)$}}
 \rput(5.5,6){\circlenode{ft0}{$\scriptstyle(\False,\True,0)$}}
 \rput(11,6){\circlenode{tt0}{$\scriptstyle(\True,\True,0)$}}

 \rput(1,9){\circlenode{ff1}{$\scriptstyle(\False,\False,1)$}}
 \rput(7,9){\circlenode{tf1}{$\scriptstyle(\True,\False,1)$}}
 \rput(5.5,14){\circlenode{ft1}{$\scriptstyle(\False,\True,1)$}}
 \rput(11,14){\circlenode{tt1}{$\scriptstyle(\True,\True,1)$}}

 \ncarc{S}{ff0}

 \ncarc{tf0}{ff0}\naput[nrot=:180]{$\scriptstyle\ol{\kw{c0\,:=\,false}}$}
 \ncarc{tf1}{ff1}\naput[nrot=:180]{$\scriptstyle\ol{\kw{c0\,:=\,false}}$}
 \ncarc{tt0}{ft0}\naput[nrot=:180]{$\scriptstyle\ol{\kw{c0\,:=\,false}}$}
 \ncarc{tt1}{ft1}\naput[nrot=:180]{$\scriptstyle\ol{\kw{c0\,:=\,false}}$}

 \ncarc{ft0}{ff0}\naput[nrot=:180]{$\scriptstyle\ol{\kw{c1\,:=\,false}}$}
 \ncarc{ft1}{ff1}\naput[nrot=:180]{$\scriptstyle\ol{\kw{c1\,:=\,false}}$}
 \ncarc{tt0}{tf0}\nbput[nrot=:180]{$\scriptstyle\ol{\kw{c1\,:=\,false}}$}
 \ncarc{tt1}{tf1}\naput[nrot=:180]{$\scriptstyle\ol{\kw{c1\,:=\,false}}$}

 \ncarc{ff0}{tf0}\naput[nrot=:0]{$\scriptstyle\ol{\kw{c0\,:=\,true}}$}
 \ncarc{ff1}{tf1}\naput[nrot=:0]{$\scriptstyle\ol{\kw{c0\,:=\,true}}$}
 \ncarc{ft0}{tt0}\naput[nrot=:0]{$\scriptstyle\ol{\kw{c0\,:=\,true}}$}
 \ncarc{ft1}{tt1}\naput[nrot=:0]{$\scriptstyle\ol{\kw{c0\,:=\,true}}$}

 \ncarc{ff0}{ft0}\naput[nrot=:0]{$\scriptstyle\ol{\kw{c1\,:=\,true}}$}
 \ncarc{ff1}{ft1}\naput[nrot=:0]{$\scriptstyle\ol{\kw{c1\,:=\,true}}$}
 \ncarc{tf0}{tt0}\naput[nrot=:0]{$\scriptstyle\ol{\kw{c1\,:=\,true}}$}
 \ncarc{tf1}{tt1}\naput[nrot=:0]{$\scriptstyle\ol{\kw{c1\,:=\,true}}$}

 \ncarc{ff1}{ff0}\naput[nrot=:0]{$\scriptstyle\ol{\kw{t\,:=\,0}}$}
 \ncarc{ft1}{ft0}\naput[nrot=:0]{$\scriptstyle\ol{\kw{t\,:=\,0}}$}
 \ncarc{tf1}{tf0}\naput[nrot=:0]{$\scriptstyle\ol{\kw{t\,:=\,0}}$}
 \ncarc{tt1}{tt0}\naput[nrot=:0]{$\scriptstyle\ol{\kw{t\,:=\,0}}$}

 \ncarc{ff0}{ff1}\naput[nrot=:0]{$\scriptstyle\ol{\kw{t\,:=\,1}}$}
 \ncarc{ft0}{ft1}\naput[nrot=:0]{$\scriptstyle\ol{\kw{t\,:=\,1}}$}
 \ncarc{tf0}{tf1}\naput[nrot=:0]{$\scriptstyle\ol{\kw{t\,:=\,1}}$}
 \ncarc{tt0}{tt1}\naput[nrot=:0]{$\scriptstyle\ol{\kw{t\,:=\,1}}$}

  \nccircle[angleA=135,angleB=315]{ff0}{.5}
  \rput(1,-1){\rnode{Lff0}{$\begin{array}{c}
  \scriptstyle\ol{\kw{c0\,:=\,false}} \\
  \scriptstyle\ol{\kw{c1\,:=\,false}} \\
  \scriptstyle\ol{\kw{t\,:=\,0}} \\
  \scriptstyle\ol{\kw{true}} \\
  \scriptstyle\ol{\kw{c1\,=\,false\,OR\,t\,=\,1}} \\
  \scriptstyle\ol{\kw{c0\,=\,false\,OR\,t\,=\,0}}
  \end{array}$}}

  \nccircle[angleA=225,angleB=45]{tf0}{.5}
  \rput(8.5,0){\rnode{Lff0}{$\begin{array}{c}
  \scriptstyle\ol{\kw{c0\,:=\,true}} \\
  \scriptstyle\ol{\kw{c1\,:=\,false}} \\
  \scriptstyle\ol{\kw{t\,:=\,0}} \\
  \scriptstyle\ol{\kw{true}} \\
  \scriptstyle\ol{\kw{c1\,=\,false\,OR\,t\,=\,1}} \\
  \scriptstyle\ol{\kw{c0\,=\,false\,OR\,t\,=\,0}}
  \end{array}$}}

  \nccircle[angleA=65]{ft0}{.5}
  \rput(4,6.5){\rnode{Lff0}{$\begin{array}{c}
  \scriptstyle\ol{\kw{c0\,=\,false\,OR\,t\,=\,0}} \\
  \scriptstyle\ol{\kw{NOT\,(c1\,=\,false\,OR\,t\,=\,1)}} \\
  \scriptstyle\ol{\kw{c0\,:=\,false}} \\
  \scriptstyle\ol{\kw{c1\,:=\,true}} \\
  \scriptstyle\ol{\kw{t\,:=\,0}} \\
  \scriptstyle\ol{\kw{true}}
  \end{array}$}}

  \nccircle[angleA=225,angleB=45]{tt0}{.5}
  \rput(10.75,4){\rnode{Lff0}{$\begin{array}{c}
  \scriptstyle\ol{\kw{true}} \\
  \scriptstyle\ol{\kw{t\,:=\,0}} \\
  \scriptstyle\ol{\kw{c0\,:=\,true}} \\
  \scriptstyle\ol{\kw{c1\,:=\,true}} \\
  \scriptstyle\ol{\kw{c0\,=\,false\,OR\,t\,=\,0}} \\
  \scriptstyle\ol{\kw{NOT\,(c1\,=\,false\,OR\,t\,=\,1)}}
  \end{array}$}}

  \nccircle[angleA=60]{ff1}{.5}
  \rput(1,11){\rnode{Lff0}{$\begin{array}{c}
  \scriptstyle\ol{\kw{c1\,=\,false\,OR\,t\,=\,1}} \\
  \scriptstyle\ol{\kw{c0\,=\,false\,OR\,t\,=\,0}} \\
  \scriptstyle\ol{\kw{c0\,:=\,false}} \\
  \scriptstyle\ol{\kw{c1\,:=\,false}} \\
  \scriptstyle\ol{\kw{t\,:=\,1}} \\
  \scriptstyle\ol{\kw{true}}
  \end{array}$}}

  \nccircle[angleA=250]{tf1}{.5}
  \rput(8.5,8.5){\rnode{Lff0}{$\begin{array}{c}
  \scriptstyle\ol{\kw{c0\,:=\,true}} \\
  \scriptstyle\ol{\kw{c1\,:=\,false}} \\
  \scriptstyle\ol{\kw{t\,:=\,1}} \\
  \scriptstyle\ol{\kw{true}} \\
  \scriptstyle\ol{\kw{c1\,=\,false\,OR\,t\,=\,1}} \\
  \scriptstyle\ol{\kw{NOT\,(c0\,=\,false\,OR\,t\,=\,0)}}
  \end{array}$}}

  \nccircle[angleA=0]{ft1}{.5}
  \rput(4.4,15){\rnode{Lff0}{$\begin{array}{c}
  \scriptstyle\ol{\kw{c1\,=\,false\,OR\,t\,=\,1}} \\
  \scriptstyle\ol{\kw{c0\,=\,false\,OR\,t\,=\,0}} \\
  \scriptstyle\ol{\kw{c0\,:=\,false}} \\
  \scriptstyle\ol{\kw{c1\,:=\,true}} \\
  \scriptstyle\ol{\kw{t\,:=\,1}} \\
  \scriptstyle\ol{\kw{true}}
  \end{array}$}}

  \nccircle[angleA=315,angleB=45]{tt1}{.5}
  \rput(10.75,16){\rnode{Lff0}{$\begin{array}{c}
  \scriptstyle\ol{\kw{c0\,:=\,true}} \\
  \scriptstyle\ol{\kw{c1\,:=\,true}} \\
  \scriptstyle\ol{\kw{t\,:=\,1}} \\
  \scriptstyle\ol{\kw{true}} \\
  \scriptstyle\ol{\kw{c1\,=\,false\,OR\,t\,=\,1}} \\
  \scriptstyle\ol{\kw{NOT\,(c0\,=\,false\,OR\,t\,=\,0)}}
  \end{array}$}}

\end{pspicture}
\caption{Transition system for the machine executing Peterson's
  protocol}
\label{fig-peterson-mach}
\end{figure}
We call this transition system $T_M$.
The behaviour of the procedures \verb#p0# and \verb#p1# upon parallel
execution on a machine can now be described as follows:
\begin{ldispl}
 \begin{geqns}
\encap{H}
 ((T_\mathtt{p0} \parcs{\commf'} T_\mathtt{p1}) \parcs{\commf} T_M)
 \end{geqns}
\end{ldispl}
where
\begin{ldispl}
\displstretch
 \begin{aeqns}
 H & = & 
 \acts(T_\mathtt{p0}) \union \acts(T_\mathtt{p1}) \union \acts(T_M)\;,
 \end{aeqns}
\end{ldispl}
the communication function $\commf'$ is undefined for any two actions,
and the communication function $\commf$ is defined such that
\begin{ldispl}
 \begin{geqns}
 \commf(a,\ol{a}) = \commf(\ol{a},a) = a^*
 \end{geqns}
\end{ldispl}
for all actions 
$a \in \acts(T_\mathtt{p0}) \union \acts(T_\mathtt{p1})$, 
and it is undefined otherwise.

Notice that most procedures, written in the same programming language  
as \verb#p0# and \verb#p1#, cannot be executed on the machine of which 
the behaviour is described by the transition system $T_M$ presented 
above.
This machine can only deal with actions that can possibly be performed 
by the procedures \verb#p0# and \verb#p1#.
However, because all actions of the machine are prevented from being 
performed on their own, $T_M$ can safely be replaced by a transition 
system for a machine that can also deal with actions that can possibly 
be performed by other procedures.
\end{example}

\section{Example: Alternating bit protocol}%
\index{alternating bit protocol}
\label{sect-interaction-abp}

Here is a fairly typical example of the use of parallel composition and
encapsulation in describing the behaviour of systems composed of
components that act concurrently and interact which each other.
The example concerns the ABP (Alternating Bit Protocol), a data 
communication protocol first introduced in~\cite{BSW69}.

\begin{figure}
\sidecaption
\setlength{\unitlength}{.15em}
\begin{picture}(150,41.5)(7,-0.5)
\put(10,20){\line(1,0){10}}
\put(10,25){\makebox(0,0){1}}
\put(30,20){\circle{20}}
\put(30,20){\makebox(0,0){S}}
\put(38,25){\line(2,1){22}}
\put(48,35){\makebox(0,0){3}}
\put(75,36){\oval(30,10)}
\put(75,36){\makebox(0,0){K}}
\put(90,36){\line(2,-1){22}}
\put(102,35){\makebox(0,0){4}}
\put(38,15){\line(2,-1){22}}
\put(48,5){\makebox(0,0){5}}
\put(75,4){\oval(30,10)}
\put(75,4){\makebox(0,0){L}}
\put(90,4){\line(2,1){22}}
\put(102,5){\makebox(0,0){6}}
\put(120,20){\circle{20}}
\put(120,20){\makebox(0,0){R}}
\put(130,20){\line(1,0){10}}
\put(140,25){\makebox(0,0){2}}
\end{picture}
\caption{Configuration of the ABP}
\label{fig-abp-conf}
\end{figure}
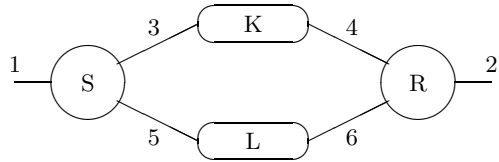

The ABP is a simple data communication protocol based on positive and 
negative acknowledgements.
Data are labeled with an alternating bit from $B = \set{0,1}$.
The sender either transmits a new datum or retransmits the most recent
datum depending on an acknowledgement represented by a bit.
The alternating bit used with the most recent datum is considered to be
a positive acknowledgement.
The configuration of the ABP is shown in Fig.~\ref{fig-abp-conf}.
We have a sender process $S$, a receiver process $R$ and two channels
$K$ and $L$.
The process $S$ waits until a datum $d$ is offered at an external port
(port~$1$).
When a datum is offered at this port, $S$ consumes it, packs it with an
alternating bit $b$ in a frame $\tup{d,b}$, and then delivers the frame
at an internal port used for sending (port~$3$).
Next, $S$ waits until a bit $b'$ is offered at an internal port used
for receiving (port~$5$).
When a bit is offered and it is the alternating bit $b$, $S$ goes back 
to waiting for a datum.
When a bit is offered and it is not the alternating bit $b$, $S$
delivers the same frame again and goes back to waiting for a bit.
The process $S$ behaves the same when an error value is offered
instead of a bit.
The process $R$ waits until a frame with a datum and an alternating bit
$\tup{d,b}$ is offered at an internal port used for receiving
(port~$4$).
When a frame is offered at this port, $R$ consumes it, unpacks it, and
then delivers the datum $d$ at an external port (port~$2$) if the
alternating bit $b$ is the right one and in any case the alternating
bit $b$ at an internal port for sending (port~$6$).
When instead an error value is offered, $R$ delivers the wrong bit.
After that, $R$ goes back to waiting for a frame, but the right bit
changes if the alternating bit was the right one.
The processes $K$ and $L$ pass on frames from an internal port of $S$
to an internal port of $R$ and bits from an internal port of $R$
to an internal port of $S$, respectively.
The processes $K$ and $L$ may corrupt frames and acknowledgements,
respectively.
In the case where this happens, $K$ and $L$ deliver an error value.

We assume a set of data $D$.
Let $F = D \x B$ be the set of frames.
For $d \in D$ and $b \in B$, we write $d,b$ for the frame $\tup{d,b}$.
For $b \in B$, we write $\ol{b}$ for the bit $1-b$.
We use the standardized notation for handshaking communication
introduced in Sect.~\ref{sect-interaction-formal}.

\begin{figure}
\begin{pspicture}(0,0)(7,5)

 \psset{arrows=->}

 \pnode(4,5){S}
 \rput(4,4){\ovalnode{*00}{$(*,0,0)$}}
 \rput(7,4){\ovalnode{d01}{$(\rd,0,1)$}}
 \rput(7,1){\ovalnode{d02}{$(\rd,0,2)$}}
 \rput(4,1){\ovalnode{*10}{$(*,1,0)$}}
 \rput(1,1){\ovalnode{d11}{$(\rd,1,1)$}}
 \rput(1,4){\ovalnode{d12}{$(\rd,1,2)$}}

 \ncline{S}{*00}
 \ncline{*00}{d01}\naput[nrot=:0]{$\kw{r}_1(\rd)$}
 \ncarc{d01}{d02}\naput[nrot=:0]{$\kw{s}_3(\rd,0)$}
 \ncarc{d02}{d01}\naput[nrot=:0]{$\begin{array}{l}
                                  \kw{r}_5(1) \\
                                  \kw{r}_5(*)
                                  \end{array}$}
 \ncline{d02}{*10}\naput[nrot=:180]{$\kw{r}_5(0)$}
 \ncline{*10}{d11}\naput[nrot=:180]{$\kw{r}_1(\rd)$}
 \ncarc{d11}{d12}\naput[nrot=:0]{$\kw{s}_3(\rd,1)$}
 \ncarc{d12}{d11}\naput[nrot=:0]{$\begin{array}{c}
                                  \kw{r}_5(0) \\
                                  \kw{r}_5(*)
                                  \end{array}$}
 \ncline{d12}{*00}\naput[nrot=:0]{$\kw{r}_5(1)$}

\end{pspicture}
\caption{Transition system for the sender}
\label{fig-ABP-sender}
\end{figure}
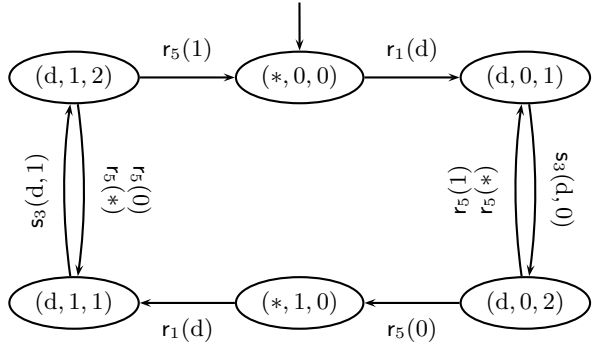

The behaviour of the sender $S$ is described by a transition system 
as follows.
As states of the sender, we have triples $\tup{d,b,i}$, where 
$d \in D \union \set{\und}$, $b \in B$ and $i \in \set{0,1,2}$,
satisfying $d = \und$ if and only if $i = 0$.
State $\tup{d,b,i}$ is roughly a state in which the datum being passed
on from the sender to the receiver is $d$ and the alternating bit is 
$b$.
If $d = \und$, no such datum is available.
The initial state is $\tup{\und,0,0}$.
As actions, we have 
$\kw{r}_1(d)$ for each $d \in D$,
$\kw{s}_3(f)$ for each $f \in F$, and
$\kw{r}_5(b)$ for each $b \in B \union \set{\und}$.
As transitions of the sender, we have the following:
\begin{iteml}
\item
for each datum $d \in D$ and bit $b \in B$:
\begin{iteml}
\item
a transition $\astep{\tup{\und,b,0}}{\kw{r}_1(d)}{\tup{d,b,1}}$,
\item
a transition $\astep{\tup{d,b,1}}{\kw{s}_3(d,b)}{\tup{d,b,2}}$,
\item
a transition $\astep{\tup{d,b,2}}{\kw{r}_5(b)}{\tup{\und,\ol{b},0}}$,
\item
a transition $\astep{\tup{d,b,2}}{\kw{r}_5(\ol{b})}{\tup{d,b,1}}$,
\item
a transition $\astep{\tup{d,b,2}}{\kw{r}_5(\und)}{\tup{d,b,1}}$.
\end{iteml}
\end{iteml}
The transition system for the sender is represented graphically in 
Fig.~\ref{fig-ABP-sender} for the case where only one datum, say $d$, 
is involved.

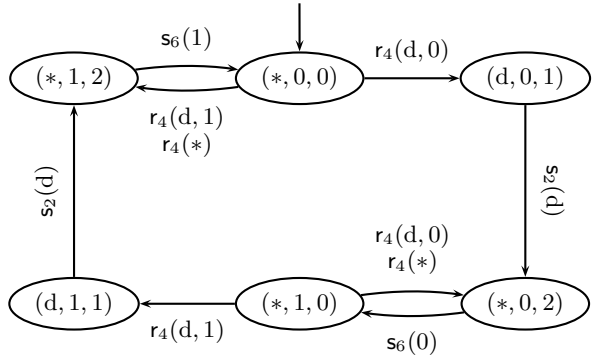
\begin{figure}
\begin{pspicture}(0,0)(7,5)

 \psset{arrows=->}

 \pnode(4,5){S}
 \rput(4,4){\ovalnode{*00}{$(*,0,0)$}}
 \rput(7,4){\ovalnode{d01}{$(\rd,0,1)$}}
 \rput(7,1){\ovalnode{*02}{$(*,0,2)$}}
 \rput(4,1){\ovalnode{*10}{$(*,1,0)$}}
 \rput(1,1){\ovalnode{d11}{$(\rd,1,1)$}}
 \rput(1,4){\ovalnode{*12}{$(*,1,2)$}}

 \ncline{S}{*00}
 \ncline{*00}{d01}\naput[nrot=:0]{$\kw{r}_4(\rd,0)$}
 \ncline{d01}{*02}\naput[nrot=:0]{$\kw{s}_2(\rd)$}
 \ncarc{*02}{*10}\naput[nrot=:180]{$\kw{s}_6(0)$}
 \ncarc{*10}{*02}\naput[nrot=:0]{$\begin{array}{c}
                                  \kw{r}_4(\rd,0) \\
                                  \kw{r}_4(*)
                                  \end{array}$}
 \ncline{*10}{d11}\naput[nrot=:180]{$\kw{r}_4(\rd,1)$}
 \ncline{d11}{*12}\naput[nrot=:0]{$\kw{s}_2(\rd)$}
 \ncarc{*12}{*00}\naput[nrot=:0]{$\kw{s}_6(1)$}
 \ncarc{*00}{*12}\naput[nrot=:180]{$\begin{array}{c}
                                    \kw{r}_4(\rd,1) \\
                                    \kw{r}_4(*)
                                    \end{array}$}
\end{pspicture}
\caption{Transition system for the receiver}
\label{fig-ABP-receiver}
\end{figure}

The behaviour of the receiver $R$ is described by a transition system 
as follows.
As states of the receiver, we have triples $\tup{d,b,i}$ where 
$d \in D \union \set{\und}$, $b \in B$ and $i \in \set{0,1,2}$,
satisfying $d = \und$ if and only if $i \neq 1$.
State $\tup{d,b,i}$ is roughly a state in which the datum to be
delivered is $d$ and the right bit is $b$.
If $d = \und$, no such datum is available.
The initial state is $\tup{\und,0,0}$.
As actions, we have 
$\kw{s}_2(d)$ for each $d \in D$,
$\kw{r}_4(f)$ for each $f \in F \union \set{\und}$, and
$\kw{s}_6(b)$ for each $b \in B$.
As transitions of the receiver, we have the following:
\begin{iteml}
\item
for each datum $d \in D$ and bit $b \in B$:
\begin{iteml}
\item
a transition 
$\astep{\tup{\und,b,0}}{\kw{r}_4(d,b)}{\tup{d,b,1}}$,
\item
a transition 
$\astep{\tup{\und,b,0}}{\kw{r}_4(d,\ol{b})}{\tup{\und,\ol{b},2}}$,
\item
a transition 
$\astep{\tup{d,b,1}}{\kw{s}_2(d)}{\tup{\und,b,2}}$;
\end{iteml}
\item
for each bit $b \in B$:
\begin{iteml}
\item
a transition 
$\astep{\tup{\und,b,0}}{\kw{r}_4(\und)}{\tup{\und,\ol{b},2}}$,
\item
a transition 
$\astep{\tup{\und,b,2}}{\kw{s}_6(b)}{\tup{\und,\ol{b},0}}$.
\end{iteml}
\end{iteml}
The transition system for the receiver is represented graphically in 
Fig.~\ref{fig-ABP-receiver} for the case where only one datum, say $d$,
is involved.

\begin{figure}
\begin{pspicture}(0,0)(12,6)

 \psset{arrows=->}

 \pnode(5,4){S}
 \rput(5,3){\ovalnode{*0}{$(*,0)$}}

 \rput(1,3){\ovalnode{d01}{$((\rd,0),1)$}}
 \rput(1,5){\ovalnode{d02}{$((\rd,0),2)$}}
 \rput(1,1){\ovalnode{d03}{$((\rd,0),3)$}}
 \rput(9,3){\ovalnode{d11}{$((\rd,1),1)$}}
 \rput(9,5){\ovalnode{d12}{$((\rd,1),2)$}}
 \rput(9,1){\ovalnode{d13}{$((\rd,1),3)$}}

 \ncline{S}{*0}
 \ncline{*0}{d01}\nbput[nrot=:180]{$\kw{r}_3(\rd,0)$}
 \ncline{*0}{d11}\naput[nrot=:0]{$\kw{r}_3(\rd,1)$}
 \ncline{d01}{d02}\naput{$\kw{i}$}
 \ncline{d01}{d03}\nbput{$\kw{i}$}
 \ncline{d11}{d12}\nbput{$\kw{i}$}
 \ncline{d11}{d13}\naput{$\kw{i}$}
 \ncline{d02}{*0}\naput[nrot=:0]{$\kw{s}_4(\rd,0)$}
 \ncline{d03}{*0}\nbput[nrot=:0]{$\kw{s}_4(*)$}
 \ncline{d12}{*0}\nbput[nrot=:180]{$\kw{s}_4(\rd,1)$}
 \ncline{d13}{*0}\naput[nrot=:180]{$\kw{s}_4(*)$}

\end{pspicture}
\caption{Transition system for the data transmission channel}
\label{fig-ABP-data-channel}
\end{figure}
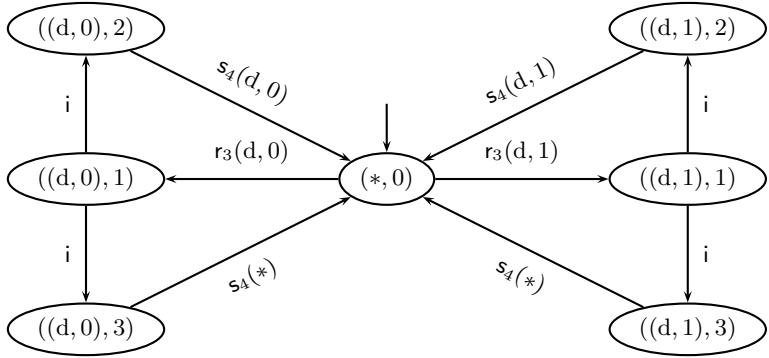

The behaviour of the data transmission channel $K$ is described by a 
transition system as follows.
As states of the channel, we have pairs $\tup{f,i}$, where
$f \in F \union \set{\und}$ and $i \in \set{0,1,2,3}$, satisfying
$f = \und$ if and only if $i = 0$.
State $\tup{f,i}$ is roughly a state in which the frame to be
transmitted is $f$.
If $f = \und$, no such frame is available.
The initial state is $\tup{\und,0}$.
As actions, we have 
$\kw{i}$,
$\kw{r}_3(f)$ for each $f \in F$, and
$\kw{s}_4(f)$ for each $f \in F \union \set{\und}$.
As transitions of the channel, we have the following:
\begin{iteml}
\item
for each frame $f \in F$: 
\begin{iteml}
\item
a transition $\astep{\tup{\und,0}}{\kw{r}_3(f)}{\tup{f,1}}$,
\item
a transition $\astep{\tup{f,2}}{\kw{s}_4(f)}{\tup{\und,0}}$,
\item
a transition $\astep{\tup{f,3}}{\kw{s}_4(\und)}{\tup{\und,0}}$;
\end{iteml}
\item
for each frame $f \in F$ and $i \in \set{2,3}$: 
\begin{iteml}
\item
a transition $\astep{\tup{f,1}}{\kw{i}}{\tup{f,i}}$.
\end{iteml}
\end{iteml}
Note that this transition system is not determinate:
for each frame $f$ we have both
$\gstep{\tup{\und,0}}{\kw{r}_3(f)\; \kw{i}}{\tup{f,2}}$ and
$\gstep{\tup{\und,0}}{\kw{r}_3(f)\; \kw{i}}{\tup{f,3}}$, but
the actions that can be performed from $\tup{f,2}$ and $\tup{f,3}$ 
are different.
The action $\kw{i}$ is an internal action that cannot be performed
synchronously with any other action.
Thus, the channel cannot be forced to leave all frames uncorrupted.
The transition system for channel $K$ is represented graphically in 
Fig.~\ref{fig-ABP-data-channel} for the case where only one datum, 
say $d$, is involved.

\begin{figure}
\begin{pspicture}(0,0)(12,6)

 \psset{arrows=->}

 \pnode(5,4){S}
 \rput(5,3){\ovalnode{*0}{$(*,0)$}}

 \rput(1,3){\ovalnode{01}{$(0,1)$}}
 \rput(1,5){\ovalnode{02}{$(0,2)$}}
 \rput(1,1){\ovalnode{03}{$(0,3)$}}
 \rput(9,3){\ovalnode{11}{$(1,1)$}}
 \rput(9,5){\ovalnode{12}{$(1,2)$}}
 \rput(9,1){\ovalnode{13}{$(1,3)$}}

 \ncline{S}{*0}
 \ncline{*0}{01}\nbput[nrot=:180]{$\kw{r}_6(0)$}
 \ncline{*0}{11}\naput[nrot=:0]{$\kw{r}_6(1)$}
 \ncline{01}{02}\naput{$\kw{i}$}
 \ncline{01}{03}\nbput{$\kw{i}$}
 \ncline{11}{12}\nbput{$\kw{i}$}
 \ncline{11}{13}\naput{$\kw{i}$}
 \ncline{02}{*0}\naput[nrot=:0]{$\kw{s}_5(0)$}
 \ncline{03}{*0}\nbput[nrot=:0]{$\kw{s}_5(*)$}
 \ncline{12}{*0}\nbput[nrot=:180]{$\kw{s}_5(1)$}
 \ncline{13}{*0}\naput[nrot=:180]{$\kw{s}_5(*)$}

\end{pspicture}
\caption{Transition system for the acknowledgement transmission channel}
\label{fig-ABP-ack-channel}
\end{figure}
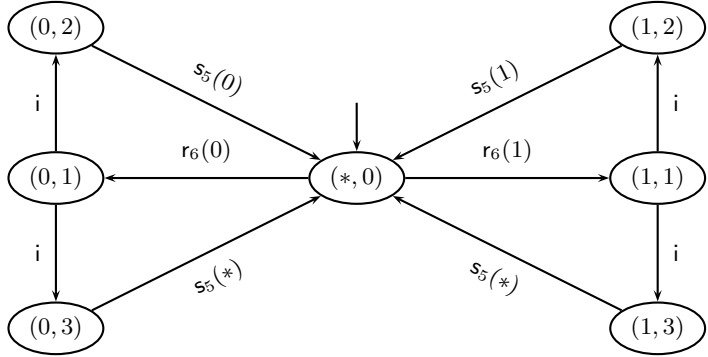

The behaviour of the acknowledgement transmission channel $L$ is 
described by a transition system as follows.
As states of the channel, we have pairs $\tup{b,i}$, where
$b \in B \union \set{\und}$ and $i \in \set{0,1,2,3}$, satisfying
$b = \und$ if and only if $i = 0$.
State $\tup{b,i}$ is roughly a state in which the bit to be
transmitted is $b$.
If $b = \und$, no such bit is available.
The initial state is $\tup{\und,0}$.
As actions, we have 
$\kw{i}$,
$\kw{s}_5(b)$ for each $b \in B \union \set{\und}$, and
$\kw{r}_6(b)$ for each $b \in B$. 
As transitions of the channel, we have the following:
\begin{iteml}
\item
for each bit $b \in B$: 
\begin{iteml}
\item
a transition $\astep{\tup{\und,0}}{\kw{r}_6(b)}{\tup{b,1}}$,
\item
a transition $\astep{\tup{b,2}}{\kw{s}_5(b)}{\tup{\und,0}}$,
\item
a transition $\astep{\tup{b,3}}{\kw{s}_5(\und)}{\tup{\und,0}}$;
\end{iteml}
\item
for each bit $b \in B$ and $i \in \set{2,3}$: 
\begin{iteml}
\item
a transition $\astep{\tup{b,1}}{\kw{i}}{\tup{b,i}}$.
\end{iteml}
\end{iteml}
Just as the transition system for channel $K$, the transition system 
for channel $L$ is not determinate:
for each bit $b$ we have both
$\gstep{\tup{\und,0}}{\kw{r}_6(b)\; \kw{i}}{\tup{b,2}}$ and
$\gstep{\tup{\und,0}}{\kw{r}_6(b)\; \kw{i}}{\tup{b,3}}$, but
the actions that can be performed from $\tup{b,2}$ and $\tup{b,3}$ 
are different.
Like in the case of channel $K$, channel $L$ cannot be forced to leave 
all acknowledgements uncorrupted.
The transition system for channel $L$ is represented graphically in 
Fig.~\ref{fig-ABP-ack-channel}.

The behaviour of the whole system is described as follows:
\begin{ldispl}
 \begin{geqns}
\encap{H}(S \parcs{\commf} K \parcs{\commf} L \parcs{\commf} R)
 \end{geqns}
\end{ldispl}
where
\begin{ldispl}
\displstretch
 \begin{aeqns}
 H & = &
  \set{\kw{s}_3(f), \kw{r}_3(f) \where f \in F} 
     \union
  \set{\kw{s}_4(f), \kw{r}_4(f) \where f \in F \union \set{\und}} 
     \\ & {} \union &
  \set{\kw{s}_5(b), \kw{r}_5(b) \where b \in B \union \set{\und}} 
     \union
  \set{\kw{s}_6(b), \kw{r}_6(b) \where b \in B}
 \end{aeqns}
\end{ldispl}
and the communication function $\commf$ is defined in the standard way
for handshaking communication (see Sect.~\ref{sect-interaction-formal}).

\begin{figure}
\begin{pspicture}(0.5,1.5)(12,13.5)

 \psset{arrows=->,radius=.2cm, xunit=1.3, yunit=1.3, runit=1.3}

  \pnode(0.5,10.5){S}

  \Cnode(1,10){*00*0*0*00} 
  \Cnode(3,10){d01*0*0*00} 
  \Cnode(6,7){d01*0*0*10} 
  \Cnode(9,8){d02*0*0*02} 
  \Cnode(4,7){d02*0*0*12} 
  \Cnode(9,10){d02*0*0d01} 
  \Cnode(9,6){d02*001*10} 
 \Cnode(9,4){d02*002*10} 
  \Cnode(7,6){d02*003*10} 
  \Cnode(3,8){d02*011*00} 
  \Cnode(4,9){d02*012*00} 
  \Cnode(2,9){d02*013*00} 
  \Cnode(5,10){d02d01*0*00} 
  \Cnode(7,8){d02d01*0*10} 
  \Cnode(7,10){d02d02*0*00} 
  \Cnode(8,7){d02d02*0*10}  
  \Cnode(5,8){d02d03*0*00} 
  \Cnode(8,9){d02d03*0*10} 

 \Cnode(9,2){*10*0*0*10} 
 \Cnode(4,5){d11*0*0*00} 
 \Cnode(7,2){d11*0*0*10} 
 \Cnode(6,5){d12*0*0*02} 
 \Cnode(1,4){d12*0*0*12} 
 \Cnode(1,2){d12*0*0d11} 
 \Cnode(7,4){d12*001*10} 
 \Cnode(6,3){d12*002*10} 
 \Cnode(8,3){d12*003*10} 
 \Cnode(1,6){d12*011*00} 
 \Cnode(1,8){d12*012*00} 
 \Cnode(3,6){d12*013*00} 
 \Cnode(3,4){d12d11*0*00} 
 \Cnode(5,2){d12d11*0*10} 
 \Cnode(2,3){d12d12*0*00} 
 \Cnode(3,2){d12d12*0*10} 
 \Cnode(2,5){d12d13*0*00} 
 \Cnode(5,4){d12d13*0*10} 

 \ncline{S}{*00*0*0*00}
 \ncline{*00*0*0*00}{d01*0*0*00}\naput[nrot=:0]{$\kw{r}_1(\rd)$}
 \ncline{*10*0*0*10}{d11*0*0*10}\naput[nrot=:180]{$\kw{r}_1(\rd)$}
 \ncline{d01*0*0*00}{d02d01*0*00}\naput[nrot=:0]{$\kw{c}_3(\rd,0)$}
 \ncline{d11*0*0*10}{d12d11*0*10}\naput[nrot=:180]{$\kw{c}_3(\rd,1)$}
 \ncline{d02d02*0*00}{d02*0*0d01}\naput[nrot=:0]{$\kw{c}_4(\rd,0)$}
 \ncline{d12d12*0*10}{d12*0*0d11}\naput[nrot=:180]{$\kw{c}_4(\rd,1)$}
 \ncline{d02*0*0d01}{d02*0*0*02}\naput[nrot=:0]{$\kw{s}_2(\rd)$}
 \ncline{d12*0*0d11}{d12*0*0*12}\naput[nrot=:0]{$\kw{s}_2(\rd)$}
 \ncline{d02*0*0*02}{d02*001*10}\naput[nrot=:0]{$\kw{c}_6(0)$}
 \ncline{d12*0*0*12}{d12*011*00}\naput[nrot=:0]{$\kw{c}_6(1)$}

 \ncline{d02*002*10}{*10*0*0*10}\naput[nrot=:0]{$\kw{c}_5(0)$}
 \ncline{d12*012*00}{*00*0*0*00}\naput[nrot=:0]{$\kw{c}_5(1)$}
 \ncline{d02d03*0*00}{d02*0*0*12}\naput[nrot=:180]{$\kw{c}_4(*)$} %
 \ncline{d12d13*0*10}{d12*0*0*02}\naput[nrot=:0]{$\kw{c}_4(*)$} %
 \ncline{d02*0*0*12}{d02*011*00}\naput[nrot=:180]{$\kw{c}_6(1)$}
 \ncline{d12*0*0*02}{d12*001*10}\naput[nrot=:0]{$\kw{c}_6(0)$}

 \ncline{d02*012*00}{d01*0*0*00}\naput[nrot=:180]{$\kw{c}_5(0)$}
 \ncline{d12*002*10}{d11*0*0*10}\naput[nrot=:0]{$\kw{c}_5(1)$}
 \ncline{d02*013*00}{d01*0*0*00}\naput[nrot=:0]{$\kw{c}_5(*)$}
 \ncline{d12*003*10}{d11*0*0*10}\naput[nrot=:180]{$\kw{c}_5(*)$}
 \ncline{d02*003*10}{d01*0*0*10}\naput[nrot=:180]{$\kw{c}_5(*)$}
 \ncline{d12*013*00}{d11*0*0*00}\naput[nrot=:0]{$\kw{c}_5(*)$}
 \ncline{d01*0*0*10}{d02d01*0*10}\naput[nrot=:0]{$\kw{c}_3(\rd,0)$}
 \ncline{d11*0*0*00}{d12d11*0*00}\naput[nrot=:180]{$\kw{c}_3(\rd,1)$}

 \ncline{d02d02*0*10}{d02*0*0*02}\naput[nrot=:0]{$\kw{c}_4(\rd,0)$}
 \ncline{d12d12*0*00}{d12*0*0*12}\nbput[nrot=:180]{$\kw{c}_4(\rd,1)$}
 \ncline{d02d03*0*10}{d02*0*0*02}\naput[nrot=:0]{$\kw{c}_4(*)$}
 \ncline{d12d13*0*00}{d12*0*0*12}\nbput[nrot=:180]{$\kw{c}_4(*)$}

 \ncline{d02*001*10}{d02*002*10}\naput[nrot=:0]{$\kw{i}$}
 \ncline{d12*011*00}{d12*012*00}\naput[nrot=:0]{$\kw{i}$}
 \ncline{d02d01*0*00}{d02d02*0*00}\naput[nrot=:0]{$\kw{i}$}
 \ncline{d12d11*0*10}{d12d12*0*10}\naput[nrot=:180]{$\kw{i}$}
 \ncline{d02d01*0*00}{d02d03*0*00}\naput[nrot=:0]{$\kw{i}$}
 \ncline{d12d11*0*10}{d12d13*0*10}\naput[nrot=:0]{$\kw{i}$}
 \ncline{d02*011*00}{d02*012*00}\nbput[nrot=:0]{$\kw{i}$}
 \ncline{d12*001*10}{d12*002*10}\nbput[nrot=:180]{$\kw{i}$}

 \ncline{d02*011*00}{d02*013*00}\naput[nrot=:180]{$\kw{i}$}
 \ncline{d12*001*10}{d12*003*10}\naput[nrot=:0]{$\kw{i}$}
 \ncline{d02*001*10}{d02*003*10}\naput[nrot=:180]{$\kw{i}$}
 \ncline{d12*011*00}{d12*013*00}\naput[nrot=:0]{$\kw{i}$}
 \ncline{d02d01*0*10}{d02d03*0*10}\naput[nrot=:0]{$\kw{i}$}
 \ncline{d12d11*0*00}{d12d13*0*00}\nbput[nrot=:180]{$\kw{i}$}
 \ncline{d02d01*0*10}{d02d02*0*10}\nbput[nrot=:0]{$\kw{i}$}
 \ncline{d12d11*0*00}{d12d12*0*00}\naput[nrot=:180]{$\kw{i}$}
\end{pspicture}
\caption{Transition system for the ABP}
\label{fig-ABP-protocol}
\end{figure}
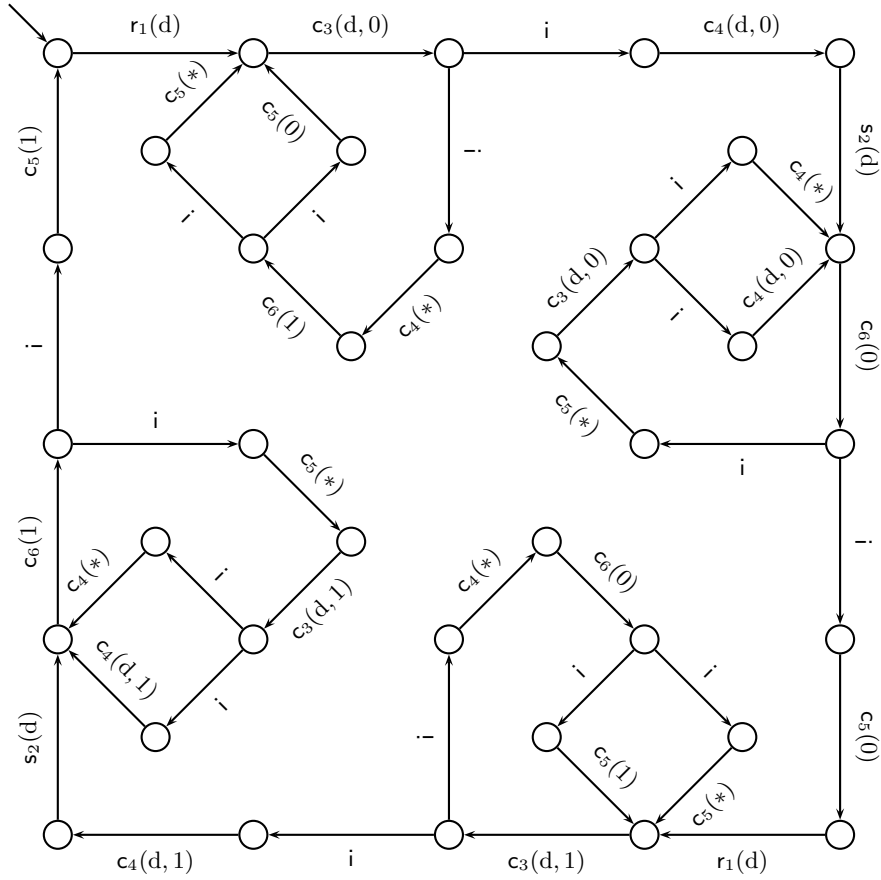

Parallel composition and encapsulation of the transition systems of 
$S$, $K$, $L$ and $R$ as described above results in the following 
transition system.
As states, we have quadruples $\tup{s,k,l,r}$, where $s$, $k$, $l$ and 
$r$ are states of $S$, $K$, $L$ and $R$, respectively.
As initial state, we have
$\tup{\tup{\und,0,0},\tup{\und,0},\tup{\und,0},\tup{\und,0,0}}$.
As actions, we have
$\kw{r}_1(d)$ and $\kw{s}_2(d)$ for each $d \in D$,
$\kw{c}_3(f)$ for each $f \in F$,
$\kw{c}_4(f)$ for each $f \in F \union \set{\und}$,
$\kw{c}_5(b)$ for each $b \in B \union \set{\und}$, 
$\kw{c}_6(b)$ for each $b \in B$, and 
$\kw{i}$.
As transitions, we have the following:
\begin{iteml}
\small
\item
for each datum $d \in D$ and bit $b \in B$: 
\begin{iteml}
\item
$\astep
 {\tup{\tup{\und,b,0},\tup{\und,0},\tup{\und,0},\tup{\und,b,0}}}
 {\kw{r}_1(d)}
 {\tup{\tup{d,b,1},\tup{\und,0},\tup{\und,0},\tup{\und,b,0}}}$,
\item
$\astep
 {\tup{\tup{d,b,1},\tup{\und,0},\tup{\und,0},\tup{\und,b,0}}}
 {\kw{c}_3(d,b)}
 {\tup{\tup{d,b,2},\tup{\tup{d,b},1},\tup{\und,0},\tup{\und,b,0}}}$,
\item
$\astep
 {\tup{\tup{d,b,2},\tup{\tup{d,b},1},\tup{\und,0},\tup{\und,b,0}}}
 {\kw{i}}
 {\tup{\tup{d,b,2},\tup{\tup{d,b},2},\tup{\und,0},\tup{\und,b,0}}}$,
\item
$\astep
 {\tup{\tup{d,b,2},\tup{\tup{d,b},2},\tup{\und,0},\tup{\und,b,0}}}
 {\kw{c}_4(d,b)}
 {\tup{\tup{d,b,2},\tup{\und,0},\tup{\und,0},\tup{d,b,1}}}$,
\item
$\astep
 {\tup{\tup{d,b,2},\tup{\und,0},\tup{\und,0},\tup{d,b,1}}}
 {\kw{s}_2(d)}
 {\tup{\tup{d,b,2},\tup{\und,0},\tup{\und,0},\tup{\und,b,2}}}$,
\item
$\astep
 {\tup{\tup{d,b,2},\tup{\und,0},\tup{\und,0},\tup{\und,b,2}}}
 {\kw{c}_6(b)}
 {\tup{\tup{d,b,2},\tup{\und,0},\tup{b,1},\tup{\und,\ol{b},0}}}$,
\item
$\astep
 {\tup{\tup{d,b,2},\tup{\und,0},\tup{b,1},\tup{\und,\ol{b},0}}}
 {\kw{i}}
 {\tup{\tup{d,b,2},\tup{\und,0},\tup{b,2},\tup{\und,\ol{b},0}}}$,
\item
$\astep
 {\tup{\tup{d,b,2},\tup{\und,0},\tup{b,2},\tup{\und,\ol{b},0}}}
 {\kw{c}_5(b)}
 {\tup{\tup{\und,\ol{b},0},\tup{\und,0},\tup{\und,0},\tup{\und,\ol{b},0}}}$,
\end{iteml}
\mbox{} \\[-2ex]
\begin{iteml}
\item
$\astep
 {\tup{\tup{d,b,2},\tup{\tup{d,b},1},\tup{\und,0},\tup{\und,b,0}}}
 {\kw{i}}
 {\tup{\tup{d,b,2},\tup{\tup{d,b},3},\tup{\und,0},\tup{\und,b,0}}}$,
\item
$\astep
 {\tup{\tup{d,b,2},\tup{\tup{d,b},3},\tup{\und,0},\tup{\und,b,0}}}
 {\kw{c}_4(\und)}
 {\tup{\tup{d,b,2},\tup{\und,0},\tup{\und,0},\tup{\und,\ol{b},2}}}$,
\item
$\astep
 {\tup{\tup{d,b,2},\tup{\und,0},\tup{\und,0},\tup{\und,\ol{b},2}}}
 {\kw{c}_6(\ol{b})}
 {\tup{\tup{d,b,2},\tup{\und,0},\tup{\ol{b},1},\tup{\und,b,0}}}$,
\item
$\astep
 {\tup{\tup{d,b,2},\tup{\und,0},\tup{\ol{b},1},\tup{\und,b,0}}}
 {\kw{i}}
 {\tup{\tup{d,b,2},\tup{\und,0},\tup{\ol{b},2},\tup{\und,b,0}}}$,
\item
$\astep
 {\tup{\tup{d,b,2},\tup{\und,0},\tup{\ol{b},2},\tup{\und,b,0}}}
 {\kw{c}_5(\ol{b})}
 {\tup{\tup{d,b,1},\tup{\und,0},\tup{\und,0},\tup{\und,b,0}}}$,
\item
$\astep
 {\tup{\tup{d,b,2},\tup{\und,0},\tup{\ol{b},1},\tup{\und,b,0}}}
 {\kw{i}}
 {\tup{\tup{d,b,2},\tup{\und,0},\tup{\ol{b},3},\tup{\und,b,0}}}$,
\item
$\astep
 {\tup{\tup{d,b,2},\tup{\und,0},\tup{\ol{b},3},\tup{\und,b,0}}}
 {\kw{c}_5(\und)}
 {\tup{\tup{d,b,1},\tup{\und,0},\tup{\und,0},\tup{\und,b,0}}}$,
\end{iteml}
\mbox{} \\[-2ex]
\begin{iteml}
\item
$\astep
 {\tup{\tup{d,b,2},\tup{\und,0},\tup{b,1},\tup{\und,\ol{b},0}}}
 {\kw{i}}
 {\tup{\tup{d,b,2},\tup{\und,0},\tup{b,3},\tup{\und,\ol{b},0}}}$,
\item
$\astep
 {\tup{\tup{d,b,2},\tup{\und,0},\tup{b,3},\tup{\und,\ol{b},0}}}
 {\kw{c}_5(\und)}
 {\tup{\tup{d,b,1},\tup{\und,0},\tup{\und,0},\tup{\und,\ol{b},0}}}$,
\item
$\astep
 {\tup{\tup{d,b,1},\tup{\und,0},\tup{\und,0},\tup{\und,\ol{b},0}}}
 {\kw{c}_3(d,b)}
 {\tup{\tup{d,b,2},\tup{\tup{d,b},1},\tup{\und,0},\tup{\und,\ol{b},0}}}$,
\item
$\astep
 {\tup{\tup{d,b,2},\tup{\tup{d,b},1},\tup{\und,0},\tup{\und,\ol{b},0}}}
 {\kw{i}}
 {\tup{\tup{d,b,2},\tup{\tup{d,b},2},\tup{\und,0},\tup{\und,\ol{b},0}}}$,
\item
$\astep
 {\tup{\tup{d,b,2},\tup{\tup{d,b},2},\tup{\und,0},\tup{\und,\ol{b},0}}}
 {\kw{c}_4(d,b)}
 {\tup{\tup{d,b,2},\tup{\und,0},\tup{\und,0},\tup{\und,b,2}}}$,
\item
$\astep
 {\tup{\tup{d,b,2},\tup{\tup{d,b},1},\tup{\und,0},\tup{\und,\ol{b},0}}}
 {\kw{i}}
 {\tup{\tup{d,b,2},\tup{\tup{d,b},3},\tup{\und,0},\tup{\und,\ol{b},0}}}$,
\item
$\astep
 {\tup{\tup{d,b,2},\tup{\tup{d,b},3},\tup{\und,0},\tup{\und,\ol{b},0}}}
 {\kw{c}_4(\und)}
 {\tup{\tup{d,b,2},\tup{\und,0},\tup{\und,0},\tup{\und,b,2}}}$.
\end{iteml}
\end{iteml}
The transition system for the whole protocol is represented graphically
in Fig.~\ref{fig-ABP-protocol} for the case where only one datum, say 
$d$, is involved.
This transition system does not reflect the configuration of the 
protocol, but is useful for analysis of the protocol.
The transition system for the whole protocol shows, for example, that
data are delivered in the order in which they were offered, without any
loss, if it is assumed that cycles of communication actions at internal
ports and the action $\kw{i}$ are eventually left.

\section{Petri nets and parallel composition}
\label{sect-interaction-conn-nets}

For a better understanding of the notion of parallel composition of
transition systems, we looked in a previous section into its
connections with the familiar notion of parallel execution of
programs.
Is there a corresponding notion for nets as well?
For the interested reader, we now show that parallel composition can 
also be defined on nets.
\begin{definition}[Parallel composition]%
\index{parallel composition!of nets}
Let $N = \tup{P,A,\step{},m_0}$ and $N' = \tup{P',A',\step{}',m_0'}$ be 
nets such that $P \inter P' = \emptyset$.
Let $\commf$ be a communication function on a set of actions that
includes $A \union A'$.
The \emph{parallel composition} of $N$ and $N'$ under $\commf$,
written $N \parcs{\commf} N'$, is the net
$\tup{P'',A'',\step{}'',m''_0}$ where
\begin{iteml}
\item
$P'' = P \union P'$;
\item
$A'' = 
A \union A' \union 
\set{\commf(a,a') \where 
     a \in A, a' \in A', \commf(a,a') \;\mathrm{is}\;\mathrm{defined}}$;
\item
$\step{}''$ is the smallest subset of 
$(\fsetof{(P'')} \diff \emptyset) 
  \x A'' \x 
 (\fsetof{(P'')} \diff \emptyset)$ such that: 
\begin{iteml}
\item
if $\astep{Q_1}{a}{Q_2}$, then
$\asteppp{Q_1}{a}{Q_2}$;
\item
if $\astepp{Q'_1}{b}{Q'_2}$, then
$\asteppp{Q'_1}{b}{Q'_2}$;
\item
if $\astep{Q_1}{a}{Q_2}$, $\astepp{Q'_1}{b}{Q'_2}$ and 
$\commf(a,b)$ is defined, then
$\asteppp{Q_1 \union Q'_1}{\commf(a,b)}{Q_2 \union Q'_2}$;
\end{iteml}
\item
$ m''_0 = m_0 \union m'_0$.
\end{iteml}
\end{definition}
Let us give a definition of encapsulation on nets as well.
\begin{definition}[Encapsulation]\index{encapsulation!of nets}
Let $N = \tup{P,A,\step{},m_0}$ be a net.
Let $H \subseteq A$.
The \emph{encapsulation} of $N$ with respect to $H$, 
written $\encap{H}(N)$, is the net 
$\tup{P',A',\step{}',m_0}$ where
\begin{iteml}
\item
$P' =
 \set{p \in P \where 
      \Exists{\sigma \in \seqof{(A \diff H)},\funct{m}{P}{\Nat}}
             {\gstep{m_0}{\sigma}{m},m(p) \neq 0}}$;
\item
$A' = 
 \set{a \in A \diff H \where 
      \Exists{\sigma \in \seqof{(A \diff H)},\funct{m}{P}{\Nat}}
             {\gstep{m_0}{\sigma\, a}{m}}}$;
\item
$\step{}'$ is the smallest subset of 
$(\fsetof{(P')} \diff \emptyset) 
  \x A' \x 
 (\fsetof{(P')} \diff \emptyset)$ such that: 
\begin{iteml}
\item
if $\astep{Q_1}{a}{Q_2}$, $Q_1 \subseteq P'$ and $a \not\in H$, 
then $\astepp{Q_1}{a}{Q_2}$.
\end{iteml}
\end{iteml}
\end{definition}

Here is an example of the use of parallel composition of nets and 
encapsulation of nets in describing process behaviour.
\begin{example}[Milner's scheduling problem]%
\index{Milner's scheduling problem}
\label{exa-scheduler-parallel}
We consider again the system of scheduled processes from 
Example~\ref{exa-scheduler}.
It consists of processes $P_1,\ldots,P_n$ ($n > 1$), each wishing to 
perform a certain task repeatedly, and a scheduler ensuring that they 
start their task in cyclic order, beginning with $P_1$.

The behaviour of process $P_i$, for $1 \leq i \leq n$, can be described 
by a net as follows.
As places of $P_i$, we have the pairs $\tup{i,\kw{idle}}$ and 
$\tup{i,\kw{busy}}$.
As initial marking, we have $\set{\tup{i,\kw{idle}}}$.
As actions, we have $\kw{request}(i)$ (request to start task $i$) and 
$\kw{finish}(i)$.
As transitions, we have the following:
\begin{ldispl}
\astep{\set{\tup{i,\kw{idle}}}}{\kw{request}(i)}
  {\set{\tup{i,\kw{busy}}}}, \\
\astep{\set{\tup{i,\kw{busy}}}}{\kw{finish}(i)}
  {\set{\tup{i,\kw{idle}}}}.
\end{ldispl}
The behaviour of scheduler $S$ can be described by a net as follows.
As places of the scheduler, we have the pairs $\tup{i,\kw{sch}}$ for 
$1 \leq i \leq n$. 
As initial marking, we have $\set{\tup{1,\kw{sch}}}$.
As actions, we have $\kw{grant}(i)$ (grant to start task $i$) for 
$1 \leq i \leq n$.
As transitions, we have the following (for $1 \leq i \leq n$):
\begin{ldispl}
\astep{\set{\tup{i,\kw{sch}}}}{\kw{grant}(i)}
  {\set{\tup{\nm{nxt}(i),\kw{sch}}}},
\end{ldispl}
where $\nm{nxt}(i) = i + 1$ if $i < n$ and $\nm{nxt}(n) = 1$.
The behaviour of the whole system is described as follows:
\begin{ldispl}
 \begin{geqns}
\encap{H}(P_1 \parcs{\commf} \ldots \parcs{\commf} P_n \parcs{\commf} S)
 \end{geqns}
\end{ldispl}
where
\begin{ldispl}
\displstretch
 \begin{aeqns}
 H & = &
  \set{\kw{request}(i), \kw{grant}(i) \where 1 \leq i \leq n}\;.
 \end{aeqns}
\end{ldispl}
and the communication function $\commf$ is defined such that
\begin{ldispl}
 \begin{geqns}
 \commf(\kw{request}(i),\kw{grant}(i)) = 
 \commf(\kw{grant}(i),\kw{request}(i)) = 
 \kw{start}(i)
 \end{geqns}
\end{ldispl}
for $1 \leq i \leq n$, and it is undefined otherwise.

The net obtained from the nets $P_1$, \ldots, $P_n$ and $S$ by parallel
composition and encapsulation as described above is the same as the
net described in Example~\ref{exa-scheduler}.
\end{example}

In Sect.~\ref{sect-basics-conn-nets}, we associated a transition
system $\TrSy(N)$ with each net $N$.
It happens that this association is useful in showing the close
connection between parallel composition of nets and parallel 
composition of transition systems, and between encapsulation of nets 
and encapsulation of transition systems.
\begin{property}
Let $N = \tup{P,A,\step{},m_0}$ and $N' = \tup{P',A',\step{}',m_0'}$ be 
nets such that $P \inter P' = \emptyset$,
let $\commf$ be a communication function on a set of actions that
includes $A \union A'$, and 
let $H \subseteq A$.
Then we have that
\begin{ldispl}
 \begin{geqns}
\TrSy(N \parcs{\commf} N') \bisim \TrSy(N) \parcs{\commf} \TrSy(N')\;,
\\
\TrSy(\encap{H}(N))  \bisim \encap{H}(\TrSy(N))\;.
 \end{geqns}
\end{ldispl}
In words, the transition system associated with a parallel composition
of nets is up to bisimulation equivalence the same as the parallel
composition of the transition systems associated with those nets; and
analogously for encapsulation.
\end{property}
\begin{example}[Milner's scheduling problem]%
\index{Milner's scheduling problem}
\label{exa-scheduler-ts}
We consider once again the system of scheduled processes from 
Examples~\ref{exa-scheduler} and~\ref{exa-scheduler-parallel}.
Associating a transition system with the net describing the behaviour
of process $P_i$ ($1 \leq i \leq n$) is trivial because only singleton
sets occur as pre- and postsets of transitions.
The resulting transition system, $\TrSy(P_i)$, can be described as
follows.
As states, we have the singleton sets of pairs
$\set{\tup{i,\kw{idle}}}$ and $\set{\tup{i,\kw{busy}}}$.
As initial state, we have $\set{\tup{i,\kw{idle}}}$.
As actions, we have $\kw{request}(i)$ and $\kw{finish}(i)$.
As transitions, we have the following:
\begin{ldispl}
\astep{\set{\tup{i,\kw{idle}}}}{\kw{request}(i)}
  {\set{\tup{i,\kw{busy}}}}, \\
\astep{\set{\tup{i,\kw{busy}}}}{\kw{finish}(i)}
  {\set{\tup{i,\kw{idle}}}}.
\end{ldispl}
Associating a transition system with the net describing the behaviour
of the scheduler $S$ is equally trivial.
The resulting transition system, $\TrSy(S)$, can be described as
follows.
As states, we have the singleton sets of pairs $\set{\tup{i,\kw{sch}}}$
for $1 \leq i \leq n$.
As initial state, we have $\set{\tup{1,\kw{sch}}}$.
As actions, we have $\kw{grant}(i)$ for $1 \leq i \leq n$.
As transitions, we have the following (for $1 \leq i \leq n$):
\begin{ldispl}
\astep{\set{\tup{i,\kw{sch}}}}{\kw{grant}(i)}
  {\set{\tup{\nm{nxt}(i),\kw{sch}}}}.
\end{ldispl}
So, the transition systems associated with the nets $P_1$, \ldots, 
$P_n$ and $S$ are simply obtained by taking the singleton sets of 
places as states.
In other words, those nets are essentially transition systems.
However, their parallel composition as nets yields the net from
Example~\ref{exa-scheduler}, which is not quite a transition system
-- because non-singleton sets of places occur as pre- and postsets of
transitions.
The transition system described by
\begin{ldispl}
 \begin{geqns}
\encap{H}
 (\TrSy(P_1) \parcs{\commf} \ldots \parcs{\commf} \TrSy(P_n) 
   \parcs{\commf} 
  \TrSy(S))\;,
 \end{geqns}
\end{ldispl}
where $H$ and $\commf$ are as in Example~\ref{exa-scheduler-parallel},
is bisimulation equivalent to the transition system associated with the
net from Example~\ref{exa-scheduler}.
\end{example}

\section{Bisimulation and trace equivalence}
\label{sect-interaction-eqv}

An important property of parallel composition and encapsulation of 
transition systems is that they preserve bisimulation equivalence, by 
which we mean the following.
\begin{property}[Preservation of bisimulation equivalence]%
\index{bisimulation equivalence!preservation of}
Let $T_1$ and $T_2$ be transition systems with $A$ as set of actions,
let $T'_1$ and $T'_2$ be transition systems with $A'$ as set of
actions, and 
let $\commf$ be a communication function on a set of actions that 
includes $A \union A'$.
Then the following holds:
\begin{ldispl}
 \begin{geqns}
\mbox
 {if $T_1 \bisim T_2$ and $T'_1 \bisim T'_2$, then
  $T_1 \parcs{\commf} T'_1 \bisim T_2 \parcs{\commf} T'_2$;}
\\
\mbox
{if $T_1 \bisim T_2$, then $\encap{H}(T_1) \bisim \encap{H}(T_2)$.}
 \end{geqns}
\end{ldispl}
\end{property}

Hence, a parallel composition of transition systems is bisimulation
equivalent to a parallel composition of transition systems obtained by
replacing the constituent transition systems by ones that are
bisimulation equivalent.
This property is actually what justifies such replacements.
It underlies many techniques for the analysis of process behaviour.

Parallel composition and encapsulation of transition systems also 
preserve trace equivalence.
\begin{property}[Preservation of trace equivalence]%
\index{trace equivalence!preservation of}
Let $T_1$ and $T_2$ be transition systems with $A$ as set of actions,
let $T'_1$ and $T'_2$ be transition systems with $A'$ as set of
actions, and 
let $\commf$ be a communication function on a set  of actions that
includes $A \union A'$.
Then the following holds:
\begin{ldispl}
 \begin{geqns}
\mbox
 {if $T_1 \treqv T_2$ and $T'_1 \treqv T'_2$, then
  $T_1 \parcs{\commf} T'_1 \treqv T_2 \parcs{\commf} T'_2$;}
\\
\mbox
{if $T_1 \treqv T_2$, then 
$\encap{H}(T_1) \treqv \encap{H}(T_2)$.}
 \end{geqns}
\end{ldispl}
\end{property}

If an equivalence is preserved by an operation, the equivalence is
called a \emph{congruence}\index{congruence} with respect to the 
operation.
Let us now illustrate how the congruence properties can be used.
\begin{example}[Split and merge connections]%
\index{split connection}\index{merge connection}
\label{exa-connections}
We consider again the split and merge connections from
Examples~\ref{exa-split} and~\ref{exa-merge}.
Both kinds of connections are used as connections between nodes in 
networks.
Suppose that the behaviour of a particular network is described by
\begin{ldispl}
 \begin{geqns}
\encap{H}
(T_1 \parcs{\commf} \ldots \parcs{\commf} T_k \parcs{\commf}
 T_{k+1} \parcs{\commf} \ldots \parcs{\commf} T_n)\;,
 \end{geqns}
\end{ldispl}
where $T_1$, \ldots, $T_n$ are transition systems describing the
behaviour of the nodes and connections that occur in the network.
Suppose further that $T_k$ is the first transition system for a merge
connection given in Example~\ref{exa-merge} and that $T_k'$ is the
second transition system for a merge connection given in
Example~\ref{exa-merge}.
Recall that the two transition systems for a merge connection are
bisimulation equivalent.
Hence, replacement of $T_k$ by $T_k'$ yields a network that is
bisimulation equivalent to the original network.

Now, suppose instead that $T_k$ is the transition system for the split
connection given in Example~\ref{exa-split} and that $T_k'$ is the
transition system for the split-like connection given in
Example~\ref{exa-split}.
Recall that those two transition systems are trace equivalent, but not
bisimulation equivalent.
Hence, replacement of $T_k$ by $T_k'$ yields a network that is trace 
equivalent to the original network.
However, the networks are not bisimulation equivalent because the 
replacement causes a premature choice of an output port.
Such changes remain unnoticed under trace equivalence, because trace 
equivalence does not tell us anything about the stages at which the 
choices of different possibilities occur.
\end{example}
The following properties of parallel composition hold because of the 
conditions imposed on the communication function 
(see Def.~\ref{def-commf}).
\begin{property}[Commutativity and associativity of parallel
 composition]%
\index{parallel composition!commutativity and associativity of}
Let $T_1$, $T_2$ and $T_3$ be transition systems with $A_1$, $A_2$ and 
$A_3$, respectively, as set of actions.
Let $\commf$ be a communication function on a set of actions that 
includes $A_1 \union A_2 \union A_3$.
Then the following holds:
\begin{ldispl}
 \begin{geqns}
  T_1 \parcs{\commf} T_2 \bisim T_2 \parcs{\commf} T_1\;,
\\
  (T_1 \parcs{\commf} T_2) \parcs{\commf} T_3 
  \bisim 
  T_1 \parcs{\commf} (T_2 \parcs{\commf} T_3)\;.
 \end{geqns}
\end{ldispl}
\end{property}


\chapter{Abstraction}
\label{ch-abstraction}

Preferably, the design of a complex system starts from a description of
its behaviour at a high level of abstraction, i.e.\ a description 
serving as a specification of the system to be developed, and ends in 
a description of the behaviour at a low level of abstraction together
with a proof that the behaviour described at the start is essentially
the same as the behaviour described at the end after abstraction from
actions that have been added during the design process.
This chapter deals with this issue of abstraction by introducing the
notions of abstraction from internal actions and branching bisimulation 
equivalence.
First of all, we explain informally what abstraction from internal 
actions is and what branching bisimulation equivalence is, and give a 
simple example of their use in comparing descriptions of process 
behaviour (Sect.~\ref{sect-abstraction-informal}).
After that, we define the notions of abstraction from internal actions
and branching bisimulation equivalence in a mathematically precise way
(Sect.~\ref{sect-abstraction-formal}).
We also use abstraction from internal actions and branching 
bisimulation equivalence to show that a merge connection with a 
feedback wire behaves as a sink 
(Sect.~\ref{sect-abstraction-merge-and-wire}), 
and to show that the simple data communication protocol from 
Sect.~\ref{sect-interaction-abp} behaves as a buffer of capacity one 
(Sect.~\ref{sect-abstraction-abp}).
For the interested reader, we define the notions of abstraction from 
internal actions and branching bisimulation equivalence for nets
(Sect.~\ref{sect-abstraction-conn-nets}).
Finally, we look at some miscellaneous issues
(Sect.~\ref{sect-abstraction-misc}).

\section{Informal explanation}
\label{sect-abstraction-informal}

Abstraction from internal actions is an important notion.  
Frequently, the behaviour of a system is first described at a high 
level of abstraction, and then as a system composed of interacting 
components.  
It should be shown that the two descriptions are equivalent after
abstraction from actions added for the interactions between the
components.
The need for abstraction from certain actions became already apparent
in the preceding chapter, while analyzing systems described using
transition systems.

Abstraction from internal actions is a means to express that certain
actions must be considered to be unobservable.
It turns actions from a certain set into a special action, denoted by
$\tau$, which is called the silent step.
Unlike other actions, the act of performing a silent step is
considered to be unobservable. 
Let us give an example of the use of abstraction.
\begin{example}[Bounded buffers]\index{buffer!bounded}
\label{exa-bbuffer-abstract}
We consider again the system composed of two bounded buffers from
Example~\ref{exa-bbuffer-parallel}.
In that example, parallel composition and encapsulation of the two
buffers, buffer~1 and buffer~2, resulted in the following transition
system.
As states, we have pairs $\tup{\sigma_1,\sigma_2}$ where $\sigma_i$
($i = 1,2$) is a sequence of data of which the length is not greater 
than $l_i$. 
As initial state, we have $\tup{\epsilon,\epsilon}$.
As actions, we have $\kw{add}_1(d)$, $\kw{rem}_2(d)$ and $\kw{trf}(d)$
for each datum $d$.
As transitions, we have the following:
\begin{iteml}
\item
for each datum $d$ and each state $\tup{\sigma_1,\sigma_2}$ with the 
length of $\sigma_1$ less than $l_1$, a transition 
$\astep{\tup{\sigma_1,\sigma_2}}{\kw{add}_1(d)}
       {\tup{d\, \sigma_1,\sigma_2}}$; 
\item
for each datum $d$ and each state $\tup{\sigma_1,\sigma_2\, d}$, a 
transition 
$\astep{\tup{\sigma_1,\sigma_2\, d}}{\kw{rem}_2(d)}
       {\tup{\sigma_1,\sigma_2}}$;
\item
for each datum $d$ and each state $\tup{\sigma_1\, d,\sigma_2}$ with 
the length of $\sigma_2$ less than $l_2$, a 
transition 
$\astep{\tup{\sigma_1\, d,\sigma_2}}{\kw{trf}(d)}
       {\tup{\sigma_1,d\, \sigma_2}}$.
\end{iteml}
At the end of Example~\ref{exa-bbuffer-parallel}, there was a need to
abstract from the internal transfer actions $\kw{trf}(d)$.
The following transition system is the result of abstraction from the
actions $\kw{trf}(d)$ for $d \in D$.
We have the same states as before.
As actions, we have $\kw{add}_1(d)$ and $\kw{rem}_2(d)$ for each datum
$d$.
As transitions, we have the following:
\begin{iteml}
\item
for each datum $d$ and each state $\tup{\sigma_1,\sigma_2}$ with the 
length of $\sigma_1$ less than $l_1$, a transition 
$\astep{\tup{\sigma_1,\sigma_2}}{\kw{add}_1(d)}
       {\tup{d\, \sigma_1,\sigma_2}}$; 
\item
for each datum $d$ and each state $\tup{\sigma_1,\sigma_2\, d}$, a 
transition 
$\astep{\tup{\sigma_1,\sigma_2\, d}}{\kw{rem}_2(d)}
       {\tup{\sigma_1,\sigma_2}}$;
\item
for each datum $d$ and each state $\tup{\sigma_1\, d,\sigma_2}$ with 
the length of $\sigma_2$ less than $l_2$, a 
transition 
$\astep{\tup{\sigma_1\, d,\sigma_2}}{\tau}
       {\tup{\sigma_1,d\, \sigma_2}}$.
\end{iteml}
This transition system is represented graphically in 
Fig.~\ref{fig-bbuffer-abstr} for the case where $l_1 = l_2 = 1$ and 
the only data involved are the natural numbers $0$ and $1$.
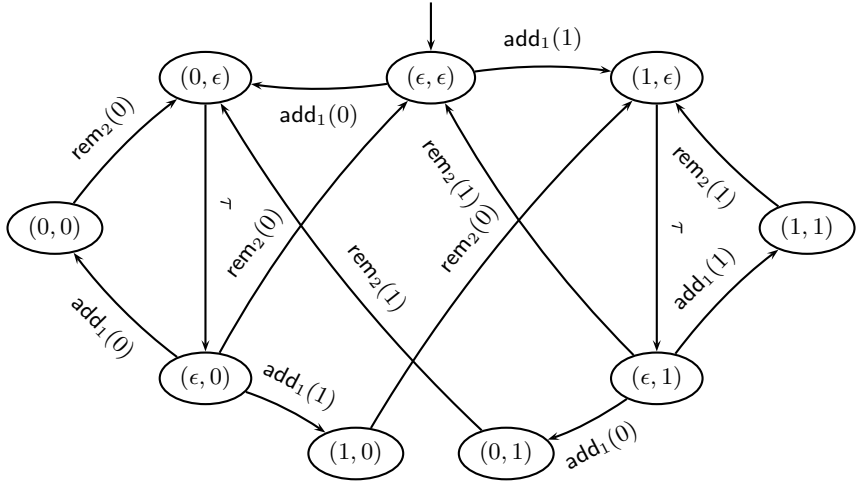
\begin{figure}
\begin{pspicture}(0,0)(12,6)

 \psset{arrows=->}

 \pnode(6,6){S}
 \rput(6,5){\ovalnode{ee}{$(\epsilon,\epsilon)$}}
 \rput(3,5){\ovalnode{0e}{$(0,\epsilon)$}}
 \rput(9,5){\ovalnode{1e}{$(1,\epsilon)$}}
 \rput(3,1){\ovalnode{e0}{$(\epsilon,0)$}}
 \rput(9,1){\ovalnode{e1}{$(\epsilon,1)$}}
 \rput(1,3){\ovalnode{00}{$(0,0)$}}
 \rput(7,0){\ovalnode{01}{$(0,1)$}}
 \rput(5,0){\ovalnode{10}{$(1,0)$}}
 \rput(11,3){\ovalnode{11}{$(1,1)$}}

 \ncline{S}{ee}

 \ncarc{ee}{0e}\naput[nrot=:180]{$\kw{add}_1(0)$}
 \ncarc{e0}{00}\naput[nrot=:180]{$\kw{add}_1(0)$}
 \ncarc{e1}{01}\naput[nrot=:180]{$\kw{add}_1(0)$}
 \ncarc{ee}{1e}\naput[nrot=:0]{$\kw{add}_1(1)$}
 \ncarc{e0}{10}\naput[nrot=:0]{$\kw{add}_1(1)$}
 \ncarc{e1}{11}\naput[nrot=:0]{$\kw{add}_1(1)$}

 \ncarc{e0}{ee}\naput[nrot=:0]{$\kw{rem}_2(0)\qquad\qquad$}
 \ncarc{00}{0e}\naput[nrot=:0]{$\kw{rem}_2(0)$}
 \ncarc{10}{1e}\naput[nrot=:0]{$\kw{rem}_2(0)$}
 \ncarc{e1}{ee}\naput[nrot=:180]{$\kw{rem}_2(1)\qquad\qquad\qquad\qquad$}
 \ncarc{01}{0e}\nbput[nrot=:180]{$\qquad\kw{rem}_2(1)$}
 \ncarc{11}{1e}\naput[nrot=:180]{$\kw{rem}_2(1)$}

 \ncline{0e}{e0}\naput[nrot=:0]{$\tau\qquad$}
 \ncline{1e}{e1}\naput[nrot=:0]{$\tau$}

\end{pspicture}
\caption{Transition system for abstraction of two parallel bounded 
  buffers}
\label{fig-bbuffer-abstr}
\end{figure}
\end{example}

As mentioned above, the act of performing a silent step is considered
to be unobservable.
However, the act of performing a silent step can sometimes be inferred
because a process may proceed as a different process after performing a
silent step.
In other words, the capabilities of a transition system may change by 
performing a silent step.
Let us look at an example of this phenomenon.
\begin{example}[Non-inert silent step]\index{silent step!non-inert}
\label{exa-tau-1}
We consider the following two transition systems, of which the second
is actually a split connection (see Example~\ref{exa-split}).
We assume a set of data $D$.
As actions of both transition systems, we have $\kw{r}_1(d)$,
$\kw{s}_2(d)$ and $\kw{s}_3(d)$ for each $d \in D$.
As states of the first transition system, we have pairs $\tup{d,i}$,
where $d \in D \union \set{\und}$ and $i \in \set{0,1,2}$, with 
$\tup{\und,0}$ as initial state.
As transitions of the first transition system, we have the following:
\begin{iteml}
\item
for each $d \in D$:
\begin{iteml}
\item
a transition
$\astep{\tup{\und,0}}{\kw{r}_1(d)}{\tup{d,1}}$, 
\item
a transition
$\astep{\tup{d,1}}{\kw{s}_2(d)}{\tup{\und,0}}$, 
\item
a transition
$\astep{\tup{d,1}}{\tau}{\tup{d,2}}$, 
\item
a transition
$\astep{\tup{d,2}}{\kw{s}_3(d)}{\tup{\und,0}}$. 
\end{iteml}
\end{iteml}
As states of the second transition system, we have pairs $\tup{d,i}$,
where $d \in D \union \set{\und}$ and $i \in \set{0,1}$, with 
$\tup{\und,0}$ as initial state.
As transitions of the second transition system, we have the following:
\begin{iteml}
\item
for each $d \in D$:
\begin{iteml}
\item
a transition
$\astep{\tup{\und,0}}{\kw{r}_1(d)}{\tup{d,1}}$, 
\item
a transition
$\astep{\tup{d,1}}{\kw{s}_2(d)}{\tup{\und,0}}$, 
\item
a transition
$\astep{\tup{d,1}}{\kw{s}_3(d)}{\tup{\und,0}}$. 
\end{iteml}
\end{iteml}
The transition systems given in this example are represented
graphically in Fig.~\ref{fig-tau-1}, for the case where 
$D = \set{0,1}$.
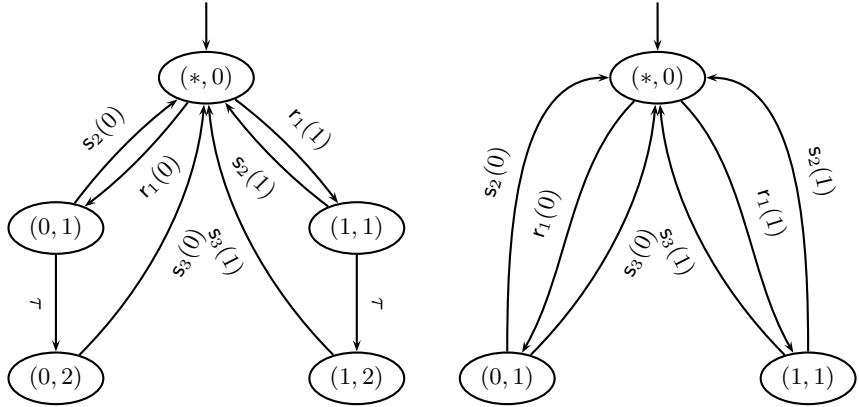
\begin{figure}
\begin{pspicture}(0,0.5)(12,6)

 \psset{arrows=->}

 \pnode(3,6){S}
 \rput(3,5){\ovalnode{*0}{$(*,0)$}}
 \rput(1,3){\ovalnode{01}{$(0,1)$}}
 \rput(5,3){\ovalnode{11}{$(1,1)$}}
 \rput(1,1){\ovalnode{02}{$(0,2)$}}
 \rput(5,1){\ovalnode{12}{$(1,2)$}}

 \ncline{S}{*0}

 \ncarc{*0}{01}\naput[nrot=:180]{$\kw{r}_1(0)$}
 \ncarc{01}{*0}\naput[nrot=:0]{$\kw{s}_2(0)$}
 \ncline{01}{02}\nbput[nrot=:0]{$\tau$}
 \nccurve[angleA=45,angleB=265]{02}{*0}\nbput[nrot=:0]{$\kw{s}_3(0)$}

 \ncarc{*0}{11}\naput[nrot=:0]{$\kw{r}_1(1)$}
 \ncarc{11}{*0}\naput[nrot=:180]{$\kw{s}_2(1)$}
 \ncline{11}{12}\naput[nrot=:0]{$\tau$}
 \nccurve[angleA=135,angleB=275]{12}{*0}\naput[nrot=:180]{$\kw{s}_3(1)$}

 \pnode(9,6){S}
 \rput(9,5){\ovalnode{*0}{$(*,0)$}}
 \rput(7,1){\ovalnode{01}{$(0,1)$}}
 \rput(11,1){\ovalnode{11}{$(1,1)$}}

 \ncline{S}{*0}

 \nccurve[angleA=225,angleB=60]{*0}{01}\nbput[nrot=:180]{$\kw{r}_1(0)$}
 \nccurve[angleA=90,angleB=180]{01}{*0}\naput[nrot=:0]{$\kw{s}_2(0)$}
 \nccurve[angleA=45,angleB=265]{01}{*0}\nbput[nrot=:0]{$\kw{s}_3(0)$}

 \nccurve[angleA=315,angleB=120]{*0}{11}\naput[nrot=:0]{$\kw{r}_1(1)$}
 \nccurve[angleA=90,angleB=0]{11}{*0}\nbput[nrot=:180]{$\kw{s}_2(1)$}
 \nccurve[angleA=135,angleB=275]{11}{*0}\naput[nrot=:180]{$\kw{s}_3(1)$}
\end{pspicture}
\caption{Transition systems of Example~\protect\ref{exa-tau-1}}
\label{fig-tau-1}
\end{figure}
The first transition system has a state, viz.\ state $\tup{d,2}$,
in which it is able to perform action $\kw{s}_3(d)$ without being able
to perform action $\kw{s}_2(d)$ instead; whereas the second transition 
system does not have such a state.
This means the following for the observable behaviour of these 
transition system. 
In the case of the first transition system, after $\kw{r}_1(d)$ has 
been performed, two observations are possible.
The act of performing $\kw{s}_2(d)$ can be observed and, after $\tau$ 
has been performed, the act of performing $\kw{s}_3(d)$ can be 
observed.
However, before anything has been observed, it may have become 
impossible to observe the act of performing $\kw{s}_2(d)$.
In the case of the second transition system, it remains possible to
observe the act of performing $\kw{s}_2(d)$ so long as nothing has been
observed.
So the observable behaviour of the two transition systems differ.
\end{example}

The purpose of abstraction from internal actions is to be able to
identify transition systems that have the same observable behaviour.
The preceding example shows that an equivalence based on the idea to
simply leave out all unobservable actions does not work.
Still, in many cases, the act of performing a silent step cannot be
inferred, because the process concerned proceeds as the same process
after performing a silent step.
In such cases, we sometimes say that the silent step is inert.
Here is an example of an inert silent step.
\begin{example}[Inert silent step]\index{silent step!inert}
\label{exa-tau-2}
We consider the following two transition systems.
We assume a set of data $D$.
As actions of both transition system, we have $\kw{r}_1(d)$ and
$\kw{s}_2(d)$ for each $d \in D$.
As states of the first transition system, we have pairs $\tup{d,i}$,
where $d \in D \union \set{\und}$ and $i \in \set{0,1,2}$, with 
$\tup{\und,0}$ as initial state.
As transitions of the first transition system, we have the following:
\begin{iteml}
\item
for each $d \in D$:
\begin{iteml}
\item
a transition
$\astep{\tup{\und,0}}{\kw{r}_1(d)}{\tup{d,1}}$, 
\item
a transition
$\astep{\tup{d,1}}{\tau}{\tup{d,2}}$, 
\item
a transition
$\astep{\tup{d,2}}{\kw{s}_2(d)}{\tup{\und,0}}$. 
\end{iteml}
\end{iteml}
As states of the second transition system, we have pairs $\tup{d,i}$,
where $d \in D \union \set{\und}$ and $i \in \set{0,1}$, with 
$\tup{\und,0}$ as initial state.
As transitions of the second transition system, we have the following:
\begin{iteml}
\item
for each $d \in D$:
\begin{iteml}
\item
a transition
$\astep{\tup{\und,0}}{\kw{r}_1(d)}{\tup{d,1}}$, 
\item
a transition
$\astep{\tup{d,1}}{\kw{s}_2(d)}{\tup{\und,0}}$. 
\end{iteml}
\end{iteml}
The transition systems given in this example are represented
graphically in Fig.~\ref{fig-tau-2}, for the case where 
$D = \set{0,1}$.
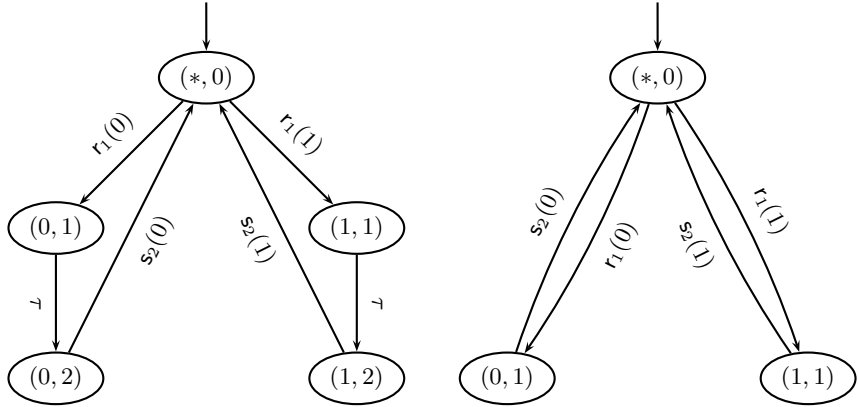
\begin{figure}
\begin{pspicture}(0,0.5)(12,6)

 \psset{arrows=->}

 \pnode(3,6){S}
 \rput(3,5){\ovalnode{*0}{$(*,0)$}}
 \rput(1,3){\ovalnode{01}{$(0,1)$}}
 \rput(5,3){\ovalnode{11}{$(1,1)$}}
 \rput(1,1){\ovalnode{02}{$(0,2)$}}
 \rput(5,1){\ovalnode{12}{$(1,2)$}}

 \ncline{S}{*0}

 \ncline{*0}{01}\nbput[nrot=:180]{$\kw{r}_1(0)$}
 \ncline{01}{02}\nbput[nrot=:0]{$\tau$}
 \ncline{02}{*0}\nbput[nrot=:0]{$\kw{s}_2(0)$}

 \ncline{*0}{11}\naput[nrot=:0]{$\kw{r}_1(1)$}
 \ncline{11}{12}\naput[nrot=:0]{$\tau$}
 \ncline{12}{*0}\naput[nrot=:180]{$\kw{s}_2(1)$}

 \pnode(9,6){S}
 \rput(9,5){\ovalnode{*0}{$(*,0)$}}
 \rput(7,1){\ovalnode{01}{$(0,1)$}}
 \rput(11,1){\ovalnode{11}{$(1,1)$}}

 \ncline{S}{*0}

 \ncarc{*0}{01}\naput[nrot=:180]{$\kw{r}_1(0)$}
 \ncarc{01}{*0}\naput[nrot=:0]{$\kw{s}_2(0)$}

 \ncarc{*0}{11}\naput[nrot=:0]{$\kw{r}_1(1)$}
 \ncarc{11}{*0}\naput[nrot=:180]{$\kw{s}_2(1)$}

\end{pspicture}
\caption{Transition systems of Example~\protect\ref{exa-tau-2}}
\label{fig-tau-2}
\end{figure}
Initially, only the act of performing $\kw{r}_1(d)$ can be observed.
After this has been observed, only the act of performing $\kw{s}_2(d)$ 
can be observed.
There is no way to infer the act of performing the silent step in
between.
So the observable behavior of these transition systems is the same.
\end{example}

What we understand from the preceding two examples is that a silent
step can only be left out if no capabilities get lost by performing
it.
According to this understanding, we adapt the notion of bisimulation 
equivalence as follows. 
Two transition systems $T$ and $T'$ are branching bisimulation
equivalent if their states can be related such that:
\begin{iteml}
\item
the initial states are related; 
\item
if states $s_1$ and $s_1'$ are related and in $T$ a transition with
label $a$ is possible from $s_1$ to some $s_2$, then 
\begin{iteml}
\item
either $a$ is the silent step and $s_2$ and $s_1'$ are related,
\item
or in $T'$ a transition with label $a$ is possible from some $s_1''$ 
to some $s_2'$ such that a generalized transition with a sequence of 
zero or more silent steps as label is possible from $s_1'$ to $s_1''$, 
$s_1$ and $s_1''$ are related, and $s_2$ and $s_2'$ are related;
\end{iteml}
\item
likewise, with the role of $T$ and $T'$ reversed.
\end{iteml}
We could have required $s_1$ to be related to all states between
$s_1'$ and $s_1''$ as well, but that turns out to be equivalent.
Let us return for a while to the preceding two examples. 
\begin{example}[Non-inert silent step]\index{silent step!non-inert}
\label{exa-tau-3}
We consider again the transition systems of Example~\ref{exa-tau-1}.
Are those transition systems identified by branching bisimulation
equivalence?
No, they are not. 
In order to be able to relate, as required, the state $\tup{\und,0}$ of
the first transition system to the state $\tup{\und,0}$ of the second
transition system, the states $\tup{d,1}$ and $\tup{d,2}$ ($d \in D$)
of the first transition system have to be related to states of the
second transition system as well.
However, we cannot relate state $\tup{d,2}$ because the second
transition system has no state from which only a transition with label
$\kw{s}_3(d)$ is possible.
\end{example}
\begin{example}[Inert silent step]\index{silent step!inert}
\label{exa-tau-4}
We also consider again the transition systems of 
Example~\ref{exa-tau-2}.
Are those transition systems identified by branching bisimulation
equivalence?
Yes, they are: relate state $\tup{\und,0}$ of the first transition
system to state $\tup{\und,0}$ of the second transition system, and for
each $d \in D$, relate the states $\tup{d,1}$ and $\tup{d,2}$ of the 
first transition system to state $\tup{d,1}$ of the second transition 
system.
In this way, the states of the two transition systems are related as
required for branching bisimulation equivalence.
\end{example}

\section{Formal definitions}
\label{sect-abstraction-formal}

With the previous section, we have prepared the way for the formal
definitions of the notions of abstraction from internal actions and
branching bisimulation equivalence.
However, we have to adapt the definitions of the notions of a transition 
system, a communication function, parallel composition and encapsulation 
from Chaps.~\ref{ch-basics} and~\ref{ch-interaction} to the presence of 
the silent step first.
In the adapted definitions, we write $A_\tau$ for 
$A \union \set{\tau}$.

\begin{definition}[Transition system]\index{transition system}
A \emph{transition system} $T$ is a quadruple $\tup{S,A,\step{},s_0}$ 
where
\begin{iteml}
\item
$S$ is a set of \emph{states}\index{state};
\item
$A$ is a set of \emph{actions}\index{action};
\item 
${\step{}} \subseteq S \x A_\tau \x S$ is a set of \emph{transitions}%
\index{transition};
\item
$s_0 \in S$ is the \emph{initial state}\index{state!initial}.
\end{iteml}
The set ${\gstep{}{}{}} \subseteq S \x \seqof{A} \x S$ of 
\emph{generalized transitions}\index{transition!generalized} of $T$ is 
the smallest subset of $S \x \seqof{A} \x S$ satisfying:
\begin{iteml}
\item
$\astep{s}{\epsilon}{s}$ for each $s \in S$;
\item
if $\astep{s}{\tau}{s'}$, then $\gstep{s}{\epsilon}{s'}$;
\item
if $\astep{s}{a}{s'}$, then $\gstep{s}{a}{s'}$;
\item
if $\gstep{s}{\sigma}{s'}$ and $\gstep{s'}{\sigma'}{s''}$, then 
$\gstep{s}{\sigma\, \sigma'}{s''}$.
\end{iteml}
A state $s \in S$ is called a \emph{reachable}\index{state!reachable} 
state of $T$ if there is a $\sigma \in \seqof{A}$ such that 
${\gstep{s_0}{\sigma}{s}}$.
A state $s \in S$ is called a \emph{terminal}\index{state!terminal} 
state of $T$ if there is no $a \in A$ and $s' \in S$ such that 
$\astep{s}{a}{s'}$.
\end{definition}
Notice that transitions labeled with the silent step may be included in
the set of transitions of a transition system, although the silent step
is never included in the set of actions.
\begin{definition}[Communication function]\index{communication function}
\label{def-commf-tau}
Let $A$ be a set of actions.
A \emph{communication function} on $A$ is a partial function 
$\funct{\commf}{A_\tau \x A_\tau}{A}$ satisfying for 
$a,b,c \in A_\tau$:
\begin{iteml}
\item
$\commf(a,\tau)$ and $\commf(\tau,a)$ are undefined;
\item
if $\commf(a,b)$ is defined, 
then $\commf(b,a)$ is defined and $\commf(a,b) = \commf(b,a)$;
\item
if $\commf(a,b)$ and $\commf(\commf(a,b),c)$ are defined,
then $\commf(b,c)$ and $\commf(a,\commf(b,c))$ are defined and
$\commf(\commf(a,b),c) = \commf(a,\commf(b,c))$.
\end{iteml}
\end{definition}
Notice that we consider the silent step to be an action that cannot be  
performed synchronously with other actions.
The reason for this is that it would otherwise be observable.
\begin{definition}[Parallel composition]\index{parallel composition}
Let $T = \tup{S,A,\step{},s_0}$ and $T' = \tup{S',A',\step{}',s'_0}$ be 
transition systems.
Let $\commf$ be a communication function on a set of actions that
includes $A \union A'$.
The \emph{parallel composition} of $T$ and $T'$ under $\commf$,
written $T \parcs{\commf} T'$, is the transition system 
$\tup{S'',A'',\step{}'',s''_0}$ where
\begin{iteml}
\item
$S'' = S \x S'$;
\item
$A'' = 
A \union A' \union 
\set{\commf(a,a') \where 
     a \in A, a' \in A', \commf(a,a') \;\mathrm{is}\;\mathrm{defined}}$;
\item
$\step{}''$ is the smallest subset of $S'' \x {A''}_\tau \x S''$ such 
that: 
\begin{iteml}
\item
if $\astep{s_1}{a}{s_2}$ and $s' \in S'$, then
$\asteppp{\tup{s_1,s'}}{a}{\tup{s_2,s'}}$;
\item
if $\astepp{s'_1}{b}{s'_2}$ and $s \in S$, then
$\asteppp{\tup{s,s'_1}}{b}{\tup{s,s'_2}}$;
\item
if $\astep{s_1}{a}{s_2}$, $\astepp{s'_1}{b}{s'_2}$ and 
$\commf(a,b)$ is defined, then
$\asteppp{\tup{s_1,s'_1}}{\commf(a,b)}{\tup{s_2,s'_2}}$;
\end{iteml}
\item
$ s''_0 = \tup{s_0,s'_0}$.
\end{iteml}
\end{definition}
\begin{definition}[Encapsulation]\index{encapsulation}
Let $T = \tup{S,A,\step{},s_0}$ be a transition system.
Let $H \subseteq A$.
The \emph{encapsulation} of $T$ with respect to $H$, 
written $\encap{H}(T)$, is the transition system 
$\tup{S',A',\step{}',s_0}$ where
\begin{iteml}
\item
$S' = 
 \set{s \where 
      \Exists{\sigma \in \seqof{(A \diff H)}}{\gstep{s_0}{\sigma}{s}}}$;
\item
$A' = 
 \set{a \in A \diff H \where 
      \Exists{s_1,s_2 \in S'}{\astep{s_1}{a}{s_2}}}$;
\item
$\step{}'$ is the smallest subset of $S' \x A'_\tau \x S'$ such that: 
\begin{iteml}
\item
if $\astep{s_1}{a}{s_2}$, $s_1 \in S'$ and $a \not\in H$, 
then $\astepp{s_1}{a}{s_2}$.
\end{iteml}
\end{iteml}
\end{definition}
The definitions of parallel composition and encapsulation are, just 
like the definition of transition system above, nothing else but simple
adjustments of the earlier definitions to cover transitions labeled 
with the silent step.
Here is an example of silent steps in parallel composition and 
encapsulation.
\begin{example}[Silent steps in parallel composition and encapsulation]%
\index{silent step}
\label{exa-tau-parc}
We consider the following two transition systems.
As actions of the first transition system, we have $\kw{s}_1(0)$ and
$\kw{s}_2(0)$.
As states of the first transition system, we have natural numbers 
$i \in \set{0,1,2,3}$, with $0$ as initial state.
As transitions of the first transition system, we have the following:
\begin{ldispl}
\astep{0}{\tau}{1},\;
\astep{0}{\tau}{2},\;
\astep{1}{\kw{s}_1(0)}{3},\;
\astep{2}{\kw{s}_2(0)}{3}.
\end{ldispl}
As actions of the second transition system, we have only $\kw{r}_1(0)$.
As states of the second transition system, we have natural numbers 
$i \in \set{0,1}$, with $0$ as initial state.
As transitions of the second transition system, we have the following:
\begin{ldispl}
\astep{0}{\kw{r}_1(0)}{1}.
\end{ldispl}
The following transition system is the result of the parallel 
composition of these two transitions systems and the subsequent 
encapsulation with respect to actions $\kw{s}_1(0)$, $\kw{r}_1(0)$, 
$\kw{s}_2(0)$ and $\kw{r}_2(0)$.
As actions, we have only $\kw{c}_1(0)$.
As states, we have the pairs $\tup{0,0}$, $\tup{1,0}$, $\tup{2,0}$ and 
$\tup{3,1}$, with $\tup{0,0}$ as initial state.
As transitions, we have the following:
\begin{ldispl}
\astep{\tup{0,0}}{\tau}{\tup{1,0}},\;
\astep{\tup{0,0}}{\tau}{\tup{2,0}},\;
\astep{\tup{1,0}}{\kw{c}_1(0)}{\tup{3,1}}.
\end{ldispl}
This transition system is capable of either first performing a
silent step, next performing a communication action, and by doing so 
getting in a terminal state or first performing a silent step and by 
doing so getting in a terminal state.
In the case where the send actions of the first transition system were
not preceded by a silent step, the resulting transition system would
only have the first alternative.
The parallel composition and the subsequent encapsulation are 
represented graphically in Fig.~\ref{fig-tau-parc} 
and~\ref{fig-tau-parc-enc}, respectively.
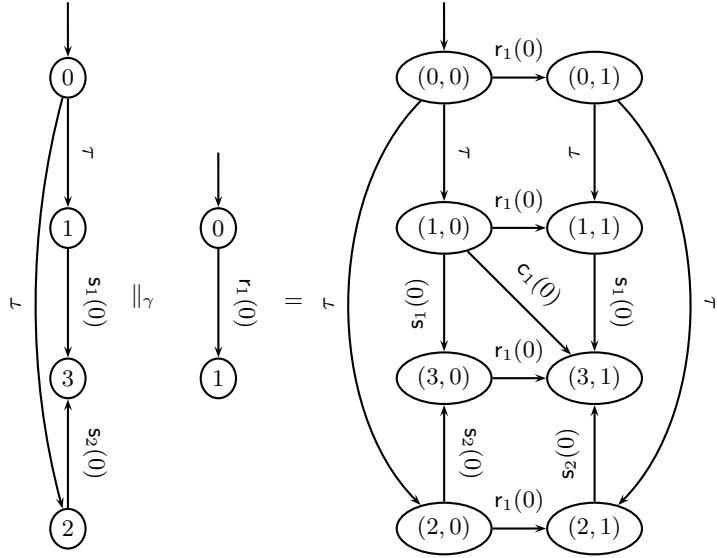
\begin{figure}
\[
\begin{array}{l}
\begin{pspicture}(0,0.5)(2,8)

 \psset{arrows=->}

 \pnode(1,8){S}
 \rput(1,7){\ovalnode{0}{$0$}}
 \rput(1,5){\ovalnode{1}{$1$}}
 \rput(1,1){\ovalnode{2}{$2$}}
 \rput(1,3){\ovalnode{3}{$3$}}

 \ncline{S}{0}
 \ncline{0}{1}\naput[nrot=:0]{$\tau$}
 \nccurve[angleA=255,angleB=105]{0}{2}\nbput[nrot=:180]{$\tau$}
 \ncline{1}{3}\naput[nrot=:0]{$\kw{s}_1(0)$}
 \ncline{2}{3}\nbput[nrot=:180]{$\kw{s}_2(0)$}
 \end{pspicture}

 \begin{pspicture}(0,0.5)(0,8)
 \rput(0,4){\Rnode{A}{$\parcs{\commf}$}}
 \end{pspicture}

\begin{pspicture}(2,0.5)(4,8)
 \psset{arrows=->}
 \pnode(3,6){S2}
 \rput(3,5){\ovalnode{01}{$0$}}
 \rput(3,3){\ovalnode{11}{$1$}}

 \ncline{S2}{01}
 \ncline{01}{11}\naput[nrot=:0]{$\kw{r}_1(0)$}
\end{pspicture}

\begin{pspicture}(0,0.5)(0,8)
 \rput(0,4){\Rnode{A}{$=$}}
\end{pspicture}

\begin{pspicture}(5,0.5)(11,8)

 \psset{arrows=->}
 \pnode(7,8){S}
 \rput(7,7){\ovalnode{00}{$(0,0)$}}
 \rput(7,5){\ovalnode{10}{$(1,0)$}}
 \rput(7,1){\ovalnode{20}{$(2,0)$}}
 \rput(7,3){\ovalnode{30}{$(3,0)$}}

 \rput(9,7){\ovalnode{01}{$(0,1)$}}
 \rput(9,5){\ovalnode{11}{$(1,1)$}}
 \rput(9,1){\ovalnode{21}{$(2,1)$}}
 \rput(9,3){\ovalnode{31}{$(3,1)$}}

 \ncline{S}{00}
 \ncline{00}{10}\naput[nrot=:0]{$\tau$}
 \nccurve[angleA=225,angleB=135]{00}{20}\nbput[nrot=:180]{$\tau$}
 \ncline{10}{30}\nbput[nrot=:180]{$\kw{s}_1(0)$}
 \ncline{20}{30}\nbput[nrot=:180]{$\kw{s}_2(0)$}

 \ncline{01}{11}\nbput[nrot=:180]{$\tau$}
 \nccurve[angleA=315,angleB=45]{01}{21}\naput[nrot=:0]{$\tau$}
 \ncline{11}{31}\naput[nrot=:0]{$\kw{s}_1(0)$}
 \ncline{21}{31}\naput[nrot=:0]{$\kw{s}_2(0)$}

 \ncline{00}{01}\naput[nrot=:0]{$\kw{r}_1(0)$}
 \ncline{10}{11}\naput[nrot=:0]{$\kw{r}_1(0)$}
 \ncline{20}{21}\naput[nrot=:0]{$\kw{r}_1(0)$}
 \ncline{30}{31}\naput[nrot=:0]{$\kw{r}_1(0)$}

 \ncline{10}{31}\naput[nrot=:0]{$\kw{c}_1(0)$}
\end{pspicture}
\end{array}
\]
\caption{Parallel composition of transition systems from
         Example~\protect\ref{exa-tau-parc}}
\label{fig-tau-parc}
\end{figure}
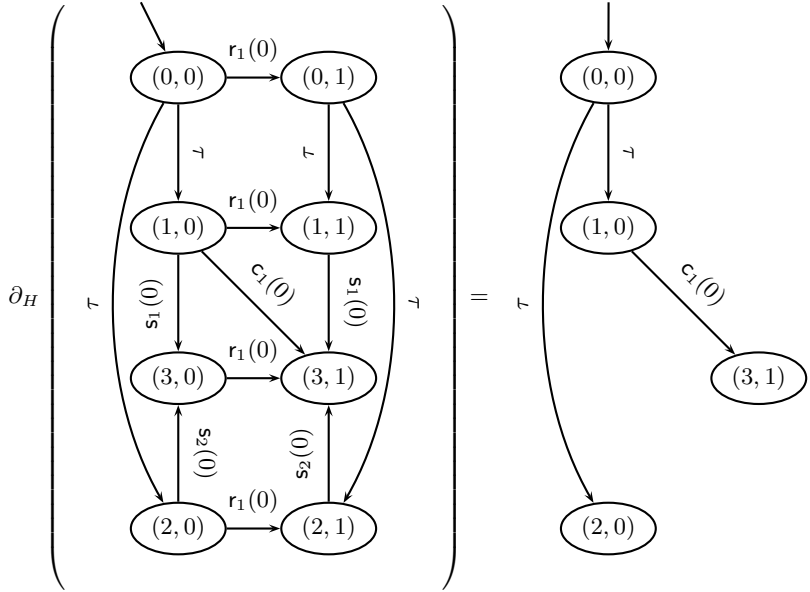
\begin{figure}
\[
\begin{array}{l}
\encap{H} \left(
\begin{pspicture}(0,4)(5,8)

 \psset{arrows=->}

 \pnode(1,8){S}
 \rput(1.5,7){\ovalnode{00}{$(0,0)$}}
 \rput(1.5,5){\ovalnode{10}{$(1,0)$}}
 \rput(1.5,1){\ovalnode{20}{$(2,0)$}}
 \rput(1.5,3){\ovalnode{30}{$(3,0)$}}

 \rput(3.5,7){\ovalnode{01}{$(0,1)$}}
 \rput(3.5,5){\ovalnode{11}{$(1,1)$}}
 \rput(3.5,1){\ovalnode{21}{$(2,1)$}}
 \rput(3.5,3){\ovalnode{31}{$(3,1)$}}

 \ncline{S}{00}
 \ncline{00}{10}\naput[nrot=:0]{$\tau$}
 \nccurve[angleA=240,angleB=120]{00}{20}\nbput[nrot=:180]{$\tau$}
 \ncline{10}{30}\nbput[nrot=:180]{$\kw{s}_1(0)$}
 \ncline{20}{30}\nbput[nrot=:180]{$\kw{s}_2(0)$}

 \ncline{01}{11}\nbput[nrot=:180]{$\tau$}
 \nccurve[angleA=300,angleB=60]{01}{21}\naput[nrot=:0]{$\tau$}
 \ncline{11}{31}\naput[nrot=:0]{$\kw{s}_1(0)$}
 \ncline{21}{31}\naput[nrot=:0]{$\kw{s}_2(0)$}

 \ncline{00}{01}\naput[nrot=:0]{$\kw{r}_1(0)$}
 \ncline{10}{11}\naput[nrot=:0]{$\kw{r}_1(0)$}
 \ncline{20}{21}\naput[nrot=:0]{$\kw{r}_1(0)$}
 \ncline{30}{31}\naput[nrot=:0]{$\kw{r}_1(0)$}

 \ncline{10}{31}\naput[nrot=:0]{$\kw{c}_1(0)$}
 \end{pspicture}

 \right) =

\begin{pspicture}(7.5,4)(12.5,8)

 \psset{arrows=->}

 \pnode(9,8){S}
 \rput(9,7){\ovalnode{00}{$(0,0)$}}
 \rput(9,5){\ovalnode{10}{$(1,0)$}}
 \rput(9,1){\ovalnode{20}{$(2,0)$}}
 \rput(11,3){\ovalnode{31}{$(3,1)$}}

 \ncline{S}{00}
 \ncline{00}{10}\naput[nrot=:0]{$\tau$}
 \nccurve[angleA=240,angleB=120]{00}{20}\nbput[nrot=:180]{$\tau$}
 \ncline{10}{31}\naput[nrot=:0]{$\kw{c}_1(0)$}

\end{pspicture}
\end{array}
\]
\caption{Encapsulation of transition system from
         Example~\protect\ref{exa-tau-parc}}
\label{fig-tau-parc-enc}
\end{figure}
\end{example}

Let us now look at the formal definitions of abstraction from internal
actions and branching bisimulation equivalence.
\begin{definition}[Abstraction]\index{abstraction}
Let $T = \tup{S,A,\step{},s_0}$ be a transition system.
Let $I \subseteq A$.
The \emph{abstraction} of $T$ with respect to $I$, 
written $\abstr{I}(T)$, is the transition system 
$\tup{S,A',\step{}',s_0}$ where
\begin{iteml}
\item
$A' = A \diff I$;
\item
$\step{}'$ is the smallest subset of $S \x {A'}_\tau \x S$ such that:
\begin{iteml}
\item
if $\astep{s_1}{a}{s_2}$ and $a \in I$, 
then $\astepp{s_1}{\tau}{s_2}$,
\item
if $\astep{s_1}{a}{s_2}$ and $a \not\in I$, 
then $\astepp{s_1}{a}{s_2}$.
\end{iteml}
\end{iteml}
\end{definition}
In the definition of branching bisimulation equivalence, we write 
$s \silent s'$ for ${\gstep{s}{\epsilon}{s'}}$. 
In other words, $s \silent s'$ indicates that state $s'$ is reachable 
from state $s$ by performing zero or more silent steps. 
\begin{definition}[Branching bisimulation]\index{branching bisimulation} 
Let $T = \tup{S,A,\step{},s_0}$ and $T' = \tup{S',A',\step{}',s_0'}$ 
be transition systems such that $A = A'$.
Then a \emph{branching bisimulation} $B$ between $T$ and $T'$ is a 
binary relation $B \subseteq S \x S'$ such that the following 
conditions hold:
\begin{enuml}
\item
$B(s_0,s_0')$;
\item 
whenever $B(s_1,s_1')$ and $\astep{s_1}{a}{s_2}$, 
then either $a = \tau$ and $B(s_2,s_1')$ or there are states 
$s_1'',s_2'$ such that $s_1' \silentp \astepp{s_1''}{a}{s_2'}$ and 
$B(s_1,s_1'')$ and $B(s_2,s_2')$;
\item 
whenever $B(s_1,s_1')$ and $\astepp{s_1'}{a}{s_2'}$, 
then either $a = \tau$ and $B(s_1,s_2')$ or there are states 
$s_1'',s_2$ such that $s_1 \silent \astep{s_1''}{a}{s_2}$ and 
$B(s_1',s_1'')$ and $B(s_2,s_2')$.
\end{enuml}
The two transition systems $T$ and $T'$ are 
\emph{branching bisimulation equivalent}%
\index{branching bisimulation equivalence}, 
written $T \bisim_\mathrm{b} T'$, if there exists a branching 
bisimulation $B$ between $T$ and $T'$.
A branching bisimulation between $T$ and $T$ is called a 
\emph{branching autobisimulation}\index{branching autobisimulation} on 
$T$.
\end{definition}
Here is an example of transition systems that are branching 
bisimulation equivalent.
\begin{example}[Bounded buffers]\index{buffer!bounded}
\label{exa-bbuffer-bbisim}
We consider again the transition system presented at the end of 
Example~\ref{exa-bbuffer-abstract} concerning abstraction of two 
encapsulated parallel bounded  buffers.
As states, we have pairs $\tup{\sigma_1,\sigma_2}$ where $\sigma_i$
($i = 1,2$) is a sequence of data of which the length of is not 
greater than $l_i$. 
As actions, we have $\kw{add}_1(d)$ and $\kw{rem}_2(d)$ for each datum
$d$.
As transitions, we have the following:
\begin{iteml}
\item
for each datum $d$ and each state $\tup{\sigma_1,\sigma_2}$ with the 
length of $\sigma_1$ less than $l_1$, a transition 
$\astep{\tup{\sigma_1,\sigma_2}}{\kw{add}_1(d)}
       {\tup{d\, \sigma_1,\sigma_2}}$; 
\item
for each datum $d$ and each state $\tup{\sigma_1,\sigma_2\, d}$, a 
transition 
$\astep{\tup{\sigma_1,\sigma_2\, d}}{\kw{rem}_2(d)}
       {\tup{\sigma_1,\sigma_2}}$;
\item
for each datum $d$ and each state $\tup{\sigma_1\, d,\sigma_2}$ with 
the length of $\sigma_2$ less than $l_2$, a 
transition 
$\astep{\tup{\sigma_1\, d,\sigma_2}}{\tau}
       {\tup{\sigma_1,d\, \sigma_2}}$.
\end{iteml}
Next, we consider the following transition system.
As states, we have sequences of data of which the length is not greater
than $l_1 + l_2$.
We have the same actions as before.
As transitions, we have the following:
\begin{iteml}
\item
for each datum $d$ and each state $\sigma$ with the length of $\sigma$ 
less than $l_1 + l_2$, 
a transition $\astep{\sigma}{\kw{add}_1(d)}{d\, \sigma}$; 
\item
for each datum $d$ and each state $\sigma\, d$, 
a transition $\astep{\sigma\, d}{\kw{rem}_2(d)}{\sigma}$.
\end{iteml}
These two transition systems are branching bisimulation equivalent.
Take the following relation:
\begin{ldispl}
 \begin{aeqns}
B & = &
\set{\tup{\tup{\sigma_1,\sigma_2},\sigma_1\,\sigma_2} \where 
     |\sigma_1| \leq l_1, |\sigma_2| \leq l_2}\;.
  \end{aeqns}
\end{ldispl}
It is easy to see that $B$ is a branching bisimulation.
The important point here is that, for each transition 
$\astep{\tup{\sigma_1\, d,\sigma_2}}
 {\tau}{\tup{\sigma_1,d\, \sigma_2}}$ of the first transition system,
the conditions imposed on a branching bisimulation permit that the 
states $\tup{\sigma_1\, d,\sigma_2}$ and $\tup{\sigma_1,d\, \sigma_2}$ 
are both related to the state $\sigma_1\, d\, \sigma_2$ of the second
transition system.
\end{example}

Just as bisimulation equivalence, branching bisimulation equivalence is 
preserved by parallel composition and encapsulation. 
Moreover, it is preserved by abstraction.
\begin{property}[Preservation of branching bisimulation equivalence]%
\index{branching bisimulation equivalence!preservation of}
Let $T_1$ and $T_2$ be transition systems with $A$ as set of actions,
let $T'_1$ and $T'_2$ be transition systems with $A'$ as set of
actions, and 
let $\commf$ be a communication function on a set of actions that 
includes $A \union A'$.
Then the following holds:
\begin{ldispl}
 \begin{geqns}
\mbox
 {if $T_1 \bisim_\mathrm{b} T_2$ and $T'_1 \bisim_\mathrm{b} T'_2$, 
  then
  $T_1 \parcs{\commf} T'_1 \bisim_\mathrm{b} T_2 \parcs{\commf} T'_2$;}
\\
\mbox
{if $T_1 \bisim_\mathrm{b} T_2$, 
 then $\encap{H}(T_1) \bisim_\mathrm{b} \encap{H}(T_2)$ 
 and $\abstr{I}(T_1) \bisim_\mathrm{b} \abstr{I}(T_2)$.}
 \end{geqns}
\end{ldispl}
\end{property}

The definition of the notion of determinacy of a transition system
has to be adapted to the presence of the silent step as well.
\begin{definition}[Determinacy]\index{transition system!determinacy of} 
Let $T = \tup{S,A,\step{},s_0}$ be a transition system.
Then  $T$ is \emph{determinate} if the following condition holds:
\begin{enuml}
\item[]
whenever $\gstep{s_0}{\sigma}{s}$ and $\gstep{s_0}{\sigma}{s'}$, then
there is a branching autobisimulation\index{branching autobisimulation} 
$B$ on $T$ such that $B(s,s')$.
\end{enuml}
\end{definition}

\section{Example: Merge connection with feedback wire}%
\index{merge connection}\index{feedback wire}
\label{sect-abstraction-merge-and-wire}

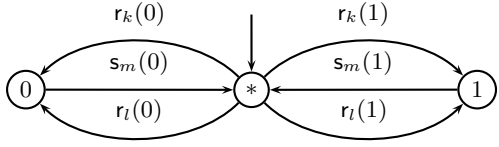
\begin{figure}
\begin{pspicture}(0,0)(7,3)

 \psset{arrows=->}

 \pnode(4,2){S}
 \rput(4,1){\circlenode{*}{$*$}}
 \rput(1,1){\circlenode{0}{$0$}}
 \rput(7,1){\circlenode{1}{$1$}}

 \ncline{S}{*}
 \ncline{0}{*}\naput{$\kw{s}_m(0)$}
 \nccurve[angleA=45,angleB=135]{*}{1}\naput{$\kw{r}_k(1)$}
 \nccurve[angleA=315,angleB=225]{*}{1}\naput{$\kw{r}_l(1)$}
 \ncline{1}{*}\nbput{$\kw{s}_m(1)$}
 \nccurve[angleA=135,angleB=45]{*}{0}\nbput{$\kw{r}_k(0)$}
 \nccurve[angleA=225,angleB=315]{*}{0}\nbput{$\kw{r}_l(0)$}

\end{pspicture}
\caption{Transition system for the merge connection}
\label{fig-merge-alt-rep}
\end{figure}

We consider again the merge connections from Example~\ref{exa-merge}.
Two transition systems describing the behaviour of a merge connection 
are given in that example.
For clearness' sake, the second one is given here again.
We assume a set of data $D$.
The behaviour of a merge connection with input ports $k$ and $l$ and 
output port $m$, $\nm{Merge}^{kl,m}$, is described by the following 
transition system.
As states, we have $\und$ and the data $d \in D$, with $\und$ as 
initial state.
As actions, we have $\kw{s}_i(d)$ and $\kw{r}_i(d)$ for $i = k,l,m$ and
$d \in D$.
As transitions, we have the following:
\begin{iteml}
\item
for each $d \in D$: 
$\astep{\und}{\kw{r}_k(d)}{d}$,
$\astep{\und}{\kw{r}_l(d)}{d}$,
$\astep{d}{\kw{s}_m(d)}{\und}$.
\end{iteml}
This transition system is represented graphically in 
Fig.~\ref{fig-merge-alt-rep} for the case where $D = \set{0,1}$.

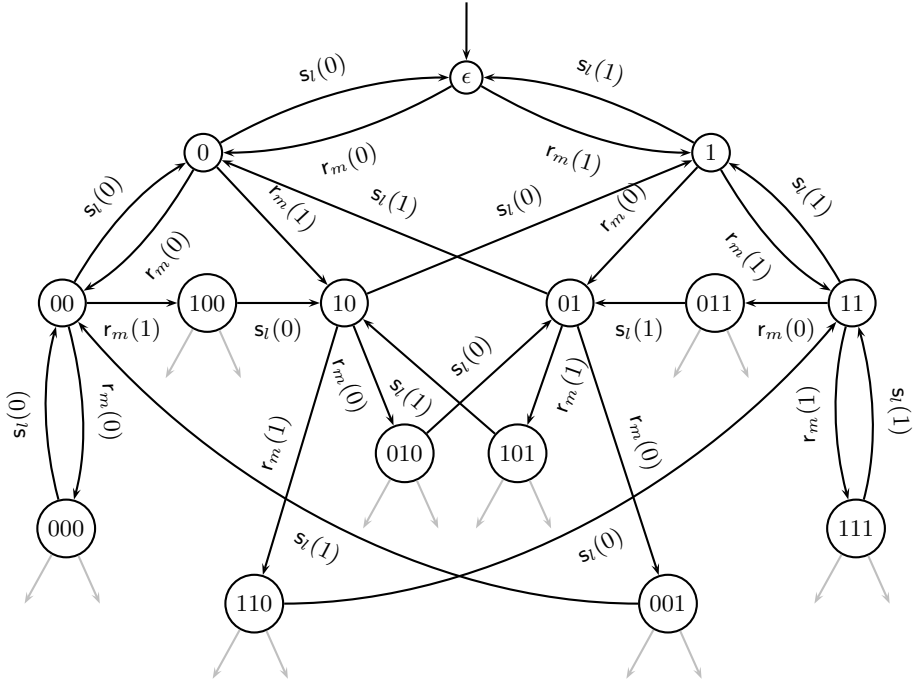
\begin{figure}
\begin{pspicture}(0,-2.5)(7,7)

 \psset{arrows=->}

 \pnode(5.875,7){S}
 \rput(5.875,6){\circlenode{e}{$\epsilon$}}
 \rput(2.375,5){\circlenode{0}{$0$}}
 \rput(9.125,5){\circlenode{1}{$1$}}
 \rput(0.5,3){\circlenode{00}{$00$}}
 \rput(4.25,3){\circlenode{10}{$10$}}
 \rput(7.25,3){\circlenode{01}{$01$}}
 \rput(11,3){\circlenode{11}{$11$}}

 \rput(0.5,0){\circlenode{000}{$000$}}
 \pnode(0,-1){E000l}
 \pnode(1,-1){E000r}
 \rput(2.375,3){\circlenode{100}{$100$}}
 \pnode(1.875,2){E100l}
 \pnode(2.875,2){E100r}
 \rput(5,1){\circlenode{010}{$010$}}
 \pnode(4.5,0){E010l}
 \pnode(5.5,0){E010r}
 \rput(3,-1){\circlenode{110}{$110$}}
 \pnode(2.5,-2){E110l}
 \pnode(3.5,-2){E110r}
 \rput(8.5,-1){\circlenode{001}{$001$}}
 \pnode(8,-2){E001l}
 \pnode(9,-2){E001r}
 \rput(6.5,1){\circlenode{101}{$101$}}
 \pnode(6,0){E101l}
 \pnode(7,0){E101r}
 \rput(9.125,3){\circlenode{011}{$011$}}
 \pnode(8.625,2){E011l}
 \pnode(9.625,2){E011r}
 \rput(11,0){\circlenode{111}{$111$}}
 \pnode(10.5,-1){E111l}
 \pnode(11.5,-1){E111r}

 \ncline{S}{e}
 \nccurve[angleA=210,angleB=0]{e}{0}\naput[nrot=:180]{$\kw{r}_m(0)$}
 \nccurve[angleA=330,angleB=180]{e}{1}\nbput[nrot=:0]{$\kw{r}_m(1)$}
 \nccurve[angleA=240,angleB=30]{0}{00}\naput[nrot=:180]{$\kw{r}_m(0)$}
 \ncline{0}{10}\naput[nrot=:0]{$\kw{r}_m(1)$}
 \ncline{1}{01}\nbput[nrot=:180]{$\kw{r}_m(0)$}
 \nccurve[angleA=300,angleB=150]{1}{11}\nbput[nrot=:0]{$\kw{r}_m(1)$}

 \nccurve[angleA=30,angleB=180]{0}{e}\naput[nrot=:0]{$\kw{s}_l(0)$}
 \nccurve[angleA=150,angleB=0]{1}{e}\nbput[nrot=:180]{$\kw{s}_l(1)$}
 \nccurve[angleA=60,angleB=210]{00}{0}\naput[nrot=:0]{$\kw{s}_l(0)$}
 \ncline{01}{0}\nbput[nrot=:180]{$\kw{s}_l(1)$}
 \ncline{10}{1}\naput[nrot=:0]{$\kw{s}_l(0)$}
 \nccurve[angleA=120,angleB=330]{11}{1}\nbput[nrot=:180]{$\kw{s}_l(1)$}

 \nccurve[angleA=285,angleB=75]{00}{000}\naput[nrot=:0]{$\kw{r}_m(0)$}
 \ncline{00}{100}\nbput[nrot=:0]{$\kw{r}_m(1)$}
 \ncline{01}{001}\naput[nrot=:0]{$\kw{r}_m(0)$}
 \ncline{01}{101}\naput[nrot=:180]{$\kw{r}_m(1)$}
 \ncline{10}{010}\nbput[nrot=:0]{$\kw{r}_m(0)$}
 \ncline{10}{110}\nbput[nrot=:180]{$\kw{r}_m(1)$}
 \ncline{11}{011}\naput[nrot=:180]{$\kw{r}_m(0)$}
 \nccurve[angleA=255,angleB=105]{11}{111}\nbput[nrot=:180]{$\kw{r}_m(1)$}

 \nccurve[angleA=105,angleB=255]{000}{00}\naput[nrot=:0]{$\kw{s}_l(0)$}
 \nccurve[angleA=180,angleB=310]{001}{00}\naput[nrot=:180]{$\kw{s}_l(1)$}
 \ncline{010}{01}\naput[nrot=:0]{$\kw{s}_l(0)$}
 \ncline{011}{01}\naput[nrot=:180]{$\kw{s}_l(1)$}
 \ncline{100}{10}\nbput[nrot=:0]{$\kw{s}_l(0)$}
 \ncline{101}{10}\naput[nrot=:180]{$\kw{s}_l(1)$}
 \nccurve[angleA=0,angleB=230]{110}{11}\nbput[nrot=:0]{$\kw{s}_l(0)$}
 \nccurve[angleA=75,angleB=285]{111}{11}\nbput[nrot=:180]{$\kw{s}_l(1)$}


\ncline[linecolor=lightgray]{000}{E000l}
\ncline[linecolor=lightgray]{100}{E100l}
\ncline[linecolor=lightgray]{110}{E110l}
\ncline[linecolor=lightgray]{010}{E010l}
\ncline[linecolor=lightgray]{101}{E101l}
\ncline[linecolor=lightgray]{001}{E001l}
\ncline[linecolor=lightgray]{011}{E011l}
\ncline[linecolor=lightgray]{111}{E111l}

\ncline[linecolor=lightgray]{000}{E000r}
\ncline[linecolor=lightgray]{100}{E100r}
\ncline[linecolor=lightgray]{110}{E110r}
\ncline[linecolor=lightgray]{010}{E010r}
\ncline[linecolor=lightgray]{101}{E101r}
\ncline[linecolor=lightgray]{001}{E001r}
\ncline[linecolor=lightgray]{011}{E011r}
\ncline[linecolor=lightgray]{111}{E111r}
\end{pspicture}
\caption{Transition system for the wire}
\label{fig-wire}
\end{figure}

Wires, which were not mentioned before, constitute another important 
kind of connection used between nodes in networks. 
A wire is reminiscent of a buffer with unbounded capacity.
The behaviour of a wire with input port $m$ and output port $l$,
$\nm{Wire}^{m,l}$, is described by the following transition system.
As states, we have all sequences $\sigma \in \seqof{D}$, with 
$\epsilon$ as initial state.
As actions, we have $\kw{r}_m(d)$ and $\kw{s}_l(d)$ for each $d \in D$.
As transitions of a wire, we have the following:
\begin{iteml}
\item
for each $d \in D$ and $\sigma \in \seqof{D}$: 
$\astep{\sigma}{\kw{r}_m(d)}{d\, \sigma}$, 
$\astep{\sigma\, d}{\kw{s}_l(d)}{\sigma}$.
\end{iteml}
This transition system is represented graphically in 
Fig.~\ref{fig-wire} for the case where $D = \set{0,1}$.%
\footnote
{In graphical representations of transition systems, we use grey tones 
 to indicate an infinite progression.
}

Let us look at the following transition system:
\begin{ldispl}
\begin{geqns}
\abstr{I}(\encap{H}(\nm{Merge}^{kl,m} \parcs{\commf} \nm{Wire}^{m,l}))
\end{geqns}
\end{ldispl}
where
\begin{ldispl}
\displstretch
 \begin{aeqns}
 H & = &
  \set{\kw{s}_i(d), \kw{r}_i(d) \where i \in \set{m,l}, d \in D}\;,
 \end{aeqns}
\end{ldispl}
\begin{ldispl}
\displstretch
 \begin{aeqns}
 I & = &
  \set{\kw{c}_i(d) \where i \in \set{m,l}, d \in D}
 \end{aeqns}
\end{ldispl}
and the communication function $\commf$ is defined in the standard way
for handshaking communication (see Sect.~\ref{sect-interaction-formal}),
i.e.\ such that
\begin{ldispl}
 \begin{geqns}
 \commf(\kw{s}_i(d),\kw{r}_i(d)) = \commf(\kw{r}_i(d),\kw{s}_i(d)) = 
 \kw{c}_i(d)
 \end{geqns}
\end{ldispl}
for all $d \in D$, and it is undefined otherwise.
Thus, the data delivered by the merge connection at port $m$ is feed
back to one of its input port, viz.\ $l$.

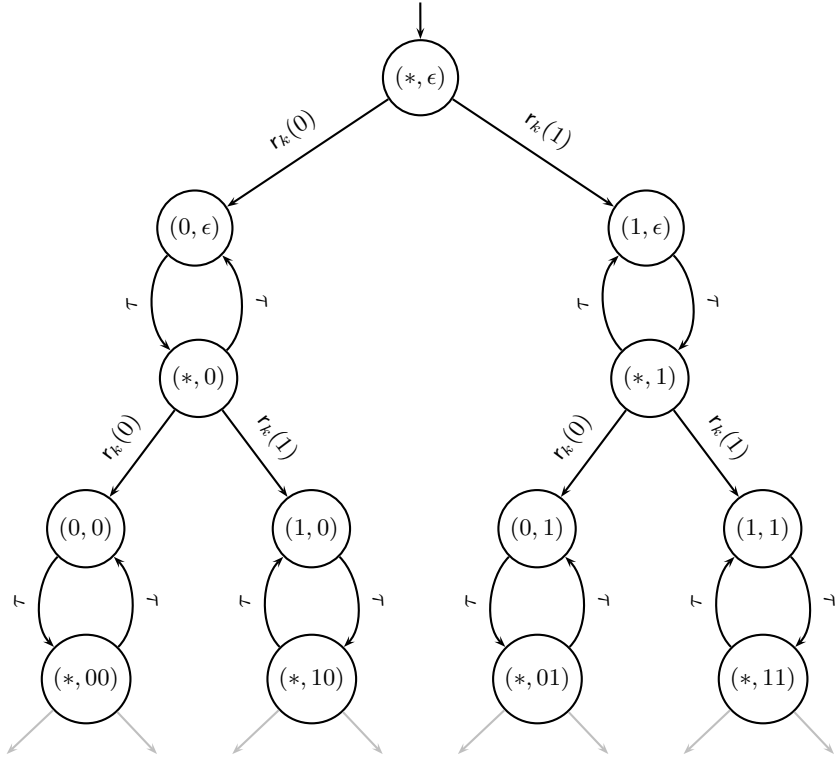
\begin{figure}
\begin{pspicture}(0,0)(8,10)

 \psset{arrows=->}

 \pnode(5.5,10){S}
 \rput(5.5,9){\circlenode{*e}{$(*,\epsilon)$}}
 \rput(2.5,7){\circlenode{0e}{$(0,\epsilon)$}}
 \rput(8.5,7){\circlenode{1e}{$(1,\epsilon)$}}
 \rput(2.5,5){\circlenode{*0}{$(*,0)$}}
 \rput(8.5,5){\circlenode{*1}{$(*,1)$}}
 \rput(1,3){\circlenode{00}{$(0,0)$}}
 \rput(4,3){\circlenode{10}{$(1,0)$}}
 \rput(7,3){\circlenode{01}{$(0,1)$}}
 \rput(10,3){\circlenode{11}{$(1,1)$}}
 \rput(1,1){\circlenode{*00}{$(*,00)$}}
 \rput(4,1){\circlenode{*10}{$(*,10)$}}
 \rput(7,1){\circlenode{*01}{$(*,01)$}}
 \rput(10,1){\circlenode{*11}{$(*,11)$}}
 \pnode(0,0){E1}
 \pnode(2,0){E2}
 \pnode(3,0){E3}
 \pnode(5,0){E4}
 \pnode(6,0){E5}
 \pnode(8,0){E6}
 \pnode(9,0){E7}
 \pnode(11,0){E8}

 \ncline{S}{*e}
 \ncline{*e}{0e}\nbput[nrot=:180]{$\kw{r}_k(0)$}
 \ncline{*e}{1e}\naput[nrot=:0]{$\kw{r}_k(1)$}
 \nccurve[angleA=225,angleB=135]{0e}{*0}\nbput[nrot=:180]{$\tau$}
 \nccurve[angleA=315,angleB=45]{1e}{*1}\naput[nrot=:0]{$\tau$}
 \nccurve[angleA=45,angleB=315]{*0}{0e}\nbput[nrot=:180]{$\tau$}
 \nccurve[angleA=135,angleB=225]{*1}{1e}\naput[nrot=:0]{$\tau$}
 \ncline{*0}{00}\nbput[nrot=:180]{$\kw{r}_k(0)$}
 \ncline{*0}{10}\naput[nrot=:0]{$\kw{r}_k(1)$}
 \ncline{*1}{01}\nbput[nrot=:180]{$\kw{r}_k(0)$}
 \ncline{*1}{11}\naput[nrot=:0]{$\kw{r}_k(1)$}
 \nccurve[angleA=225,angleB=135]{00}{*00}\nbput[nrot=:180]{$\tau$}
 \nccurve[angleA=315,angleB=45]{10}{*10}\naput[nrot=:0]{$\tau$}
 \nccurve[angleA=225,angleB=135]{01}{*01}\nbput[nrot=:180]{$\tau$}
 \nccurve[angleA=315,angleB=45]{11}{*11}\naput[nrot=:0]{$\tau$}
 \nccurve[angleA=45,angleB=315]{*00}{00}\nbput[nrot=:180]{$\tau$}
 \nccurve[angleA=135,angleB=225]{*10}{10}\naput[nrot=:0]{$\tau$}
 \nccurve[angleA=45,angleB=315]{*01}{01}\nbput[nrot=:180]{$\tau$}
 \nccurve[angleA=135,angleB=225]{*11}{11}\naput[nrot=:0]{$\tau$}

 \ncline[linecolor=lightgray]{*00}{E1}
 \ncline[linecolor=lightgray]{*10}{E3}
 \ncline[linecolor=lightgray]{*01}{E5}
 \ncline[linecolor=lightgray]{*11}{E7}
 \ncline[linecolor=lightgray]{*00}{E2}
 \ncline[linecolor=lightgray]{*10}{E4}
 \ncline[linecolor=lightgray]{*01}{E6}
 \ncline[linecolor=lightgray]{*11}{E8}

\end{pspicture}
\caption{Transition system for the merge connection with feedback wire}
\label{fig-merge-and-wire}
\end{figure}

Parallel composition, encapsulation and abstraction of the transition
systems $\nm{Merge}^{kl,m}$ and $\nm{Wire}^{m,l}$ as described above
results in the following transtion system.
As states, we have pairs $\tup{d,\sigma}$ where $d$ and $\sigma$ are
states of $\nm{Merge}^{kl,m}$ and $\nm{Wire}^{m,l}$, respectively.
As initial state, we have $\tup{\und,\epsilon}$.
As actions we have $\kw{r}_k(d)$ for each $d \in D$.
As transitions we have the following:
\begin{iteml}
\item
for each $d \in D$ and $\sigma \in \seqof{D}$:  
$\astep{\tup{\und,\sigma}}{\kw{r}_k(d)}{\tup{d,\sigma}}$,
$\astep{\tup{d,\sigma}}{\tau}{\tup{\und,d\, \sigma}}$, and
$\astep{\tup{\und,\sigma\, d}}{\tau}{\tup{d,\sigma}}$.
\end{iteml}
This transition system is represented graphically in 
Fig.~\ref{fig-merge-and-wire} for the case where $D = \set{0,1}$.

Let us also look at the following transition system.
As states, we have only $\und$.
Consequently, $\und$ is the initial state.
As actions we have $\kw{r}_k(d)$ for each $d \in D$.
As transitions we have the following:
\begin{iteml}
\item
for each $d \in D$, a transition $\astep{\und}{\kw{r}_k(d)}{\und}$.
\end{iteml}
This transition system described the behaviour of a special node in a
network, viz.\ a \emph{sink}.
A sink consumes data, but does not deliver it anywhere.

These two transition systems are branching bisimulation equivalent.
Take the following relation:
\begin{ldispl}
 \begin{aeqns}
B & = &
\set{\tup{\tup{d,\sigma},\und} \where 
     d \in D \union \set{\und},\sigma \in \seqof{D}}\;.
  \end{aeqns}
\end{ldispl}
It is easy to see that $B$ is a branching bisimulation.
The important point here is that, for each transition 
$\astep{\tup{d,\sigma}}{\tau}{\tup{\und,d\, \sigma}}$ of the first 
transition system, the conditions imposed on a branching bisimulation 
permit that the states $\tup{d,\sigma}$ and $\tup{\und,d\, \sigma}$ 
are both related to the state $\und$ of the second transition system;
and for each transition 
$\astep{\tup{\und,\sigma\, d}}{\tau}{\tup{d,\sigma}}$ of the first 
transition system, the conditions imposed on a branching bisimulation 
permit that the states $\tup{\und,d\, \sigma}$ and $\tup{d,\sigma}$ 
are both related to the state $\und$ of the second transition system.

\section{Example: Alternating bit protocol}%
\index{alternating bit protocol}
\label{sect-abstraction-abp}

We continue with the example of Sect.~\ref{sect-interaction-abp} 
concerning the ABP.
At the end of that section, we presented the transition system that
was the result of parallel composition and encapsulation of the 
transition systems of the sender $S$, the data transmission channel 
$K$, the acknowledgement transmission channel $L$ and the receiver $R$ 
as described earlier in that section. 

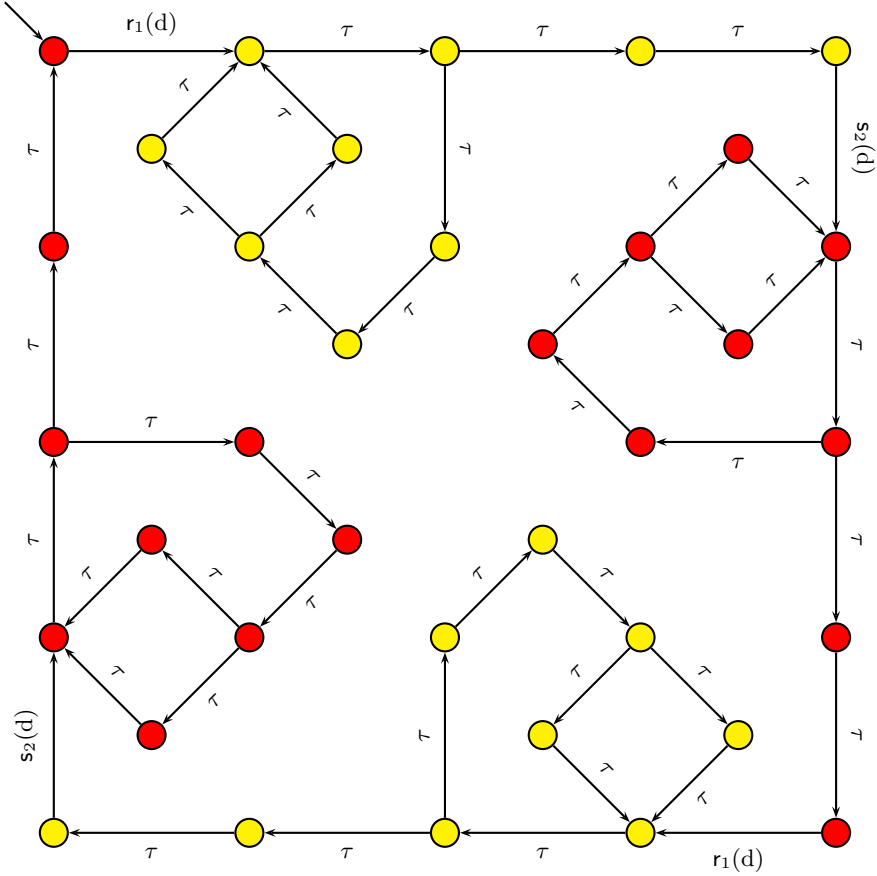
\begin{figure}
\begin{pspicture}(0.5,1.5)(12,13.5)

 \psset{arrows=->,radius=.2cm, xunit=1.3, yunit=1.3, runit=1.3,fillstyle=solid}

  \pnode(0.5,10.5){S}

  \Cnode[fillcolor=red](1,10){*00*0*0*00} 
  \Cnode[fillcolor=yellow](3,10){d01*0*0*00} 
  \Cnode[fillcolor=red](6,7){d01*0*0*10} 
  \Cnode[fillcolor=red](9,8){d02*0*0*02} 
  \Cnode[fillcolor=yellow](4,7){d02*0*0*12} 
  \Cnode[fillcolor=yellow](9,10){d02*0*0d01} 
  \Cnode[fillcolor=red](9,6){d02*001*10} 
 \Cnode[fillcolor=red](9,4){d02*002*10} 
  \Cnode[fillcolor=red](7,6){d02*003*10} 
  \Cnode[fillcolor=yellow](3,8){d02*011*00} 
  \Cnode[fillcolor=yellow](4,9){d02*012*00} 
  \Cnode[fillcolor=yellow](2,9){d02*013*00} 
  \Cnode[fillcolor=yellow](5,10){d02d01*0*00} 
  \Cnode[fillcolor=red](7,8){d02d01*0*10} 
  \Cnode[fillcolor=yellow](7,10){d02d02*0*00} 
  \Cnode[fillcolor=red](8,7){d02d02*0*10}  
  \Cnode[fillcolor=yellow](5,8){d02d03*0*00} 
  \Cnode[fillcolor=red](8,9){d02d03*0*10} 

 \Cnode[fillcolor=red](9,2){*10*0*0*10} 
 \Cnode[fillcolor=red](4,5){d11*0*0*00} 
 \Cnode[fillcolor=yellow](7,2){d11*0*0*10} 
 \Cnode[fillcolor=yellow](6,5){d12*0*0*02} 
 \Cnode[fillcolor=red](1,4){d12*0*0*12} 
 \Cnode[fillcolor=yellow](1,2){d12*0*0d11} 
 \Cnode[fillcolor=yellow](7,4){d12*001*10} 
 \Cnode[fillcolor=yellow](6,3){d12*002*10} 
 \Cnode[fillcolor=yellow](8,3){d12*003*10} 
 \Cnode[fillcolor=red](1,6){d12*011*00} 
 \Cnode[fillcolor=red](1,8){d12*012*00} 
 \Cnode[fillcolor=red](3,6){d12*013*00} 
 \Cnode[fillcolor=red](3,4){d12d11*0*00} 
 \Cnode[fillcolor=yellow](5,2){d12d11*0*10} 
 \Cnode[fillcolor=red](2,3){d12d12*0*00} 
 \Cnode[fillcolor=yellow](3,2){d12d12*0*10} 
 \Cnode[fillcolor=red](2,5){d12d13*0*00} 
 \Cnode[fillcolor=yellow](5,4){d12d13*0*10} 

 \ncline{S}{*00*0*0*00}
 \ncline{*00*0*0*00}{d01*0*0*00}\naput[nrot=:0]{$\kw{r}_1(\rd)$}
 \ncline{*10*0*0*10}{d11*0*0*10}\naput[nrot=:180]{$\kw{r}_1(\rd)$}
 \ncline{d01*0*0*00}{d02d01*0*00}\naput[nrot=:0]{$\tau$}
 \ncline{d11*0*0*10}{d12d11*0*10}\naput[nrot=:180]{$\tau$}
 \ncline{d02d02*0*00}{d02*0*0d01}\naput[nrot=:0]{$\tau$}
 \ncline{d12d12*0*10}{d12*0*0d11}\naput[nrot=:180]{$\tau$}
 \ncline{d02*0*0d01}{d02*0*0*02}\naput[nrot=:0]{$\kw{s}_2(\rd)$}
 \ncline{d12*0*0d11}{d12*0*0*12}\naput[nrot=:0]{$\kw{s}_2(\rd)$}
 \ncline{d02*0*0*02}{d02*001*10}\naput[nrot=:0]{$\tau$}
 \ncline{d12*0*0*12}{d12*011*00}\naput[nrot=:0]{$\tau$}

 \ncline{d02*002*10}{*10*0*0*10}\naput[nrot=:0]{$\tau$}
 \ncline{d12*012*00}{*00*0*0*00}\naput[nrot=:0]{$\tau$}
 \ncline{d02d03*0*00}{d02*0*0*12}\naput[nrot=:180]{$\tau$} %
 \ncline{d12d13*0*10}{d12*0*0*02}\naput[nrot=:0]{$\tau$} %
 \ncline{d02*0*0*12}{d02*011*00}\naput[nrot=:180]{$\tau$}
 \ncline{d12*0*0*02}{d12*001*10}\naput[nrot=:0]{$\tau$}

 \ncline{d02*012*00}{d01*0*0*00}\naput[nrot=:180]{$\tau$}
 \ncline{d12*002*10}{d11*0*0*10}\naput[nrot=:0]{$\tau$}
 \ncline{d02*013*00}{d01*0*0*00}\naput[nrot=:0]{$\tau$}
 \ncline{d12*003*10}{d11*0*0*10}\naput[nrot=:180]{$\tau$}
 \ncline{d02*003*10}{d01*0*0*10}\naput[nrot=:180]{$\tau$}
 \ncline{d12*013*00}{d11*0*0*00}\naput[nrot=:0]{$\tau$}
 \ncline{d01*0*0*10}{d02d01*0*10}\naput[nrot=:0]{$\tau$}
 \ncline{d11*0*0*00}{d12d11*0*00}\naput[nrot=:180]{$\tau$}

 \ncline{d02d02*0*10}{d02*0*0*02}\naput[nrot=:0]{$\tau$}
 \ncline{d12d12*0*00}{d12*0*0*12}\nbput[nrot=:180]{$\tau$}
 \ncline{d02d03*0*10}{d02*0*0*02}\naput[nrot=:0]{$\tau$}
 \ncline{d12d13*0*00}{d12*0*0*12}\nbput[nrot=:180]{$\tau$}

 \ncline{d02*001*10}{d02*002*10}\naput[nrot=:0]{$\tau$}
 \ncline{d12*011*00}{d12*012*00}\naput[nrot=:0]{$\tau$}
 \ncline{d02d01*0*00}{d02d02*0*00}\naput[nrot=:0]{$\tau$}
 \ncline{d12d11*0*10}{d12d12*0*10}\naput[nrot=:180]{$\tau$}
 \ncline{d02d01*0*00}{d02d03*0*00}\naput[nrot=:0]{$\tau$}
 \ncline{d12d11*0*10}{d12d13*0*10}\naput[nrot=:0]{$\tau$}
 \ncline{d02*011*00}{d02*012*00}\nbput[nrot=:0]{$\tau$}
 \ncline{d12*001*10}{d12*002*10}\nbput[nrot=:180]{$\tau$}

 \ncline{d02*011*00}{d02*013*00}\naput[nrot=:180]{$\tau$}
 \ncline{d12*001*10}{d12*003*10}\naput[nrot=:0]{$\tau$}
 \ncline{d02*001*10}{d02*003*10}\naput[nrot=:180]{$\tau$}
 \ncline{d12*011*00}{d12*013*00}\naput[nrot=:0]{$\tau$}
 \ncline{d02d01*0*10}{d02d03*0*10}\naput[nrot=:0]{$\tau$}
 \ncline{d12d11*0*00}{d12d13*0*00}\nbput[nrot=:180]{$\tau$}
 \ncline{d02d01*0*10}{d02d02*0*10}\nbput[nrot=:0]{$\tau$}
 \ncline{d12d11*0*00}{d12d12*0*00}\naput[nrot=:180]{$\tau$}
\end{pspicture}
\caption{Transition system for the ABP after abstraction from internal
  actions}
\label{fig-ABP-protocol-abstr}
\end{figure}

Most transitions of that transition system concern internal actions.
The behaviour of the ABP after abstraction from the internal actions
is described as follows:
\begin{ldispl}
 \begin{geqns}
\abstr{I}
 (\encap{H}(S \parcs{\commf} K \parcs{\commf} L \parcs{\commf} R))
 \end{geqns}
\end{ldispl}
where
\begin{ldispl}
\displstretch
 \begin{aeqns}
 I & = &
  \set{\kw{c}_3(f) \where f \in F} \union
  \set{\kw{c}_4(f) \where f \in F \union \set{\und}} \\ & {} \union &
  \set{\kw{c}_5(b) \where b \in B \union \set{\und}} \union
  \set{\kw{c}_6(b) \where b \in B} \union \set{\kw{i}}
 \end{aeqns}
\end{ldispl}
and $H$ and $\commf$ are as in Section~\ref{sect-interaction-abp}.
Parallel composition, encapsulation and abstraction of the transition 
systems of $S$, $K$, $L$ and $R$ as described above results in the 
following transition system.
We have the same states as before.
As actions, we have
$\kw{r}_1(d)$ and $\kw{s}_2(d)$ for each $d \in D$.
As transitions, we have:
\begin{iteml}
\small
\item
for each datum $d \in D$ and bit $b \in B$: 
\begin{iteml}
\item
$\astep
 {\tup{\tup{\und,b,0},\tup{\und,0},\tup{\und,0},\tup{\und,b,0}}}
 {\kw{r}_1(d)}
 {\tup{\tup{d,b,1},\tup{\und,0},\tup{\und,0},\tup{\und,b,0}}}$,
\item
$\astep
 {\tup{\tup{d,b,1},\tup{\und,0},\tup{\und,0},\tup{\und,b,0}}}
 {\tau}
 {\tup{\tup{d,b,2},\tup{\tup{d,b},1},\tup{\und,0},\tup{\und,b,0}}}$,
\item
$\astep
 {\tup{\tup{d,b,2},\tup{\tup{d,b},1},\tup{\und,0},\tup{\und,b,0}}}
 {\tau}
 {\tup{\tup{d,b,2},\tup{\tup{d,b},2},\tup{\und,0},\tup{\und,b,0}}}$,
\item
$\astep
 {\tup{\tup{d,b,2},\tup{\tup{d,b},2},\tup{\und,0},\tup{\und,b,0}}}
 {\tau}
 {\tup{\tup{d,b,2},\tup{\und,0},\tup{\und,0},\tup{d,b,1}}}$,
\item
$\astep
 {\tup{\tup{d,b,2},\tup{\und,0},\tup{\und,0},\tup{d,b,1}}}
 {\kw{s}_2(d)}
 {\tup{\tup{d,b,2},\tup{\und,0},\tup{\und,0},\tup{\und,b,2}}}$,
\item
$\astep
 {\tup{\tup{d,b,2},\tup{\und,0},\tup{\und,0},\tup{\und,b,2}}}
 {\tau}
 {\tup{\tup{d,b,2},\tup{\und,0},\tup{b,1},\tup{\und,\ol{b},0}}}$,
\item
$\astep
 {\tup{\tup{d,b,2},\tup{\und,0},\tup{b,1},\tup{\und,\ol{b},0}}}
 {\tau}
 {\tup{\tup{d,b,2},\tup{\und,0},\tup{b,2},\tup{\und,\ol{b},0}}}$,
\item
$\astep
 {\tup{\tup{d,b,2},\tup{\und,0},\tup{b,2},\tup{\und,\ol{b},0}}}
 {\tau}
 {\tup{\tup{\und,\ol{b},0},\tup{\und,0},\tup{\und,0},\tup{\und,\ol{b},0}}}$,
\end{iteml}
\mbox{} \\[-2ex]
\begin{iteml}
\item
$\astep
 {\tup{\tup{d,b,2},\tup{\tup{d,b},1},\tup{\und,0},\tup{\und,b,0}}}
 {\tau}
 {\tup{\tup{d,b,2},\tup{\tup{d,b},3},\tup{\und,0},\tup{\und,b,0}}}$,
\item
$\astep
 {\tup{\tup{d,b,2},\tup{\tup{d,b},3},\tup{\und,0},\tup{\und,b,0}}}
 {\tau}
 {\tup{\tup{d,b,2},\tup{\und,0},\tup{\und,0},\tup{\und,\ol{b},2}}}$,
\item
$\astep
 {\tup{\tup{d,b,2},\tup{\und,0},\tup{\und,0},\tup{\und,\ol{b},2}}}
 {\tau}
 {\tup{\tup{d,b,2},\tup{\und,0},\tup{\ol{b},1},\tup{\und,b,0}}}$,
\item
$\astep
 {\tup{\tup{d,b,2},\tup{\und,0},\tup{\ol{b},1},\tup{\und,b,0}}}
 {\tau}
 {\tup{\tup{d,b,2},\tup{\und,0},\tup{\ol{b},2},\tup{\und,b,0}}}$,
\item
$\astep
 {\tup{\tup{d,b,2},\tup{\und,0},\tup{\ol{b},2},\tup{\und,b,0}}}
 {\tau}
 {\tup{\tup{d,b,1},\tup{\und,0},\tup{\und,0},\tup{\und,b,0}}}$,
\item
$\astep
 {\tup{\tup{d,b,2},\tup{\und,0},\tup{\ol{b},1},\tup{\und,b,0}}}
 {\tau}
 {\tup{\tup{d,b,2},\tup{\und,0},\tup{\ol{b},3},\tup{\und,b,0}}}$,
\item
$\astep
 {\tup{\tup{d,b,2},\tup{\und,0},\tup{\ol{b},3},\tup{\und,b,0}}}
 {\tau}
 {\tup{\tup{d,b,1},\tup{\und,0},\tup{\und,0},\tup{\und,b,0}}}$,
\end{iteml}
\mbox{} \\[-2ex]
\begin{iteml}
\item
$\astep
 {\tup{\tup{d,b,2},\tup{\und,0},\tup{b,1},\tup{\und,\ol{b},0}}}
 {\tau}
 {\tup{\tup{d,b,2},\tup{\und,0},\tup{b,3},\tup{\und,\ol{b},0}}}$,
\item
$\astep
 {\tup{\tup{d,b,2},\tup{\und,0},\tup{b,3},\tup{\und,\ol{b},0}}}
 {\tau}
 {\tup{\tup{d,b,1},\tup{\und,0},\tup{\und,0},\tup{\und,\ol{b},0}}}$,
\item
$\astep
 {\tup{\tup{d,b,1},\tup{\und,0},\tup{\und,0},\tup{\und,\ol{b},0}}}
 {\tau}
 {\tup{\tup{d,b,2},\tup{\tup{d,b},1},\tup{\und,0},\tup{\und,\ol{b},0}}}$,
\item
$\astep
 {\tup{\tup{d,b,2},\tup{\tup{d,b},1},\tup{\und,0},\tup{\und,\ol{b},0}}}
 {\tau}
 {\tup{\tup{d,b,2},\tup{\tup{d,b},2},\tup{\und,0},\tup{\und,\ol{b},0}}}$,
\item
$\astep
 {\tup{\tup{d,b,2},\tup{\tup{d,b},2},\tup{\und,0},\tup{\und,\ol{b},0}}}
 {\tau}
 {\tup{\tup{d,b,2},\tup{\und,0},\tup{\und,0},\tup{\und,b,2}}}$,
\item
$\astep
 {\tup{\tup{d,b,2},\tup{\tup{d,b},1},\tup{\und,0},\tup{\und,\ol{b},0}}}
 {\tau}
 {\tup{\tup{d,b,2},\tup{\tup{d,b},3},\tup{\und,0},\tup{\und,\ol{b},0}}}$,
\item
$\astep
 {\tup{\tup{d,b,2},\tup{\tup{d,b},3},\tup{\und,0},\tup{\und,\ol{b},0}}}
 {\tau}
 {\tup{\tup{d,b,2},\tup{\und,0},\tup{\und,0},\tup{\und,b,2}}}$.
\end{iteml}
\end{iteml}
The transition system for the whole protocol is represented graphically
in Fig.~\ref{fig-ABP-protocol-abstr} for the case where only one datum 
is involved.

Next, we consider the following transition system.
As states, we have $d \in D \union \set{\und}$.
The initial state is $\und$.
As actions, we have $\kw{r}_1(d)$ and $\kw{s}_2(d)$ for each $d \in D$.
As transitions, we have the following:
\begin{iteml}
\item
for each datum $d \in D$:
\begin{iteml}
\item
a transition $\astep{\und}{\kw{r}_1(d)}{d}$,
\item
a transition $\astep{d}{\kw{s}_2(d)}{\und}$.
\end{iteml}
\end{iteml}
This transition system describes the behaviour of a bounded buffer with 
capacity $1$, but the actions of a bounded buffer as introduced in 
Example~\ref{exa-bbuffer} have been renamed.

The two transition systems presented above are branching bisimulation 
equivalent.
For each $d \in D$, we define the sets $R(d)$ and $S(d)$ of states of 
the first transition system to be related to states $d$ and $\und$,
respectively, of the second transition system: 

\begin{small}
\begin{ldispl}
 \begin{aeqns}
R(d) \\ {} =
\{
\tup{\tup{d,b,1},\tup{\und,0},\tup{\und,0},\tup{\und,b,0}},
\tup{\tup{d,b,2},\tup{\tup{d,b},1},\tup{\und,0},\tup{\und,b,0}}, \\
\phantom{{} = \{}
\tup{\tup{d,b,2},\tup{\tup{d,b},2},\tup{\und,0},\tup{\und,b,0}},
\tup{\tup{d,b,2},\tup{\und,0},\tup{\und,0},\tup{d,b,1}}, \\
\phantom{{} = \{}
\tup{\tup{d,b,2},\tup{\tup{d,b},1},\tup{\und,0},\tup{\und,b,0}},
\tup{\tup{d,b,2},\tup{\tup{d,b},3},\tup{\und,0},\tup{\und,b,0}}, \\
\phantom{{} = \{}
\tup{\tup{d,b,2},\tup{\und,0},\tup{\und,0},\tup{\und,\ol{b},2}},
\tup{\tup{d,b,2},\tup{\und,0},\tup{\ol{b},1},\tup{\und,b,0}}, \\
\phantom{{} = \{}
\tup{\tup{d,b,2},\tup{\und,0},\tup{\ol{b},2},\tup{\und,b,0}},
\tup{\tup{d,b,2},\tup{\und,0},\tup{\ol{b},1},\tup{\und,b,0}}, \\
\phantom{{} = \{}
\tup{\tup{d,b,2},\tup{\und,0},\tup{\ol{b},3},\tup{\und,b,0}} \}\;;
 \end{aeqns}
\end{ldispl}
\begin{ldispl}
 \begin{aeqns}
S(d) \\ {} =
\{
\tup{\tup{d,b,2},\tup{\und,0},\tup{\und,0},\tup{\und,b,2}},
\tup{\tup{d,b,2},\tup{\und,0},\tup{b,1},\tup{\und,\ol{b},0}}, \\
\phantom{{} = \{}
\tup{\tup{d,b,2},\tup{\und,0},\tup{b,2},\tup{\und,\ol{b},0}},
\tup{\tup{\und,b,0},\tup{\und,0},\tup{\und,0},\tup{\und,b,0}}, \\
\phantom{{} = \{}
\tup{\tup{d,b,2},\tup{\und,0},\tup{b,1},\tup{\und,\ol{b},0}},
\tup{\tup{d,b,2},\tup{\und,0},\tup{b,3},\tup{\und,\ol{b},0}}, \\
\phantom{{} = \{}
\tup{\tup{d,b,1},\tup{\und,0},\tup{\und,0},\tup{\und,\ol{b},0}},
\tup{\tup{d,b,2},\tup{\tup{d,b},1},\tup{\und,0},\tup{\und,\ol{b},0}}, \\
\phantom{{} = \{}
\tup{\tup{d,b,2},\tup{\tup{d,b},2},\tup{\und,0},\tup{\und,\ol{b},0}},
\tup{\tup{d,b,2},\tup{\tup{d,b},1},\tup{\und,0},\tup{\und,\ol{b},0}}, \\
\phantom{{} = \{}
\tup{\tup{d,b,2},\tup{\tup{d,b},3},\tup{\und,0},\tup{\und,\ol{b},0}} 
\}\;.
  \end{aeqns}
\end{ldispl}
\end{small}

\noindent
In Fig.~\ref{fig-ABP-protocol-abstr}, the states from the sets $R(d)$
and $S(d)$ are coloured light (yellow) and dark (red), respectively.
Next we define the relation $B$ as follows:
\begin{ldispl}
 \begin{aeqns}
  B & = &
  \set{\tup{s,d} \where d \in D, s \in R(d)} \union 
  \set{\tup{s,\und} \where s \in \Union_{d \in D} S(d)}\;.
 \end{aeqns}
\end{ldispl}
It is straightforward to see that the conditions imposed on branching 
bisimulation equivalence permit that all states in $R(d)$ are related 
to state $d$ and that all states in $\Union_{d \in D} S(d)$ are related 
to state $\und$.
In other words, the two transition systems presented above are 
branching bisimulation equivalent.
This justifies the claim that, after abstraction from internal actions,
the ABP behaves the same as a bounded buffer with capacity $1$.

As a corollary, we have that the relation $B'$ defined by
\begin{ldispl}
 \begin{aeqns}
  B' & = &
  \Union_{d \in D} (R(d) \x R(d)) \union 
  (\Union_{d \in D} S(d) \x \Union_{d \in D} S(d))
 \end{aeqns}
\end{ldispl}
is a branching autobisimulation on the first transition system 
presented above.
It is easy to show by means of $B'$ that, although the transition 
systems for the channels are not determinate, the transition system for 
the whole protocol is determinate.

\section{Petri nets and abstraction}
\label{sect-abstraction-conn-nets}

For the interested reader, we now show that abstraction and branching 
bisimulation equivalence can be defined on nets as well.

Like with the definition of encapsulation on nets, the definition of 
abstraction on nets is similar to the definition of abstraction on 
transition systems.
Here is the definition concerned.
\begin{definition}[Abstraction]\index{abstraction!of nets}
Let $N = \tup{P,A,\step{},m_0}$ be a net.
Let $I \subseteq A$.
The \emph{abstraction} of $N$ with respect to $I$, 
written $\abstr{I}(N)$, is the net 
$\tup{P,A',\step{}',m_0}$ where
\begin{iteml}
\item
$A' = A \diff I$;
\item
$\step{}'$ is the smallest subset of 
$(\fsetof{(P)} \diff \emptyset) 
  \x A' \x 
 (\fsetof{(P)} \diff \emptyset)$ such that: 
\begin{iteml}
\item
if $\astep{Q_1}{a}{Q_2}$ and $a \in I$, 
then $\astepp{Q_1}{\tau}{Q_2}$,
\item
if $\astep{Q_1}{a}{Q_2}$ and $a \not\in I$, 
then $\astepp{Q_1}{a}{Q_2}$.
\end{iteml}
\end{iteml}
\end{definition}
Branching bisimulation equivalence on nets is simply defined as 
branching bisimulation equivalence on their associated transition
systems.
\begin{definition}[Branching bisimulation equivalence]%
\index{branching bisimulation equivalence!of nets} 
\sloppy
Let $N = \tup{P,A,\step{},m_0}$ and $N' = \tup{P',A',\step{}',m_0'}$ 
be nets such that $A = A'$.
Then the nets $N$ and $N'$ are \emph{branching bisimulation 
equivalent}, written $N \bisim_\mathrm{b} N'$, if 
$\TrSy(N) \bisim_\mathrm{b} \TrSy(N')$.
\end{definition}

As explained in Sect.~\ref{sect-basics-conn-nets}, different from
transition systems, nets may indicate that transitions can occur
simultaneously.
By identifying branching bisimulation equivalent nets, this aspect of 
process behaviour described by nets is no longer covered.
Let us look at an example.
\begin{example}[Bounded counter]\index{counter!bounded}
\label{exa-bcounter-net}
We consider the bounded counter from Example~\ref{exa-bcounter}.
In this example, we focus on the behaviour of a bounded counter with 
bound $2$.
It can simply be described by the following net.
As places of the counter with bound $2$, we have the natural numbers 
$0$ to $2$.
As initial marking, we have $\set{0}$.
As actions, we have $\kw{inc}$ and $\kw{dec}$.
As transitions, we have the following:
\begin{iteml}
\item
for each place $i \in \set{0,1}$:
$\astep{\set{i}}{\kw{inc}}{\set{i+1}}$ and
$\astep{\set{i+1}}{\kw{dec}}{\set{i}}$.
\end{iteml}
It is easy to see that all reachable markings of this net are singleton 
sets.
The marking $\set{i}$ indicates that the value of the counter is $i$.
Next, we consider a net that is branching bisimulation equivalent to 
the one just presented.
As places, we have the natural numbers $0$ to $3$.
As initial marking, we have $\set{0,2}$.
As actions, we still have $\kw{inc}$ and $\kw{dec}$.
As transitions, we have the following:
\begin{iteml}
\item
$\astep{\set{0}}{\kw{inc}}{\set{1}}$,
$\astep{\set{1,2}}{\tau}{\set{0,3}}$, 
$\astep{\set{3}}{\kw{dec}}{\set{2}}$.
\end{iteml}
Unlike the first net, the second net indicates that if both $\kw{inc}$ 
and $\kw{dec}$ can occur, they can occur simultaneously.
Nevertheless, these nets are branching bisimulation equivalent.
\end{example}

The association of a transition system $\TrSy(N)$ with each net $N$ is 
also useful in showing the close connection between abstraction of nets 
and abstraction of transition systems.
\begin{property}
Let $N = \tup{P,A,\step{},m_0}$ be a net, and 
let $I \subseteq A$.
Then we have that
\begin{ldispl}
 \begin{geqns}
\TrSy(\abstr{I}(N))  \bisim \abstr{I}(\TrSy(N))\;.
 \end{geqns}
\end{ldispl}
\end{property}
Similar connections were already shown for parallel composition and
encapsulation in Sect.~\ref{sect-interaction-conn-nets}.

\section{Miscellaneous}
\label{sect-abstraction-misc}

\subsection*{Programs and abstraction}

In Sects.~\ref{sect-basics-conn-programs} 
and~\ref{sect-interaction-conn-programs}, we have seen that the 
behaviour of programs upon execution can be described in a 
straightforward way by means of transition systems and parallel 
composition of transition systems.
Is abstraction from internal actions relevant in this area as well?
In most programming languages, there are no features related to this 
kind of abstraction.
This is to be expected:
programs are primarily intended to be executed, not to be analyzed; 
whereas transition systems are primarily intended to be analyzed.

\subsection*{Trace equivalence}
In Sect.~\ref{sect-basics-trace-eqv}, trace equivalence was defined as 
follows.
Let $T = \tup{S,A,\step{},s_0}$ be a transition system.
A trace of $T$ is a sequence $\sigma \in \seqof{A}$ such that
$\gstep{s_0}{\sigma}{s}$ for some $s \in S$.
We write $\tr(T)$ for the set of all traces of $T$.
Then two transition systems $T$ and $T'$ are trace equivalent, written 
$T \treqv T'$, if $\tr(T) = \tr(T')$.
With the adapted definition of the generalized transitions of a 
transition system, this means that in the case of trace equivalence
we simply leave out all unobservable actions.
Because it does not matter in the case of trace equivalence at which 
stages choices occur, this is all right.


\chapter{Composition}
\label{ch-composition}

In Chap.~\ref{ch-interaction}, we have seen that, by means of parallel 
composition, a transition system can be composed of others that act 
concurrently and interact with each other.
This is not the only conceivable way of composition.
This chapter treats several basic ways in which transition systems can 
be composed of others that do not interact with each other.
Sequential composition is used to describe that a transition systems is
composed of two others that act successively.
Alternative composition is used to describe that a transition system is
composed of two others that act the one or the other.
Iteration is used to describe that a transition system is composed of
two others of which the first one acts repeatedly until the second one
takes over.
Many transition systems can be composed using these three ways of 
composition.
Thus, they support mastering the complexity of large transition systems.
First of all, we explain informally what alternative composition, 
sequential composition and iteration are, and give simple examples of 
their use in describing process behaviour 
(Sect.~\ref{sect-composition-informal}).
After that, we first adapt the definitions of transition system, 
parallel composition, encapsulation and abstraction from 
Chap.~\ref{ch-abstraction} to the addition of alternative composition, 
sequential composition and iteration 
(Sect.~\ref{sect-composition-formal-1}), and then define 
alternative composition, sequential composition and iteration in a 
mathematically precise way (Sect.~\ref{sect-composition-formal-2}).
We also use these operations to define the components
of the simple data communication protocol from 
Sect.~\ref{sect-interaction-abp} (Sect.~\ref{sect-composition-abp}).
Next, we have another look at bisimulation equivalence and trace 
equivalence (Sect.~\ref{sect-composition-eqv}).
Finally, we look at some miscellaneous issues
(Sect.~\ref{sect-composition-misc}).

\section{Informal explanation}
\label{sect-composition-informal}

The alternative composition of two transition systems $T$ and $T'$ is 
a transition system describing that there is a choice between the
behaviour described by $T$ and the behaviour described by $T'$.
The choice is resolved at the instant that one of them performs its
first action.
The sequential composition of two transition systems $T$ and $T'$ is a
transition system describing that the behaviour described by $T$ and
the behaviour described by $T'$ follow each other.
The notion of a transition system needs to be adapted in the presence of
sequential composition because $T'$ should only take over on successful
termination of $T$.
The iteration of transition system $T$ with exit transition system $T'$
is a transition system describing that initially there is a choice
between the behaviour described by $T$ and the behaviour described by
$T'$, and upon successful termination of $T$ there is this choice
again.
Often, we need to describe that a transition system simply acts 
repeatedly for ever.
Such a no-exit iteration can be treated as a special case of iteration 
with exit (see Sect.~\ref{sect-composition-formal-2}).
The no-exit iteration of transition system $T$ is a transition system 
describing that initially there is the behaviour described by $T$, and 
upon successful termination of $T$ the behaviour is again as initially.
Here are a couple of examples.  
\begin{example}[Simple telephone system]\index{simple telephone system}
\label{exa-bcp}
We consider the simple telephone system from 
Example~\ref{exa-telephones}.
Recall that in this telephone system each telephone is provided with a 
process, called its basic call process, to establish and maintain 
connections with other telephones.
Actions of this process include receiving an off-hook or on-hook signal
{from} the telephone, receiving a dialed number from the telephone,
sending a signal to start or to stop emitting a dial tone, ring tone or
ring-back tone to the telephone, and receiving an alert signal from
another telephone -- indicating an incoming call.
Initially, there is a choice between the following two alternatives:
\begin{iteml}
\item
receiving an off-hook signal from the telephone followed by a process
of which the first action is sending a signal to start emitting a
dial tone to the telephone;
\item
receiving an alert signal from another telephone followed by a process
of which the first action is sending a signal to start emitting a
ring tone to the telephone.
\end{iteml}
In either case the basic call process goes back to waiting for another 
off-hook or alert signal after the call is terminated.
Therefore, the behaviour of the basic call process of a telephone can 
be described as the no-exit iteration of a process that is itself the 
alternative composition of two subprocesses, one reacting to an 
off-hook signal sent to the basic call process and the other reacting 
to an alert signal sent to the basic call process.
The first one of these subprocesses first goes through a dialling phase 
and after that through a calling phase.
So, the behaviour of this process can itself be described as the 
sequential composition of a subprocess for the dialling phase and a 
subprocess for the calling phase.
And so forth.
\end{example}
\begin{example}[Telephone answering machine controller]%
\index{telephone answering machine controller}
\label{exa-tamc}
In order to control telephone answering, the control component of
an answering machine has to communicate with the recorder component of
the answering machine, the telephone network, and the telephone 
connected with the answering machine.
When an incoming call is detected, the answering is not started
immediately:
\begin{iteml}
\item
if the incoming call is broken off or the receiver of the telephone is
lifted within a certain period, answering is discontinued;
\item
otherwise, an off-hook signal is issued to the network when this period
has elapsed and after that a pre-recorded message is played.
\end{iteml}
Upon termination of the message, the recorder is started and a beep
signal is issued to the network.
The recorder is stopped when:
\begin{iteml}
\item
either the call is broken off;
\item
or a certain time period has passed in the case where the call has not 
been broken off earlier.
\end{iteml}
Thereafter, an on-hook signal is issued to the network.
The behaviour of the control component can be described as the no-exit
iteration of a process that is itself the sequential composition of 
three subprocesses, one checking whether the receiver is not lifted
when an incoming call is detected, one controlling the answering with 
the pre-recorded message, and one controlling the recording of a 
message from the caller.
Each of these subprocesses must respond properly if the call is broken
off prematurely.
Therefore, the behaviour of each of them can be described as an 
alternative composition with one of the alternatives reacting to
signals indicating that the call is broken off prematurely.
\end{example}

\section{Adjustment of earlier definitions}
\label{sect-composition-formal-1}

In the previous section, we have prepared the way for the formal 
definitions of the notions of alternative composition of transition 
systems, sequential composition of transition systems, and iteration of
transition systems.
Before we give those definitions in the next section, we first adapt 
the definitions of the notions of a transition system, parallel 
composition, encapsulation and abstraction from 
Chap.~\ref{ch-abstraction}.

We already mentioned that the notion of a transition system needs to be 
adapted, because sequential composition requires that successful 
termination is distinguished from becoming inactive.
\begin{definition}[Transition system]\index{transition system}
\label{def-ts-with-st}
A \emph{transition system} $T$ is a quintuple 
$\tup{S,A,\step{},\term{},s_0}$ 
where
\begin{iteml}
\item
$S$ is a set of \emph{states}\index{state};
\item
$A$ is a set of \emph{actions}\index{action};
\item 
${\step{}} \subseteq S \x A_\tau \x S$ is a set of \emph{transitions}%
\index{transition};
\item
${\term{}} \subseteq S$, with $s \in {\term{}}$ only if there are no 
$a \in A_\tau$ and $s' \in S$  such that $\astep{s}{a}{s'}$, is a set 
of \emph{successfully terminating states}%
\index{state!successfully terminating};
\item
$s_0 \in S \diff {\term{}}$ is the \emph{initial state}%
\index{state!initial}.
\end{iteml}
We write $\term{s}$ instead $s \in {\term{}}$.
The set ${\gstep{}{}{}} \subseteq S \x \seqof{A} \x S$ of 
\emph{generalized transitions}\index{transition!generalized} of $T$ is 
the smallest subset of $S \x \seqof{A} \x S$ satisfying:
\begin{iteml}
\item
$\astep{s}{\epsilon}{s}$ for each $s \in S$;
\item
if $\astep{s}{\tau}{s'}$, then $\gstep{s}{\epsilon}{s'}$;
\item
if $\astep{s}{a}{s'}$, then $\gstep{s}{a}{s'}$;
\item
if $\gstep{s}{\sigma}{s'}$ and $\gstep{s'}{\sigma'}{s''}$, then 
$\gstep{s}{\sigma\, \sigma'}{s''}$.
\end{iteml}
A state $s \in S$ is called a \emph{reachable}\index{state!reachable} 
state of $T$ if there is a $\sigma \in \seqof{A}$ such that 
${\gstep{s_0}{\sigma}{s}}$.
A state $s \in S$ is called a \emph{terminal}\index{state!terminal} 
state of $T$ if there is no $a \in A$ and $s' \in S$ such that 
$\astep{s}{a}{s'}$.
\end{definition}
Notice that only terminal states may be successfully terminating states.
Moreover, the initial state may not be a successfully terminating state.
This excludes transition systems that can terminate successfully 
without performing any action.
Such transition systems are not excluded here because they are 
problematic in whatever way.
However, their inclusion would clutter up the coming definitions.

When looking at those definitions, it is important to take the 
following into account. 
A transition system can be made more intelligible by a judicious choice 
of states.
However, the identity of the states of a transition system are never 
really relevant to the behaviour described by the transition system.
For example, transition systems that differ only with respect to the 
identity of their states are bisimulation equivalent.
Hence, we can ignore the identity of the states of a transition system.
If transition systems differ only with respect to the identity of their
states, they are called isomorphic.
Here is the mathematically precise definition.
\begin{definition}[Isomorphy]\index{isomorphy} 
\sloppy
Let $T = \tup{S,A,\step{},\term{},s_0}$ 
and $T' = \tup{S',A',\step{}',\term{}',s_0'}$ 
be transition systems such that $A = A'$.
Then $T$ and $T'$ are \emph{isomorphic} if there exists a bijective 
relation $B \subseteq S \x S'$ such that the following conditions hold:
\begin{enuml}
\item
$B(s_0,s_0')$;
\item 
whenever $B(s_1,s_1')$ and $B(s_2,s_2')$, 
then $\astep{s_1}{a}{s_2}$ if and only if $\astepp{s_1'}{a}{s_2'}$;
\item
whenever $B(s,s')$, then $\term{s}$ if and only if $\termp{s'}$.
\end{enuml}
\end{definition}
We will always consider two transition systems the same if they are 
isomorphic, and write $T = T'$ if $T$ and $T'$ are isomorphic.
Because of this, the disjointness requirement on the sets of states 
that occurs in the definitions of alternative composition, sequential 
composition and iteration given below does not cause any loss of 
generality.
Moreover, it does not matter that an arbitrary fresh initial state is 
chosen in the case of alternative composition and iteration: up to 
isomorphism the result is independent of the particular choice.

Unreachable states, and transitions between them, are never really 
relevant to the behaviour described by the transition system.
For example, transition systems that differ only with respect to 
unreachable states are bisimulation equivalent.
In fact, we are only interested in connected transition systems.
\begin{definition}[Connected transition system]%
\index{transition system!connected} 
Let $T = \tup{S,A,\step{},\linebreak[2]\term{},s_0}$ be a transition 
system.
Then the set of \emph{reachable} states of $T$, written $\reach(T)$, is
$\set{s \in S \where 
      \Exists{\sigma \in \seqof{A}}{\gstep{s_0}{\sigma}{s}}}$; 
and the set of \emph{not immediately reachable} states of $T$, written 
$\reachp(T)$, is
$\set{s' \in S \where 
      \Exists 
       {s \in \reach(T) \diff \set{s_0},\sigma \in \seqof{A}}
       {\gstep{s}{\sigma}{s'}}}$.
The transition system $T$ is called a \emph{connected} transition
system if $S = \reach(T)$.
\end{definition}
All operations defined in this chapter result in connected transition 
systems if they are applied to connected transition systems.
Notice that either $\reach(T) = \reachp(T)$ or
$\reach(T) = \reachp(T) \union \set{s_0}$, depending on\linebreak[2] 
whether $s_0$ is reachable from other states than $s_0$. 

A further restriction to finitely branching or countably branching 
transition systems and/or to finite or regular transition systems is 
often made.
\begin{definition}[Classification of transition systems]
Let $T = \tup{S,A,\linebreak[2]\step{},\term{},s_0}$ be a transition 
system.
Then $T$ is a \emph{finitely branching} transition system%
\index{transition system!finitely branching} if for all 
$s \in \reach(T)$ we have that the set 
$\set{\tup{a,s'} \in A \x S \where \astep{s}{a}{s'}}$ is finite,
and $T$ is a \emph{countably branching} transition system%
\index{transition system!countably branching} if for all 
$s \in \reach(T)$ we have that the set 
$\set{\tup{a,s'} \in A \x S \where \astep{s}{a}{s'}}$ is countable.
Furthermore, $T$ is a \emph{finite} transition system%
\index{transition system!finite} if the set 
$\set{\tup{\sigma,s} \in \seqof{A} \x S \where 
      \gstep{s_0}{\sigma}{s}}$ is finite, 
and $T$ is a \emph{regular} transition system%
\index{transition system!regular} if the set 
$\reach(T)$, i.e. the set 
$\set{s \in S \where 
      \Exists{\sigma \in \seqof{A}}{\gstep{s_0}{\sigma}{s}}}$, 
is finite.
\end{definition}
Here is an example showing that it also makes sense to distinguish
successfully terminating states in  a setting without operations such
as sequential composition.
\begin{example}[Factorial and greatest common divisor programs]%
\index{factorial program}\index{greatest common divisor program}
\label{exa-fact-gcd-term}
We consider again the transition systems describing the behaviours of
PASCAL programs upon execution from Examples~\ref{exa-factorial}
and~\ref{exa-gcd}.
State $7$ of the transition system for the factorial program and state
$8$ of the transition system for the greatest common divisor program
are intended to be successfully terminating states.
However, this cannot be made explicit with the definition of transition
system from Chap.~\ref{ch-basics}.
With the definition of transition system given in this chapter, we can
designate the above-mentioned states as successfully terminating states
of those transition systems.
As an aside, we mention that the transition systems from
Examples~\ref{exa-factorial} and~\ref{exa-gcd} are connected, finitely
branching, and regular.
\end{example}

Because successfully terminating states are now distinguished from
other terminal states, we have to adapt the definitions of parallel
composition, encapsulation and abstraction from 
Chap.~\ref{ch-abstraction} as well.
The new definitions are nothing else but simple adjustments of the
earlier definitions to cover successfully terminating states.
\begin{definition}[Parallel composition]\index{parallel composition}
Let $T = \tup{S,A,\step{},\term{},s_0}$ 
and $T' = \tup{S',A',\step{}',\term{}',s'_0}$ be 
transition systems.
Let $\commf$ be a communication function on a set of actions that
includes $A \union A'$.
The \emph{parallel composition} of $T$ and $T'$ under $\commf$,
written $T \parcs{\commf} T'$, is the transition system 
$\tup{S'',A'',\step{}'',\term{}'',s''_0}$ where
\begin{iteml}
\item
$S'' = S \x S'$;
\item
$A'' = 
A \union A' \union 
\set{\commf(a,a') \where 
     a \in A, a' \in A', \commf(a,a') \;\mathrm{is}\;\mathrm{defined}}$;
\item
$\step{}''$ is the smallest subset of $S'' \x {A''}_\tau \x S''$ such 
that: 
\begin{iteml}
\item
if $\astep{s_1}{a}{s_2}$ and $s' \in S'$, then
$\asteppp{\tup{s_1,s'}}{a}{\tup{s_2,s'}}$;
\item
if $\astepp{s'_1}{b}{s'_2}$ and $s \in S$, then
$\asteppp{\tup{s,s'_1}}{b}{\tup{s,s'_2}}$;
\item
if $\astep{s_1}{a}{s_2}$, $\astepp{s'_1}{b}{s'_2}$ and 
$\commf(a,b)$ is defined, then
$\asteppp{\tup{s_1,s'_1}}{\commf(a,b)}{\tup{s_2,s'_2}}$;
\end{iteml}
\item
$\term{}''$ is the smallest subset of $S''$ such that: 
\begin{iteml}
\item
if $\term{s}$ and $\termp{s'}$, then
$\termpp{\tup{s,s'}}$;
\end{iteml}
\item
$ s''_0 = \tup{s_0,s'_0}$.
\end{iteml}
\end{definition}
What is new in this definition of parallel composition, compared with 
the definition from Chap.~\ref{ch-abstraction}, concerns successful
termination.
Notice that the parallel composition of two transition systems $T$ and 
$T'$ can only terminate successfully when both $T$ and $T'$ can 
terminate successfully.
\begin{definition}[Encapsulation]\index{encapsulation}
Let $T = \tup{S,A,\step{},\term{},s_0}$ be a transition system.
Let $H \subseteq A$.
The \emph{encapsulation} of $T$ with respect to $H$, 
written $\encap{H}(T)$, is the transition system 
$\tup{S',A',\step{}',\term{}',s_0}$ where
\begin{iteml}
\item
$S' = 
 \set{s \where 
      \Exists{\sigma \in \seqof{(A \diff H)}}{\gstep{s_0}{\sigma}{s}}}$;
\item
$A' = 
 \set{a \in A \diff H \where 
      \Exists{s_1,s_2 \in S'}{\astep{s_1}{a}{s_2}}}$;
\item
$\step{}'$ is the smallest subset of $S' \x A'_\tau \x S'$ such that: 
\begin{iteml}
\item
if $\astep{s_1}{a}{s_2}$, $s_1 \in S'$ and $a \not\in H$, 
then $\astepp{s_1}{a}{s_2}$;
\end{iteml}
\item
$\term{}' = {\term{}} \inter S'$.
\end{iteml}
\end{definition}
Like in the case of parallel composition, what is new in this 
definition of encapsulation, compared with the definition from 
Chap.~\ref{ch-abstraction}, concerns successful termination.
Notice that the encapsulation of a transition system $T$ can only 
terminate successfully when $T$ could terminate successfully.
Notice further that successfully terminating states of $T$ may become
unreachable by encapsulation.
Here is an example of successful termination in parallel composition
and encapsulation.
\begin{example}[Successful termination in parallel composition and 
encapsulation]\index{successful termination}
\label{exa-term-parc}
We consider the following two transition systems, which are closely
related to the ones of Example~\ref{exa-tau-parc}.
As actions of the first transition system, we have $\kw{s}_1(0)$ and
$\kw{s}_2(0)$.
As states of the first transition system, we have natural numbers 
$i \in \set{0,1,2,3}$, with $0$ as initial state and $3$ as only
successfully terminating state.
As transitions of the first transition system, we have the following:
\begin{ldispl}
\astep{0}{\tau}{1},\;
\astep{0}{\tau}{2},\;
\astep{1}{\kw{s}_1(0)}{3},\;
\astep{2}{\kw{s}_2(0)}{3}.
\end{ldispl}
As actions of the second transition system, we have only $\kw{r}_1(0)$.
As states of the second transition system, we have natural numbers 
$i \in \set{0,1}$, with $0$ as initial state and $1$ as only
successfully terminating state.
As transitions of the second transition system, we have the following:
\begin{ldispl}
\astep{0}{\kw{r}_1(0)}{1}.
\end{ldispl}
These two transition systems are represented graphically in 
Fig.~\ref{fig-term-parc}.%
\footnote
{In graphical representations of transition systems, we indicate the 
 successfully terminating state by an outgoing unlabeled arrow.
}
\begin{figure}
\[
\begin{array}{c}
\encap{H} \left(

\begin{pspicture}(0.5,2)(3.5,4)
 \psset{arrows=->}
 \pnode(1,4){S}
 \rput(1,3){\ovalnode{0}{$0$}}
 \rput(3,3){\ovalnode{1}{$1$}}
 \rput(1,1){\ovalnode{2}{$2$}}
 \rput(3,1){\ovalnode{3}{$3$}}
 \pnode(3,0){E}
 \ncline{S}{0}
 \ncline{0}{1}\nbput{$\tau$}
 \ncline{0}{2}\naput{$\tau$}
 \ncline{1}{3}\naput[nrot=:0]{$\kw{s}_1(0)$}
 \ncline{2}{3}\naput{$\kw{s}_2(0)$}
 \ncline{3}{E}
\end{pspicture}

\begin{pspicture}(3.5,2)(4.5,4)
\rput(4,2){\rnode{D}{$\parcs{\commf}$}}
\end{pspicture}

\begin{pspicture}(4.5,2)(5.5,4)
 \psset{arrows=->}
 \pnode(5,4){S'}
 \rput(5,3){\ovalnode{0}{$0$}}
 \rput(5,1){\ovalnode{1}{$1$}}
 \pnode(5,0){E'}
 \ncline{S'}{0}
 \ncline{0}{1}\naput[nrot=:0]{$\kw{r}_1(0)$}
 \ncline{1}{E'}
\end{pspicture}

\right)

\begin{pspicture}(5.5,2)(6.5,4)
\rput(6,2){\rnode{D}{$=$}}
\end{pspicture}

\begin{pspicture}(6.5,2)(9.5,4)
 \psset{arrows=->}
 \pnode(7,4){S''}
 \rput(7,3){\ovalnode{00}{$(0,0)$}}
 \rput(9,3){\ovalnode{10}{$(1,0)$}}
 \rput(7,1){\ovalnode{20}{$(2,0)$}}
 \rput(9,1){\ovalnode{31}{$(3,1)$}}
 \pnode(9,0){E''}
 \ncline{S''}{00}
 \ncline{00}{10}\naput{$\tau$}
 \ncline{00}{20}\naput{$\tau$}
 \ncline{10}{31}\naput[nrot=:0]{$\kw{c}_1(0)$}
 \ncline{31}{E''}
\end{pspicture}
\end{array}
\]
\caption{Transition systems of Example~\protect\ref{exa-term-parc}}
\label{fig-term-parc}
\end{figure}
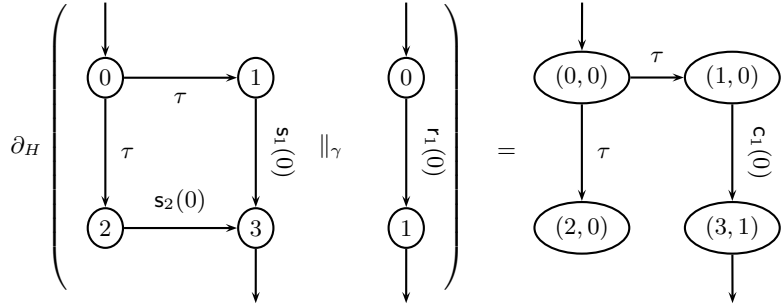
Parallel composition of these transitions systems and subsequent
encapsulation with respect to actions $\kw{s}_1(0)$, $\kw{r}_1(0)$, 
$\kw{s}_2(0)$ and $\kw{r}_2(0)$ result in the following transition
system.
As actions of the resulting transition system, we have only 
$\kw{c}_1(0)$.
As states of the resulting transition system, we have the pairs
$\tup{0,0}$, $\tup{1,0}$, $\tup{2,0}$ and $\tup{3,1}$, with $\tup{0,0}$
as initial state and $\tup{3,1}$ as only successfully terminating
state.
As transitions of the resulting transition system, we have the
following:
\begin{ldispl}
\astep{\tup{0,0}}{\tau}{\tup{1,0}},\;
\astep{\tup{0,0}}{\tau}{\tup{2,0}},\;
\astep{\tup{1,0}}{\kw{c}_1(0)}{\tup{3,1}}.
\end{ldispl}
The resulting transition system, which is also represented graphically 
in Fig.~\ref{fig-term-parc}, is capable of either first performing a 
silent step, next performing a communication action, and then 
terminating successfully or first performing a silent step and then 
becoming inactive.
In the case of Example~\ref{exa-tau-parc}, i.e.\ in the setting without
successful termination, it was not clear from the transition system
that states $\tup{3,1}$ and $\tup{2,0}$ are states of a different
nature.
\end{example}
\begin{definition}[Abstraction]\index{abstraction}
Let $T = \tup{S,A,\step{},\term{},s_0}$ be a transition system.
Let $I \subseteq A$.
The \emph{abstraction} of $T$ with respect to $I$, 
written $\abstr{I}(T)$, is the transition system 
$\tup{S,A',\step{}',\term{},s_0}$ where
\begin{iteml}
\item
$A' = A \diff I$;
\item
$\step{}'$ is the smallest subset of $S \x {A'}_\tau \x S$ such that:
\begin{iteml}
\item
if $\astep{s_1}{a}{s_2}$ and $a \in I$, 
then $\astepp{s_1}{\tau}{s_2}$,
\item
if $\astep{s_1}{a}{s_2}$ and $a \not\in I$, 
then $\astepp{s_1}{a}{s_2}$.
\end{iteml}
\end{iteml}
\end{definition}
Like in the cases of parallel composition and encapsulation, what is 
new in this definition of abstraction, compared the definition from 
Chap.~\ref{ch-abstraction}, concerns successful termination.
Notice that the abstraction of a transition system $T$ can only 
terminate successfully when $T$ could terminate successfully.
Notice further that successfully terminating states of $T$ never become
unreachable by abstraction.

\section{New definitions}
\label{sect-composition-formal-2}

Let us now look at the formal definitions of alternative composition, 
sequential composition, and (single-exit) iteration.
\begin{definition}[Alternative composition]%
\index{alternative composition}
Let $T = \tup{S,A,\step{},\term{},s_0}$ 
and $T' = \tup{S',A',\step{}',\term{}',s_0'}$ be 
transition systems such that $S \inter S' = \emptyset$.
The \emph{alternative composition} of $T$ and $T'$,
written $T \altc T'$, is the transition system 
$\tup{S'',A'',\step{}'',\term{}'',s''_0}$ where
\begin{iteml}
\item
$S'' = \set{s''_0} \union \reachp(T) \union \reachp(T')$;
\item
$A'' = A \union A'$;
\item
$\step{}''$ is the smallest subset of $S'' \x {A''}_\tau \x S''$ such 
that: 
\begin{iteml}
\item
if $\astep{s_0}{a}{s}$, then $\asteppp{s_0''}{a}{s}$;
\item
if $\astepp{s_0'}{a}{s'}$, then $\asteppp{s_0''}{a}{s'}$;
\item
if $\astep{s_1}{a}{s_2}$ and $s_1 \in S''$, 
then $\asteppp{s_1}{a}{s_2}$;
\item
if $\astepp{s_1'}{a}{s_2'}$ and $s_1' \in S''$, 
then $\asteppp{s_1'}{a}{s_2'}$;
\end{iteml}
\item
$\term{}''$ is the smallest subset of $S''$ such that: 
\begin{iteml}
\item
if $\term{s}$, then $\termpp{s}$;
\item
if $\termp{s'}$, then $\termpp{s'}$;
\end{iteml}
\item
$ s''_0 \notin S \union S'$.
\end{iteml}
\end{definition}
The following things should be noted about the definition of 
alternative composition.
The alternative composition of transition systems $T$ and $T'$ has a
fresh initial state.
This fresh initial state adopts the transitions from the initial 
state of $T$ and the transitions from the initial state of $T'$.
However, the fresh initial state does not replace the initial states of 
$T$ and $T'$.
Thus, transitions to the initial state of $T$ or $T'$ do not lead to
transitions to the fresh initial state.
The latter transitions would imply that the choice, that should be 
there only initially, could come back later.
Here is an example to illustrate that it is quite natural to look at
certain real-life processes as the alternative composition of other 
processes.
\begin{example}[Railroad crossing controller]%
\index{railroad crossing controller}
\label{exa-rrc-altc}
We consider a simple railroad crossing controller.
An approach signal is sent to the controller as soon as a train passes 
a detector placed backward from the gate.
An exit signal is sent to the controller as soon as the train passes 
another detector placed forward from the gate.
The controller is able to receive approach and exit signals from the 
train detectors at any time.
When the controller receives an approach signal, a lower signal must be 
sent to the gate.
When the controller receives an exit signal, a raise signal must be 
sent to the gate.
Suppose that $A$ and $E$ are the transition systems describing the 
behaviours of the subprocesses dedicated to receiving and handling an 
approach signal and an exit signal, respectively, in the case where the
signal is received at the beginning of a cycle of the controller, i.e.\
when there is no previous signal being handled.
Then the behaviour of one cycle of the controller is described by 
$A \altc E$.
\end{example}
Let us also give an example illustrating the details of alternative 
composition.
\begin{example}[Alternative composition]\index{alternative composition}
\label{exa-altc}
We assume a set of data $D$, and two input ports $k$ and $l$.
For $d \in D$, let $R_k(d)$ and $R_l(d)$ be the transition systems
$\tup{S,A,\step{},\term{},s_0}$ and 
$\tup{S',A',\step{}',\term{}',s_0'}$ where
\[
\begin{aeqns}
S & = & \set{\tup{k,\und},\tup{k,d}}\enspace, \\
A & = & \set{\kw{r}_k(d)}\enspace, \\
{\step{}} & = & 
 \set{\astep{\tup{k,\und}}{\kw{r}_k(d)}{\tup{k,d}}}\enspace, \\
{\term{}} & = & \set{\tup{k,d}}\enspace, \\
s_0 & = & \tup{k,\und}\enspace,
\end{aeqns}
\qquad
\begin{aeqns}
S' & = & \set{\tup{l,\und},\tup{l,d}}\enspace, \\
A' & = & \set{\kw{r}_l(d)}\enspace, \\
{\step{}'} & = & 
 \set{\astep{\tup{l,\und}}{\kw{r}_l(d)}{\tup{l,d}}}\enspace, \\
{\term{}'} & = & \set{\tup{l,d}}\enspace, \\
s_0' & = & \tup{l,\und}\enspace.
\end{aeqns}
\]
The transition system $R_i(d)$ is capable of receiving $d$ at port $i$ 
and then terminating successfully ($i = k,l$).
The alternative composition $R_k(d) \altc R_l(d)$ is the transition 
system $\tup{S'',A'',\step{}'',\term{}'',s_0''}$ where
\[
\begin{aeqns}
S'' & = & \set{\tup{\und,\und},\tup{k,d},\tup{l,d}}\enspace, \\
A'' & = & \set{\kw{r}_k(d),\kw{r}_l(d)}\enspace, \\
{\step{}''} & = & 
 \set{\astep{\tup{\und,\und}}{\kw{r}_k(d)}{\tup{k,d}},
      \astep{\tup{\und,\und}}{\kw{r}_l(d)}{\tup{l,d}}}\enspace, \\
{\term{}''} & = & \set{\tup{k,d},\tup{l,d}}\enspace, \\
s_0'' & = & \tup{\und,\und}\enspace.
\end{aeqns}
\]
This transition system is capable of receiving datum $d$ at port $k$ or 
$l$ and then terminating successfully.
The alternative composition of $R_k(d)$ and $R_l(d)$ is represented
graphically in Fig.~\ref{fig-altc}.
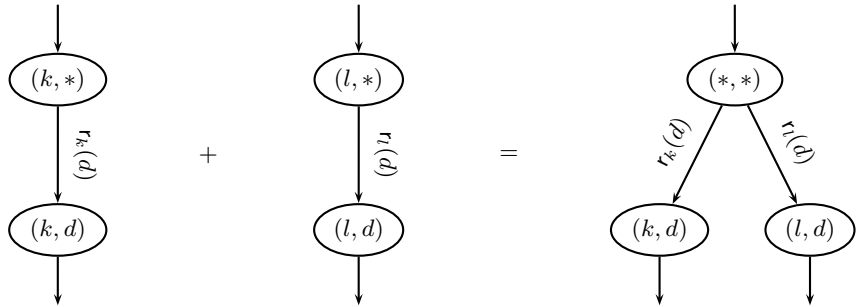
\begin{figure}
\begin{pspicture}(0,0)(12,5)

 \psset{arrows=->}

 \pnode(1,4){S}
 \rput(1,3){\ovalnode{k*}{$(k,*)$}}
 \rput(1,1){\ovalnode{kd}{$(k,d)$}}
 \pnode(1,0){E}
 \ncline{S}{k*}
 \ncline{k*}{kd}\naput[nrot=:0]{$\kw{r}_k(d)$}
 \ncline{kd}{E}

 \rput(3,2){\rnode{D}{$+$}}

 \pnode(5,4){S'}
 \rput(5,3){\ovalnode{l*}{$(l,*)$}}
 \rput(5,1){\ovalnode{ld}{$(l,d)$}}
 \pnode(5,0){E'}
 \ncline{S'}{l*}
 \ncline{l*}{ld}\naput[nrot=:0]{$\kw{r}_l(d)$}
 \ncline{ld}{E'}

 \rput(7,2){\rnode{D}{$=$}}

 \pnode(10,4){S''}
 \rput(10,3){\ovalnode{**}{$(*,*)$}}
 \rput(9,1){\ovalnode{kd}{$(k,d)$}}
 \rput(11,1){\ovalnode{ld}{$(l,d)$}}
 \pnode(9,0){E1''}
 \pnode(11,0){E2''}
 \ncline{S''}{**}
 \ncline{**}{kd}\nbput[nrot=:180]{$\kw{r}_k(d)$}
 \ncline{**}{ld}\naput[nrot=:0]{$\kw{r}_l(d)$}
 \ncline{kd}{E1''}
 \ncline{ld}{E2''}

\end{pspicture}
\caption{Alternative composition of $R_k(d)$ and $R_l(d)$}
\label{fig-altc}
\end{figure}
\end{example}
\begin{definition}[Sequential composition]\index{sequential composition}
Let $T = \tup{S,A,\step{},\term{},s_0}$ 
and $T' = \tup{S',A',\step{}',\term{}',s'_0}$ be 
transition systems such that $S \inter S' = \emptyset$.
The \emph{sequential composition} of $T$ and $T'$,
written $T \seqc T'$, is the transition system 
$\tup{S'',A'',\step{}'',\term{}',s_0}$ where
\begin{iteml}
\item
$S'' = (S \union S') \diff {\term{}}$;
\item
$A'' = A \union A'$;
\item
$\step{}''$ is the smallest subset of $S'' \x {A''}_\tau \x S''$ such 
that: 
\begin{iteml}
\item
if $\astep{s_1}{a}{s_2}$ and not $\term{s_2}$, 
then $\asteppp{s_1}{a}{s_2}$;
\item
if $\astep{s_1}{a}{s_2}$ and $\term{s_2}$, 
then $\asteppp{s_1}{a}{s_0'}$;
\item
if $\astepp{s_1'}{a}{s_2'}$, 
then $\asteppp{s_1'}{a}{s_2'}$.
\end{iteml}
\end{iteml}
\end{definition}
The definition of sequential composition is the first definition of
a way in which transition systems can be composed where successfully 
terminating states are relevant to the transitions of the resulting
transition system. 
Notice that, in the sequential composition of transition systems $T$ 
and $T'$, the initial state of $T'$ replaces all successfully 
terminating states of $T$. 
However, it does not become a successfully terminating state itself.
Here is an example to illustrate that it is quite natural to look at
certain real-life processes as the sequential composition of other 
processes.
\begin{example}[Railroad crossing controller]%
\index{railroad crossing controller}
\label{exa-rrc-seqc}
We look again at the railroad crossing controller from 
Example~\ref{exa-rrc-altc}.
Suppose that $R(\nm{appr})$ and $R(\nm{exit})$ are the transition 
systems describing the behaviours of the subprocesses dedicated to 
receiving an approach signal and an exit signal, respectively.
Suppose that $D$ and $U$ are the transition systems describing the 
behaviours of the subprocesses dedicated to handling an approach 
signal and an exit signal, respectively, that is received 
at the beginning of a cycle of the controller.
Then the behaviour of one cycle of the controller is described by
$(R(\nm{appr}) \seqc D) \altc (R(\nm{exit}) \seqc U)$.
\end{example}
Let us also give an example illustrating the details of sequential 
composition.
\begin{example}[Sequential composition]\index{sequential composition}
\label{exa-seqc}
We assume a set of data $D$ and one output port $m$.
For $d \in D$, let $S_m(d)$ be the transition system
$\tup{S',A',\step{}',\linebreak[2]\term{}',\linebreak[2]s_0'}$ where
\[
\begin{aeqns}
S' & = & \set{\tup{m,d},\tup{m,\und}}\enspace, \\
A' & = & \set{\kw{s}_m(d)}\enspace, \\
{\step{}'} & = & 
 \set{\astep{\tup{m,d}}{\kw{s}_m(d)}{\tup{m,\und}}}\enspace, \\
{\term{}'} & = & \set{\tup{m,\und}}\enspace, \\
s_0' & = & \tup{m,d}\enspace.
\end{aeqns}
\]
The transition system $S_m(d)$ is capable of sending datum $d$ at port 
$m$ and then terminating successfully.
Let $R_k(d) \altc R_l(d)$ be as defined in Example~\ref{exa-altc}.
The sequential composition $(R_k(d) \altc R_l(d)) \seqc S_m(d)$ is the 
transition system $\tup{S'',A'',\step{}'',\term{}'',s_0''}$ where
\[
\begin{aeqns}
S'' & = & \set{\tup{\und,\und},\tup{m,d},\tup{m,\und}}\enspace, \\
A'' & = & \set{\kw{r}_k(d),\kw{r}_l(d),\kw{s}_m(d)}\enspace, \\
{\step{}''} & = & 
 \set{\astep{\tup{\und,\und}}{\kw{r}_k(d)}{\tup{m,d}},
      \astep{\tup{\und,\und}}{\kw{r}_l(d)}{\tup{m,d}},
      \astep{\tup{m,d}}{\kw{s}_m(d)}{\tup{m,\und}}}\enspace, \\
{\term{}''} & = & \set{\tup{m,\und}}\enspace, \\
s_0'' & = & \tup{\und,\und}\enspace.
\end{aeqns}
\]
This transition system is capable of receiving datum $d$ at port $k$ or 
$l$, next sending datum $d$ at port $m$ and then terminating 
successfully.
The sequential composition of $R_k(d) \altc R_l(d)$ and $S_m(d)$ is 
represented graphically in Fig.~\ref{fig-seqc}.
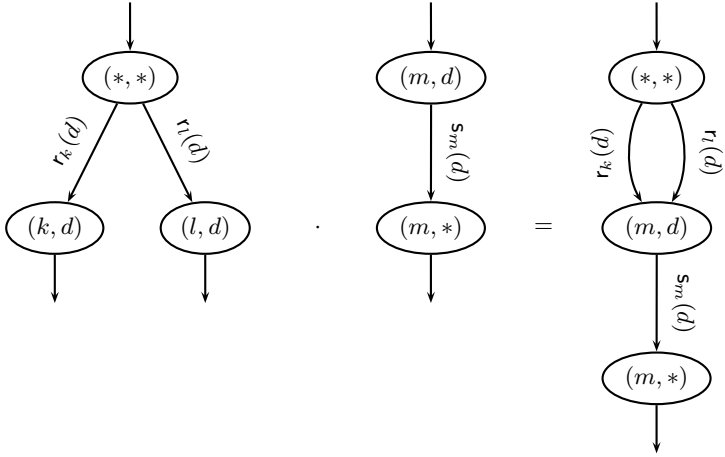
\begin{figure}
\begin{pspicture}(0,0)(10,6)

 \psset{arrows=->}

 \pnode(2,6){S''}
 \rput(2,5){\ovalnode{**}{$(*,*)$}}
 \rput(1,3){\ovalnode{kd}{$(k,d)$}}
 \rput(3,3){\ovalnode{ld}{$(l,d)$}}
 \pnode(1,2){E1''}
 \pnode(3,2){E2''}
 \ncline{S''}{**}
 \ncline{**}{kd}\nbput[nrot=:180]{$\kw{r}_k(d)$}
 \ncline{**}{ld}\naput[nrot=:0]{$\kw{r}_l(d)$}
 \ncline{kd}{E1''}
 \ncline{ld}{E2''}

 \rput(4.5,3){\rnode{D}{$\cdot$}}

 \pnode(6,6){S''}
 \rput(6,5){\ovalnode{md}{$(m,d)$}}
 \rput(6,3){\ovalnode{m*}{$(m,*)$}}
 \pnode(6,2){E}
 \ncline{S''}{md}
 \ncline{md}{m*}\naput[nrot=:0]{$\kw{s}_m(d)$}
 \ncline{m*}{E}

 \rput(7.5,3){\rnode{D}{$=$}}

\pnode(9,6){S}
 \rput(9,5){\ovalnode{**}{$(*,*)$}}
 \rput(9,3){\ovalnode{md}{$(m,d)$}}
 \rput(9,1){\ovalnode{m*}{$(m,*)$}}
 \pnode(9,0){E}
 \ncline{S}{**}
 \nccurve[angleA=240,angleB=120]{**}{md}\nbput[nrot=:180]{$\kw{r}_k(d)$}
 \nccurve[angleA=300,angleB=60]{**}{md}\naput[nrot=:0]{$\kw{r}_l(d)$}
 \ncline{md}{m*}\naput[nrot=:0]{$\kw{s}_m(d)$}
 \ncline{m*}{E}
\end{pspicture}
\caption{Sequential composition of $R_k(d) \altc R_l(d)$ and $S_m(d)$}
\label{fig-seqc}
\end{figure}
\end{example}
\begin{definition}[Iteration]\index{iteration}
Let $T = \tup{S,A,\step{},\term{},s_0}$ 
and $T' = \tup{S',A',\linebreak[2]\step{}',\linebreak[2]\term{}',s_0'}$ 
be transition systems such that $S \inter S' = \emptyset$.
The \emph{iteration} of $T$ with exit $T'$,
written $T \iter T'$, is the transition system 
$\tup{S'',A'',\step{}'',\term{}',s_0''}$ where
\begin{iteml}
\item
$S'' = \set{s_0''} \union \reachp(T) \union \reachp(T')$;
\item
$A'' = A \union A'$;
\item
$\step{}''$ is the smallest subset of $S'' \x {A''}_\tau \x S''$ such 
that: 
\begin{iteml}
\item
if $\astep{s_0}{a}{s_2}$ and not $\term{s_2}$, 
then $\asteppp{s_0''}{a}{s_2}$;
\item
if $\astep{s_0}{a}{s_2}$ and $\term{s_2}$, 
then $\asteppp{s_0''}{a}{s_0''}$;
\item
if $\astepp{s_0'}{a}{s_2'}$, 
then $\asteppp{s_0''}{a}{s_2'}$;
\item
if $\astep{s_1}{a}{s_2}$, $s_1 \in S''$ and not $\term{s_2}$, 
then $\asteppp{s_1}{a}{s_2}$;
\item
if $\astep{s_1}{a}{s_2}$, $s_1 \in S''$ and $\term{s_2}$, 
then $\asteppp{s_1}{a}{s_0''}$;
\item
if $\astepp{s_1'}{a}{s_2'}$ and $s_1' \in S''$, 
then $\asteppp{s_1'}{a}{s_2'}$;
\end{iteml}
\item
$ s''_0 \notin S \union S'$.
\end{iteml}
\end{definition}
Like in the case of alternative composition, the iteration of 
transition systems $T$ with exit transition system $T'$ has a fresh 
initial state that adopts the transitions from the initial state of $T$ 
and the transitions from the initial state of $T'$.
Again, this is needed because otherwise choices could come back 
unintentionally. 
Like in the case of sequential composition, successfully terminating 
states are relevant to the transitions of the resulting transition 
system.
In the case of iteration, the fresh initial state replaces all 
successfully terminating states of $T$.
In this way, the choice, that is there initially, will come back after 
successful termination of $T$.
Here is an example to illustrate that it is quite natural to look at
certain real-life processes as the iteration with exit of other 
processes.
\begin{example}[Railroad crossing controller]%
\index{railroad crossing controller}
\label{exa-rrc-iter}
We look once more at the railroad crossing controller from 
Examples~\ref{exa-rrc-altc} and~\ref{exa-rrc-seqc}.
In this example, we take into account that, because of fault tolerance 
considerations, approach signals should always cause the gate to go 
down, and exit signals should be ignored while the gate is going down.
Suppose that $S(\nm{lower})$ is the transition system describing the 
behaviour of the subprocess dedicated to sending a lower signal. 
The behaviour of the subprocess dedicated to handling an approach 
signal that is received at the beginning of a cycle of the controller 
is described by 
$(R(\nm{appr}) \altc R(\nm{exit})) \iter S(\nm{lower})$.
This is the transition system $D$ referred to in 
Example~\ref{exa-rrc-seqc}.
\end{example}
As mentioned in Sect.~\ref{sect-composition-informal}, no-exit 
iteration can be treated as a special case of iteration with exit.
Here follows the precise definition.
\begin{definition}[No-exit iteration]\index{iteration!no-exit}
The \emph{no-exit iteration} of $T$, written $T \neiter$, is the 
transition system $T \iter T'$, where $T'$ is the transition system 
$\tup{\set{s_0},\emptyset,\emptyset,\emptyset,s_0}$.
\end{definition}
Here is an example to illustrate that it is quite natural to look at
certain real-life processes as the no-exit iteration of other 
processes.
\begin{example}[Railroad crossing controller]%
\index{railroad crossing controller}
\label{exa-rrc-neiter}
We look again at the railroad crossing controller from 
Examples~\ref{exa-rrc-altc}, \ref{exa-rrc-seqc} and~\ref{exa-rrc-iter}.
The transition system 
$(R(\nm{appr}) \seqc D) \altc (R(\nm{exit}) \seqc U)$ from 
Example~\ref{exa-rrc-seqc} describes the behaviour of one cycle of the 
controller.
The behaviour of the controller is described by
$((R(\nm{appr}) \seqc D) \altc (R(\nm{exit}) \seqc U)) \neiter$.
\end{example}
Let us also give an example illustrating the details of (no-exit)
iteration.
\begin{example}[Merge connection]\index{merge connection}
\label{exa-iter}
Let $(R_k(d)$ and $R_l(d)) \seqc S_m(d)$ be as defined in 
Example~\ref{exa-seqc}.
The no-exit iteration $((R_k(d) \altc R_l(d)) \seqc S_m(d)) \neiter$ is
the transition system $\tup{S',A',\step{}',\term{}',s_0'}$ where
\[
\begin{aeqns}
S' & = & \set{\und,\tup{m,d}}\enspace, \\
A' & = & \set{\kw{r}_k(d),\kw{r}_l(d),\kw{s}_m(d)}\enspace, \\
{\step{}'} & = & 
 \set{\astep{\und}{\kw{r}_k(d)}{\tup{m,d}},
      \astep{\und}{\kw{r}_l(d)}{\tup{m,d}},
      \astep{\tup{m,d}}{\kw{s}_m(d)}{\und}}\enspace, \\
{\term{}'} & = & \emptyset\enspace, \\
s_0' & = & \und\enspace.
\end{aeqns}
\]
This transition system is isomorphic to the second transition system
given for a merge connection in Example~\ref{exa-merge} in the case
where $D$ is a singleton set.
The no-exit iteration of $(R_k(d) \altc R_l(d)) \seqc S_m(d)$ is 
represented graphically in Fig.~\ref{fig-neiter}
\begin{figure}
\[
\begin{array}{c}
\left(

\begin{pspicture}(0.5,3)(3.5,6)
 \psset{arrows=->}
 \pnode(2,6){S}
 \rput(2,5){\ovalnode{**}{$(*,*)$}}
 \rput(2,3){\ovalnode{md}{$(m,d)$}}
 \rput(2,1){\ovalnode{m*}{$(m,*)$}}
 \pnode(2,0){E}
 \ncline{S}{**}
 \nccurve[angleA=240,angleB=120]{**}{md}\nbput[nrot=:180]{$\kw{r}_k(d)$}
 \nccurve[angleA=300,angleB=60]{**}{md}\naput[nrot=:0]{$\kw{r}_l(d)$}
 \ncline{md}{m*}\naput[nrot=:0]{$\kw{s}_m(d)$}
 \ncline{m*}{E}
\end{pspicture}

\right)^{\omega}

\begin{pspicture}(3.5,3)(4.5,6)
\rput(4,3){\rnode{D}{$=$}}
\end{pspicture}

\begin{pspicture}(5.5,3)(7.5,6)
 \psset{arrows=->}
 \pnode(6.5,6){S}
 \rput(6.5,5){\ovalnode{**}{$~~*~~$}}
 \rput(6.5,3){\ovalnode{md}{$(m,d)$}}
 \ncline{S}{**}
 \nccurve[angleA=240,angleB=120]{**}{md}\nbput[nrot=:180]{$\kw{r}_k(d)$}
 \nccurve[angleA=300,angleB=60]{**}{md}\naput[nrot=:0]{$\kw{r}_l(d)$}
 \nccurve[angleA=0,angleB=0]{md}{**}\nbput[nrot=:180]{$\kw{s}_m(d)$}
\end{pspicture}
\end{array}
\]
\caption{No-exit iteration of $(R_k(d) \altc R_l(d)) \seqc S_m(d)$}
\label{fig-neiter}
\end{figure}
\end{example}

Before we turn to more examples of the use of alternative composition, 
sequential composition and iteration, we will introduce atomic 
transition systems, i.e.\ transition systems that are capable of first 
performing a single action and then terminating successfully, and the 
inactive transition system, i.e. the transition system that is 
incapable of performing any action and of terminating successfully.
\begin{definition}[Atomic transition system]%
\index{atomic transition system}
\sloppy
Let $a$ be an action.
The \emph{atomic} transition system performing $a$ is the transition 
system 
$\tup{\set{s_0,s_1},\set{a},\set{\astep{s_0}{a}{s_1}},\set{s_1},s_0}$ 
where $s_0$ and $s_1$ are fresh states.
The \emph{inactive} transition system is the transition system 
$\tup{\set{s_0},\emptyset,\emptyset,\emptyset,s_0}$ where $s_0$ is a 
fresh state.
If no confusion can arise, the atomic transition system performing $a$
is simply denoted by $a$.
The inactive transition system is denoted by $\dead$.
\end{definition}
Bear in mind that it does not matter that arbitrary fresh states are
chosen, as up to isomorphism the result is independent of the
particular choice.
Notice that the inactive transition system $\dead$ is used in the 
definition of no-exit iteration: $T \neiter = T \iter \dead$.

Like for parallel composition, we use the convention of association to
the left for alternative composition and sequential composition.
The need to use parentheses is further reduced by ranking the
precedence of the binary operations on transition systems.
We adhere to the following precedence rules:
\begin{iteml}
\item
the operation $\altc$ has lower precedence than all others;
\item
the operation $\seqc$ has higher precedence than all others;
\item
all other operations have the same precedence.
\end{iteml}
For example, we write $x \seqc z \altc y \seqc z$ for
$(x \seqc z) \altc (y \seqc z)$.

Here are a couple of examples of the composition of transition systems 
starting from atomic transition systems.
These examples show a way to present transition systems that is quite 
different from the way that we used before.
It looks to be a more convenient way.
We will return to this later in Chapter~\ref{ch-expressions}.
\begin{example}[Bounded buffer]\index{buffer!bounded}
\label{exa-bbuffer-composed}
We consider again the bounded buffer from Example~\ref{exa-bbuffer}.
We restrict ourselves to the case where its capacity is $1$ and
it can only keep bits, i.e.\ $D = \set{0,1}$.
Using alternative composition, sequential composition and iteration,
its behaviour can be described as follows:
\begin{ldispl}
(\kw{add}(0) \seqc \kw{rem}(0) \altc \kw{add}(1) \seqc \kw{rem}(1)) 
 \neiter\;.
\end{ldispl}
\end{example}
\begin{example}[Split and merge connections]%
\index{split connection}\index{merge connection}
\label{exa-split-merge-composed}
We consider again the split connection from Example~\ref{exa-split}
and the merge connection from Example~\ref{exa-merge}.
We restrict ourselves once more to the case where only bits are 
involved, i.e.\ $D = \set{0,1}$.
Using alternative composition, sequential composition and iteration,
the behaviour of the split connection and the merge connection can be 
described as follows:
\begin{ldispl}
(\kw{r}_k(0) \seqc (\kw{s}_l(0) \altc \kw{s}_m(0))
  \altc 
 \kw{r}_k(1) \seqc (\kw{s}_l(1) \altc \kw{s}_m(1))) \neiter
\end{ldispl}
and 
\begin{ldispl}
((\kw{r}_k(0) \altc \kw{r}_l(0)) \seqc \kw{s}_m(0)
  \altc 
 (\kw{r}_k(1) \altc \kw{r}_l(1)) \seqc \kw{s}_m(1)) \neiter\;.
\end{ldispl}
\end{example}
Here is another example, showing that the behaviour of simple PASCAL 
programs upon execution can also be described using alternative 
composition, sequential composition and iteration.
\begin{example}[Factorial program]\index{factorial program}
\label{exa-factorial-composed}
We consider again the PASCAL program to calculate factorials from 
Example~\ref{exa-factorial}.
Using alternative composition, sequential composition and iteration,
the behaviour of this program upon execution can be described as 
follows:
\begin{ldispl}
(\kw{read(n)}) \seqc (\kw{i\,:=\,0}) \seqc (\kw{f\,:=\,1}) \seqc
\\
\left(
 ((\kw{i\,<\,n}) \seqc 
  (\kw{i\,:=\,i\,+\,1}) \seqc 
  (\kw{f\,:=\,f\,*\,i})) \iter
 (\kw{NOT\,i\,<\,n})\right) \seqc (\kw{write(f)})\;.
\end{ldispl}
\end{example}
For reasons of readability, we have enclosed all atomic transition 
systems in parentheses.
We cannot directly give a transition system describing the behaviour of 
a program upon execution on a machine by means of atomic transition 
systems, alternative composition, sequential composition and iteration.
Nor we can give a transition system describing the behaviour of the
machine on which the program is executed in this way.
For the machine, as well as a category of simple programs, it is
possible if we use in addition parallel composition, encapsulation and
abstraction.
However, it requires special tricks.
The kind of tricks needed here, will be illustrated later in 
Example~\ref{exa-bcounter-expr}.
We will see later in Chapter~\ref{ch-expressions} that we can do better
if it is in addition possible to define transition systems
recursively.

\section{Example: Alternating bit protocol}%
\index{alternating bit protocol}
\label{sect-composition-abp}

We continue with the example of Sects.~\ref{sect-interaction-abp}
and~\ref{sect-abstraction-abp} concerning the ABP.
Here, we describe the behaviour of the sender $S$, the data transmission
channel $K$, the acknowledgement transmission channel $L$ and the 
receiver $R$ using alternative composition, sequential composition and 
iteration.

We restrict ourselves to the case where the set $D$ of data is finite.
Thus, we will use the following abbreviation.
Let $\mathcal{I} = \set{i_1,\ldots,i_n}$ be an index set and $T_i$ be 
a transition system for each $i \in \mathcal{I}$.
Then we write $\vAltc{i \in {\cal I}} T_i$ for
$T_{i_1} \altc \ldots \altc T_{i_n}$.
We further use the convention that $\vAltc{i \in {\cal I}} T_i$ stands
for $\dead$ if $\mathcal{I} = \emptyset$.

The behaviour of the sender $S$ can be described as follows:
\begin{ldispl}
\Bigg(
 \Altc{d \in D} 
  \kw{r}_1(d) \seqc \kw{s}_3(d,0) \seqc
  \left(
   ((\kw{r}_5(1) \altc \kw{r}_5(\und)) \seqc \kw{s}_3(d,0)) \iter 
   \kw{r}_5(0)\right) \seqc
 \\ \phantom{\Bigg(}
 \Altc{d \in D} 
  \kw{r}_1(d) \seqc \kw{s}_3(d,1) \seqc
  \left(
   ((\kw{r}_5(0) \altc \kw{r}_5(\und)) \seqc \kw{s}_3(d,1)) \iter 
   \kw{r}_5(1)\right)
\Bigg) \neiter
\end{ldispl}
The behaviour of the receiver $R$ can be described as follows:
\begin{ldispl}
\Bigg(\left(\left(\left(
    \Altc{d \in D} \kw{r}_4(d,1) \altc \kw{r}_4(\und)\right) \seqc 
   \kw{s}_6(1)\right) \iter
  \Altc{d \in D} \kw{r}_4(d,0)\right) \seqc 
 \kw{s}_2(d) \seqc \kw{s}_6(0) \seqc
 \\ \phantom{\Bigg(}
 \left(\left(\left(
    \Altc{d \in D} \kw{r}_4(d,0) \altc \kw{r}_4(\und)\right) \seqc 
   \kw{s}_6(0)\right) \iter
  \Altc{d \in D} \kw{r}_4(d,1)\right) \seqc 
 \kw{s}_2(d) \seqc \kw{s}_6(1)
\Bigg) \neiter
\end{ldispl}
The behaviour of the data transmission channel $K$ can be described as 
follows:
\begin{ldispl}
\left(
 \Altc{f \in F} 
  \kw{r}_3(f) \seqc 
  (\kw{i} \seqc \kw{s}_4(f) \altc \kw{i} \seqc \kw{s}_4(\und))
\right) \neiter
\end{ldispl}
The behaviour of the acknowledgement transmission channel $L$ can be 
described as follows:
\begin{ldispl}
\left(
 \Altc{b \in B}
  \kw{r}_6(b) \seqc 
  (\kw{i} \seqc \kw{s}_5(b) \altc \kw{i} \seqc \kw{s}_5(\und))
\right) \neiter
\end{ldispl}
The transition systems for $S$, $R$, $K$ and $L$ presented above using 
alternative composition, sequential composition and iteration are 
bisimulation equivalent to the ones presented in 
Sect.~\ref{sect-interaction-abp}.
The transition systems for $K$ and $L$ are even isomorphic to the ones
presented in Sect.~\ref{sect-interaction-abp}.

\section{Bisimulation and trace equivalence}
\label{sect-composition-eqv}

Because successfully terminating states are now distinguished from
other terminal states, the definition of branching bisimulation
equivalence needs to be adapted as well.
The new definition is nothing but a rather simple adjustment of the
earlier definition reflecting that successful termination is now a 
capability that counts as well: 
\begin{iteml}
\item
if states $s$ and $s'$ are related and $s$ is a successfully 
terminating state in $T$, then there is some successfully terminating 
state $s''$ in $T'$ such that a generalized transition with a sequence 
of zero or more silent steps as label is possible from $s'$ to $s''$, 
and $s$ and $s''$ are related;
\item
likewise, with the role of $T$ and $T'$ reversed.
\end{iteml}
\begin{definition}[Branching bisimulation]\index{branching bisimulation} 
Let $T = \tup{S,A,\step{},\term{},s_0}$ 
and $T' = \tup{S',A',\step{}',\term{}',s_0'}$ 
be transition systems such that $A = A'$.
Then a \emph{branching bisimulation} $B$ between $T$ and $T'$ is a 
binary relation $B \subseteq S \x S'$ such that the following 
conditions hold:
\begin{enuml}
\item
$B(s_0,s_0')$;
\item 
whenever $B(s_1,s_1')$ and $\astep{s_1}{a}{s_2}$, 
then either $a = \tau$ and $B(s_2,s_1')$ or there are states 
$s_1'',s_2'$ such that $s_1' \silentp \astepp{s_1''}{a}{s_2'}$ and 
$B(s_1,s_1'')$ and $B(s_2,s_2')$;
\item 
whenever $B(s_1,s_1')$ and $\astepp{s_1'}{a}{s_2'}$, 
then either $a = \tau$ and $B(s_1,s_2')$ or there are states 
$s_1'',s_2$ such that $s_1 \silent \astep{s_1''}{a}{s_2}$ and 
$B(s_1',s_1'')$ and $B(s_2,s_2')$;
\item
whenever $B(s,s')$ and $\term{s}$, 
then there is a state $s''$ such that $s' \silentp \termp{s''}$ and 
$B(s,s'')$;
\item
whenever $B(s,s')$ and $\termp{s'}$, 
then there is a state $s''$ such that $s \silent \term{s''}$ and 
$B(s',s'')$.
\end{enuml}
Two transition systems $T$ and $T'$ are \emph{branching bisimulation
equivalent}\index{branching bisimulation equivalence}, written 
$T \bisim_\mathrm{b} T'$, if there exists a branching bisimulation $B$ 
between $T$ and $T'$.
\end{definition}
What is new in this definition of branching bisimulation equivalence,
compared with the definition from Chap.~\ref{ch-abstraction}, concerns
again successful termination.

However, this generalization introduces an anomaly as we will 
demonstrate in the following example.
\begin{example}[Non-preservation of branching bisimulation equivalence]%
\index{branching bisimulation equivalence!preservation of}
\label{exa-rooted}
We consider again the transition systems from Example~\ref{exa-tau-1}.
We restrict ourselves to the case where only bits are involved, i.e.\
$D = \set{0,1}$.
Using atomic transition systems, alternative composition and sequential 
composition, they can be presented as follows:
\begin{ldispl}
\kw{r}_1(0) \seqc (\kw{s}_2(0) \altc \tau \seqc \kw{s}_3(\und)) \altc
\kw{r}_1(1) \seqc (\kw{s}_2(1) \altc \tau \seqc \kw{s}_3(\und)) 
\end{ldispl}
and
\begin{ldispl}
\kw{r}_1(0) \seqc (\kw{s}_2(0) \altc \kw{s}_3(\und)) \altc
\kw{r}_1(1) \seqc (\kw{s}_2(1) \altc \kw{s}_3(\und))\;.
\end{ldispl}
The second case is the first case with $\tau \seqc \kw{s}_3(\und)$ 
replaced by $\kw{s}_3(\und)$.
The latter two transition systems are branching bisimulation 
equivalent, but the former two are not as explained in 
Example~\ref{exa-tau-1}.
Hence, branching bisimulation equivalence fails to be a congruence
with respect to alternative composition.
\end{example}
This anomaly can simply be resolved by requiring that the initial 
states are related as in the case of standard bisimulation equivalence.
\begin{definition}[Rooted branching bisimulation]%
\index{branching bisimulation!rooted}
Let $T = \tup{S,A,\linebreak[2]\step{},\linebreak[2]\term{},s_0}$ 
and $T' = \tup{S',A',\step{}',\term{}',s_0'}$ 
be transition systems such that $A = A'$.
If $B$ is a branching bisimulation between $T$ and $T'$, 
then we say that a pair $\tup{s_1,s_1'} \in S \x S'$ satisfies the 
\emph{root condition}\index{root condition} in $B$ if the following 
conditions hold:
\begin{enuml}
\item 
whenever $\astep{s_1}{a}{s_2}$, then there is a state $s_2'$ such 
that $\astepp{s_1'}{a}{s_2'}$ and $B(s_2,s_2')$;
\item 
whenever $\astepp{s_1'}{a}{s_2'}$, then there is a state $s_2$ such 
that $\astep{s_1}{a}{s_2}$ and $B(s_2,s_2')$.
\end{enuml}
The two transition systems $T$ and $T'$ are  
\emph{rooted branching bisimulation equivalent}%
\index{branching bisimulation equivalence!rooted}, 
written $T \bisim_\mathrm{rb} T'$, if there exists a branching 
bisimulation $B$ between $T$ and $T'$ such that the pair 
$\tup{s_0,s_0'}$ satisfies the root condition in $B$.
\end{definition}

Just as branching bisimulation equivalence, rooted branching 
bisimulation equivalence is preserved by parallel composition, 
encapsulation and abstraction. 
Moreover, it is preserved by alternative composition, sequential 
composition and iteration.
\begin{property}%
[Preservation of rooted branching bisimulation equivalence]%
\index{branching bisimulation equivalence!rooted!preservation of}
Let $T_1$ and $T_2$ be transition systems with $A$ as set of actions,
let $T'_1$ and $T'_2$ be transition systems with $A'$ as set of
actions, and 
let $\commf$ be a communication function on a set of actions that 
includes $A \union A'$.
Then the following holds:
\begin{ldispl}
 \begin{geqns}
\mbox
 {if $T_1 \bisim_\mathrm{rb} T_2$ and $T'_1 \bisim_\mathrm{rb} T'_2$, 
  then
  $T_1 \altc T'_1 \bisim_\mathrm{rb} T_2 \altc T'_2$,\,}
\\ 
\mbox
 {\phantom{if}
  $T_1 \seqc T'_1 \bisim_\mathrm{rb} T_2 \seqc T'_2$,\,
  $T_1 \iter T'_1 \bisim_\mathrm{rb} T_2 \iter T'_2$ and
  $T_1 \parcs{\commf} T'_1 \bisim_\mathrm{rb} T_2 \parcs{\commf} T'_2$;}
\\
\mbox
{if $T_1 \bisim_\mathrm{rb} T_2$, 
 then $\encap{H}(T_1) \bisim_\mathrm{rb} \encap{H}(T_2)$
 and  $\abstr{I}(T_1) \bisim_\mathrm{rb} \abstr{I}(T_2)$.}
 \end{geqns}
\end{ldispl}
\end{property}

If we consider transition systems the same if they are rooted branching
bisimulation equivalent, then both parallel composition and alternative
composition are commutative and associative, and sequential composition
is associative.
\begin{property}%
[Commutativity and associativity of binary operations]%
\index{alternative composition!commutativity and associativity of}%
\index{sequential composition!associativity of}%
\index{parallel composition!commutativity and associativity of}
Let $T_1$, $T_2$ and $T_3$ be transition systems with $A_1$, $A_2$ and 
$A_3$, respectively, as set of actions.
Let $\commf$ be a communication function on a set of actions that 
includes $A_1 \union A_2 \union A_3$.
Then the following holds:
\begin{ldispl}
 \begin{geqns}
  T_1 \altc T_2 \bisim_\mathrm{rb} T_2 \altc T_1\;,
\\
  (T_1 \altc T_2) \altc T_3 
  \bisim_\mathrm{rb} 
  T_1 \altc (T_2 \altc T_3)\;,
\\
  (T_1 \seqc T_2) \seqc T_3 
  \bisim_\mathrm{rb} 
  T_1 \seqc (T_2 \seqc T_3)\;,
\\
  T_1 \parcs{\commf} T_2 \bisim_\mathrm{rb} T_2 \parcs{\commf} T_1\;,
\\
  (T_1 \parcs{\commf} T_2) \parcs{\commf} T_3 
  \bisim_\mathrm{rb} 
  T_1 \parcs{\commf} (T_2 \parcs{\commf} T_3)\;.
 \end{geqns}
\end{ldispl}
\end{property}

Transition systems can be reduced to connected transition systems as 
follows.
\begin{definition}[Reduction]\index{reduction} 
Let $T = \tup{S,A,\step{},\term{},s_0}$ be a transition system.
Then the \emph{reduction} of $T$, written $\reduct(T)$, is the 
transition system $\tup{S',A',\step{}',\term{}',s_0}$ where
\begin{iteml}
\item
$S' = \reach(T)$;
\item
$A' = \set{a \in A \where \Exists{s,s' \in S'}{\astep{s}{a}{s'}}}$;
\item
${\step{}'} = {\step{}} \inter (S' \x A' \x S')$;
\item
${\term{}'} = {\term{}} \inter S'$. 
\end{iteml}
\end{definition}
Any transition system is rooted branching bisimulation equivalent to
its reduction, which is a connected transition system.
\begin{property}
Let $T$ be a transition system.
Then the following holds:
\begin{ldispl}
 \begin{geqns}
T \bisim_\mathrm{rb} \reduct(T)\;.
 \end{geqns}
\end{ldispl}
\end{property}

The definition of trace equivalence needs to be adapted to the setting
with successful termination as well.
\begin{definition}[Trace equivalence]\index{trace equivalence}
Let $T = \tup{S,A,\step{},\term{},s_0}$ be a transition system.
A \emph{terminating} trace\index{trace!terminating} of $T$ is a sequence 
$\sigma \in \seqof{A}$ such that $\gstep{s_0}{\sigma}{s}$ and $\term{s}$ 
for some $s \in S$.
We write $\lang(T)$ for the set of all terminating traces of $T$.
Then two transition systems $T$ and $T'$ are \emph{trace equivalent},
written $T \treqv T'$, if $\tr(T) = \tr(T')$ and $\lang(T) = \lang(T')$.
\end{definition}
In those cases where only the terminating traces of a transition system
matter, an equivalence can be used that is even coarser than trace
equivalence.
\begin{definition}[Language equivalence]\index{language equivalence}
Two transition systems $T$ and $T'$ are \emph{language equivalent},
written $T \leqv T'$, if $\lang(T) = \lang(T')$.
\end{definition}
Obvious the terminology used here is based on viewing a transition 
system as an automaton by regarding its actions as symbols and its
successfully terminating states as final states, cf.\ 
Sect.~\ref{sect-basics-conn-automata}.

\section{Miscellaneous}
\label{sect-composition-misc}

We have seen in Chap.~\ref{ch-interaction} that it is slightly simpler 
to define parallel composition on nets than it is on transition systems.
On the other hand, it is fairly complicated to define alternative 
composition, sequential composition and iteration on nets.
For that reason, we will not show that alternative composition, 
sequential composition and iteration can be defined on nets as well.


\chapter{Expressions and Recursion}
\label{ch-expressions}

Transition systems describing the behaviour of real-life systems are
generally very large or even infinite.
They become lightly unintelligible.
Succinctness can be gained by using the operations introduced to 
compose transition systems of others.
We have already illustrated this in Chaps.~\ref{ch-interaction} 
and~\ref{ch-composition}.
However, the notation used there was introduced in an ad hoc and
informal way.
In order to preclude any difference of opinion about the form and 
intended meaning of the expressions concerned, called process
expressions, we give in this chapter a syntax and semantics which 
describe in a mathematically precise way how to construct process
expressions and how to assign meanings to them.
In other words, we turn the informal notation used the preceding 
chapters into a formalized language.
The formalization enables us to define transition systems, up to rooted
branching bisimulation equivalence, by means of recursive 
specifications. 
First of all, we discuss some important issues concerning process 
expressions and recursive specifications 
(Sect.~\ref{sect-expressions-informal}).
After that, we first give the syntax of process expressions 
(Sect.~\ref{sect-expressions-syntax}) and then give the semantics of 
process expressions (Sect.~\ref{sect-expressions-semantics}).
Next, we look at recursive specifications 
(Sect.~\ref{sect-expressions-rec}).
We also use recursive specification to define the components of the 
simple data communication protocol from Sect.~\ref{sect-interaction-abp}
(Sect.~\ref{sect-expressions-abp}) and to define the components of a 
workcell in a manufacturing system 
(Sect.~\ref{sect-expressions-workcell}).

\section{Introduction}
\label{sect-expressions-informal}

A main purpose of this chapter is to turn the informal notation used 
in the preceding chapters into a formalized language, and consequently
to make the intended meaning of the expressions concerned fully 
precise. 
The way in which the notation was introduced does not make it really 
fully precise.
Actually, we have used the same notation in different chapters for 
slightly different things.
In order to make the intended meaning of the expressions concerned 
fully precise, we have to make the form of the expressions fully
precise first.
The expressions given in Examples~\ref{exa-bbuffer-composed},
\ref{exa-split-merge-composed}, \ref{exa-factorial-composed}
and~\ref{exa-rooted} from Chap.~\ref{ch-composition} are all of the 
form that we will make precise in this chapter.

As already suggested after Example~\ref{exa-factorial-composed}, 
recursive specifications add to the behaviours that can be defined by 
means of process expressions.
It can be guaranteed that the solutions of recursive specifications are
unique up to rooted branching bisimulation equivalence by imposing
a weak restriction, known as guardedness, on the shape of recursive 
specifications. 
This is one of the reason why transition systems are no suitable 
candidates for the meanings of process expressions.  
The meaning of each process expression should represent all transition 
systems rooted branching bisimulation equivalent to some transition 
system. 
The solution is simply to have sets of transition systems that are 
rooted branching bisimulation equivalent to some transition system as 
meanings.
We will call these meanings processes, hence the name process 
expressions.
It is easy to lift the operations on transition systems defined in
Chap.~\ref{ch-composition} to processes because rooted branching 
bisimulation equivalence is a congruence with respect to those
operations.
\begin{example}[Process expressions]\index{process expression}
\label{exa-expr}
In the informal notation of Chap.~\ref{ch-composition},  
$(\kw{r}_k(0) \altc \kw{r}_l(0)) \seqc \kw{s}_m(0)$ and
$\kw{r}_k(0) \seqc \kw{s}_m(0) \altc \kw{r}_l(0) \seqc \kw{s}_m(0)$
denote different, but rooted branching bisimulation equivalent, 
transition systems.
As expressions of the formalized language that is introduced in this
chapter, they have the same meaning.
This is shown in detail later in Example~\ref{exa-meaning}.
\end{example}

We do not take all transition systems into consideration.
Because unreachable states and transitions are not relevant to the 
behaviour described by a transition system, we do not consider 
transition systems that are not connected.
In fact, we consider only finitely branching connected transition 
systems.
The reason for this is that, with finitely branching connected 
transition systems, we still cover all processes that are definable by 
means of process expressions or specifiable by means of guarded 
recursion.
An important thing to remember here is that the operations on 
transition systems defined in Chap.~\ref{ch-composition} result in 
connected transition systems if they are applied to connected 
transition systems.

It is interesting that there are unguarded recursive specifications of
which all solutions consist of transition systems that are not finitely 
branching.
However, we do not consider unguarded recursive specifications, because 
not all of them have a unique solution.

\section{Syntax of process expressions}
\label{sect-expressions-syntax}

In the previous section, we have prepared the way for the formal 
definition of the syntax and semantics of process expressions.
We give the definition of the syntax in this section, and the 
definition of the semantics in the next section.

We assume a fixed but arbitrary set $\Act$ of actions and a fixed but
arbitrary communication function $\funct{\commf}{\Act \x \Act}{\Act}$.
The set $\Act$ and the function $\commf$ can be regarded as parameters 
of the language, instantiated for each application of the language.
Moreover, we assume a set $\Var$ of \emph{process variables}.

We are now ready to describe in a fully precise way how to construct 
process expressions.
This is done by defining the set of process expressions inductively by 
formation rules.
\begin{definition}[Process expression]\index{process expression}
The set of \emph{process expressions} over $\Act$ is the smallest set
$\PE(\Act)$ satisfying:
\begin{iteml}
\item
$x \in \PE(\Act)$ for each variable $x \in \Var$;
\item
$a \in \PE(\Act)$ for each action $a \in \Act$;
\item
$\dead \in \PE(\Act)$;
\item
if $p \in \PE(\Act)$ and $q \in \PE(\Act)$, then 
$(p \altc q) \in \PE(\Act)$, $(p \seqc q) \in \PE(\Act)$, 
$(p \iter q) \in \PE(\Act)$, and $(p \parc q) \in \PE(\Act)$;
\item
if $p \in \PE(\Act)$, then $p \neiter \in \PE(\Act)$,
$\encap{H}(p) \in \PE(\Act)$ for each $H \subseteq \Act$, and 
$\abstr{I}(p) \in \PE(\Act)$ for each $I \subseteq \Act$.
\end{iteml}
A process expression is \emph{closed}\index{process expression!closed} 
if it does not contain variables.
We write $\CPE(\Act)$ for the set 
$\set{p \in \PE(\Act) \where p\; \mathrm{is}\;\mathrm{closed}}$.
\end{definition}
If the set of actions is clear or irrelevant, we write $\CPE$ and $\PE$ 
instead of $\CPE(\Act)$ and $\PE(\Act)$, respectively.
Let us give an example of the construction of process expressions.
\begin{example}[Process expression]\index{process expression}
\label{exa-form}
We consider the expression 
$((\kw{r}_k(0) \altc \kw{r}_l(0)) \seqc \kw{s}_m(0))$.
Let $\kw{r}_k(0),\kw{r}_l(0),\kw{s}_m(0) \in \Act$.
Then $\kw{r}_k(0),\kw{r}_l(0) \in \PE(\Act)$.
Hence, $(\kw{r}_k(0) \altc \kw{r}_l(0)) \in \PE(\Act)$.
Because $\kw{s}_m(0) \in \PE(\Act)$ as well, 
$((\kw{r}_k(0) \altc \kw{r}_l(0)) \seqc \kw{s}_m(0)) \in \PE(\Act)$.
In other words, $((\kw{r}_k(0) \altc \kw{r}_l(0)) \seqc \kw{s}_m(0))$ 
is a process expression.
\end{example}

All $a \in \Act$ and $\dead$ are called 
\emph{constants}\index{constant}, and $\altc$, $\seqc$, $\iter$, 
$\parc$, $\neiter$, $\encap{H}$ (for $H \subseteq \Act$) and $\abstr{I}$ 
(for $I \subseteq \Act$) are called \emph{operators}\index{operator}.
Constants and operators are symbols.
That is, the nature of constants and operators is purely syntactic. 

In order to reduce the need to use parentheses, like in the case of the
informal notation used in the preceding chapter, we use the convention 
of association to the left for the operators $\altc$, $\seqc$ and 
$\parc$, and in addition the following precedence rules:
\begin{iteml}
\item
the operator $\altc$ has lower precedence than all others;
\item
the operator $\seqc$ has higher precedence than all others;
\item
all other operators have the same precedence.
\end{iteml}
Moreover, we omit the outermost parentheses.
For example, we write $x \seqc y \seqc z \altc w$ for
$(((x \seqc y) \seqc z) \altc w)$.

\section{Semantics of process expressions}
\label{sect-expressions-semantics}

We begin with defining the set $\Prc(\Act)$.
This set is the semantic domain for process expressions, which means 
that the meanings of process expressions are elements of $\Prc(\Act)$.
\begin{definition}[Process]\index{process}
We consider transition systems as defined in Def.~\ref{def-ts-with-st}.
We write $\TS(\Act)$ for the set of all finitely branching connected
transition systems of which the set of actions is a subset of $\Act$.
A \emph{process} is an equivalence class of $\TS(\Act)$ with respect to
rooted branching bisimulation.
We write $\rbbeqvc{T}$ for the process 
$\set{T' \in \TS(\Act) \where T \bisim_\mathrm{rb} T'}$.
We write $\Prc(\Act)$ for $\set{\rbbeqvc{T} \where T \in \TS(\Act)}$,
i.e.\ the set of all processes of which the set of actions is a subset 
of $\Act$.
If a transition system $T \in \TS(\Act)$ is a member of a process 
$P \in \Prc(\Act)$, then $T$ is called a 
\emph{representative}\index{process!representative of} of $P$.
A process $P$ is called a \emph{regular} process\index{process!regular} 
if $P$ has a regular representative.
\end{definition}
If the set of actions is clear or irrelevant, we write $\TS$ and $\Prc$
instead of $\TS(\Act)$ and $\Prc(\Act)$, respectively.

For process expressions that contain variables, the meanings depend on 
the meanings assigned to the variables.
This is done by means of an assignment.
\begin{definition}[Assignment]\index{assignment}
An \emph{assignment} is a function $\funct{\alpha}{\Var}{\Prc(\Act)}$.
\end{definition}

In order to assign meanings to process expressions, we have to give
an interpretation to each constant and operator.
The interpretation of each constant is a process and the interpretation 
of each operator is an operation on processes.
Those operations on processes correspond to the operations on 
transition systems defined in Chap.~\ref{ch-composition}.
In order to distinguish the operators, the operations on processes 
corresponding to the operators and the operations on transition systems 
corresponding to the operations on processes from each other, we will
write for each operator, say $o$, $o\ssup{\Prc}$ for the corresponding
operation on processes and $o\ssup{\TS}$ for the corresponding
operation on transition systems.
It is important to remember that the operation denoted in this chapter
by $o\ssup{\TS}$ was denoted by $o$ in Chap.~\ref{ch-composition}.
\begin{definition}[Interpretation of constants and operators]%
\index{interpretation}\index{constant}\index{operator}
The \emph{interpretations} of the constants and operators of $\PE(\Act)$ 
are defined as follows:
\begin{ldispl}
\begin{aeqns}
a\ssup{\Prc} & = & \rbbeqvc{a\ssup{\TS}} 
                              & (\mathrm{for\; each\;} a \in \Act)
\\
\deadi{\Prc} & = & \rbbeqvc{\deadi{\TS}}
\\
\rbbeqvc{T} \altci{\Prc} \rbbeqvc{T'} & = & \rbbeqvc{T \altci{\TS} T'}
\\
\rbbeqvc{T} \seqci{\Prc} \rbbeqvc{T'} & = & \rbbeqvc{T \seqci{\TS} T'}
\\
\rbbeqvc{T} \iteri{\Prc} \rbbeqvc{T'} & = & \rbbeqvc{T \iteri{\TS} T'}
\\
\rbbeqvc{T} \parci{\Prc} \rbbeqvc{T'} & = & 
                                  \rbbeqvc{T \parcsi{\commf}{\TS} T'}
\\
\rbbeqvc{T} \neiteri{\Prc} & = & \rbbeqvc{T \neiteri{\TS}}
\\
\encapi{H}{\Prc}(\rbbeqvc{T}) & = & \rbbeqvc{\encapi{H}{\TS}(T)}
                             & (\mathrm{for\; each\;} H \subseteq \Act)
\\
\abstri{I}{\Prc}(\rbbeqvc{T}) & = & \rbbeqvc{\abstri{I}{\TS}(T)}
                             & (\mathrm{for\; each\;} I \subseteq \Act)
\end{aeqns}
\end{ldispl}
\end{definition}
These interpretations of the operators are well-defined because rooted 
branching bisimulation equivalence is a congruence with respect to the
corresponding operations on transition systems, so the choice of a
representative from an equivalence class does not matter.
Here is an example concerning the application of operations on processes 
defined above.
\begin{example}[Application of operations on processes]
\label{exa-rbbeqvc}
We consider the following application of operations on processes defined
above:
$(\kw{r}_k(0)\ssup{\Prc} \altci{\Prc} \kw{r}_l(0)\ssup{\Prc}) 
  \seqci{\Prc}
 \kw{s}_m(0)\ssup{\Prc}$.
It corresponds to one cycle of the merge connection from
Example~\ref{exa-merge}.
We calculate the resulting process:
\begin{ldispl}
\begin{aeqns}
& &
(\kw{r}_k(0)\ssup{\Prc} \altci{\Prc} \kw{r}_l(0)\ssup{\Prc}) 
  \seqci{\Prc}
 \kw{s}_m(0)\ssup{\Prc}
\\
& = &
(\rbbeqvc{\kw{r}_k(0)\ssup{\TS}} \altci{\Prc}
 \rbbeqvc{\kw{r}_l(0)\ssup{\TS}}) \seqci{\Prc}
\rbbeqvc{\kw{s}_m(0)\ssup{\TS}}
\\
& = &
\rbbeqvc{\kw{r}_k(0)\ssup{\TS} \altci{\TS} \kw{r}_l(0)\ssup{\TS}}
 \seqci{\Prc} 
\rbbeqvc{\kw{s}_m(0)\ssup{\TS}}
\\
& = &
\rbbeqvc{(\kw{r}_k(0)\ssup{\TS} \altci{\TS} \kw{r}_l(0)\ssup{\TS})
          \seqci{\TS} 
         \kw{s}_m(0)\ssup{\TS}}
\\
& = &
\set{(\kw{r}_k(0)\ssup{\TS} \altci{\TS} \kw{r}_l(0)\ssup{\TS})
      \seqci{\TS} 
     \kw{s}_m(0)\ssup{\TS}, 
\\
& & \phantom{\{}
     \kw{r}_k(0)\ssup{\TS} \seqci{\TS} \kw{s}_m(0)\ssup{\TS}   
      \altci{\TS} 
     \kw{r}_l(0)\ssup{\TS} \seqci{\TS} \kw{s}_m(0)\ssup{\TS}}\;.
\end{aeqns}
\end{ldispl}
The equivalence class
$\rbbeqvc{(\kw{r}_k(0)\ssup{\TS} \altci{\TS} \kw{r}_l(0)\ssup{\TS})
           \seqci{\TS} 
          \kw{s}_m(0)\ssup{\TS}}$ 
contains only two transition systems, because we consider transition 
systems to be the same if they are isomorphic.
It is clear that
$\rbbeqvc{(\kw{r}_k(0)\ssup{\TS} \altci{\TS} \kw{r}_l(0)\ssup{\TS})
           \seqci{\TS} 
          \kw{s}_m(0)\ssup{\TS}} =
 \rbbeqvc{\kw{r}_k(0)\ssup{\TS} \seqci{\TS} \kw{s}_m(0)\ssup{\TS}   
           \altci{\TS} 
          \kw{r}_l(0)\ssup{\TS} \seqci{\TS} \kw{s}_m(0)\ssup{\TS}}$.
In Example~\ref{exa-meaning}, we will see that this is important.
\end{example}

We are now in the position to describe in a fully precise way how to
assign meanings to process expressions.
This is done by defining evaluation functions, one for each 
assignment.
\begin{definition}[Meaning of process expressions]%
\index{process expression!meaning of}
\label{def-eval}
Let $\alpha$ be an assignment.
The \emph{evaluation} function for $\alpha$, 
$\funct{\M{\alpha}}{\PE(\Act)}{\Prc(\Act)}$, is recursively defined as 
follows:
\begin{ldispl}
\begin{aeqns}
\M{\alpha}(x) & = & \alpha(x)
\\
\M{\alpha}(a) & = & a\ssup{\Prc} & (\mathrm{for\; each\;} a \in \Act)
\\
\M{\alpha}(\dead) & = & \deadi{\Prc} 
\\
\M{\alpha}(p \altc q) & = & \M{\alpha}(p) \altci{\Prc} \M{\alpha}(q)
\\
\M{\alpha}(p \seqc q) & = & \M{\alpha}(p) \seqci{\Prc} \M{\alpha}(q)
\\
\M{\alpha}(p \iter q) & = & \M{\alpha}(p) \iteri{\Prc} \M{\alpha}(q)
\\
\M{\alpha}(p \parc q) & = & \M{\alpha}(p) \parci{\Prc} \M{\alpha}(q)
\\
\M{\alpha}(p \neiter) & = & \M{\alpha}(p) \neiteri{\Prc}
\\
\M{\alpha}(\encap{H}(p)) & = & \encapi{H}{\Prc}(\M{\alpha}(p))
                             & (\mathrm{for\; each\;} H \subseteq \Act)
\\
\M{\alpha}(\abstr{I}(p)) & = & \abstri{I}{\Prc}(\M{\alpha}(p))
                             & (\mathrm{for\; each\;} I \subseteq \Act)
\end{aeqns}
\end{ldispl}
We say that $p = q$ \emph{holds} iff $\M{\alpha}(p) = \M{\alpha}(q)$ 
for all assignments $\alpha$.
\end{definition}
Clearly, the meanings of closed process expressions do not depend on
the assignment concerned.
Process expressions that contain variables are essential for recursive
specification of processes.
An important thing to note about process expressions is that all of 
them denote regular processes.
Recursively specified processes need not be regular.
We will return to this in Section~\ref{sect-expressions-rec}.
Here is an example of the evaluation of process expressions.
\begin{example}[Meaning of process expressions]%
\index{process expression!meaning of}
\label{exa-meaning}
We consider the process expression 
$(\kw{r}_k(0) \altc \kw{r}_l(0)) \seqc \kw{s}_m(0)$ from 
Example~\ref{exa-form}.
We assign a meaning to this process expression as follows:
\begin{ldispl}
\begin{aeqns}
& &
\M{\alpha}((\kw{r}_k(0) \altc \kw{r}_l(0)) \seqc \kw{s}_m(0))
\\
& = &
(\M{\alpha}(\kw{r}_k(0)) \altci{\Prc} \M{\alpha}(\kw{r}_l(0)))
 \seqci{\Prc} 
\M{\alpha}(\kw{s}_m(0))
\\
& = &
(\kw{r}_k(0)\ssup{\Prc} \altci{\Prc} \kw{r}_l(0)\ssup{\Prc}) 
  \seqci{\Prc}
 \kw{s}_m(0)\ssup{\Prc}
\\
& = &
\rbbeqvc{(\kw{r}_k(0)\ssup{\TS} \altci{\TS} \kw{r}_l(0)\ssup{\TS})
          \seqci{\TS} 
         \kw{s}_m(0)\ssup{\TS}}
\\
& = &
\set{(\kw{r}_k(0)\ssup{\TS} \altci{\TS} \kw{r}_l(0)\ssup{\TS})
      \seqci{\TS} 
     \kw{s}_m(0)\ssup{\TS}, 
\\
& & \phantom{\{}
     \kw{r}_k(0)\ssup{\TS} \seqci{\TS} \kw{s}_m(0)\ssup{\TS}   
      \altci{\TS} 
     \kw{r}_l(0)\ssup{\TS} \seqci{\TS} \kw{s}_m(0)\ssup{\TS}}\;.
\end{aeqns}
\end{ldispl}
From the third step, we made use of the calculations made in 
Example~\ref{exa-rbbeqvc}.
As for any closed process expression, we see that the meaning assigned 
to $(\kw{r}_k(0) \altc \kw{r}_l(0)) \seqc \kw{s}_m(0)$ does not depend 
on the assignment concerned.
Similarly, we obtain:
\begin{ldispl}
\begin{aeqns}
& &
\M{\alpha}
 (\kw{r}_k(0) \seqc \kw{s}_m(0) \altc \kw{r}_l(0) \seqc \kw{s}_m(0))
\\
& = &
\rbbeqvc{\kw{r}_k(0)\ssup{\TS} \seqci{\TS} \kw{s}_m(0)\ssup{\TS}   
          \altci{\TS} 
         \kw{r}_l(0)\ssup{\TS} \seqci{\TS} \kw{s}_m(0)\ssup{\TS}}
\\
& = &
\set{(\kw{r}_k(0)\ssup{\TS} \altci{\TS} \kw{r}_l(0)\ssup{\TS})
      \seqci{\TS} 
     \kw{s}_m(0)\ssup{\TS}, 
\\
& & \phantom{\{}
     \kw{r}_k(0)\ssup{\TS} \seqci{\TS} \kw{s}_m(0)\ssup{\TS}   
      \altci{\TS} 
     \kw{r}_l(0)\ssup{\TS} \seqci{\TS} \kw{s}_m(0)\ssup{\TS}}\;.
\end{aeqns}
\end{ldispl}
Thus, as to be expected, the process expressions
$(\kw{r}_k(0) \altc \kw{r}_l(0)) \seqc \kw{s}_m(0)$ and
$\kw{r}_k(0) \seqc \kw{s}_m(0) \altc \kw{r}_l(0) \seqc \kw{s}_m(0)$ are
assigned the same meaning for all assignments.
This means that 
$(\kw{r}_k(0) \altc \kw{r}_l(0)) \seqc \kw{s}_m(0) =
 \kw{r}_k(0) \seqc \kw{s}_m(0) \altc \kw{r}_l(0) \seqc \kw{s}_m(0)$
holds.
\end{example}
In the preceding example, the meaning of process expressions is given 
in terms of atomic transition systems and operations on transition 
systems.
In the following two examples, the meaning of process expressions is
given directly in terms of transition systems. 
\begin{example}[Milner's scheduling problem]%
\index{Milner's scheduling problem}
\label{exa-scheduler-expr}
We consider again the system of scheduled processes from 
Examples~\ref{exa-scheduler}, \ref{exa-scheduler-parallel}
and~\ref{exa-scheduler-ts}.
It is easy to see that the process that is the meaning of the process 
expression
\begin{ldispl}
(\kw{request}(i) \seqc \kw{finish}(i)) \neiter
\end{ldispl}
has the transition system for $P_i$ ($1 \leq i \leq n$) given in 
Example~\ref{exa-scheduler-ts} as a representative. 
It is also easy to see that the process that is the meaning of the 
process expression
\begin{ldispl}
(\kw{grant}(1) \seqc \ldots \seqc \kw{grant}(n)) \neiter
\end{ldispl}
has the transition system for $S$ given in 
Example~\ref{exa-scheduler-ts} as a representative. 
\end{example}
\begin{example}[Binary memory cell]\index{binary memory cell}
\label{exa-binvar-expr}
We consider again the binary memory cell from Example~\ref{exa-binvar}.
It is easy to see that the process that is the meaning of the process 
expression
\begin{ldispl}
(((\kw{rtr}(0) \altc \kw{sto}(0)) \iter \kw{sto}(1)) \seqc
 ((\kw{rtr}(1) \altc \kw{sto}(1)) \iter \kw{sto}(0))) \neiter
\end{ldispl}
has the transition system for the binary memory cell given at the end
of Example~\ref{exa-binvar} as a representative.
\end{example}

\section{Recursive specification}
\label{sect-expressions-rec}

In this section, we first explain what a recursive specification is and
after that we define it in a mathematically precise way.

An equation of the form $X = t$, where $X$ is a process variable and 
$t$ is a process expression that contains no variables other than $X$, 
determines a process, i.e. has a unique solution, if it satisfies a 
criterion known as guardedness.
Roughly speaking, this means that $X$ is always preceded by an action
in $t$.
An equation of the above-mentioned form is called a recursive
specification.
A recursive specification that satisfies the guardedness criterion is
called a guarded recursive specification.
A solution for $X$ of a recursive specification $X = t$ is a process
$p$ such that $X = t$ holds if $X$ stands for $p$.
In case $X = t$ is a guarded recursive specification, it has a unique
solution for $X$.
The capabilities of that solution can be approximated to any finite
depth $n$ by taking $t$ and replacing $n$ times all occurrences of $X$
by $t$.
It is easy to see that in the case where $X = t$ is not guarded, there 
are occurrences of $X$ that will inhibit a definite answer about a part 
of the capabilities, even to depth $1$.
Here are a couple of examples about guardedness and uniqueness of 
solutions.
\begin{example}[Uniqueness of solutions]%
\index{recursive specification!solution of!uniqueness of}
\label{exa-unguarded}
For any action $a \in \Act$, the recursive specifications $X = X$ and 
$Y = a \altc Y$ are unguarded.
Each process is a solution of $X = X$.
Replacing in the right-hand side of this equation the occurrences of 
$X$ by the right-hand side, even repeatedly, does not reveal anything 
about the capabilities of a solution.
Each process that has the option to perform action $a$ and then to 
terminate successfully is a solution of $Y = a \altc Y$.
Replacing in the right-hand side of this equation the occurrences of 
$Y$ by the right-hand side, even repeatedly, only confirms what was 
already known, viz.\ that a solution must be capable of performing 
action $a$ and then terminating successfully.
Let us now look at the guarded recursive specification $Z = a \seqc Z$.
Its only solution is the process that keeps performing action $a$
forever.
Replacing in the right-hand side of this equation the occurrences of 
$Z$ by the right-hand side $n$ times reveals that a solution must be 
capable of performing action $a$ $n + 1$ times and then proceeding as 
$Z$. 
\end{example}
\begin{example}[Existence of solutions]%
\index{recursive specification!solution of!existence of}
\label{exa-inf-branching}
For any action $a \in \Act$, the recursive specification 
$X = a \altc X \seqc a$ is unguarded.
A solution of this recursive specification must be capable of either
performing action $a$ once and then terminating successfully, or 
performing action $a$ twice and then terminating successfully, or 
performing action $a$ three times and then terminating successfully, 
etc.
Hence, a solution cannot be finitely branching.
This means that $X = a \altc X \seqc a$ has no solution, because only 
finitely branching transition systems are considered.
\end{example}
All of this extends from one equation to a set of equations where the 
left-hand sides of the equations are process variables and the 
right-hand sides of the equations are process expressions that contain 
only process variables that are among the ones on the left-hand sides of 
the equations.
This allows a number of processes to be defined in terms of each other,
which is known as mutual recursion.
Let us give an example of mutual recursion.
\begin{example}[Bounded counter]\index{counter!bounded}
\label{exa-mutual-recursion}
We consider once more the bounded counter from 
Example~\ref{exa-bcounter}.
We give a recursive specification for the case where the bound is 
$2$:
\begin{ldispl}
\begin{aeqns}
C2_0 & = &
\kw{inc} \seqc C2_1\;,
\eqnsep
C2_1 & = &
\kw{dec} \seqc C2_0 \altc 
\kw{inc} \seqc \kw{dec} \seqc C2_1\;.
\end{aeqns}
\end{ldispl}
The counter of which the value is $0$ ($C2_0$) is defined in terms of
the counter of which the value is $1$ ($C2_1$); and
the counter of which the value is $1$ ($C2_1$) is defined in terms of 
both counters ($C2_0$ and $C2_1$).
\end{example}

Let us now turn to the precise definitions of the notions of a recursive
specification, a solution of a recursive specification, and guardedness
of a recursive specification.
\begin{definition}[Recursive specification]%
\index{recursive specification}
A \emph{recursive specification} is a set of \emph{recursive}
equations $E = \set{X = t_X \where X \in V}$ where $V$ is a set of
process variables and each $t_X$ is a process expression that only
contains variables from $V$.
We denote the variables that occur in a recursive specification by
$X,X',Y,Y',...$.
Let $E$ be a recursive specification.
Then we write $\vars(E)$ for the set of all variables that occur on the
left-hand side of an equation in $E$.
\end{definition}
Notice that infinite sets of recursive equations are not excluded.
\begin{definition}[Solution of recursive specification]%
\index{recursive specification!solution of}
A \emph{solution} of a recursive specification $E$ is a set of 
processes $\set{p_X \in \Prc(\Act) \where X \in \vars(E)}$ such that
$\M{\alpha}(X) = \M{\alpha}(t_X)$ for all equations
$X = t_X \in E$ if $\alpha$ is an assignment such that
$\alpha(X) = p_X$ for all $X \in \vars(E)$.
\end{definition}
\begin{definition}[Guarded recursive specification]%
\index{recursive specification!guarded}
Let $t$ be a process expression containing a variable $X$.
We call an occurrence of $X$ in $p$ \emph{guarded} if $p$ has a 
subexpression of the form $a \seqc q$, where $a \in \Act$, with $q$ a 
process expression containing this occurrence of $X$.  
A recursive specification is called a \emph{guarded} recursive
specification if all occurrences of variables on the right-hand sides
of its equations are guarded or it can be rewritten to such a recursive
specification using equations that hold and the equations of the 
recursive specification.
\end{definition}
It is important to remember that guarded recursive specifications have 
unique solutions.
Let us look at an example of guarded recursive specifications of 
processes.
\begin{example}[Split and merge connections]%
\index{split connection}\index{merge connection}
\label{exa-split-merge-rec}
We consider again the split connection from Example~\ref{exa-split} and
the merge connection from Example~\ref{exa-merge}. 
As in Example~\ref{exa-split-merge-composed}, we restrict ourselves to
the case where only bits are involved, i.e.\ $D = \set{0,1}$.
The split connection and the merge connection can be recursively
specified as follows:
\begin{ldispl}
\begin{aeqns}
\nm{Split}^{k,lm} 
\\ \quad {} =
(\kw{r}_k(0) \seqc (\kw{s}_l(0) \altc \kw{s}_m(0)) \altc
 \kw{r}_k(1) \seqc (\kw{s}_l(1) \altc \kw{s}_m(1))) \seqc 
\nm{Split}^{k,lm}
\end{aeqns}
\end{ldispl}
and 
\begin{ldispl}
\begin{aeqns}
\nm{Merge}^{kl,m}
\\ \quad {} =
((\kw{r}_k(0) \altc \kw{r}_l(0)) \seqc \kw{s}_m(0) \altc
 (\kw{r}_k(1) \altc \kw{r}_l(1)) \seqc \kw{s}_m(1)) \seqc 
\nm{Merge}^{kl,m}\;.
\end{aeqns}
\end{ldispl}
The processes denoted by the process expressions given in
Example~\ref{exa-split-merge-composed} are the solutions of these
recursive specifications.
\end{example}
Here is another example of guarded recursive specifications of 
processes.
\begin{example}[Bounded buffer]\index{buffer!bounded}
\label{exa-bbuffer-rec}
We consider once more the bounded buffer from Example~\ref{exa-bbuffer}.
Like in Example~\ref{exa-bbuffer-composed}, we restrict ourselves to 
the case where it can only keeps bits, i.e.\ $D = \set{0,1}$.
We give guarded recursive specifications for the cases where its 
capacity is $1$ and $2$.
The buffer with capacity $1$ can be recursively specified as follows:
\begin{ldispl}
\begin{aeqns}
B1 & = &
(\kw{add}(0) \seqc \kw{rem}(0) \altc \kw{add}(1) \seqc \kw{rem}(1)) 
 \seqc
B1\;.
\end{aeqns}
\end{ldispl}
The solution of this guarded recursive specifications is the process 
denoted by the process expression given in
Example~\ref{exa-bbuffer-composed}. 
The buffer with capacity $2$ can be recursively specified as follows:
\begin{ldispl}
\begin{aeqns}
B2 & = &
\kw{add}(0) \seqc B2'_0 \altc \kw{add}(1) \seqc B2'_1\;,
\eqnsep
B2'_d & = &
\kw{rem}(d) \seqc B2 \altc 
\kw{add}(0) \seqc \kw{rem}(d) \seqc B2'_0 \altc 
\kw{add}(1) \seqc \kw{rem}(d) \seqc B2'_1 
\cndsep
\cnd{(\mathrm{for\; every}\; d \in \set{0,1}).}
\end{aeqns}
\end{ldispl}
The solution of this guarded recursive specification can be denoted by 
a process expression as well, but it is very clumsy.
\end{example}

It is not the case that the solution of each guarded recursive 
specification can be denoted by a process expression.
In the following couple of examples, we give guarded recursive 
specifications of which the solution cannot be denoted by process 
expressions.
\begin{example}[Unbounded counter]\index{counter!unbounded}
\label{exa-counter-rec}
We consider an unbounded counter.
The difference with a bounded counter is that its value can always be 
incremented.
The unbounded counter can be recursively specified as follows:
\begin{ldispl}
\begin{aeqns}
C & = & \kw{inc} \seqc C' \seqc C\;,
\eqnsep
C' & = &
\kw{dec} \altc \kw{inc} \seqc C' \seqc C'\;.
\end{aeqns}
\end{ldispl}
The solution of this guarded recursive specification cannot be denoted 
by a process expression.
A representative of the solution of this guarded recursive 
specification is represented graphically in Fig.~\ref{fig-counter-rec}.
\begin{figure}
\begin{pspicture}(0,0)(7,2)

 \psset{arrows=->}

 \pnode(0,1){S}
 \rput(1,1){\circlenode{0}{0}}
 \rput(3,1){\circlenode{1}{1}}
 \rput(5,1){\circlenode{2}{2}}
 \pnode(7,1.1){3} \pnode(7,0.9){4}

 \ncline{S}{0}
 \nccurve[angleA=30,angleB=150]{0}{1}\naput{$\kw{inc}$}
 \nccurve[angleA=210,angleB=330]{1}{0}\naput{$\kw{dec}$}
 \nccurve[angleA=30,angleB=150]{1}{2}\naput{$\kw{inc}$}
 \nccurve[angleA=210,angleB=330]{2}{1}\naput{$\kw{dec}$}
 \nccurve[angleA=30,angleB=150,linecolor=lightgray]{2}{3}
 \nccurve[angleA=210,angleB=330,linecolor=lightgray]{4}{2}
\end{pspicture}
\caption{Transition systems for the unbounded counter}
\label{fig-counter-rec}
\end{figure}
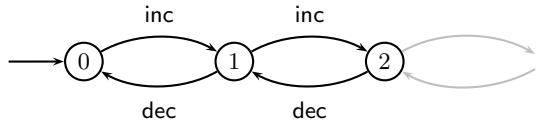
\end{example}
\begin{example}[Unbounded buffer]\index{buffer!unbounded}
\label{exa-buffer-rec}
We consider an unbounded buffer.
The difference with a bounded buffer is that new data can always be 
added to the data that it keeps.
Like in Example~\ref{exa-bbuffer-rec}, we restrict ourselves to the 
case where it can only keeps bits, i.e.\ $D = \set{0,1}$.
The unbounded buffer can be recursively specified as follows:
\begin{ldispl}
\begin{aeqns}
B & = & B'_\epsilon
\eqnsep
B'_\epsilon & = &
\kw{add}(0) \seqc B'_0 \altc \kw{add}(1) \seqc B'_1\;,
\eqnsep
B'_{\sigma\, d} & = &
\kw{rem}(d) \seqc B'_\sigma \altc 
\kw{add}(0) \seqc B'_{0\, \sigma\, d} \altc 
\kw{add}(1) \seqc B'_{1\, \sigma\, d} 
\cndsep
\cnd{(\mathrm{for\; every}\; \sigma \in \seqof{\set{0,1}}\;
      \mathrm{and}\; d \in \set{0,1}).}
\end{aeqns}
\end{ldispl}
The solution of this guarded recursive specification cannot be denoted 
by a process expression either.
A representative of the solution of this guarded recursive 
specification is represented graphically in Fig.~\ref{fig-buffer-rec}.
\begin{figure}
\begin{pspicture}(0,-2.5)(7,7)

 \psset{arrows=->}

 \pnode(5.875,7){S}
 \rput(5.875,6){\circlenode{e}{$\epsilon$}}
 \rput(2.375,5){\circlenode{0}{$0$}}
 \rput(9.125,5){\circlenode{1}{$1$}}
 \rput(0.5,3){\circlenode{00}{$00$}}
 \rput(4.25,3){\circlenode{10}{$10$}}
 \rput(7.25,3){\circlenode{01}{$01$}}
 \rput(11,3){\circlenode{11}{$11$}}

 \rput(0.5,0){\circlenode{000}{$000$}}
 \pnode(0,-1){E000l}
 \pnode(1,-1){E000r}
 \rput(2.375,3){\circlenode{100}{$100$}}
 \pnode(1.875,2){E100l}
 \pnode(2.875,2){E100r}
 \rput(5,1){\circlenode{010}{$010$}}
 \pnode(4.5,0){E010l}
 \pnode(5.5,0){E010r}
 \rput(3,-1){\circlenode{110}{$110$}}
 \pnode(2.5,-2){E110l}
 \pnode(3.5,-2){E110r}
 \rput(8.5,-1){\circlenode{001}{$001$}}
 \pnode(8,-2){E001l}
 \pnode(9,-2){E001r}
 \rput(6.5,1){\circlenode{101}{$101$}}
 \pnode(6,0){E101l}
 \pnode(7,0){E101r}
 \rput(9.125,3){\circlenode{011}{$011$}}
 \pnode(8.625,2){E011l}
 \pnode(9.625,2){E011r}
 \rput(11,0){\circlenode{111}{$111$}}
 \pnode(10.5,-1){E111l}
 \pnode(11.5,-1){E111r}

 \ncline{S}{e}
 \nccurve[angleA=210,angleB=0]{e}{0}\naput[nrot=:180]{$\kw{add}(0)$}
 \nccurve[angleA=330,angleB=180]{e}{1}\nbput[nrot=:0]{$\kw{add}(1)$}
 \nccurve[angleA=240,angleB=30]{0}{00}\naput[nrot=:180]{$\kw{add}(0)$}
 \ncline{0}{10}\naput[nrot=:0]{$\kw{add}(1)$}
 \ncline{1}{01}\nbput[nrot=:180]{$\kw{add}(0)$}
 \nccurve[angleA=300,angleB=150]{1}{11}\nbput[nrot=:0]{$\kw{add}(1)$}

 \nccurve[angleA=30,angleB=180]{0}{e}\naput[nrot=:0]{$\kw{rem}(0)$}
 \nccurve[angleA=150,angleB=0]{1}{e}\nbput[nrot=:180]{$\kw{rem}(1)$}
 \nccurve[angleA=60,angleB=210]{00}{0}\naput[nrot=:0]{$\kw{rem}(0)$}
 \ncline{01}{0}\nbput[nrot=:180]{$\kw{rem}(1)$}
 \ncline{10}{1}\naput[nrot=:0]{$\kw{rem}(0)$}
 \nccurve[angleA=120,angleB=330]{11}{1}\nbput[nrot=:180]{$\kw{rem}(1)$}

 \nccurve[angleA=285,angleB=75]{00}{000}\naput[nrot=:0]{$\kw{add}(0)$}
 \ncline{00}{100}\nbput[nrot=:0]{$\kw{add}(1)$}
 \ncline{01}{001}\naput[nrot=:0]{$\kw{add}(0)$}
 \ncline{01}{101}\naput[nrot=:180]{$\kw{add}(1)$}
 \ncline{10}{010}\nbput[nrot=:0]{$\kw{add}(0)$}
 \ncline{10}{110}\nbput[nrot=:180]{$\kw{add}(1)$}
 \ncline{11}{011}\naput[nrot=:180]{$\kw{add}(0)$}
 \nccurve[angleA=255,angleB=105]{11}{111}\nbput[nrot=:180]{$\kw{add}(1)$}

 \nccurve[angleA=105,angleB=255]{000}{00}\naput[nrot=:0]{$\kw{rem}(0)$}
 \nccurve[angleA=180,angleB=310]{001}{00}\naput[nrot=:180]{$\kw{rem}(1)$}
 \ncline{010}{01}\naput[nrot=:0]{$\kw{rem}(0)$}
 \ncline{011}{01}\naput[nrot=:180]{$\kw{rem}(1)$}
 \ncline{100}{10}\nbput[nrot=:0]{$\kw{rem}(0)$}
 \ncline{101}{10}\naput[nrot=:180]{$\kw{rem}(1)$}
 \nccurve[angleA=0,angleB=230]{110}{11}\nbput[nrot=:0]{$\kw{rem}(0)$}
 \nccurve[angleA=75,angleB=285]{111}{11}\nbput[nrot=:180]{$\kw{rem}(1)$}

\ncline[linecolor=lightgray]{000}{E000l}
\ncline[linecolor=lightgray]{100}{E100l}
\ncline[linecolor=lightgray]{110}{E110l}
\ncline[linecolor=lightgray]{010}{E010l}
\ncline[linecolor=lightgray]{101}{E101l}
\ncline[linecolor=lightgray]{001}{E001l}
\ncline[linecolor=lightgray]{011}{E011l}
\ncline[linecolor=lightgray]{111}{E111l}

\ncline[linecolor=lightgray]{000}{E000r}
\ncline[linecolor=lightgray]{100}{E100r}
\ncline[linecolor=lightgray]{110}{E110r}
\ncline[linecolor=lightgray]{010}{E010r}
\ncline[linecolor=lightgray]{101}{E101r}
\ncline[linecolor=lightgray]{001}{E001r}
\ncline[linecolor=lightgray]{011}{E011r}
\ncline[linecolor=lightgray]{111}{E111r}
\end{pspicture}
\caption{Transition systems for the unbounded buffer}
\label{fig-buffer-rec}
\end{figure}
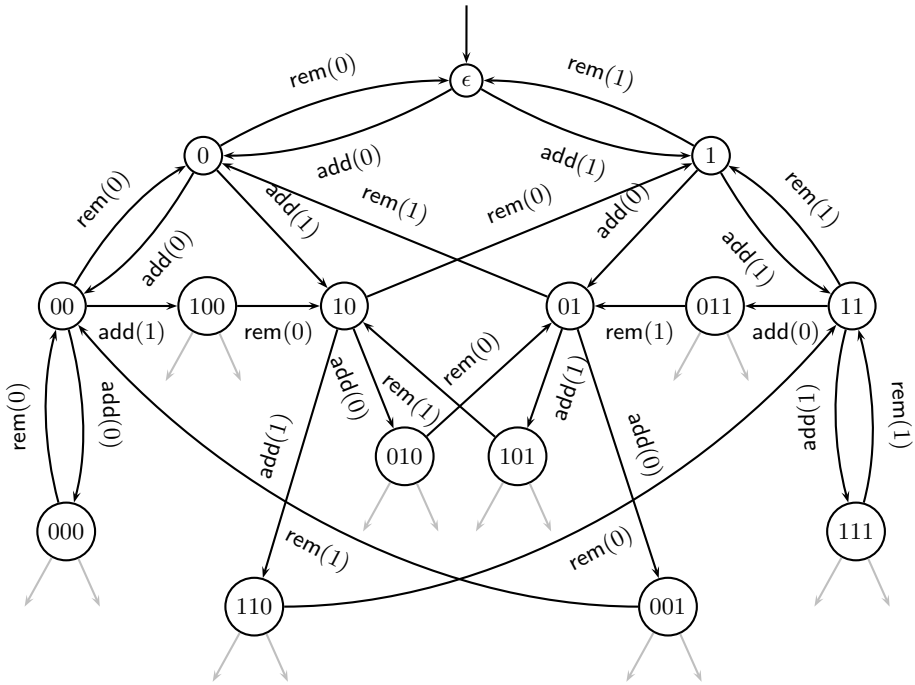
\end{example}
All process expressions introduced at the beginning of this chapter 
denote regular processes.
On the other hand, the solutions of guarded recursive specifications 
are not necessarily regular processes.
In other words, the processes that are specifiable by means of guarded
recursion include processes that are not definable by means of process
expressions.
For example, we have seen that the unbounded counters and buffers from 
Examples~\ref{exa-counter-rec} and~\ref{exa-buffer-rec} cannot be 
denoted by process expressions. 
That is because unbounded counters and buffers are not regular.

In the recursive specifications given in the preceding examples, 
parallel composition and encapsulation do not appear.
However, this is not excluded.
Let us give an example.
\begin{example}[Unbounded counter]\index{counter!unbounded}
\label{exa-counter-rec-alt}
We consider once more the unbounded counter from 
Example~\ref{exa-counter-rec}.
It can also be recursively specified as follows:
\begin{ldispl}
\begin{aeqns}
C & = & \kw{inc} \seqc (\kw{dec} \parc C)\;.
\end{aeqns}
\end{ldispl}
\end{example}

Example~\ref{exa-buffer-rec} is the first occasion where a guarded 
recursive specification with a (countably) infinite number of equations
is given.
It is surprising that, if our language of process expressions is 
extended with operators for the renaming of actions, it becomes 
possible to specify the unbounded buffer with two equations.
Parallel composition and encapsulation has to be used in that case as 
well.

A more advanced example of guarded recursive specification of processes
is given in the next section.

\section{Example: Alternating bit protocol}%
\index{alternating bit protocol}
\label{sect-expressions-abp}

We continue with the example of Sects.~\ref{sect-interaction-abp},
\ref{sect-abstraction-abp} and~\ref{sect-composition-abp} concerning
the ABP.
Here, we give guarded recursive specifications of the sender $S$, the 
data transmission channel $K$, the acknowledgement transmission channel 
$L$ and the receiver $R$.

We restrict ourselves again to the case where the set $D$ of data is 
finite.
Like in Sect.~\ref{sect-composition-abp}, we will use the sum
notation.
Let $\mathcal{I} = \set{i_1,\ldots,i_n}$ be an index set and $p_i$ be 
a process expression for each $i \in \mathcal{I}$.
Then we write $\vAltc{i \in {\cal I}} t_i$ for
$t_{i_1} \altc \ldots \altc t_{i_n}$.
We also use the convention that $\vAltc{i \in {\cal I}} t_i$ stands for 
$\dead$ if $\mathcal{I} = \emptyset$.
As in Sects.~\ref{sect-interaction-abp} and \ref{sect-abstraction-abp},
we write $\ol{b}$ for $1 - b$.

The guarded recursive specification of the sender $S$ consists of the 
following recursive equations:
\begin{ldispl}
\begin{aeqns}
S    & = & S'_0\;,
\eqnsep
S'_b & = &
\Altc{d \in D} \kw{r}_1(d) \seqc \kw{s}_3(d,b) \seqc  S''_{d,b}
\cndsep
\cnd{(\mathrm{for\; every}\; b \in B),}
\eqnsep
S''_{d,b} & = &
(\kw{r}_5(\ol{b}) \altc \kw{r}_5(\und)) \seqc 
                     \kw{s}_3(d,b) \seqc S''_{d,b} \altc
\kw{r}_5(b) \seqc S'_{\ol{b}}
\cndsep
\cnd{(\mathrm{for\; every}\; d \in D \; \mathrm{and}\; b \in B).}
\end{aeqns}
\end{ldispl}
The guarded recursive specification of the receiver $R$ consists of the 
following recursive equations:
\begin{ldispl}
\begin{aeqns}
R    & = & R'_0\;,
\eqnsep
R'_b & = &
(\Altc{d \in D} \kw{r}_4(d,\ol{b}) \altc \kw{r}_4(\und)) \seqc 
                                      \kw{s}_6(\ol{b}) \seqc R'_b 
\\ & {} \altc &
\Altc{d \in D} \kw{r}_4(d,b) \seqc 
 \kw{s}_2(d) \seqc \kw{s}_6(b) \seqc R'_{\ol{b}}
\cndsep
\cnd{(\mathrm{for\; every}\; b \in B).}
\end{aeqns}
\end{ldispl}
The guarded recursive specification of the data transmission channel
$K$ consists of the following recursive equation:
\begin{ldispl}
\begin{aeqns}
K & = &
\Altc{f \in F} 
 \kw{r}_3(f) \seqc 
 (\kw{i} \seqc \kw{s}_4(f) \altc \kw{i} \seqc \kw{s}_4(\und)) \seqc K\;.
\end{aeqns}
\end{ldispl}
The guarded recursive specification of the acknowledgement transmission 
channel $L$ consists of the following recursive equation:
\begin{ldispl}
\begin{aeqns}
L & = &
\Altc{b \in B}
 \kw{r}_6(b) \seqc 
 (\kw{i} \seqc \kw{s}_5(b) \altc \kw{i} \seqc \kw{s}_5(\und)) \seqc L\;.
\end{aeqns}
\end{ldispl}
The processes denoted by the process expressions given in
Sect.~\ref{sect-composition-abp} are the solutions of these guarded 
recursive specifications.

\section{Example: Workcell}\index{workcell}
\label{sect-expressions-workcell}

Here is another example of the use of guarded recursion in describing
the behaviour of systems.
The example concerns a workcell.
CIM (Computer Intergrated Manufacturing) systems are usually 
constructed from several workcells connected to each other via some
transport service, and controlled by some supervisor.
A workcell is itself constructed from various connected components,
including a workcell controller.
The workcell described in this section is the same as the workcell with 
quality check described in~\cite{Mau90}.

The main purpose of this example, which  is to illustrate that it is also 
possible to describe the behaviour of processes whose actions differ 
from those commonly found in pure software systems.
This is important because many systems are composed of both software 
and hardware components.
In this example, processes are involved that do not only send and 
receive messages, but also accept and deliver products.
Another thing to note about this example is the following.
The size and complexity of the system concerned exceed those of systems 
treated in preceding examples.
A corresponding description at the level of transition systems would
be fairly unintelligible.

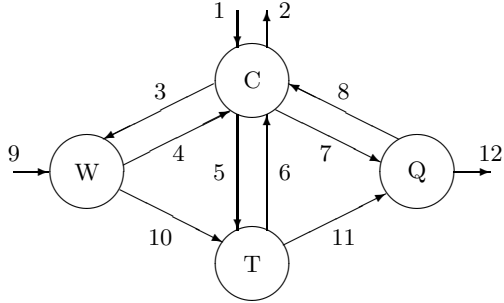
\begin{figure}
 
\setlength{\unitlength}{.15em}
\begin{picture}(150,90)(7,-0.5)
\put(10,40){\vector(1,0){9.5}}
\put(10,45){\makebox(0,0){9}}
\put(30,40){\circle{20}}
\put(30,40){\makebox(0,0){W}}
\put(64.5,64){\vector(-2,-1){30}}
\put(50,62){\makebox(0,0){3}}
\put(40,42){\vector(2,1){29}}
\put(55,45){\makebox(0,0){4}}
\put(71,84){\vector(0,-1){10}}
\put(66,84){\makebox(0,0){1}}
\put(79,74){\vector(0,1){10}}
\put(84,84){\makebox(0,0){2}}
\put(75,65){\circle{20}}
\put(75,65){\makebox(0,0){C}}
\put(71,55.5){\vector(0,-1){31.5}}
\put(66,40){\makebox(0,0){5}}
\put(79,24){\vector(0,1){31.5}}
\put(84,40){\makebox(0,0){6}}
\put(115,49){\vector(-2,1){30}}
\put(100,62){\makebox(0,0){8}}
\put(81.5,57){\vector(2,-1){28.5}}
\put(95,45){\makebox(0,0){7}}
\put(39,35){\vector(2,-1){28}}
\put(50,22){\makebox(0,0){10}}
\put(75,15){\circle{20}}
\put(75,15){\makebox(0,0){T}}
\put(83.5,20){\vector(2,1){28}}
\put(100,22){\makebox(0,0){11}}
\put(120,40){\circle{20}}
\put(120,40){\makebox(0,0){Q}}
\put(130,40){\vector(1,0){10}}
\put(140,45){\makebox(0,0){12}}
\end{picture}
\caption{Configuration of the workcell}
\label{fig-workcell-conf}
\end{figure}

The simple workcell described in this section consists of four 
components: a workstation, a transport service, a quality checker, and
a workcell controller.
The workstation accepts products, processes them, and delivers 
processed products of which the quality is either good or bad.
The transport service accepts products at the one end, transports them,
and delivers the transported products at the other end.
The quality checker determines whether the processed products are good.
A good product is passed, while a bad product is removed.
When a product is removed, this is signalled to the workcell 
controller.
The workcell controller controls the workcell.
It receives instructions to process a certain number of products.
When an instruction is received, it directs the workcell to do so.
While the processing is going on, the workcell controller counts the
number of products removed by the quality checker.
When the processing is completed, the workcell controller directs the
workcell to process again a number of products to compensate for the
removed products.
The configuration of the workcell is shown in 
Fig.~\ref{fig-workcell-conf}.
The four components are connected to each other and the environment by
12 ports.
Ports 3 to 8, 10 and 11 are internal ports and ports 1, 2, 9 and 12 are
external ports.
Ports 1 to 8 are used to communicate data and ports 9 to 12 are used to 
exchange products.

Along ports 1, 3, 5 and 7 a message $\nm{produce}(n)$ can be sent to
indicate to the receiver that the workcell has to produce $n$ products.
Along ports 2, 4, 6 and 8 a message $\nm{ready}$ can be sent back to
indicate that the component has fulfilled its part of the task.
Along port 8 a message $\nm{reject}$ can be sent back as well.
This message indicates that a product has not been passed to the 
environment. 
At port 9 unprocessed products are exchanged.
At ports 10, 11 and 12 processed products are exchanged.

We assume a finite set of unprocessed products $P_\mathrm{in}$.
Moreover, we assume that for each $p \in P_\mathrm{in}$ there are a 
processed product of good quality denoted by $\nm{proc}(p,\nm{ok})$ and 
a processed product of bad quality denoted by $\nm{proc}(p,\nm{nok})$.
We write 
$P_\mathrm{out}^\mathrm{ok}$ for the set
$\set{\nm{proc}(p,\nm{ok}) \where p \in P_\mathrm{in}}$, 
$P_\mathrm{out}^\mathrm{nok}$ for the set
$\set{\nm{proc}(p,\nm{nok}) \where p \in P_\mathrm{in}}$, and
$P_\mathrm{out}$ for the set
$P_\mathrm{out}^\mathrm{ok} \union P_\mathrm{out}^\mathrm{nok}$.
We also assume that there is a bound $N$ on the number of products that
the workcell can be requested to produce.

The workstation waits until a message $\nm{produce}(n)$ is received 
from the workcell controller.
When such a message is received, it accepts, processes and delivers
$n$ products, and then sends the message $\nm{ready}$ to the workcell 
controller.
The workstation may deliver products of good quality as well as 
products of bad quality.
After the message $\nm{ready}$ has been sent, the workstation goes back
to waiting for a message from the workcell controller.
The guarded recursive specification of the workstation $W$ consists of
the following recursive equations:
\begin{ldispl}
\begin{aeqns}
W    & = & \Altc{n \leq N} \kw{r}_3(\nm{produce}(n)) \seqc W'_n\;,
\eqnsep
W'_0 & = & \kw{s}_1(\nm{ready}) \seqc W\;,
\eqnsep
W'_{n+1} & = & 
\Altc{p \in P_\mathrm{in}} 
 \kw{r}_9(p) \seqc 
 (\kw{i} \seqc \kw{r}_{10}(\nm{proc}(p,\nm{ok})) \altc
  \kw{i} \seqc \kw{r}_{10}(\nm{proc}(p,\nm{nok}))) \seqc 
W'_n
\cndsep
\cnd{(\mathrm{for\; every}\; n < N).}
\end{aeqns}
\end{ldispl}
The action $\kw{i}$ is again an internal action that cannot be 
performed synchronously with any other action.
Thus, the workstation cannot be forced to produce products of good
quality only.

The transport service waits until a message $\nm{produce}(n)$ is 
received from the workcell controller.
When such a message is received, it accepts, transports and delivers
$n$ products, and then sends the message $\nm{ready}$ to the workcell 
controller.
The transport service may have to accept products from the workstation
while there are accepted products that it could not deliver to the
quality checker yet.
After the message $\nm{ready}$ has been sent, the transport service 
goes back to waiting for a message from the workcell controller.
The guarded recursive specification of the transport service $T$
consists of the following recursive equations:
\begin{ldispl}
\begin{aeqns}
T\phantom{{}'_{n+1,\sigma\, q}} & = & 
\Altc{n \leq N} \kw{r}_5(\nm{produce}(n)) \seqc T'_{n,\epsilon}\;,
\eqnsep
T'_{0,\epsilon}   & = & \kw{s}_6(\nm{ready}) \seqc T\;,
\eqnsep
T'_{n+1,\epsilon} & = & 
\Altc{q \in P_\mathrm{out}} \kw{r}_{10}(q) \seqc T'_{n,q}
\cndsep
\cnd{(\mathrm{for\; every}\; n < N),}
\end{aeqns}
\end{ldispl}
\begin{ldispl}
\begin{aeqns}
T'_{0,\sigma\, q}   & = & \kw{s}_{11}(q) \seqc T'_{0,\sigma}
\cndsep
\cnd{(\mathrm{for\; every}\; q \in P_\mathrm{out}\;
      \mathrm{and}\; \sigma \in \seqof{P_\mathrm{out}}),}
\eqnsep
T'_{n+1,\sigma\, q} & = & 
\Altc{q' \in P_\mathrm{out}} 
 \kw{r}_{10}(q') \seqc T'_{n,q'\, \sigma\, q} \altc
\kw{s}_{11}(q) \seqc T'_{n+1,\sigma}
\cndsep
\cnd{(\mathrm{for\; every}\; n < N,\, q \in P_\mathrm{out}\;
      \mathrm{and}\; \sigma \in \seqof{P_\mathrm{out}}).}
\end{aeqns}
\end{ldispl}

The quality checker waits until a message $\nm{produce}(n)$ is 
received from the workcell controller.
When such a message is received, it checks $n$ products and then sends 
the message $\nm{ready}$ to the workcell controller.
After the message $\nm{ready}$ has been sent, the quality checker 
goes back to waiting for a message from the workcell controller.
Checking a product includes accepting the product and delivering the
product if its quality is good.
Each time that the quality checker encounters a product of which the 
quality is bad, it sends the message $\nm{reject}$ to the workcell 
controller.
The guarded recursive specification of the quality checker $Q$
consists of the following recursive equations:
\begin{ldispl}
\begin{aeqns}
Q    & = & \Altc{n \leq N} \kw{r}_7(\nm{produce}(n)) \seqc Q'_n\;,
\eqnsep
Q'_0 & = & \kw{s}_8(\nm{ready}) \seqc Q\;,
\eqnsep
Q'_{n+1} & = & 
\Altc{q \in P_\mathrm{out}^\mathrm{ok}} 
 \kw{r}_{11}(q) \seqc \kw{s}_{12}(q) \seqc Q'_n \altc
\Altc{q \in P_\mathrm{out}^\mathrm{nok}} 
 \kw{r}_{11}(q) \seqc \kw{s}_8(\nm{reject}) \seqc Q'_n
\cndsep
\cnd{(\mathrm{for\; every}\; n < N).}
\end{aeqns}
\end{ldispl}

The workcell controller waits until a message $\nm{produce}(n)$ is 
received from the environment.
When such a message is received, it sends the same message to the 
quality checker, the transport service and workstation, in that order.
After that, the workcell controller waits for reception of the message 
$\nm{ready}$ from the quality checker, the transport service and 
workstation, again in that order.
If it receives instead the message $\nm{reject}$ from the quality 
checker, it increments a counter of rejections and goes back to 
waiting for the ready messages.
When the ready messages are received, the workcell controller continues
as follows.
In the case where there are rejections, say $n'$ ($0 < n' \leq n$), it
first sends the message $\nm{produce}(n')$ to the quality checker, the
transport service and workstation, in that order, and then goes back to
waiting for ready messages.
In the case where there are no rejections, it sends the message
$\nm{ready}$ to the environment.
The guarded recursive specification of the workcell controller $C$
consists of the following recursive equations:
\begin{ldispl}
\begin{aeqns}
C\phantom{{}'_{n+1}} & = & 
\Altc{n \leq N} \kw{r}_1(\nm{produce}(n)) \seqc C'_n\;,
\eqnsep
C'_0 & = & 
\kw{s}_2(\nm{ready}) \seqc C\;,
\end{aeqns}
\end{ldispl}
\begin{ldispl}
\begin{aeqns}
C'_{n+1} & = & 
\kw{s}_7(\nm{produce}(n\!+\!1)) \seqc 
\kw{s}_5(\nm{produce}(n\!+\!1)) \seqc
\kw{s}_3(\nm{produce}(n\!+\!1)) \seqc C''_0
\cndsep
\cnd{(\mathrm{for\; every}\; n < N),}
\eqnsep
C''_n & = & 
\kw{r}_8(\nm{ready}) \seqc \kw{r}_6(\nm{ready}) \seqc
\kw{r}_4(\nm{ready}) \seqc C'_n \altc
\kw{r}_8(\nm{reject}) \seqc C''_{n+1}
\cndsep
\cnd{(\mathrm{for\; every}\; n < N),}
\eqnsep
C''_N & = & 
\kw{r}_8(\nm{ready}) \seqc \kw{r}_6(\nm{ready}) \seqc
\kw{r}_4(\nm{ready}) \seqc C'_N\;.
\end{aeqns}
\end{ldispl}
The whole workcell is described by
\begin{ldispl}
\begin{geqns}
\abstr{I}(\encap{H}(C \parc W \parc T \parc Q))
\end{geqns}
\end{ldispl}
where
\begin{ldispl}
\displstretch
 \begin{aeqns}
 H & = &
 \set{\kw{s}_i(m),\kw{r}_i(m) \where 
     i \in \set{3,5,7}, m \in \set{\nm{produce}(n) \where n \in N}}
\\ & {} \union &
 \set{\kw{s}_i(m),\kw{r}_i(m) \where 
     i \in \set{4,6,8}, m \in \set{\nm{ready},\nm{reject}}}
\\ & {} \union &
 \set{\kw{s}_i(p),\kw{r}_i(p) \where 
     i \in \set{10,11}, p \in P_\mathrm{out}}
 \end{aeqns}
\end{ldispl}
and
\begin{ldispl}
\displstretch
 \begin{aeqns}
 I & = &
 \set{\kw{c}_i(m) \where 
     i \in \set{3,5,7}, m \in \set{\nm{produce}(n) \where n \in N}}
\\ & {} \union &
 \set{\kw{c}_i(m) \where 
     i \in \set{4,6,8}, m \in \set{\nm{ready},\nm{reject}}}
\\ & {} \union &
 \set{\kw{c}_i(p) \where 
     i \in \set{10,11}, p \in P_\mathrm{out}} 
        \union 
 \set{\kw{i}}\;.
 \end{aeqns}
\end{ldispl}

The workcell is considered to be correct if it behaves as follows in 
the case where there is a supplier that delivers an unlimited number of 
any one unprocessed product.
The workcell, together with the supplier, waits until a message 
$\nm{produce}(n)$ is received from the environment.
When such a message is received, it processes and delivers $n$ 
products, and then sends the message $\nm{ready}$ to the environment.
After the message $\nm{ready}$ has been sent, the workcell goes back
to waiting for a message from the environment.

Let $p_0$ be a fixed but arbitrary member of $P_\mathrm{in}$.
The guarded recursive specification of the supplier $S$ consists of the 
following recursive equation:
\begin{ldispl}
\begin{aeqns}
S & = & \kw{s}_9(p_0) \seqc S\;.
\end{aeqns}
\end{ldispl}
The workcell together with the supplier is described by
\begin{ldispl}
\begin{geqns}
\abstr{I'}
 (\encap{H'}(S \parc \abstr{I}(\encap{H}(C \parc W \parc T \parc Q))))
\end{geqns}
\end{ldispl}
where
\begin{ldispl}
\displstretch
 \begin{aeqns}
 H' & = & \set{\kw{s}_9(p),\kw{r}_9(p) \where p \in P_\mathrm{in}}
 \end{aeqns}
\end{ldispl}
and
\begin{ldispl}
\displstretch
 \begin{aeqns}
 I' & = &
 \set{\kw{c}_9(p) \where p \in P_\mathrm{in}}\;.
 \end{aeqns}
\end{ldispl}
We can show that this process is the solution of the following guarded 
recursive specification:
\begin{ldispl}
\begin{aeqns}
V & = & \Altc{n \leq N} \kw{r}_1(\nm{produce}(n)) \seqc V'_n \seqc V\;,
\eqnsep
V'_0     & = & \kw{s}_2(\nm{ready})\;,
\eqnsep
V'_{n+1} & = & \kw{s}_{12}(\nm{proc}(p_0,\nm{ok})) \seqc V'_n
\cndsep
\cnd{(\mathrm{for\; every}\; n < N).}
\end{aeqns}
\end{ldispl}
This guarded recursive specification describes exactly what is 
considered in the preceding paragraph to be the correct behaviour of
the workcell.


\chapter{Selected topics}
\label{ch-topics}

There are many interesting topics related to process expressions and 
guarded recursive specifications which are not treated in 
Chap.~\ref{ch-expressions}.
This chapter treats some selected topics.
First of all, we give the semantics of closed process expressions in  
an alternative way known as structural operational semantics 
(Sect.~\ref{sect-topics-sos}).
After that, we give equational laws that hold
(Sect.~\ref{sect-topics-eqns}).
We also look briefly at the expressive power of process expressions
(Sect.~\ref{sect-topics-power}) and an interesting restricted form of 
guarded recursive specification (Sect.~\ref{sect-topics-linear}).

\section{Structural operational semantics}
\label{sect-topics-sos}

We still assume a fixed but arbitrary set $\Act$ of actions and a fixed 
but arbitrary communication function 
$\funct{\commf}{\Act \x \Act}{\Act}$.

We associate a transition system with a closed process expression $p$
by taking the closed process expressions as states, with $p$ as initial
state, and by defining the transitions using transition rules in the
style of Plotkin.
The way of giving semantics adopted is called 
\emph{structural operational semantics}.%
\footnote
{A lot of theory has been developed about structural operational 
 semantics (see e.g.~\cite{AFV01,MRG07}).}
The transition rules used to define the transitions have the form
\begin{ldispl}
 \Rule{\phi_1,\ldots,\phi_m}{\psi}\;,
\end{ldispl}
which is to be read as
\begin{ldispl}
\mbox{if $\phi_1$ and \ldots\ and $\phi_m$, then $\psi$.}
\end{ldispl}
As customary, $\phi_1,\ldots,\phi_m$ and $\psi$ are called the 
\emph{premises} and the \emph{conclusion}, respectively.
The premises and conclusions of a transition rule are of the form
$\astep{t}{a}{t'}$, where $t \in \PE(\Act)$ and 
$t' \in \PE(\Act) \union \set{\surd}$.
The transition rules used (see Table~\ref{tbl-rules}) are actually 
transition rule schemas: $a$, $b$ and $c$ are placeholders for 
arbitrary members of $\Actt$, and $H$ and $I$ are placeholders for 
arbitrary subsets of $\Act$.
A side-condition is added to some of them to restrict the members of
$\Actt$ for which $a$, $b$ and $c$ are placeholders.
In applying the transition rules, the process variables $x$, $x'$, $y$ 
and $y'$ may be replaced by any process expression, but not by $\surd$.

Let ${\step{}}$ be the smallest subset of
$\CPE(\Act) \x \Act \x (\CPE(\Act) \union \set{\surd})$ 
satisfying the transition rules from Table~\ref{tbl-rules}.
\begin{table}
\caption{Transition rules for process expressions}
\label{tbl-rules}
\begin{ruletbl}
\Rule
{\phantom{\astep{x}{a}{x'}}}
{\astep{a}{a}{\surd}}
\quad
\\
\Rule
{\astep{x}{a}{x'}}
{\astep{x \altc y}{a}{x'}}
\quad
\Rule
{\astep{y}{a}{y'}}
{\astep{x \altc y}{a}{y'}}
\quad
\Rule
{\astep{x}{a}{\surd}}
{\astep{x \altc y}{a}{\surd}}
\quad
\Rule
{\astep{y}{a}{\surd}}
{\astep{x \altc y}{a}{\surd}}
\\
\Rule
{\astep{x}{a}{x'}}
{\astep{x \seqc y}{a}{x' \seqc y}}
\quad
\Rule
{\astep{x}{a}{\surd}}
{\astep{x \seqc y}{a}{y}}
\\
\Rule
{\astep{x}{a}{x'}}
{\astep{x \iter y}{a}{x' \seqc (x \iter y)}}
\quad
\Rule
{\astep{x}{a}{\surd}}
{\astep{x \iter y}{a}{x \iter y}}
\quad
\Rule
{\astep{y}{a}{y'}}
{\astep{x \iter y}{a}{y'}}
\quad
\Rule
{\astep{y}{a}{\surd}}
{\astep{x \iter y}{a}{\surd}}
\\
\Rule
{\astep{x}{a}{x'}}
{\astep{x \neiter}{a}{x' \seqc x \neiter}}
\quad
\Rule
{\astep{x}{a}{\surd}}
{\astep{x \neiter}{a}{x \neiter}}
\\
\Rule
{\astep{x}{a}{x'}}
{\astep{x \parc y}{a}{x' \parc y}}
\quad
\Rule
{\astep{y}{a}{y'}}
{\astep{x \parc y}{a}{x \parc y'}}
\quad
\Rule
{\astep{x}{a}{\surd}}
{\astep{x \parc y}{a}{y}}
\quad
\Rule
{\astep{y}{a}{\surd}}
{\astep{x \parc y}{a}{x}}
\\
\RuleC
{\astep{x}{a}{x'},\; \astep{y}{b}{y'}}
{\astep{x \parc y}{c}{x' \parc y'}}
{\commf(a,b) = c}
\quad
\RuleC
{\astep{x}{a}{x'},\; \astep{y}{b}{\surd}}
{\astep{x \parc y}{c}{x'}}
{\commf(a,b) = c}
\\
\RuleC
{\astep{x}{a}{\surd},\; \astep{y}{b}{y'}}
{\astep{x \parc y}{c}{y'}}
{\commf(a,b) = c}
\quad
\RuleC
{\astep{x}{a}{\surd},\; \astep{y}{b}{\surd}}
{\astep{x \parc y}{c}{\surd}}
{\commf(a,b) = c}
\\
\RuleC
{\astep{x}{a}{x'}}
{\astep{\encap{H}(x)}{a}{\encap{H}(x')}}
{a \not\in H}
\quad
\RuleC
{\astep{x}{a}{\surd}}
{\astep{\encap{H}(x)}{a}{\surd}}
{a \not\in H}
\\
\RuleC
{\astep{x}{a}{x'}}
{\astep{\abstr{I}(x)}{a}{\abstr{I}(x')}}
{a \not\in I}
\quad
\RuleC
{\astep{x}{a}{\surd}}
{\astep{\abstr{I}(x)}{a}{\surd}}
{a \not\in I}
\\
\RuleC
{\astep{x}{a}{x'}}
{\astep{\abstr{I}(x)}{\tau}{\abstr{I}(x')}}
{a \in I}
\quad
\RuleC
{\astep{x}{a}{\surd}}
{\astep{\abstr{I}(x)}{\tau}{\surd}}
{a \in I}
\end{ruletbl}
\end{table}
We can look at the members of ${\step{}}$ as follows:
\begin{iteml}
\item
a member of the form $\astep{p}{a}{p'}$, where $p' \neq \surd$,  
indicates that the process denoted by $p$ is capable of first 
performing action $a$ and then proceeding as the process denoted by 
$p'$; 
\item
a member of the form $\astep{p}{a}{\surd}$ indicates that the process 
denoted by $p$ is capable of first performing action $a$ and then 
terminating successfully.
\end{iteml}
So, $\surd$ is introduced to represent successful termination.
Notice that ${\step{}}$ has no members of the form 
$\astep{\surd}{a}{p'}$.

The transition rules from Table~\ref{tbl-rules} provide an alternative 
way to assign meanings to process expressions.
\begin{definition}[Meaning induced by the transition rules]%
\index{process expression!meaning of}\index{transition rule}
Let $p \in \CPE(\Act)$.
Then the meaning of $p$ induced by the transition rules from
Table~\ref{tbl-rules}, written $\cM'(p)$, is 
$\rbbeqvc{\reduct\tup{S,A,\step{},\term{},s_0}}$ where
\begin{iteml}
\item
$S = \CPE(\Act) \union \set{\surd}$;
\item
$A = \Act$;
\item
$\step{}$ is the smallest subset of
$\CPE(\Act) \x \Act \x (\CPE(\Act) \union \set{\surd})$ 
satisfying the transition rules from Table~\ref{tbl-rules};
\item
${\term{}} = \set{\surd}$; 
\item
$s_0 = p$.
\end{iteml}
\end{definition}
Recall that $\reduct$ reduces any transition system to a connected
transition system that is rooted branching bisimulation equivalent.

The meaning of a closed process expression induced by the transition 
rules coincides with its meaning according to Def.~\ref{def-eval}.
\begin{property}[Equality of meanings]%
\index{process expression!meaning of}\index{transition rule}
For every $p \in \CPE(\Act)$, we have that $\cM'(p) = \M{\alpha}(p)$ 
for all assignments $\alpha$.
\end{property}

\section{Equational laws}
\label{sect-topics-eqns}

In Table~\ref{tbl-axioms}, a number of equations that hold are given.
\begin{table}
\caption{Equational laws for process expressions}
\label{tbl-axioms}
\begin{eqntbl}
\begin{axcol}
x \altc y = y \altc x                               & \axiom{A1} \\
(x \altc y) \altc z = x \altc (y \altc z)           & \axiom{A2} \\
x \altc x = x                                       & \axiom{A3} \\
(x \altc y) \seqc z = x \seqc z \altc y\seqc z      & \axiom{A4} \\
(x \seqc y) \seqc z = x \seqc (y \seqc z)           & \axiom{A5} \\
x \altc \dead = x                                   & \axiom{A6} \\
\dead \seqc x = \dead                               & \axiom{A7} \\
{} \\
a \commm b = c\;               \mif \commf(a,b) = c & \axiom{CF1} \\
a \commm b = \dead\;
              \mif \commf(a,b)\; \mathrm{undefined} & \axiom{CF2} \\
{} \\
x \parc y =
   (x \leftm y \altc y \leftm x) \altc x \commm y   & \axiom{CM1} \\
a \leftm x = a \seqc x                              & \axiom{CM2} \\
a \seqc x \leftm y = a \seqc (x \parc y)            & \axiom{CM3} \\
(x \altc y) \leftm z = x \leftm z \altc y \leftm z  & \axiom{CM4} \\
a \seqc x \commm b = (a \commm b) \seqc x           & \axiom{CM5} \\
a \commm b \seqc x = (a \commm b) \seqc x           & \axiom{CM6} \\
a \seqc x \commm b \seqc y =
                     (a \commm b) \seqc (x \parc y) & \axiom{CM7} \\
(x \altc y) \commm z = x \commm z \altc y \commm z  & \axiom{CM8} \\
x \commm (y \altc z) = x \commm y \altc x \commm z  & \axiom{CM9} 
\end{axcol}
\qquad
\begin{axcol}
x \seqc \tau = x                                    & \axiom{B1} \\
x \seqc (\tau \seqc (y \altc z) \altc y) = x \seqc (y \altc z)  
                                                    & \axiom{B2} \\
{} \\
x \iter y = x \seqc (x \iter y) \altc y             & \axiom{BKS1} \\
x \iter (y \seqc z) = (x \iter y) \seqc z           & \axiom{BKS2} \\
(x \altc y) \iter z = \eqnbreak
    x \iter (y \seqc ((x \altc y) \iter z) \altc z) & \axiom{BKS3} \\
{} \\
x \neiter = x \iter \dead                           & \axiom{NEI} \\
{} \\
{} \\
\encap{H}(a) = a\;                 \mif a \not\in H & \axiom{D1}  \\
\encap{H}(a) = \dead\;                 \mif a \in H & \axiom{D2}  \\
\encap{H}(x \altc y) =
                    \encap{H}(x) \altc \encap{H}(y) & \axiom{D3}  \\
\encap{H}(x \seqc y) =
                    \encap{H}(x) \seqc \encap{H}(y) & \axiom{D4}  \\
{} \\
\abstr{I}(a) = a\; \mif a \not\in I                    & \axiom{TI1} \\
\abstr{I}(a) = \tau\; \mif a \in I                     & \axiom{TI2} \\
\abstr{I}(x \altc y) = \abstr{I}(x) \altc \abstr{I}(y) & \axiom{TI3} \\
\abstr{I}(x \seqc y) = \abstr{I}(x) \seqc \abstr{I}(y) & \axiom{TI4} 
\end{axcol}
\end{eqntbl}
\end{table}
Many equations are actually equation schemas: $a$, $b$ and $c$ are
placeholders for arbitrary members of $\Actt \union \set{\dead}$, and
$H$ and $I$ are placeholders for arbitrary subsets of $\Act$.
A side-condition is added to some of them to restrict the members of
$\Actt \union \set{\dead}$ for which $a$, $b$ and $c$ are
placeholders.
Notice that, unlike in the transition rules from Table~\ref{tbl-rules},
$a$, $b$ and $c$ are also placeholders for $\dead$ in the equations
from Table~\ref{tbl-axioms}.
Two auxiliary operators appear in Table~\ref{tbl-axioms}: $\leftm$ and
$\commm$.
The operator $\leftm$ is interpreted as left merge, which is the same
as parallel composition except that the left merge of $p_1$ and $p_2$
starts with performing an action of $p_1$.
The operator $\commm$  is interpreted as communication merge, which is
the same as parallel composition except that the communication merge of
$p_1$ and $p_2$ starts with performing an action of $p_1$ and an action
of $p_2$ synchronously.
These interpretations are clearly reflected by the additional
transition rules for $\leftm$ and $\commm$ given in
Table~\ref{tbl-add-rules}.
\begin{table}
\caption{Additional transition rules for $\leftm$ and $\commm$}
\label{tbl-add-rules}
\begin{ruletbl}
\Rule
{\astep{x}{a}{x'}}
{\astep{x \leftm y}{a}{x' \parc y}}
\quad
\Rule
{\astep{x}{a}{\surd}}
{\astep{x \leftm y}{a}{y}}
\\
\RuleC
{\astep{x}{a}{x'},\; \astep{y}{b}{y'}}
{\astep{x \commm y}{c}{x' \parc y'}}
{\commf(a,b) = c}
\quad
\RuleC
{\astep{x}{a}{x'},\; \astep{y}{b}{\surd}}
{\astep{x \commm y}{c}{x'}}
{\commf(a,b) = c}
\\
\RuleC
{\astep{x}{a}{\surd},\; \astep{y}{b}{y'}}
{\astep{x \commm y}{c}{y'}}
{\commf(a,b) = c}
\quad
\RuleC
{\astep{x}{a}{\surd},\; \astep{y}{b}{\surd}}
{\astep{x \commm y}{c}{\surd}}
{\commf(a,b) = c}
\end{ruletbl}
\end{table}
From the equations given in Table~\ref{tbl-axioms}, we can derive many 
other equations that hold.
Actually, we can derive all equations between closed process
expressions in which only the operators $\altc$, $\seqc$, $\parc$,
$\encap{H}$ and $\abstr{I}$ occur.
Let us illustrate by means of a simple example what can be done with
equational laws for process expressions.
\begin{example}[Merge connection]\index{merge connection}
\label{exa-calc}
We consider once more the merge connection from 
Example~\ref{exa-merge}.
Let $T_1$ and $T_2$ be the first and second transition system from
Example~\ref{exa-merge}, for the case where $D = \set{0,1}$.
According to Definition~\ref{def-eval}, we assign to the process
expressions
\begin{ldispl}
(\kw{r}_k(0) \seqc \kw{s}_m(0) \altc \kw{r}_l(0) \seqc \kw{s}_m(0)
  \altc 
 \kw{r}_k(1) \seqc \kw{s}_m(1) \altc \kw{r}_l(1) \seqc \kw{s}_m(1))
 \neiter
\end{ldispl}
and
\begin{ldispl}
((\kw{r}_k(0) \altc \kw{r}_l(0)) \seqc \kw{s}_m(0)
  \altc 
 (\kw{r}_k(1) \altc \kw{r}_l(1)) \seqc \kw{s}_m(1)) \neiter
\end{ldispl}
the meanings $\rbbeqvc{T_1}$ and $\rbbeqvc{T_2}$, respectively.
The simplest way to show that $\rbbeqvc{T_1}$ equals $\rbbeqvc{T_2}$, 
is by applying equation A4 from Table~\ref{tbl-axioms}.
We do not have to construct a bisimulation, like in
Example~\ref{exa-merge}, to prove this.
\end{example}

The equations given in Table~\ref{tbl-axioms} constitute the axiom 
system ACP$^{\tau{*}}$ from~\cite{BP01}.

\section{Expressive power of process expressions}
\label{sect-topics-power}

All regular processes can be denoted by process expressions.
\begin{property}[Expressive power]%
\index{process expression!expressive power of}
Let $\alpha$ be an arbitrary assignment.
Then, for every $P \in \Prc(\Act)$ that is regular, there exists a 
$\Act' \supseteq \Act$ and a $p \in \CPE(\Act')$ such that
$\M{\alpha}(p) = P$.
\end{property}
Although regular processes can be denoted by process expressions, it
may easily become very clumsy.
This is nicely illustrated in the following example.
\begin{example}[Bounded counter]\index{counter!bounded}
\label{exa-bcounter-expr}
We consider once more the bounded counter from
Example~\ref{exa-bcounter}.
In Example~\ref{exa-bcounter}, a regular transition system was given
for the bounded counter in a direct way.
The corresponding recursive specifications was given in Example~\ref{exa-mutual-recursion} for the case where the bound is $2$.
In this example, we give corresponding process expressions for the
cases where the bound is $1$ and $2$.
In the case where the bound is $1$, the process expression is:
\begin{ldispl}
\begin{geqns}
(\kw{inc} \seqc \kw{dec}) \neiter\;.
\end{geqns}
\end{ldispl}
However, in the case where the bound is $2$, the simplest process
expression is:
\begin{ldispl}
\begin{geqns}
\abstr{\set{\kw{i}}}
 (\encap{\set{\kw{inc}',\kw{dec}'}}
  ((\kw{inc} \seqc \kw{dec}') \neiter \parc 
   (\kw{inc}' \seqc \kw{dec}) \neiter))
\end{geqns}
\end{ldispl}
where the communication function $\commf$ is defined such that
$\commf(\kw{inc}',\kw{dec}') = \commf(\kw{dec}',\kw{inc}') = \kw{i}$.

\end{example}

\section{Linear recursive specifications}
\label{sect-topics-linear}

Given a finitely branching transition system, we can easily construct a
guarded recursive specification that has the process of which that
transition system is a representative as its solution.
For every reachable state $s$, we introduce a corresponding process
variable $X_s$.
The right-hand side of the recursive equation for $X_s$ is an
alternative composition with an alternative $a \seqc X_{s'}$ for each
transition $\astep{s}{a}{s'}$ and an alternative $a$ for each
transition $\astep{s}{a}{\surd}$.
Here are a couple of examples.
\begin{example}[Binary memory cell]\index{binary memory cell}
\label{exa-binvar-rec}
We consider again the binary memory cell from Example~\ref{exa-binvar}.
In that example, a transition system was given for the binary memory 
cell in a direct way.
The corresponding recursive specification is as follows:
\begin{ldispl}
\begin{aeqns}
M & = & M'_0\;,
\eqnsep
M'_b & = & 
\kw{rtr}(b) \seqc M'_b \altc
\kw{sto}(b) \seqc M'_b \altc
\kw{sto}(1-b) \seqc M'_{1-b}
\cndsep
\cnd{(\mathrm{for\; every}\; b \in \set{0,1}).}
\end{aeqns}
\end{ldispl}
\end{example}
\begin{example}[Calculator]\index{calculator}
\label{exa-calculator-rec}
We consider again the simple calculator from
Example~\ref{exa-calculator}.
In that example, a transition system was given for the calculator in a 
direct way.
The corresponding recursive specification is as follows:
\begin{ldispl}
\begin{aeqns}
C & = & C'_{\tup{\und,\und}}\;,
\eqnsep
C'_{\tup{\und,\und}} & = & 
\Altc{i \in \set{i \where \nm{min} \leq i \leq \nm{max}}} 
 \kw{rd}(i) \seqc C'_{\tup{i,\und}}\;,
\eqnsep
C'_{\tup{i,\und}} & = & 
\Altc{o \in \set{\kw{clr},\kw{eq},\kw{add},\kw{sub},\kw{mul},\kw{div}}} 
 \kw{rd}(o) \seqc C'_{\tup{i,o}}\;,
\eqnsep
C'_{\tup{i,\kw{clr}}} & = & 
\kw{wr}(0) \seqc C'_{\tup{\und,\und}}\;,
\eqnsep
C'_{\tup{i,\kw{eq}}} & = & 
\kw{wr}(i) \seqc C'_{\tup{i,\und}}\;,
\eqnsep
C'_{\tup{i,\kw{add}}} & = & 
\Altc{j \in \set{j \where \nm{min} \leq i+j \leq \nm{max}}} 
 \kw{rd}(j) \seqc C'_{\tup{i+j,\und}}\;,
\eqnsep
C'_{\tup{i,\kw{sub}}} & = & 
\Altc{j \in \set{j \where \nm{min} \leq i-j \leq \nm{max}}} 
 \kw{rd}(j) \seqc C'_{\tup{i-j,\und}}\;,
\eqnsep
C'_{\tup{i,\kw{mul}}} & = & 
\Altc{j \in \set{j \where \nm{min} \leq i \cdot j \leq \nm{max}}} 
 \kw{rd}(j) \seqc C'_{\tup{i \cdot j,\und}}\;,
\eqnsep
C'_{\tup{i,\kw{div}}} & = & 
\Altc
 {j \in \set{j \where \nm{min} \leq i \div j \leq \nm{max}, j \neq 0}} 
  \kw{rd}(j) \seqc C'_{\tup{i \div j,\und}}\;.
\end{aeqns}
\end{ldispl}
We refrained from mentioning after each equation schema that there is
an instance for every $i$ such that $\nm{min} \leq i \leq \nm{max}$.
\end{example}

Conversely, given a guarded recursive specification consisting of
equations whose right-hand sides are alternative compositions of which
the alternatives are of the form $a$ or $a \seqc X$, we can construct 
a finitely branching transition system that is a representative of the
process that is the solution of that guarded recursive specification.
Here is an example.
\begin{example}[Unbounded buffer]\index{buffer!unbounded}
\label{exa-buffer-ts}
We consider again the unbounded buffer from 
Example~\ref{exa-buffer-rec}.
In that example, a guarded recursive specification was given for an 
unbounded buffer that can only keep bits, i.e.\ $D = \set{0,1}$.
The corresponding transition system is as follows.
As states of the unbounded buffer, we have all sequences 
$\sigma \in \seqof{D}$, with $\epsilon$ as initial state.
There are no successfully terminating states.
As actions, we have $\kw{add}(d)$ and $\kw{rem}(d)$ for each $d \in D$.
As transitions of an unbounded buffer, we have the following:
\begin{iteml}
\item
for each $d \in D$, 
a transition $\astep{\epsilon}{\kw{add}(d)}{d}$;
\item
for each $d \in D$ and $\sigma \in \seqof{D}$, 
a transition $\astep{\sigma\, d}{\kw{rem}(d)}{\sigma}$;
\item
for each $d,d' \in D$ and $\sigma \in \seqof{D}$, 
a transition $\astep{\sigma\, d}{\kw{add}(d')}{d'\, \sigma\, d}$.
\end{iteml}
\end{example}

A guarded recursive specification consisting of equations whose
right-hand sides are alternative compositions of which the alternatives
are of the form $a$ or $a \seqc X$ is called a \emph{linear} recursive specification.
The examples given above show the close connection between linear
recursive specification and finitely branching transition systems.

\appendix

\chapter{Set theoretical preliminaries}
\label{ch-prelims}

In this appendix, we give a brief summary of facts from set theory used 
in these lecture notes.
This will at least serve to establish the terminology and notation
concerning sets.
First of all, we treat elementary sets (Appendix~\ref{sect-sets}).
After that, we look at relations, functions 
(Appendix~\ref{sect-rels-funcs}) and sequences
(Appendix~\ref{sect-seqs}).

\section{Sets}
\label{sect-sets}

A \emph{set}\index{set} is a collection of things which are said to be 
the \emph{members}\index{member} of the set.
A set is completely determined by its members.
That is, if two sets $A$ and $A'$ have the same members, then $A = A'$.
We write $a \in A$ to indicate that $a$ is a member of the set $A$, and
$a \not\in A$ to indicate that $a$ is not a member of the set $A$.
A set $A$ is a \emph{subset}\index{subset} of a set $A'$, written 
$A \subseteq A'$ or $A' \supseteq A$, if for all $x$, $x \in A$ implies 
$x \in A'$.

If a set has a finite number of members $a_1$, \ldots, $a_n$, then the 
set is written as follows:
\begin{ldispl}
\set{a_1,\ldots,a_n}\;.
\end{ldispl}
Let $P(x)$ be the statement that $x$ has property $P$.
Then the set whose members are exactly the things that have property 
$P$, if such a set exists, is written as follows:
\begin{ldispl}
\set{x \where P(x)}\;.
\end{ldispl}

If $A$ is a set and $P(x)$ is the statement that $x$ has property $P$,
then there exists a subset of $A$ of which the members are exactly the 
members of $A$ that have property $P$.
This set is denoted by $\set{x \in A \where P(x)}$:
\begin{ldispl}
\begin{aeqns}
\set{x \in A \where P(x)} & = & 
\set{x \where x \in A \;\mathrm{and}\; P(x)}\;.
\end{aeqns}
\end{ldispl}

If $A$ is a set, then there exists a set of which the members are 
exactly the subsets of $A$.
This set is called the \emph{powerset}\index{powerset} of $A$ and is 
denoted by $\setof{(A)}$:
\begin{ldispl}
\begin{aeqns}
\setof{(A)} & = & \set{x \where x \subseteq A}\;.
\end{aeqns}
\end{ldispl}

If $\cA$ is a set of sets, then there exists a set of which the members 
are exactly the members of the subsets of $\cA$.
This set is called the \emph{union}\index{union} of $\cA$ and is denoted 
by $\Union \cA$:
\begin{ldispl}
\begin{aeqns}
\Union \cA & = & \set{x \where \Exists{A \in \cA}{x \in A}}\;.
\end{aeqns}
\end{ldispl}

There exists a set with no members.
This set is called the \emph{empty set}\index{empty set} and is denoted 
by $\emptyset$: 
\begin{ldispl}
\begin{aeqns}
\emptyset & = & \set{x \where x \neq x}\;.
\end{aeqns}
\end{ldispl}

Let $A$ and $A'$ be sets.
Then the usual set operations \emph{union}\index{union} ($\union$), 
\emph{intersection}\index{intersection} ($\inter$) and 
\emph{difference}\index{difference} ($\diff$) are 
defined as follows:
\begin{ldispl}
\begin{aeqns}
A \union A' & = & \set{x \where x \in A \;\mathrm{or}\;  x \in A'}\;,
\\
A \inter A' & = & \set{x \where x \in A \;\mathrm{and}\; x \in A'}\;,
\\
A \diff  A' & = & 
              \set{x \where x \in A \;\mathrm{and}\; x \not\in A'}\;.
\end{aeqns}
\end{ldispl}

If $A$ and $A'$ are sets, then there exists a set of which the members 
are exactly $A$ and $A'$.
This set is called the \emph{unordered pair}\index{unordered pair} of 
$A$ and $A'$ and is denoted by $\set{A,A'}$.
Let $A$ be a set, $a \in A$ and $a' \in A$.
Then the \emph{ordered pair}\index{ordered pair}, or shortly pair, with 
\emph{first element} $a$ and \emph{second element} $a'$, written 
$\tup{a,a'}$, is the set defined as follows:
\begin{ldispl}
\begin{aeqns}
\tup{a,a'} & = & \set{\set{a},\set{a,a'}}\;.
\end{aeqns}
\end{ldispl}
Let $A$ and $A'$ be sets.
Then the set operation \emph{cartesian product}\index{cartesian product} 
($\x$) is defined as follows:
\begin{ldispl}
\begin{aeqns}
A \x  A' & = & 
\set{\tup{x,x'} \where x \in A \;\mathrm{and}\; x' \in A'}\;.
\end{aeqns}
\end{ldispl}
This is extended in the obvious way to the cartesian product of more
than two sets.
An \emph{ordered} $n$-\emph{tuple}\index{tuple} ($n > 2$), or shortly 
$n$-\emph{tuple}, with \emph{first element} $a_1$, \ldots, 
\emph{$n$th element} $a_n$, written $\tup{a_1,\ldots,a_n}$, is the set
defined as follows:
\begin{ldispl}
\begin{aeqns}
\tup{a_1,\ldots,a_n} & = & \tup{\tup{a_1,\ldots,a_{n-1}},a_n}\;.
\end{aeqns}
\end{ldispl} 
A pair is sometimes also called a $2$-tuple.
Let $A_1$, \ldots, $A_n$ be sets.
Then the \emph{cartesian product} of more than two sets is defined as 
follows:
\begin{ldispl}
\begin{aeqns}
A_1 \x  \ldots \x A_n & = & 
\set{\tup{x_1,\ldots,x_n} \where x_1 \in A_1,\ldots,x_n \in A_n}\;.
\end{aeqns}
\end{ldispl}

If a set has a finite number of members, the set is said to be 
\emph{finite}\index{set!finite}.
We use the following abbreviation.
We write $\fsetof{(A)}$ for 
$\set{x \in \setof{(A)} \where x \;\mathrm{is}\;\mathrm{finite}}$,
the set of all finite subsets of $A$.

As usual, we write $\Nat$ to denote the set of all natural numbers, and
$\Bool$ to denote the set $\set{\True,\False}$ of all boolean values.

\section{Relations and functions}
\label{sect-rels-funcs}

Let $A_1$, \ldots, $A_n$ be sets.
An \emph{$n$-ary relation}\index{relation} $R$ \emph{between} 
$A_1$, \ldots, $A_n$ is a subset of $A_1 \x  \ldots \x A_n$.
If $A_1 = \ldots = A_n$, $R$ is called an $n$-ary relation \emph{on}
$A_1$.
We often write $R(a_1,\ldots,a_n)$ for $\tup{a_1,\ldots,a_n} \in R$.

Let $A$ be a set and $R$ be a binary relation on $A$.
Then we define the following:
\begin{iteml}
\item
$R$ is \emph{reflexive}\index{reflexivity} if $R(x,x)$ for all $x \in A$;
\item
$R$ is \emph{symmetric}\index{symmetry} if $R(x,y)$ implies $R(y,x)$;
\item
$R$ is \emph{transitive}\index{transitivity} if $R(x,y)$ and $R(y,z)$ 
implies $R(x,z)$;
\item 
$R$ is an \emph{equivalence relation}\index{equivalence relation} on $A$ 
if $R$ is reflexive, symmetric and transitive.
\end{iteml}

Let $A$ be a set and $R$ be an equivalence relation on $A$.
Then, for each $a \in A$, the set $\set{x \in A \where R(a,x)}$ is
called an \emph{equivalence class}\index{equivalence class} with respect 
to $R$.
The members of an equivalence class are said to be 
\emph{representatives}\index{equivalence class!representative of} of the 
equivalence class.

Let $A$ and $A'$ be sets.
Then a \emph{function from}\index{function} $A$ \emph{to} $A'$ is a 
relation $f$ between $A$ and $A'$ such that for all $x \in A$ there 
exists a unique $x' \in A'$ with $\tup{x,x'} \in f$.
This $x'$ is called the \emph{value of}\index{function!value of} 
$f$ \emph{at} $x$.
We write $\funct{f}{A}{A'}$ to indicate that $f$ is a function from $A$ 
to $A'$, and we write $f(x)$ for the value of $f$ at $x$.

If $A$, $A'$ and $A''$ are sets, $A \subseteq A'$ and 
$\funct{f}{A'}{A''}$, then there exists a set of which the members are 
exactly the values of $f$ at the members of $A$.
This set is denoted by $\set{f(x) \where x \in A}$:
\begin{ldispl}
\begin{aeqns}
\set{f(x) \where x \in A} & = & 
\set{x' \where \Exists{x \in A}{f(x) = x'}}\;.
\end{aeqns}
\end{ldispl}

Let $\cI$ be a set, $\cA$ be a set of sets.
Then a \emph{family}\index{family} indexed by $\cI$ is a function
$\funct{A}{\cI}{\cA}$.
The set $\cI$ is called the \emph{index}\index{index} set of the family.
We write $A_i$ for $A(i)$.
If $A$ is a family indexed by $\cI$, then we write 
$\Union_{i \in \cI} A_i$ for $\Union \set{A_i \where i \in \cI}$.

We also use the following abbreviation.
We write $\set{f(x) \where x \in A, P(x)}$ for 
$\set{f(x) \where x \in \set{x' \in A \where P(x')}}$.

Let $A$ and $A'$ be sets.
Then a \emph{partial function from}\index{function!partial} $A$ 
\emph{to} $A'$ is a relation $f$ between $A$ and $A'$ such that 
there exist a set $B \subseteq A$ for which $\funct{f}{B}{A'}$.
For $x \in A$, $f(x)$ is said to be \emph{defined}\index{definedness} if 
$x$ is a member of the unique set $B \subseteq A$ for which 
$\funct{f}{B}{A'}$ and $f(x)$ is said to be \emph{undefined} otherwise.

\section{Sequences}
\label{sect-seqs}

Let $A$ be a set and $n \in \Nat$.
Then a (finite) \emph{sequence}\index{sequence} over $A$ of 
\emph{length} $n$, is a function 
$\funct{\sigma}{\set{i \in \Nat \where 1 \leq i \leq n}}{A}$.
If $n > 0$ and $\sigma(1) = a_1$, \ldots, $\sigma(n) = a_n$, then the 
sequence is written as follows:
\begin{ldispl}
a_1 \,\ldots\, a_n\;.
\end{ldispl}
The sequence of length $0$ is called the \emph{empty} 
sequence\index{empty sequence} and is denoted by $\epsilon$.

Let $A$ be a set.
Then the set of all sequences over $A$ is denoted by $\seqof{A}$, and
the set of all nonempty sequences over $A$ is denoted by
$\neseqof{A}$.
For each $\sigma \in \seqof{A}$, we write $|\sigma|$ for the length of
$\sigma$.

Let $A$ be a set, and $\sigma,\sigma' \in \seqof{A}$.
Then the sequence operation \emph{concatenation}\index{concatenation} 
($\conc$) is defined as follows.
$\sigma \conc \sigma'$ is the unique sequence 
$\sigma'' \in \seqof{A}$ with $|\sigma''| = |\sigma| + |\sigma'|$ 
such that:
\begin{ldispl}
\begin{aeqns}
\sigma''(i) & = & \sigma(i) & \;\mathrm{if}\; 1 \leq i \leq |\sigma|\;,
\\
\sigma''(i) & = & \sigma'(i - |\sigma|) & 
\;\mathrm{if}\; |\sigma| + 1 \leq i \leq |\sigma| + |\sigma'|\;.
\end{aeqns}
\end{ldispl}
We usually write $\sigma \, \sigma'$ for 
$\sigma \conc \sigma'$.

Let $A$ be a set, and $\sigma,\sigma' \in \seqof{A}$.
Then $\sigma'$ is a \emph{prefix}\index{prefix} of $\sigma$, written 
$\sigma' \preceq \sigma$, if there exists a $\sigma'' \in \seqof{A}$
such that $\sigma' \, \sigma'' = \sigma$; and 
$\sigma'$ is a \emph{proper prefix} of $\sigma$, written 
$\sigma' \prec \sigma$, if $\sigma' \preceq \sigma$ and 
$\sigma' \neq \sigma$.

\bibliographystyle{plain}                                 
\addcontentsline{toc}{chapter}{\numberline{}References} 
\bibliography{PT}                                         
\cleardoublepage                                          
\addcontentsline{toc}{chapter}{\numberline{}Index}        
\flushbottom                                              
\printindex                                               

\end{document}